\documentclass[fontsize=12pt,a4paper,headings=normal,
twoside=false,leqno,parskip=half-,abstract=true,bibliography=totoc]{scrartcl}
\usepackage[english]{babel}
\usepackage{bm}
\usepackage[utf8]{inputenc}
\usepackage{hyperref}
\hypersetup{
	pdftitle={Stable Heteroclinic Chains in HL},
	pdfauthor={Hauke Sprink and Phillipo Lappicy},
	colorlinks=true,
	linkcolor=blue,
	citecolor=blue,
	filecolor=blue,
	urlcolor=blue}

\usepackage{graphicx}
\usepackage[format=plain,labelfont=bf,font=small]{caption}
\usepackage{subfig}
\usepackage{xcolor}
\usepackage[arrow, matrix, curve]{xy}
\usepackage{float}
\usepackage{enumitem}

\usepackage{caption}
\usepackage{comment}
\usepackage[textwidth=2cm,textsize=small,backgroundcolor=none]{todonotes}

\usepackage{xfrac}

\usepackage{tabulary}
\usepackage{array}

\usepackage{amsmath,amsthm}
\usepackage{amssymb} 
\usepackage{latexsym}

\usepackage{tikz}
\usetikzlibrary{arrows.meta, arrows, shapes}
\usepackage{pgfplots}

\usetikzlibrary{intersections,through,backgrounds}
\usetikzlibrary{calc, bending}

\newcommand{\shortquad}{\hspace{0.5em}}

\usepackage{wasysym}

\newtheorem{thm}{Theorem}[section]
\newtheorem{lem}[thm]{Lemma}
\newtheorem{prop}[thm]{Proposition}
\newtheorem{cor}[thm]{Corollary}

\newtheorem*{remark}{Remark}

\usepackage{aliascnt}
\newaliascnt{lemm}{theorem}

\aliascntresetthe{lemm}

\newaliascnt{coro}{theorem}

\aliascntresetthe{coro}

\newaliascnt{propo}{theorem}

\aliascntresetthe{propo}

\usepackage[notref,notcite,color,final %make final instead of draft, by % %
]{showkeys}
\DeclareOldFontCommand{\bf}{\normalfont\bfseries}{\mathbf}
\DeclareOldFontCommand{\it}{\normalfont\itshape}{\textit}

\newcommand{\eps}{\varepsilon}
\newcommand{\RR}{\mathbb R}
\newcommand{\ZZ}{\mathbb Z}
\newcommand{\QQ}{\mathbb Q}
\newcommand{\NN}{\mathbb N}

\newcommand{\BIIa}{\mathcal{B}_{N_1}}
\newcommand{\BIIb}{\mathcal{B}_{N_2}}
\newcommand{\BIIc}{\mathcal{B}_{N_3}}

\newcommand{\BIIalpha}{\mathcal{B}_{N_\alpha}}

\newcommand{\Cantor}{\mathcal{C}}

\newcommand{\sg}{\lambda(v)}

\newcommand{\N}{\cal N}

\newcommand{\A}{\mathrm{{A}}}

\newcommand{\KC}{\mathrm{K}^{\ocircle}}

\newcommand{\Q}{{\bf \mathrm{Q}}}
\newcommand{\T}{{\bf \mathrm{T}}}

\newcommand{\Tb}{{\bf \mathrm{T}_2}}

\newcommand{\KM}{\mathcal{K}}

\renewcommand{\S}{\mathcal{S}}

\newcommand{\KI}{{\mathcal{K}}}

\newcommand{\tab}{{\bf \mathrm{t}_{12}}}
\newcommand{\tac}{{\bf \mathrm{t}_{13}}}

\newcommand{\tca}{{\bf \mathrm{t}_{31}}}

\newcommand{\talphabeta}{{\bf \mathrm{t}_{\alpha\beta}}}
\newcommand{\tbetaalpha}{{\bf \mathrm{t}_{\beta\alpha}}}
\newcommand{\talphagamma}{{\bf \mathrm{t}_{\alpha\gamma}}}

\newcommand{\eva}{\mu_1}
\newcommand{\evb}{\mu_2}
\newcommand{\evc}{\mu_3}
\newcommand{\evi}{\mu_\alpha}
\newcommand{\evj}{\mu_\beta}
\newcommand{\evk}{\mu_\gamma}

\newcommand{\ssev}{\mu_{ss}}
\newcommand{\sev}{\mu_{s}}
\newcommand{\sssev}{\mu_{s,ss}}
\newcommand{\uev}{\mu_{u}}

\newcommand{\tsssev}{\tilde{\mu}_{s,ss}}

\newcommand{\insection}{S^{{\normalfont \text{\,in}}}}
\newcommand{\outsection}{S^{{\normalfont \text{out}}}}
\newcommand{\tinsection}{\tilde{S}^{{\normalfont \text{\,in}}}}
\newcommand{\toutsection}{\tilde{S}^{{\normalfont \text{out}}}}

\renewcommand{\Psi}{\Pi}
\newcommand{\Psiloc}{\Pi^{\normalfont \text{loc}}}
\newcommand{\Psiglob}{\Pi^{\normalfont \text{glob}}}
\newcommand{\tPsiglob}{\tilde{\Pi}^{\normalfont \text{glob}}}

\newcommand{\xin}{x^{\normalfont \text{\,in}}}
\newcommand{\xuin}{x_u^{\normalfont \text{\,in}}}
\newcommand{\xcin}{x_c^{\normalfont \text{\,in}}}
\newcommand{\xsin}{x_s^{\normalfont \text{\,in}}}
\newcommand{\xssin}{x_{ss}^{\normalfont \text{\,in}}}
\newcommand{\xsssin}{x_{s,ss}^{\normalfont \text{\,in}}}
\newcommand{\xusin}{x_{u,s}^{\normalfont \text{\,in}}}

\newcommand{\yin}{y^{\normalfont \text{\,in}}}
\newcommand{\yuin}{y_u^{\normalfont \text{\,in}}}

\newcommand{\ysin}{y_s^{\normalfont \text{\,in}}}
\newcommand{\yssin}{y_{ss}^{\normalfont \text{\,in}}}
\newcommand{\ysssin}{y_{s,ss}^{\normalfont \text{\,in}}}

\newcommand{\txin}{\tilde{x}^{\normalfont \text{\,in}}}
\newcommand{\txuin}{\tilde{x}_u^{\normalfont \text{\,in}}}
\newcommand{\txcin}{\tilde{x}_c^{\normalfont \text{\,in}}}
\newcommand{\txsin}{\tilde{x}_s^{\normalfont \text{\,in}}}

\newcommand{\txsssin}{\tilde{x}_{s,ss}^{\normalfont \text{\,in}}}
\newcommand{\txusin}{\tilde{x}_{u,s}^{\normalfont \text{\,in}}}

\newcommand{\xout}{x^{\normalfont \text{out}}}
\newcommand{\xuout}{x_u^{\normalfont \text{out}}}
\newcommand{\xcout}{x_c^{\normalfont \text{out}}}
\newcommand{\xsout}{x_s^{\normalfont \text{out}}}
\newcommand{\xssout}{x_{ss}^{\normalfont \text{out}}}
 \newcommand{\xsssout}{x_{s,ss}^{\normalfont \text{out}}}

\newcommand{\yout}{y^{\normalfont \text{out}}}
\newcommand{\yuout}{y_u^{\normalfont \text{out}}}

\newcommand{\ysout}{y_s^{\normalfont \text{out}}}
\newcommand{\yssout}{y_{ss}^{\normalfont \text{out}}}
\newcommand{\ysssout}{y_{s,ss}^{\normalfont \text{out}}}

\newcommand{\txout}{\tilde{x}^{\normalfont \text{out}}}

\newcommand{\txcout}{\tilde{x}_c^{\normalfont \text{out}}}

\newcommand{\txsssout}{\tilde{x}_{s,ss}^{\normalfont \text{out}}}

\newcommand{\x}{x}
\newcommand{\xu}{x_u}
\newcommand{\xc}{x_c}
\newcommand{\xs}{x_s}
\newcommand{\xss}{x_{ss}}
\newcommand{\xsss}{x_{s,ss}}
\newcommand{\xus}{x_{u,s}}

\newcommand{\tx}{\tilde{x}}
\newcommand{\txu}{\tilde{x}_u}
\newcommand{\txc}{\tilde{x}_c}
\newcommand{\txsss}{\tilde{x}_{s,ss}}
\newcommand{\txus}{\tilde{x}_{u,s}}

\newcommand{\ttzero}{\tilde{t_0}}

\newcommand{\y}{y}
\newcommand{\yu}{y_u}
\newcommand{\yc}{y_c}
\newcommand{\ys}{y_s}
\newcommand{\yss}{y_{ss}}
\newcommand{\ysss}{y_{s,ss}}
\newcommand{\yus}{y_{u,s}}

\newcommand{\tyus}{\tilde{y}_{u,s}}

\newcommand{\talpha}{\tilde{\alpha}}

\newcommand{\tout}{t^{\normalfont \text{out}}}

\newcommand{\Cglob}{M^\text{glob}}
\newcommand{\Cret}{M^\text{ret}}

\newcommand{\Cc}[2]{C^c_{#1}(#2)}
\newcommand{\Cus}[2]{C^{u,s}_{#1}(#2)}

\newcommand{\B}{B}

\newcommand{\ex}[1]{e^{#1}}

\newcommand{\lip}{\zeta}
\newcommand{\tlip}{\tilde{\zeta}}
\newcommand{\lipstar}{\zeta^\star}

\newcommand{\graph}{\text{graph}}

\newcommand{\Mid}{\mathrm{A}^{\text{m}}(v)}
\newcommand{\Stab}{\mathrm{A}^{\text{s}}(v)}
\newcommand{\Unstab}{\mathrm{A}^{\text{u}}(v)}

\begin{document}
	
	\title{
            %\LARGE{Stable oscillations in Bianchi cosmologies }
            %\\or\\
            \LARGE{Stable manifolds of heteroclinic chains \\ in Ho\v{r}ava-Lifshitz cosmology }
            }

	\author{
		\\
		{~}\\
		Hauke Sprink* and Phillipo Lappicy${}^\circ$\\
		\vspace{2cm}}
	
	\date{ }
	\maketitle
	\thispagestyle{empty}
	
	\vfill
	
	$\ast$\\
	Institute for Mathematics, Otto-von-Guericke Universität, Magdeburg, Germany\\
	%Arnimallee 3, 14195 Berlin, Germany\\

	$\circ$\\
	Universidad Complutense de Madrid and ICMAT, Madrid, Spain\\
	%ICMC, Universidade de S\~ao Paulo\\
	%Avenida trabalhador são-carlense 400, 13566-590, São Carlos, SP, Brazil\\
	
	%%%%%%%%%%%%%%%%%%%%%%%%%%%%%%%%%%%%%%%%%%%%%%%%%%%%%%%%%%%
	
	\newpage
	\pagestyle{plain}
	\pagenumbering{arabic}
	\setcounter{page}{1}
	
	\begin{abstract}
		\noindent
		We consider vacuum spatially homogeneous models in Ho\v{r}ava-Lifshitz gravity, which are expected to describe generic singularities for these modified gravity theories. These models perturb the well-known Bianchi models in general relativity (GR) by a parameter $v\in (0,1)$, in which GR occurs at $v=1/2$. 
        We show that all periodic heteroclinic chains of Bianchi type II orbits are robust for small perturbations of GR, i.e. they exist for any $v\approx 1/2$.  
		Moreover, we prove the existence of a stable manifold for each heteroclinic chain within a Cantor set for the supercritical case, $v\in (1/2,1)$. % for all $v\in [0,1]$.
		In particular, we discuss the heteroclinic chains of periods three and four in detail. %We also describe the dynamics for the extremal parameter cases $v=0$ and $v=1$.
		

        \medskip \noindent \textbf{MSC 2020:} 34D45, 37G35, 37N20, 83C20, 83C75, 83D05.  
%34D45      Attractors of solutions to ordinary differential equations
%37C70      Attractors and repellers of smooth dynamical systems and their topological structure
%37G35  	Dynamical aspects of attractors and their bifurcations
%37N20      Dynamical systems in other branches of physics (quantum mechanics, general relativity, laser physics)
%83C05      Einstein’s equations (general structure, canonical formalism, Cauchy problems)
%83C10       Equations of motion in general relativity and gravitational theory 
%83C20       Classes of solutions; algebraically special solutions, metrics with symmetries for problems in general relativity and gravitational theory
%83C75       Space-time singularities, cosmic censorship, etc.
%83D05       Relativistic gravitational theories other than Einstein’s, including asymmetric field theories 
		%\textbf{Keywords:} ...
	\end{abstract}

\tableofcontents

		\numberwithin{equation}{section}
		\numberwithin{figure}{section}
		\numberwithin{table}{section}

\pagebreak

\section{Introduction}\label{sec:intro}
The Einstein equations are ten coupled partial differential equations (PDEs) that relate the curvature of spacetime
%, in terms the Ricci tensor $R_{\mu \nu}$ and scalar curvature $R =g^{\mu \nu} R_{\mu \nu} $
to the presence of matter.
%, encoded in the stress-energy tensor $T_{\mu\nu}$.
%The unknown in these equations is the metric of spacetime.
%metric tensor  $g_{\mu \nu}$.
%
%
Six of them are evolution equations, whereas the remaining four constrain the initial data.
%Six of them describe how the metric evolves in time and four of them constrain the initial data.
%
%
%With the "mathematical rescaling" the EFE in vacuum $(T_{\mu\nu} = 0)$ become\\
%\begin{equation}\label{EFE}
%R_{\mu \nu} -\frac 12 R g_{\mu \nu} = 0
%\end{equation}
%
An important feature of these equations is that they possess singularities, e.g. \cite{Penrose65,HawkingPenrose70,Geroch66,HawkingEllis73,Senovilla15}, however, little is known about the dynamics of generic singularities.

A heuristic approach to describe generic singularities was provided by Belinski, Khalatnikov and Lifshitz (BKL), which is called the \textit{BKL conjecture}.
According to their work \cite{bkl70}, generic singularities, for example the Big Bang, should possess the following properties:
\begin{itemize}
	\item [(i)] \textit{Vacuum domination:} Solutions in a cosmological model that contains matter converge to the vacuum model, i.e. ``matter does not matter".
	\item [(ii)] \textit{Locality:} Spatial points evolve independently of their neighbors. Hence terms in the Einstein equations containing spatial derivatives are negligible and the PDEs are well approximated by a system of ordinary differential equations (ODEs).
	\item [(iii)] \textit{Oscillatory:} The behavior of general solutions is governed by simpler solutions possessing oscillatory character. In particular, the dynamics of the Bianchi type VIII and IX models should be governed by the Bianchi type I and II models through an induced discrete map, the so-called \emph{Kasner map}, which is chaotic.
	%	 Solutions of type I form a circle of equilibria in phase space, the so-called Kasner circle and solutions of type II are heteroclinic solutions connecting each two equilibria on the Kasner circle. The discrete map generating the sequences of heteroclinic solutions is called Kasner map.
\end{itemize}
%
%The spatially homogeneous \emph{Bianchi cosmologies} are models of the universe such that all points behave equivalently.
%
Due to (i)-(ii), vacuum spatially homogeneous cosmologies, the Bianchi models, are expected to play a key role in the asymptotic behavior of generic singularities\footnote{We emphasize that the BKL picture remains quite controversial, as there are cases where there are no oscillations, see the discussion in \cite{AnderssonRendall,RodSpeck,FourRodSpeck,Nutzi,RingstromData} and references therein. }, see \cite{mis69a,bkl70,bkl82,Hobill94,Mixmaster,ugg13a,ugg13b}. These models can be written using a dynamical systems formulation, see \cite{Wainwright_Hsu,Dynamical_systems_in_cosmology}.
The six evolutionary PDEs become six dimensionless expansion-normalized ODEs, where one of them decouples from the others; whereas the four constraints reduce to one algebraic constraint, the (vacuum) Gauss constraint.

Regarding the oscillatory behaviour (iii), which concerns the dynamics of the Bianchi ODEs, a major achievement is the Bianchi type IX attractor theorem by H. Ringstr\"om, see  \cite{Ringstrom, heiugg09b}. It states that the $\omega$-limit set of generic solutions of Bianchi type IX is contained in the space of solutions of Bianchi type I and II, known as the \emph{Mixmaster attractor}.
An analogous theorem for Bianchi type VIII solutions was proven by B. Brehm in \cite{Bernhard}.
However, these two results do not provide information on how solutions approach the Mixmaster attractor, i.e. if the asymptotic behaviour is related to the dynamics restricted to the Mixmaster. %, which in turn is described by the Kasner map.
It is expected that the concatenation of heteroclinic orbits of Bianchi type II and the induced map of Bianchi type I (the so-called \textit{Kasner map}) play a key role in the dynamics.
%More rigorous results can be found, for example, in \cite{Beguin,Liebscher,Dutilleul}. 
A state~of~the~art overview is provided in \cite{Hobill94,Mixmaster}.

The fact that the induced dynamics of the Mixmaster, which is described by the discrete dynamics of the Kasner map, plays a fundamental role in the continuous dynamics of the Bianchi models was achieved for example in \cite{Beguin,Liebscher,Dutilleul}. Mathematically, they prove that for suitable sequences 
%$(h_k)_{k \in \NN}$
of Bianchi type II solutions obtained by the Kasner map, a \emph{heteroclinic chain}, there exists a codimension-one (three-dimensional) stable manifold consisting of Bianchi type VIII and IX solutions that approach such a heteroclinic chain as $t\to \infty$. 
%This aspect is called \emph{shadowing} and 

Since oscillation proofs in GR are quite demanding, recently there has been an effort
to try to understand the GR dynamics by means of perturbation theory, see \cite{HellLappicyUggla}. Indeed, we consider a Ho\v{r}ava-Lifshitz theory, which perturbs the Bianchi ODEs by a parameter $v\in (0,1)$, where GR occurs for $v = 1/2$. Our present goal is to establish stable oscillations (by means of stable heteroclinic chains) for the perturbed model, $v\neq 1/2$, as a first step towards understanding how they carry over to the limiting case, when $v = 1/2$. We emphasize that even though modified gravity theories remain speculative, they may provide suitable modifications that allow one to obtain rigorous proofs in GR.

The outcome is the following ODE system in $\RR^5$ with a parameter $v \in (0,1)$:
%
% formulation of Einstein's equation for homogeneous non-isotropic universes/spacetimes in
%What is left is the following system in $\RR^5$:
\vspace{-0.05cm}
\begin{subequations}\label{full:subs}
	\begin{align}
		N_1^\prime 			  & \,=\, -2(2v\Sigma^2 - 2\Sigma_+) \, N_1 \label{full:N1} \\
		N_2^\prime 			  &\,=\, -2(2v\Sigma^2 + \Sigma_+ + \sqrt{3}\Sigma_-)N_2 \label{full:N2} \\
		N_3^\prime 			  &\,=\, -2(2v\Sigma^2 + \Sigma_+ - \sqrt{3}\Sigma_- )N_3 \label{full:N3}  \\
		\Sigma_+^\prime  &\,=\, 4v(1-\Sigma^2)\Sigma_+ +  {\cal S}_+ \label{full:Sigma+}\\
		\Sigma_-^\prime   &\,=\, 4v(1-\Sigma^2)\Sigma_- + {\cal S}_- \label{full:Sigma-}
	\end{align}
\end{subequations}
%
% \begin{subequations}\label{full_alt:subs}
% 	\begin{align}
% 		N_1^\prime 			  & \,=\, -2(\Sigma^2 - 2\Sigma_+) \, N_1 \label{full_alt:N1} \\
% 		N_2^\prime 			  & \,=\, -2(\Sigma^2 + \Sigma_+ + \sqrt{3}\Sigma_-)N_2 \label{full_alt:N2} \\
% 		N_3^\prime 			  & \,=\, -2(\Sigma^2 + \Sigma_+ - \sqrt{3}\Sigma_- )N_3 \label{full_alt:N3} \\
% 		\Sigma_+^\prime  & \,=\, 2(1-\Sigma^2)\Sigma_+ +  {\cal S}_+ \label{full_alt:Sigma+}\\
% 		\Sigma_-^\prime   & \,=\, 2(1-\Sigma^2)\Sigma_- +  {\cal S}_- \label{full_alt:Sigma-}
% 	\end{align}
% \end{subequations}
where
\begin{subequations}
	\begin{equation}
		\begin{aligned}
			\Sigma^2  &	:= \Sigma_+^2 + \Sigma_-^2 \label{Sigma}\\
			{\cal S}_+ &  := 2 \left[\left(N_2-N_3\right)^2 - N_1\left(2N_1-N_2-N_3\right) \right]\\
			{\cal S}_- &  := 2\sqrt{3} \left(N_2-N_3\right)\left(N_2+N_3-N_1\right).
		\end{aligned}
	\end{equation}
\end{subequations}
The Gauss constraint restricts the phase-space $\RR^5$ to a four-dimensional manifold given by
\begin{equation} \label{constraint}
	1 =  \Sigma^2 + N_1^2 + N_2^2 + N_3^2 - 2N_1N_2 - 2N_2N_3 - 2N_3N_1.
\end{equation}
The derivative $(.)^\prime=d/dt$ is with respect to a time variable $t$ which is directed towards the physical past. Thus the Big Bang singularity occurs as $t \rightarrow \infty$.

The model \eqref{full:subs} was derived in \cite{HellLappicyUggla} within the context of Ho{\v{r}}ava-Lifshitz gravity, see \cite{hor09a,hor09b, HL_status_report}, which considers higher order terms in the Einstein-Hilbert action. Depending on which terms are considered, the system $\eqref{full:subs}$ is expected to describe the asymptotic dynamics of the Big Bang for a fixed $v \in (0,1)$.
In particular, the parameter $v$ perturbs general relativity, which corresponds to $v=1/2$.
%Moreover, the  are expected to describe generic singularities, see \cite{Bakas10,Miso11,Kamenshchik,HellLappicyUggla}.º
In \cite[Appendix A]{HellLappicyUggla}, it is argued that there is an `asymptotically dominant' curvature term in the Einstein-Hilbert action towards the singularity, yielding a respective dominant model \eqref{full:subs} for some $v\in (0,1)$, see also \cite{Bakas10,Miso11,Kamenshchik}. 
The analogue of  Ringstr\"om's attractor theorem was obtained in \cite{Beatriz}.
For the existence of periodic orbits in the extremal case $v=0$, see \cite{Church}.

%All variables in \eqref{full_alt:subs} are dimensionless and expansion-normalized. In these coordinates, $ N = \left(N_1, N_2, N_3\right)$ relates to the intrinsic curvature of a spacelike hypersurface, whereas $ \Sigma = \left(\Sigma_+, \Sigma_-\right)$ relates to the extrinsic curvature of space, i.e. ``how is space embedded in spacetime".
%
%Mathematically, the $N$-variables are directly related to structure constants of the Lie algebra, whereas the $\Sigma$-variables correspond to the components of the traceless second fundamental form. For more details, see, e.g. \cite{Dynamical_systems_in_cosmology}.

We recall that the Bianchi type of a solution of system \eqref{full:subs} is determined by the signs of the $N$-variables, which are invariant under the flow. This provides a stratification of phase-space by invariant sets of increasing dimension.
Bianchi type $\mathrm{I}$ consists of all $N_\alpha , \alpha=1,2,3,$ being zero (constituting a circle of equilibria), Bianchi type $\mathrm{II}$ consists of a single non-zero $N_\alpha$ (yielding a hemisphere containing heteroclinic orbits between the equilibria of type $\mathrm{I}$), Bianchi types $\mathrm{VI}_0,\mathrm{VII}_0$ consist of two non-zero $N_\alpha$, and Bianchi types $\mathrm{VIII},\mathrm{IX}$ consist of three non-zero $N_\alpha$.
%
%Thus, the solutions of type $\mathrm{VIII}$ and $\mathrm{IX}$ form open sets in phase space, whereas the solutions of the lower types reside on the lower dimensional invariant boundaries of these open sets, 
See Table~\ref{table}.
%

%The most general solutions of class A and therefore of particular interest are the solutions of type VIII and IX. BKL and Misner \cite{mis69a} suggested that the asymptotic dynamics of Bianchi IX solutions is described by a sequence of solutions of type I, which constitute equilibria in phase space. We derive the induced discrete map, the Kasner map, from which the sequence of equilibria is obtained.
%We describe the solutions of Bianchi type I and II in more detail. 
%Wainwright: Moreover, surprising that Kasner vacuum solutions are important
%
	\begin{figure}[H]
		\centering
			\begin{tikzpicture}[scale=1]
				%
%\draw [thick, -Latex] (0,-0.1)--(0,3.5);
%\draw [thick] (0.15,0)--(-0.15,0);
%\draw [thick] (0.15,1)--(-0.15,1);
%\draw [thick] (0.15,2)--(-0.15,2);
%\draw [thick] (0.15,3)--(-0.15,3);

%
%\draw (0,3) circle (0pt) node [anchor = east] {$D$};
%\draw (0,2.5) circle (0pt) node [anchor = east] {$4$};
%\draw (0,2) circle (0pt) node [anchor = east] {$4$};
%\draw (0,1.5) circle (0pt) node [anchor = east] {$3$};
%\draw (0,1) circle (0pt) node [anchor = east] {$3$};
%\draw (0,0.5) circle (0pt) node [anchor = east] {$2$};
%\draw (0,0) circle (0pt) node [anchor = east] {$1$};

\draw [] (-0.9,2.9)--(4.8,2.9);

\draw (0,3.1) circle (0pt) node [anchor = east] {Type};
%\draw (3,3.1) circle (0pt) node [anchor = west] {Lie algebra};

%\draw (3,2.5) circle (0pt) node [anchor = west] {$\mathfrak{su}(2)$};
%\draw (3,2) circle (0pt) node [anchor = west] {$\mathfrak{sl}(2,\RR)$};
%\draw (3,1.5) circle (0pt) node [anchor = west] {$\mathfrak{isom}(\RR^2)$};
%\draw (3,1) circle (0pt) node [anchor = west] {$\mathfrak{isom}($Min$_2)$};
%\draw (3,0.5) circle (0pt) node [anchor = west] {Heisenberg};
%\draw (3,0) circle (0pt) node [anchor = west] {$\RR^3$};

\draw (-0.1,2.5) circle (0pt) node [anchor = east] {I};
\draw (-0.1,2) circle (0pt) node [anchor = east] {II};
\draw (-0.1,1.5) circle (0pt) node [anchor = east] {VI$_0$};
\draw (-0.1,1) circle (0pt) node [anchor = east] {VII};
\draw (-0.1,0.5) circle (0pt) node [anchor = east] {VIII};
\draw (-0.1,0) circle (0pt) node [anchor = east] {IX};

\draw (1,3.1) circle (0pt) node [anchor = east] {$N_\alpha$};
\draw (1.7,3.1) circle (0pt) node [anchor = east] {$N_\beta$};
\draw (2.4,3.1) circle (0pt) node [anchor = east] {$N_\gamma$};
\draw (4.8,3.1) circle (0pt) node [anchor = east] {Dimension};

\draw (1,2.5) circle (0pt) node [anchor = east] {$0$};
\draw (1.7,2.5) circle (0pt) node [anchor = east] {$0$};
\draw (2.4,2.5) circle (0pt) node [anchor = east] {$0$};
\draw (4,2.5) circle (0pt) node [anchor = east] {$1$};

\draw (1,2) circle (0pt) node [anchor = east] {$+$};
\draw (1.7,2) circle (0pt) node [anchor = east] {$0$};
\draw (2.4,2) circle (0pt) node [anchor = east] {$0$};
\draw (4,2) circle (0pt) node [anchor = east] {$2$};

\draw (1,1.5) circle (0pt) node [anchor = east] {$+$};
\draw (1.7,1.5) circle (0pt) node [anchor = east] {$-$};
\draw (2.4,1.5) circle (0pt) node [anchor = east] {$0$};
\draw (4,1.5) circle (0pt) node [anchor = east] {$3$};

\draw (1,1) circle (0pt) node [anchor = east] {$+$};
\draw (1.7,1) circle (0pt) node [anchor = east] {$+$};
\draw (2.4,1) circle (0pt) node [anchor = east] {$0$};
\draw (4,1) circle (0pt) node [anchor = east] {$3$};

\draw (1,0.5) circle (0pt) node [anchor = east] {$+$};
\draw (1.7,0.5) circle (0pt) node [anchor = east] {$+$};
\draw (2.4,0.5) circle (0pt) node [anchor = east] {$-$};
\draw (4,0.5) circle (0pt) node [anchor = east] {$4$};

\draw (1,0) circle (0pt) node [anchor = east] {$+$};
\draw (1.7,0) circle (0pt) node [anchor = east] {$+$};
\draw (2.4,0) circle (0pt) node [anchor = east] {$+$};
\draw (4,0) circle (0pt) node [anchor = east] {$4$};
			\end{tikzpicture}\captionof{table}{The Bianchi type of a solution of \eqref{full:subs} is described by the signs of its $N$-variables, for a permutation $(\alpha\beta\gamma)$ of  $(123)$. Note there is an equivalent representation where all signs are reversed.}\label{table} 	\end{figure}

For $v\neq 1/2$, some of the features of GR persist whereas some new phenomena emerge.
In particular, %the Bianchi hierarchy of invariant sets in Table~\ref{table} remains valid depending on whether the variables $N_\alpha$ are nonzero. 
solutions of Bianchi type I, the Kasner circle of equilibria, are independent of the parameter $v\in (0,1)$, although their linearization is not. The heteroclinic solutions of Bianchi type II joining each two Bianchi type I solutions are perturbed and thus the Kasner map is affected by the parameter. 
Notably, the infinite heteroclinic chains, for $v\in (1/2,1)$, form a Cantor set denoted by $\mathcal{C}(v)$.
The investigation of this parameter dependence is provided in \cite{HellLappicyUggla}, which shows that general relativity is a critical case where a bifurcation occurs and chaos becomes generic. The relationship between the discrete dynamics of the Kasner map and the continuous dynamics of $\eqref{full:subs}$, however, is only conjectured.

In this paper, we prove Conjectures 7.1 and 7.2 of \cite{HellLappicyUggla}, which state that all infinite heteroclinic chains in the supercritical case, $v\in (1/2,1)$, admit a three-dimensional stable manifold. 
The summary of the main result, which is stated fully in Theorem~\ref{manifold_theorem}, can be stated as:
%In this case, the Kasner map is chaotic only on a Cantor set of measure zero. %We prove Conjecture 7.2 in this thesis.
%More precisely,
\begin{thm}\label{intro:thm}
    For $v\in (1/2,1)$, all Bianchi type $\mathrm{II}$ heteroclinic chains of \eqref{full:subs} that are contained in the Cantor set $\mathcal{C}(v)$ possess a codimension-one Lipschitz stable manifold of Bianchi types $\mathrm{VIII}$ and $\mathrm{IX}$ solutions. 
    Moreover, the union of all such stable manifolds amounts to a continuous foliation for $p\in \mathcal{C}(v)$, denoted by $\mathcal{W}^s(\mathcal{C}(v)):=\cup_{p\in \mathcal{C}(v)} W^s(p)$. 
\end{thm} 
%
%In contrast to the result that the generic attractor of the equation~\eqref{full_alt:subs} is contained in the Mixmaster, see~\cite{Ringstrom, heiugg09b}, Theorem~\ref{intro:thm} proves that all heteroclinic chains within the Mixmaster (not containing a Taub point) attract a solution of~\eqref{full_alt:subs}. This settles the \emph{Mixmaster attractor conjecture} in \cite{Mixmaster}, up to solutions which have a Taub point in its $\omega$-limit.
%For a discussion on chains that contain a Taub point, see~\autoref{chap:conclusion}.
%
This Theorem indicates that the set $\mathcal{W}^s(\mathcal{C}(v))$ could be a solenoidal manifold, see \cite{Verjovsky} and references therein.
Our proof is achieved by adapting~\cite{Liebscher,Dutilleul} for the perturbed equation~\eqref{full:subs}, which unfolds the Taub points and their pre-images under the Kasner map into a uniformly normally hyperbolic Cantor set for $v>1/2$. 
For this reason, Theorem~\ref{intro:thm} does not have any restriction on the heteroclinic chain, such as periodicity or an algebraic condition on its continued fraction expansion when $v=1/2$, as in~\cite{ReitererTrubowitz,Beguin,Liebscher,Buchner,Dutilleul}. 
%
%Opposed to these results, note that Theorem~\ref{intro:thm} does not provide any further property of the stable set: our perturbation arguments loses track if it is a graph with of certain dimensionality and . See~\autoref{chap:conclusion} for further discussions.
%
%These algebraic conditions on the continued fractions amount to heteroclinic chains that do not get too close to the Taub points too fast. In this regard, the unfolding of the Taub points which amounts to the normally hyperbolic Cantor set yields a simpler analysis for $v>1/2$ without any algebraic condition in the continued fractions, in comparison with \cite{Dutilleul,ReitererTrubowitz}, as \emph{all} points in the Cantor set are normally hyperbolic.

%\textcolor{red}{ADD PARAGRAPH SAYING THAT THIS IS BETTER THAN BEGUIN, REITERER AND TRUBOWITZ: WE HAVE NO ALGEBRAIC CONDITION (without any assumption on how close a chain is to the Taubp points, or how fast it approaches the Taub points).}

The remainder of this paper is organized as follows. 
In \autoref{chap:background}, we provide a background of known results.
Then, \autoref{sec:warm-up} deals with periodic heteroclinic chains of Bianchi type II solutions (i.e. periodic orbits of the Kasner map). 
In \autoref{chap:eigenvalues} we give a complete characterization of the linearization structure for all $v\in (0,1)$, and in particular, we show that the crucial assumption in \cite{Liebscher} that concerns the ordering of the eigenvalues at the Bianchi type I solutions is always satisfied in the supercritical case. %, whereas the subcritical case $(0<v<\frac 12)$ is more complicated. Roughly speaking, stability along heteroclinic chains decreases as $v$ decreases. 
In \autoref{sec:periods}, we illustrate the heteroclinic chains of periods three and four. 
More generally, in \autoref{sec:periodic}, we prove the robustness of periodic orbits of the Kasner map: they persist under small perturbations of GR (i.e., for $v\approx 1/2$). 
We emphasize that all heteroclinic chains within the Cantor set for $v>1/2$ exist in the limit $v\to 1/2$, except a period 2 chain that converges to the Taub point for $v=1/2$; see~\cite{HellLappicyUggla}. 
In \autoref{chap:shadowing} we prove the existence of aforementioned stable manifolds using the techniques presented in \cite{Liebscher,Liebscher_magnetic}.
In \autoref{chap:conclusion} we provide a broad discussion on our results and remaining open issues.

%In section \ref{sec:per4} we discuss the heteroclinic chains of period four, which exist for all $v\in [0,1]$, but changes its stability character and its dynamical behaviour for some value $v_*\in (0,1/2)$.
%Sections \ref{sec:extremal0} and \ref{sec:extremal1} display results on the dynamics of Bianchi type I, II, VI and VII solutions for the extremal cases $v=0$ and $v=1$, respectively.

\section{Background}\label{chap:background}

\subsection{$D_3 \times \ZZ_2$ equivariance}

Equivariance is evident in Misner variables,
%\begin{align}\label{6dim_coordinates}
	$\Sigma_1 = -2  \Sigma_+,
	\Sigma_2 =  \Sigma_+ + \sqrt{3} \Sigma_-,
	\Sigma_3 =  \Sigma_+ - \sqrt{3} \Sigma_-$. 
%\end{align}
%with no change in the $N$-variables. 
For a cyclic permutation $(\alpha \beta \gamma)$ of $(123)$, we obtain a system in $\RR^6$
\begin{subequations}\label{6dim_system}
	\begin{align}
		N_\alpha^\prime &= -2\left(2v |\Sigma|^2 + \Sigma_\alpha\right)N_\alpha, \\ \Sigma_\alpha^\prime &= 4v\left(1-|\Sigma|^2\right)\Sigma_\alpha -4 \,\mathcal{S}_\alpha,
	\end{align}
\end{subequations}
where $\alpha=1,2,3$, $\mathcal{S}_\alpha := \left(N_\beta - N_\gamma\right)^2 - N_\alpha \left(2N_\alpha - N_\beta - N_\gamma\right)$, and $|\Sigma|^2:=\frac{1}{6}\left(\Sigma_1^2+\Sigma_2^2+\Sigma_3^2\right) $. Thus, $|\Sigma|^2 = \Sigma^2$ and the Gauss constraint \eqref{constraint} remains unchanged. To counteract the additional variable we establish the algebraic constraint $\Sigma_1+ \Sigma_2+ \Sigma_3 = 0$.

Equation \eqref{6dim_system} remains unchanged by permuting indices in both $\Sigma$- and $N$-variables, and thus the $D_3$-equivariance. The same applies to the $\ZZ_2$-action, $N \mapsto -N$. %, with no change in $\Sigma$-variables.
A cyclic permutation of the $N$-variables corresponds to a rotation by $\pm 2\pi/3$ in $(\Sigma_+,\Sigma_-)$-space, whereas a non-cyclic permutation corresponds to a reflection along certain axes, see \autoref{figure_circle_sectors}.
\subsection{Bianchi type I}
\begin{figure}[H]
		\centering
		\begin{tikzpicture}[scale=1.25]
				\input{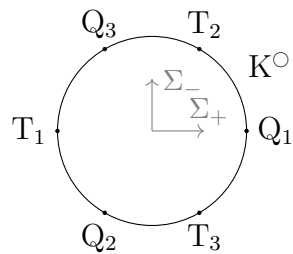}
			\end{tikzpicture}
			\caption{Solutions of Bianchi type I form the Kasner circle $\KC$, which is divided in six equivalent sectors, each of which is bounded by one of the Taub points and one of the $\Q$-points as defined in \eqref{Taub_points} and \eqref{Q_points}.}\label{figure_circle_sectors}
	\end{figure}

%\textcolor{red}{difference Kasner - Bianchi type I}\\
Solutions of Bianchi type I are characterized by $N_\alpha=0$ for $\alpha=1,2,3$. The constraint \eqref{constraint} reduces the phase space to the  \textit{Kasner circle} of equilibria for all parameters $v \in (0,1)$:
\begin{equation}\label{defKC}
	\KC := \left\{ (N,\Sigma) \in \RR^5 : N = 0, \Sigma^2 = 1 \right\} \cong \left\{ (\Sigma_+,\Sigma_-) \in \RR^2 : \Sigma_+^2 + \Sigma_-^2 = 1 \right\} .
\end{equation}
%We use the same symbol for both the circle in full phase space as a subset of $\RR^5$ and for the circle in $\Sigma$-space as a subset of $\RR^2$.
%(In General relativity $(v= \frac 12)$)
%
There are three special points corresponding to the Taub representation of flat (Minkowski) spacetime, called the \emph{Taub points}, \cite{taub}. %These points represent metrics with exactly one $p_\alpha=1$ and therefore $p_\beta = p_\gamma = 0$.
These points are given by
\begin{equation}\label{Taub_points}
	\T_1 = \left(-1,0\right), \qquad \T_2 = \left(\frac12, \frac{\sqrt3}{2}\right), \qquad \T_3 = \left(\frac12, -\frac{\sqrt3}{2}\right).
\end{equation}
It is also useful to introduce the three antipodal points %\vspace{-0.1cm}
\begin{equation}\label{Q_points}
	\Q_1 = \left(1,0\right), \qquad \Q_2 = \left(-\frac12, -\frac{\sqrt3}{2}\right), \qquad \Q_3 = \left(-\frac12, \frac{\sqrt3}{2}\right).
\end{equation}
The Taub points and the $\Q$-points are distributed alternately at equal distance of $\frac\pi3$ along the circle. Moreover, the $\T_\alpha\Q_\alpha$-axes are precisely the axes of reflection for which system \eqref{full:subs} has $D_3$-symmetry.
Thus, the equivalence classes of the Taub points and the $\Q$-points consist of three points, whereas each point in the open sector between two such points is a representative of an equivalence class of six points, see \autoref{figure_circle_sectors}. Points in the same equivalence class represent physically equivalent states.

%	\begin{figure}
	%	\centering
	%	\subfloat[The Lie algebras of Bianchi class A with the signs of the $N$-variables of the coresponding solutions of system \eqref{full_alt:subs} for a permutation $(\alpha\beta\gamma)$ of  $(123)$. There exists an equivalent representation where all signs are reversed.]{
		%		\begin{tikzpicture}[scale=1.1]
			%			\input{figures/table}
			%		\end{tikzpicture} \label{table}} \qquad\qquad
	%	\subfloat[Solutions of Bianchi type I form the Kasner circle $\KC$, which is divided in six equivalent sectors, each of which is bouneded by one of the Taub points and one of the $\Q$-point as defined in \eqref{Taub_points} and \eqref{Q_points}.]
	%	{\begin{tikzpicture}[scale=1.5]
			%			\input{figures/circle_sectors}
			%	\end{tikzpicture}}\label{fig_table_and_kc}
	%\end{figure}
	
	Note the existence of the Kasner circle is independent of the parameter $v\in (0,1)$. Its stability, however, depends strongly on the parameter and this affects the type II solutions, which will be described next.

	\subsection{Bianchi type II}
 
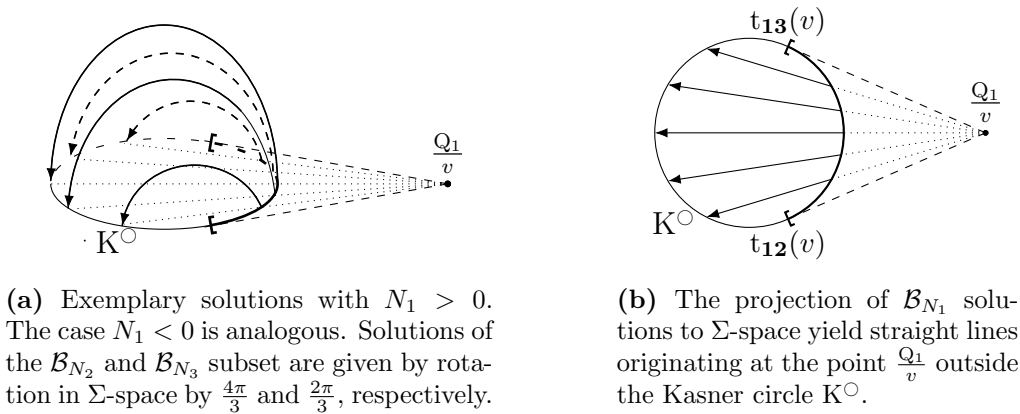
\begin{figure}[H]
		\centering
		%projected solutions
		\subfloat[Exemplary solutions with $N_1>0$. The case $N_1<0$ is analogous. Solutions of the $\BIIb$ and $\BIIc$ subset are given by rotation in $\Sigma$-space by $\frac{4\pi}{3}$ and $\frac{2\pi}{3}$, respectively.]{\begin{tikzpicture}[scale=1.5]
				\clip (-1.4,-0.8) rectangle (2.9,1.5);
 	%kasner circle
 	\draw [thin] (-1,0) arc (180:360:1cm and 0.4cm);
 	\draw [thin, loosely dashed] (1,0) arc (0:180:1cm and 0.4cm);
 	
 	\draw [-Bracket, line width= 1pt ] (1,0) arc (0:-66.4218:1cm and 0.4cm);% node[anchor=north] {\tiny$\tab$};
 	\draw [-Bracket, line width= 1pt , loosely dashed] (1,0) arc (0:66.4218:1cm and 0.4cm);% node[anchor=south] {\tiny$\tac$};
 	
 	% caption K
 	\filldraw [black] (-0.7,-0.5) circle (0.1pt) node[anchor=west] {$\KC$};

 	%%%%% mitte
 	\draw [-Latex ] (1,0) arc (0:179:1cm and 1.4cm);
 	 \draw [semithick ] (1,0) arc (0:179:1cm and 1.4cm);

   %%%%%% außen
   % vorne
	\draw [ -Latex ]  (0.86,-0.21) arc (22:174:0.64cm and 0.6cm);
	\draw [semithick ]  (0.86,-0.21) arc (22:174:0.64cm and 0.6cm);
   % hinten 	
	\draw [-Latex,  dashed]  (0.86,0.21) arc (15:150:0.65cm and 0.7cm);
	\draw [semithick,  dashed]  (0.86,0.21) arc (15:150:0.65cm and 0.7cm);

   %%%%%% mittlere
    %vorne   
  % \draw [- Latex ]  (0.96, -0.11) arc (10:175:0.88cm and 1.16cm);
    \draw [ ]  (0.68, 0.64) arc (48:8:0.93cm and 1.23cm);
    \draw [-Latex]  (0.68, 0.64) arc (50:177:0.93cm and 1.2cm);
     \draw [semithick]  (0.68, 0.64) arc (50:177:0.93cm and 1.2cm);
    %hinten
    \draw [-Latex,  dashed]  (0.96, 0.11) arc (3:170:0.90cm and 1.1cm);
     \draw [semithick,  dashed]  (0.96, 0.11) arc (3:170:0.90cm and 1.1cm);
 	
 	%projected
 	\draw[dotted] (2.5,0) -- (-1,0);
 	\draw[dotted] (2.5,0) -- (-0.8,0.22);
 	\draw[dotted] (2.5,0) -- (-0.8,-0.22);
	\draw[dotted] (2.5,0) -- (-0.35,0.356);
 	\draw[dotted] (2.5,0) -- (-0.35,-0.356);

 	% coordinate
 	%\draw[- Latex] (0,0) -- (0,1.8) node [anchor=west] {$N_1$};
 	
 	 	%tangents
 	\draw[dashed,-] (2.5,0) -- (0.3,0.39);
 	\draw[dashed,-] (2.5,0) -- (0.3,-0.39);
% 	%initial points
% 	%front
% 	\filldraw [black] (0.9,-0.17)circle (0.6pt);
% 	\draw (0.9,-0.17) circle (0.6pt);% node[anchor=east] {$\alpha(x_0)$};
% 	
% 	%middle
% 	\filldraw [black] (1,0) circle (0.6pt);
% 	\draw (1,0) circle (0.6pt);% node[anchor=east] {$\alpha(x_0)$};
% 
% 	%back
% 	\filldraw [black] (0.95,0.13) circle (0.6pt);
% 	\draw (0.95,0.13) circle (0.6pt);% node[anchor=east] {$\alpha(x_0)$};
% 	
% 	 \filldraw [black] (0.76,-0.26) circle (0.6pt);
% 	\draw (0.76,-0.26) circle (0.6pt);% node[anchor=east] {$\alpha(x_0)$};
% 	
% 	
% 	%final points
% 	%front
% 	\filldraw [black] (-0.6,-0.33) circle (0.6pt);
% 	\draw [black] (-0.6,-0.33) circle (0.6pt); % node[anchor=south west] {$\omega(x_0)$};
% 	
% 	%middle
% 	\filldraw [black] (-1,0) circle (0.6pt);
% 	%\draw (-1,0) circle (0.7pt);% node[anchor=east] {$\alpha(x_0)$};
% 	
% 	 %back
% 	\filldraw [black] (-0.75,0.26) circle (0.6pt);
% 	\draw [black] (-0.75,0.26) circle (0.6pt); % node[anchor=south west] {$\omega(x_0)$};
% 	
% 	\filldraw [black] (-0.15,-0.4) circle (0.6pt);
% 	\draw (-0.15,-0.4) circle (0.6pt);% node[anchor=east] {$\alpha(x_0)$};

 	%eta
 	\filldraw [black] (2.5,0) circle (0.7pt);
 	\draw [black] (2.5,0) circle (0.7pt) node[anchor=south] {$\frac{\Q_1}{v}$};
 	
 	% dummy
 	 	\filldraw [black] (0,-0.7) circle (0pt);
 	%x_0
 %	\filldraw [black] (-0.3,0.59) circle (0.7pt) node[anchor=east] {$x_0$};
 	
		\end{tikzpicture}}
		\hspace{1.5cm} 
		\subfloat[The projection of $\BIIa$ solutions to $\Sigma$-space yield straight lines originating at the point $\frac{\Q_1}{v}$ outside the Kasner circle $\KC$.]{\begin{tikzpicture}[scale=1.25]
				\clip (-1.4,-1.5) rectangle (2.85,1.5);

	%coordinate system
%	\draw [line width = 0.1mm] ( 0,-1.4) -- (0,1.4) ;
%	\draw [line width = 0.1mm]  (-1.4,0) -- (1.4,0) ;
	
	%circle
	\draw [thin] (1,0) arc (0:360:1cm and 1cm);	 	
	
	% A_1
	\draw [thick,Bracket-Bracket] (0.4,-0.9165) arc (-66.4218:66.4218:1cm and 1cm);	 	
	
	% Q/v
	\filldraw [black] (2.5,0) circle (0.7pt);
	\draw [black] (2.5,0) circle (0.7pt) node[anchor=south] {$\frac{\Q_1}{v}$};
	
%	%tangent points
%	\filldraw [black] (0.4,0.9165) circle (0pt) node[anchor=south] {$\tac$};
%	\filldraw [black] (0.4,-0.9165) circle (0pt) node[anchor=north] {$\tab$};

%	\draw[-Latex, dotted,thick] (2.5,0) -- (-1,0); 
%	\draw[-Latex, dotted,thick] (2.5,0) -- (-0.11,-0.98); 
%	\draw[-Latex, dotted,thick] (2.5,0) -- (-0.54,-0.83); 
%	\draw[-Latex, dotted,thick] (2.5,0) -- (-0.8,0.59); 
%	%
%	%
%	\draw[-Latex, dotted,thick] (2.5,0) -- (-0.94,0.3); 
%	\draw[-Latex, dotted,thick] (2.5,0) -- (-0.11,0.98); 
%	\draw[-Latex, dotted,thick] (2.5,0) -- (-0.54,0.83); 
%	\draw[-Latex, dotted,thick] (2.5,0) -- (-0.8,-0.59); 
%	\draw[-Latex, dotted,thick] (2.5,0) -- (-0.94,-0.3); 

\foreach \s in {0.15,0.34,0.5,0.66,0.85}
{\draw[dotted] (2.5,0) -- ({cos(42-\s*84)},{sin(42-\s*84)});}
%die äußeren und mitte funktionieren mit den s-werten
\draw[-Latex] ({cos(42-0.15*84)},{sin(42-0.15*84)}) -- ({cos(90+0.15*180)},{sin(90+0.15*180)});
\draw[-Latex] ({cos(42-0.5*84)},{sin(42-0.5*84)}) -- ({cos(90+0.5*180)},{sin(90+0.5*180)});
\draw[-Latex] ({cos(42-0.85*84)},{sin(42-0.85*84)}) -- ({cos(90+0.85*180)},{sin(90+0.85*180)});

%die funktionieren nicht
\draw[-Latex] ({cos(42-0.34*84)},{sin(42-0.34*84)}) -- ({cos(90+0.33*180)},{sin(90+0.33*180)});
\draw[-Latex] ({cos(42-0.66*84)},{sin(42-0.66*84)}) -- ({cos(90+0.67*180)},{sin(90+0.67*180)});
%	\draw[-Latex, dotted,thick] (2.5,0) -- (-1,0); 
%	\draw[-Latex, dotted,thick] (2.5,0) -- (-0.11,-0.98); 
%	\draw[-Latex, dotted,thick] (2.5,0) -- (-0.54,-0.83);
%	
%	 
%	\draw[-Latex, dotted,thick] (2.5,0) -- (-0.8,-0.59); 
%	\draw[-Latex, dotted,thick] (2.5,0) -- (-0.8,-0.59); 
%
%	
	\draw[dashed, -] (2.5,0) -- (0.41,-0.91) node[anchor = north] {\small$\tab(v)$} ;
	\draw[dashed, -] (2.5,0) -- (0.41, 0.91) node[anchor = south] {\small$\tac(v)$} ;

	\filldraw [black] (-0.8,-0.7) circle (0pt) node[anchor=north] {$\KC$}; 	
	
	
			\end{tikzpicture}\label{heteroclinics_figure:projected}}
		\caption {$\BIIalpha$ solutions lie in a hemisphere spanning over $\Sigma$-space in $N_\alpha$-direction, given by the Gauss constraint \eqref{constraint} with $N_{\beta} = N_{\gamma} = 0$. The unstable arc (bold) is $\A_1(v)$.
			%		The thick arc represents the unstable arc $\A_\alpha(v)$.
		}\label{heteroclinics_figure}
	\end{figure}

	Solutions of Bianchi type II are the two-parameter family of Taub vacuum solutions, \cite{taub}. In our phase space
	%\textcolor{red}{Wainwright: "Taub vacuum solution", similar representation as BI?}\\
	these are solutions where $N_\alpha \neq 0$ for exactly one $\alpha= 1,2,3$. The Gauss constraint \eqref{constraint} yields the three sets
	\begin{equation}
        \BIIalpha = \left\{ (N,\Sigma) \in \RR^5 : \Sigma^2 + N_\alpha^2 = 1, \, N_\beta = N_\gamma =0 \right\}.
        \end{equation}
	Each set, given $\alpha=1,2,3$, is the union of two two-dimensional hemispheres, called \emph{Kasner caps}, with $\KC$ as their common boundary. 
	The invariance of the Gauss constraint further allows one to solve system \eqref{full:subs} for $\Sigma$ only, and obtain $N_\alpha$ afterwards. The two-dimensional equation for $\Sigma$, after rescaling of time, becomes 
    \begin{equation}
        \Sigma^\prime=\Sigma - \frac {\Q_\alpha}{v}.
    \end{equation}
    Accordingly, the projected solutions lie on straight lines originating at $\frac{\Q_{\alpha}}{v}$.
	It is necessary that $\Sigma(t)^2 <1$ for all $t\in \RR$. Therefore, $\BIIalpha$ consists of heteroclinic solutions connecting two equilibria on the Kasner circle $\KC$, see \autoref{heteroclinics_figure}.
	%Solving the constraint \eqref{constraint} for $N_\alpha$ shows that $\BIIalpha$ consists of two hemispheres spanning in positive and negative $N_\alpha$-direction,
	%By $D_3 \times \ZZ_2$-equivariance there are exactly six of these hemispheres, two for each of the three $N$-variables.

In general, linearization of equation~\eqref{full:subs} at $\mathrm{K}^{\ocircle}$ results in $N'_1 = 4(\Sigma_+|_{\mathrm{K}^{\ocircle}}-v){N}_1$, and thereby the stability behaviour of $N_{1}$ changes when $\Sigma_+|_{\mathrm{K}^{\ocircle}} = v$. We define the \emph{unstable Kasner arc} to be the points in $\mathrm{K}^{\ocircle}$ that are unstable in the $N_1$ variable: %, i.e., when $\Sigma_+ > v$: %. The closure of $\mathrm{int}(A_{1}(v))$ is denoted by $A_1(v)$ and is given by
\begin{equation}\label{domain_KM_def}
		A_1(v):= \left\{ (\Sigma_+,\Sigma_-)\in \mathrm{K}^{\ocircle} \text{ $|$ }
		\Sigma_+ > v\right\}.
	\end{equation}
Note that the arc $A_1(v)$ is a symmetric portion of
$\mathrm{K}^{\ocircle}$ with the point $\mathrm{Q}_1 := -\mathrm{T}_1$ in the middle. %, where $\Sigma_+$ attains its minimum value $-2$ in $\mathrm{K}^\ocircle$.
The boundary set $\partial A_1(v)$ consists of two fixed points, which coincide with the Taub points $\mathrm{T}_2$ and $\mathrm{T}_3$ when $v=1/2$; for $v\neq 1/2$, each Taub point unfolds into two non-hyperbolic fixed points.
We denote these boundary points by $\tab(v)$ and $\tac(v)$, see \autoref{heteroclinics_figure}.
%Indeed, for $v=1/2$, the Taub point $\mathrm{T}_1$ has three center eigenvalues, whereas it has one center eigenvalue for $v\neq 1/2$ corresponding to the Kasner circle.  Each of the two unfoldings of the Taub points is a fixed point with two center directions.
Equivariance yields the arcs $A_2(v),A_3(v)$, where the respective variables $N_2,N_3$ are unstable, and leads to similar statements for the other Taub points $\mathrm{T}_2$ and $\mathrm{T}_3$.
	%		\begin{figure}
		%			\centering
		%			\begin{tikzpicture}[scale=0.78]
			%				\input{figures/vsmall_circle_defense}
			%				\input{figures/vsmall_orbit1_defense}
			%				\input{figures/vsmall_orbit2_defense}
			%			\end{tikzpicture}
		%		\end{figure}
	%
	%
	%		\begin{figure}
		%			\centering
		%			\begin{tikzpicture}[scale=0.78]
			%				\input{figures/vhalf_circle_defense}
			%				\input{figures/vhalf_orbit_defense}
			%			\end{tikzpicture}
		%			%	\caption*{$\boxed {v=\frac12}$}
		%		\end{figure}
	%
	%
	%		\begin{figure}
		%			\centering
		%			\begin{tikzpicture}[scale=0.78]
			%			\input{figures/vbig_circle_defense}
			%			\input{figures/vbig_orbit_defense}
			%			\end{tikzpicture}
		%		\end{figure}
	\subsection{The Kasner map}\label{sec:kasner}

Each point $p \in \A_1(v)$ uniquely determines the corresponding (projected) type II solution by the line passing through $\frac{\Q_1}{v}$ and $p$.
Thus, the type II solutions in $\BIIa$ induce a map from the $\alpha$-limit set to the $\omega$-limit set of heteroclinics in $\BIIa$, denoted by $\KM_1:A_1(v) \longrightarrow A_1^c(v)$,
%
%In particular, we can parametrize $A_1$ by polar coordinates $(\Sigma_+,\Sigma_-)=(\cos(\varphi),\sin(\varphi))$, which implies that the angle variable $\varphi\in A_1=[-\arccos v, \arccos v]$.
%\textcolor{blue}{The proof for period 4 as it is now makes use of the domain for the parametrization being $[-\arccos(v), 2\pi - \arccos(v)]$ But maybe it's clearer to give those details later} \textcolor{red}{ok. let us leave the way things are, for now, and then when we finish the next section, we come back and see how we write the introduction. for example, maybe defining the maps on the whole arc $A_1$ might make the notation a bit simpler, allow tha kasner map $\KM$ below be well-defined at the taub points, and might not change the proof. we can discuss that later...}
%In this coordinates, the map $\KM_1$ is given by
%\begin{equation}\label{KM_angle}
%		\KM_1(\varphi) :=
%		\pi - 2 \arctan\left(\left(\frac{1+v}{1-v} \right) \tan\left(\frac{\varphi}{2}\right)\right),
%	\end{equation}
%with derivative given by $|D\KM_1(\varphi)| = g(\varphi) := \frac{1-v^2}{ v^2 + 1 -2\cos(\varphi) v} \geq 1.$
% 
% Each point $p \in \A_1(v)$ uniquely determines the corresponding (projected) type II solution by the line passing through $\frac{\Q_1}{v}$ and $p$.
%	This line is parametrized by $$\Sigma= \eta \left(p - \frac{\Q_1}{v}\right) + \frac{\Q_1}{v}$$ with a scalar funtion $\eta$. The first intersection with $\KC$ always occurs when $\eta=1$. We denote the value for $\eta$ which corresponds to the second intersection by $g_1$ and we obtain the formula
	\begin{align} \label{KM1}
		p \, &{\longmapsto}  \,
		\KM_1(p):=g_1(p,v)\left(p+ \frac{\T_1}{v} \right)- \frac{\T_1}{v}, \hspace{1.4cm} g_1(p,v)  := \frac{1-v^2}{-2\Sigma_+ v + v^2 + 1} > 1.
	\end{align}	
	%The arc $\A_1(v)$ is bounded by points whose connecting line to $\frac{\Q_1}{v}$ is tangent to $\KC$, i.e. when the line has exactly one point of intersection with $\KC$. This occurs when $g_1(p,v)=1$, which in turn happens when $\Sigma_+ = v$. We denote these boundary points by $\tab(v)$ and $\tac(v)$, see \autoref{heteroclinics_figure}. These points are excluded in the definition of $\KM_1$, but they can be viewed as fixed points. The unstable arc $\A_1(v)$ is characterized by $g_1(p,v)>1$, which results in
	%\begin{equation}\label{domain_KM_def}
	%	A_1(v) = \left\{p \in \KC : \Sigma_+>v\right\}.
	%\end{equation} 
%
Hence $g_1(p,v)=|D\KM_1(p)|$ is the expansion rate of the map: it is maximal and equal to $(1+v)/(1-v)$ at the midpoint $\Q_1\in A_1$, and it decreases to $1$ towards the boundary of $A_1$. %, i.e. as $\Sigma_+\to v$. %, i.e. as $\varphi \rightarrow \pm \arccos(v)$. % or equivalently $\Sigma_+ \rightarrow v$.
%If $v$ decreases, the domain $A_1$ gets bigger and the derivative flattens towards the constant value $-1$, whereas if $v$ increases, the domain $A_1$ shrinks and the expansion rate increases in $\text{int} (A_1)$. %Analogous considerations hold for heteroclinics in $\BIIb$ and $\BIIc$, yielding the respective maps $\KM_2$ and $\KM_3$.
%

\begin{figure}[H]
		\centering
		\captionsetup[subfigure]{labelformat=empty}
		\subfloat[][]{
			\begin{tikzpicture}[scale=1.2]
				% v= 0.25
% labeling 
\draw (-1.7,1.3) circle(0pt) node[anchor=west] { };
\tiny
% Kasner circle
\draw[thin] (1,0) arc (0:360:1cm and 1cm);

% tangents
%1
\draw[dashed, -, thin] (0.25,0.975)--(3.5,0);
\draw[dashed, -, thin] (0.25,-0.975)--(3.5,0);
%2
\draw[dashed, -, thin] (0.71,-0.72)--(-1.75,-3.03);
\draw[dashed, -, thin] (-0.978,0.25)--(-1.75,-3.03);
%3
\draw[dashed, -, thin] (0.71,0.72)--(-1.75,3.03);
\draw[dashed, -, thin] (-0.978,-0.25)--(-1.75,3.03);

% domains
\draw [Bracket-Bracket] (0.25,-0.9682) arc (-75.5224878:75.5224878:1cm and 1cm);
\draw [Bracket-Bracket] (0.7071,0.7071) arc (45:194.5:1cm and 1cm);
\draw [Bracket-Bracket] (0.7071,-0.7071) arc (-45:-194.5:1cm and 1cm);

%%overlap
\draw [line width = 1.1 pt] (0.7071,0.7071) arc (45:75.52248:1cm and 1cm);
\draw [line width = 1.1 pt] (0.7071,-0.7071) arc (315.52248:285
:1cm and 1cm);
\draw [line width = 1.1 pt] (-0.97, 0.25) arc (165.5:194.5:1cm and 1cm);

%Q/v
\filldraw [black] (3.5,0) circle (0.5pt); \draw  (3.5,0) circle (0pt) node [anchor=west]{$\frac{\Q_1}v$};
\filldraw [black] (-1.75,-3.03) circle (0.5pt); \draw (-1.75,-3.03) circle (0.5pt)  node [anchor=east]{$\frac{\Q_2}v$};
\filldraw [black] (-1.75,3.03) circle (0.5pt); \draw (-1.75,3.03) circle (0.7pt)  node [anchor=east]{$\frac{\Q_3}v$};

%\draw (1,-2.5) circle(0pt) node [text width=3cm] {\footnotesize  $(0 < v < \frac 12$) The Kasner map is multivalued in the intersection of two unstable arcs};

\filldraw  (1.2,-2.5) circle (0pt) node [anchor=north]{\boxed{\text{\scriptsize$v\in (0,1/2)$}}};

%\draw (1.5,-3) circle(0pt) node % [text width=3cm]
%{\fontdimen2\font=2em%
%	\parbox{3.5cm}{\footnotesize \justifying \noindent The Kasner map is multivalued in the intersection of two unstable arcs.}};

%\fontdimen2\font=2em%
%\parbox{3cm}{\RaggedRight This is a ragged right and hyphenated text in a TikZ node.}

\filldraw [black] (0,3) circle (0pt);
\filldraw [black] (0,-3.2) circle (0pt);
				% projected outside circle
\draw [densely dotted,  thin] (3.5,0)--(0.997,0.0125);

\filldraw  (0.997,0.0125) circle (0.7pt) node [anchor=south west]{\scriptsize$p_{\small 0}$};

\draw [densely dotted, thin] (-1.75,3.03)--(-0.69,0.72);

\draw [densely dotted, thin] (-1.75,-3.03)--(0.10, -0.99);

%\draw [green,densely dotted,  thin] (-1.75,3.03) -- ( -0.83, 0.55);
%\draw [densely dotted, very thin] (-1.75,3.03) -- (-0.995,0.022);
%\draw [densely dotted, very thin] (-1.75,3.03) -- ( -0.02, 0.99);

\draw [densely dotted,  thin] (-1.75,-3.03) -- (-0.995,0.022);
%\draw [densely dotted, very thin] (-1.75,-3.03) -- ( -0.75, -0.68);
%\draw [red, densely dotted, very thin] (-1.75,-3.03) -- ( -0.22, -0.975);

%\draw [blue, densely dotted,  thin] (3.5,0) -- (0.86,0.53);
%\draw [densely dotted, very thin] (3.5,0) -- (0.97,-0.24);

% projected inside circle
\draw [thin] (-0.69,0.72)--(0.1,-0.99);
\draw [thin] (0.997,0.0125)--(0.1,-0.99);
\draw [thin,-latex] (0.997,0.0125)--( -0.1, 0.016);
\draw [thin] ( -0.05, 0.0159)--( -0.995, 0.022);
\draw [thin, -latex] (-0.995,0.022)-- ( -0.83, 0.55);
%\draw [thin] ( -0.83, 0.55) -- ( -0.22, -0.975);
%\draw [thin] ( -0.22, -0.975) -- (0.86,0.53);
%\draw [thin,-latex] (0.86,0.53)--( 0.5, 0.6) ;

%\draw [very thin] (-0.995,0.022)-- ( -0.75, -0.67);
%\draw [very thin] ( -0.75, -0.67)-- ( -0.02, 0.99);
%\draw [very thin] ( -0.02, 0.99)-- (0.97,-0.24);
%\draw [very thin,-stealth] (0.97,-0.24)-- ( 0.5, -0.28);
				\input{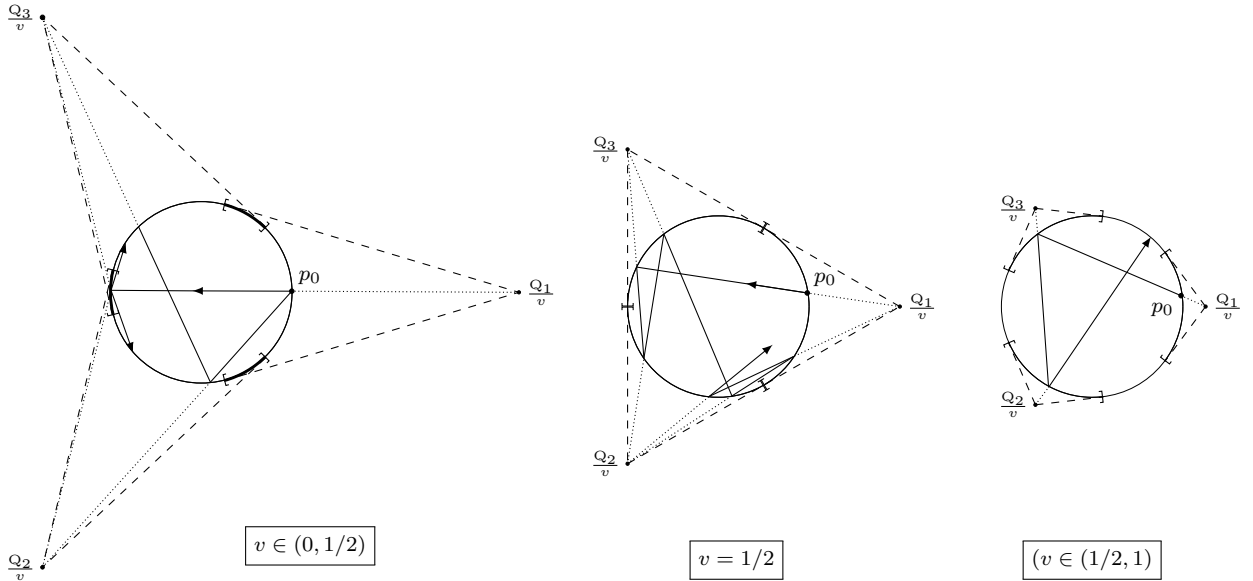}
			\end{tikzpicture} \label{figure_subcritical}
		}
		\subfloat[][]{
			\begin{tikzpicture}[scale=1.2]
				clip (-1.5,-2) rectangle (2.5,2);
% v= 1/2
\tiny 
% Kasner circle
\draw[thin] (1,0) arc (0:360:1cm and 1cm);
% tangents
\draw[thin,dashed] (2,0)--(-1,1.732) ;
\draw[thin,dashed] (-1,1.732) --(-1,-1.732) ;
\draw[thin,dashed] (-1,-1.732) --(2,0);

%\draw (-1.7,1.3) circle(0pt) node[anchor=west] { };

%% arcs

% A_1
\draw [-Bracket] (1,0) arc (0:60:1cm and 1cm);
\draw [-Bracket] (1, 0) arc (0:-60:1cm and 1cm);

% A_3
\draw [-Bracket] (-0.5,0.866) arc (120:180:1cm and 1cm); 
\draw [-Bracket] (-0.5, 0.866) arc (120:60:1cm and 1cm);

% A_2
\draw [-Bracket] (-0.5,-0.866) arc (240:180:1cm and 1cm);
\draw [-Bracket] (-0.5, -0.866) arc (240:300:1cm and 1cm);

%% Q/v
\filldraw [black] (2,0) circle (0.5pt);
\draw  (2,0) circle (0pt) node [anchor=west]{$\frac{\Q_1}v$};
\filldraw [black] (-1,-1.732) circle (0.5pt);
\draw (-1,-1.732) circle (0pt)  node [anchor=east]{$\frac{\Q_2}v$};
\filldraw [black] (-1,1.732)  circle (0.5pt);
\draw (-1,1.732)  circle (0pt)  node [anchor=east]{$\frac{\Q_3}v$};

\filldraw  (0.2,-2.5) circle (0pt) node [anchor=north]{\boxed{\text{\scriptsize$v=1/2$}}};

%\draw (0,-2.5) circle(0pt) node {\footnotesize $\mathrm{(b)} \, \, v= \frac 12$};
%\draw (0.5,-3) circle(0pt) node % [text width=3cm]
%{\fontdimen2\font=2em%
%	\parbox{3.5cm}{\footnotesize \justifying \noindent The Kasner map is well-defined. Its image $\KM(\KC)$ is a double cover of $\KC$. }};

%align 
%\filldraw [black] (-2,3) circle (0pt);
\filldraw [black] (0,2.5) circle (0pt);
\filldraw [black] (0,-3) circle (0pt);
				clip (-1.5,-2) rectangle (2.5,2);
\filldraw  (0.98,0.15) circle (0.7pt) node [anchor=south west]{\scriptsize$p_{\small 0}$};

%orbit in correct order
\draw [thin] (0.98,0.15)--(-0.9,0.435);
\draw [thin,-latex] (0.98,0.15)--(0.3,0.255);
\draw [ thin] (-0.9,0.435)--(-0.82,-0.57);
\draw [ thin] (-0.82,-0.57)--(-0.6,0.8);
\draw [ thin] (-0.6,0.8)--(0.15,-0.985);
\draw [ thin] (0.15,-0.985)--(0.83,-0.55);
\draw [ thin] (0.83,-0.55)--(-0.1,-0.99);
\draw [ thin,-latex] (-0.1,-0.99)--(0.6,-0.42);

\draw [ thin,densely dotted] (2,0)--(0.98,0.15);
\draw [ thin,densely dotted] (2,0)--(0.83,-0.55);

\draw [ thin,densely dotted] (-1,-1.732)--(-0.82,-0.57);
\draw [ thin,densely dotted] (-1,-1.732)--(-0.1,-0.99);
\draw [ thin,densely dotted] (-1,-1.732)--(0.15,-0.985);

\draw [ thin,densely dotted] (-1,1.732)--(-0.6,0.8);
\draw [ thin,densely dotted] (-1,1.732)--(-0.9,0.435);
			\end{tikzpicture} \label{figure_half}
		}
		\subfloat[][]{
			\begin{tikzpicture}[scale=1.2]
				clip (-1.5,-2) rectangle (2.5,6);

% 1/2 < v < 1    v=0.8 
\tiny 
\draw (-1.4,1.3) circle(0pt) node[anchor=west] { };

% Kasner circle
\draw[very thin] (1,0) arc (0:360:1cm and 1cm);

% arcs
%A_1
\draw [-Bracket] (1,0) arc (0:36.87:1cm and 1cm);
% node [anchor= west] {$\tac$};
\draw [-Bracket] (1, 0) arc (0:-36.87:1cm and 1cm);
% node [anchor= west] {$\tab$};

%A_3
\draw [-Bracket] (-0.5, 0.866) arc (120:156.87:1cm and 1cm);
% node [anchor= east] {$\tcb$};
\draw [-Bracket] (-0.5, 0.866) arc (120:83.13:1cm and 1cm);
% node [anchor= south] {$\tca$};

%A_2
\draw [-Bracket] (-0.5, -0.866) arc (240:276.87:1cm and 1cm);
\draw [-Bracket] (-0.5, -0.866) arc (240:203.13:1cm and 1cm);

% tangents
\draw[dashed, -, thin] (1.25,0)--(0.8,0.607);
\draw[dashed, -, thin] (1.25,0)--(0.8,-0.607);

\draw[dashed, -, thin] (-0.625,1.0825)--(-0.9257, 0.3893);
\draw[dashed, -, thin] (-0.625,1.0825)--(0.1257, 0.9963);

\draw[dashed, -, thin] (-0.625,-1.0825)--(0.1257, -0.9963);
\draw[dashed, -, thin] (-0.625,-1.0825)--(-0.9257,-0.3893);

%T/v
\filldraw  (1.25,0) circle (0.5pt) node [anchor=west]{$\frac{\Q_1}v$};
\filldraw (-0.625,-1.0825) circle (0.5pt) node [anchor=east]{$\frac{\Q_2}v$};
\filldraw (-0.625,1.0825)  circle (0.5pt)  node [anchor=east]{$\frac{\Q_3}v$};

%\draw (0,-2.5) circle(0pt) node {\footnotesize $ \mathrm{(c)} \, \, \frac 12 < v < 1$};

\filldraw  (0,-2.5) circle (0pt) node [anchor=north]{\boxed{\text{\scriptsize$(v\in (1/2,1)$}}};

%\draw (0.1,-3) circle(0pt) node % [text width=3cm]
%{\fontdimen2\font=2em%
%	\parbox{3.2cm}{\footnotesize \justifying \noindent  The Kasner map is well-defined. Generic orbits eventually enter $\S$.}};

%align 
%\filldraw [black] (-2,3) circle (0pt);
\filldraw [black] (0,3) circle (0pt);
\filldraw [black] (0,-3) circle (0pt);
				\draw[densely dotted,  thin] (1.25,0)--(0.98, 0.12);
\draw[densely dotted, thin]  (-0.625,1.0825)--(-0.6, 0.8);
\draw[densely dotted, thin]  (-0.625,-1.0825)--(-0.48,-0.89);

\draw[ thin] (0.98, 0.12)--(-0.6, 0.8);
\draw[ thin]  (-0.6, 0.8)--(-0.48,-0.89);
\draw[ thin,-latex]  (-0.48,-0.89)--(0.65,0.77);

\filldraw  (0.98, 0.12) circle (0.7pt) node [anchor=north east]{\scriptsize$p_{\small 0}$};
			\end{tikzpicture} \label{figure_supercritical}
		}
		\caption{Orbits of a point $p_0 \in \A_1$ under the Kasner map $\KM$ for different parameter values $v \in (0,1)$. } \label{figures_KM}
	\end{figure}

By $D_3$-equivariance, analogous constructions in $\BIIb$ and $\BIIc$ yield the maps $\KM_2$ and $\KM_3$ with the same properties. Altogether, they
 %Therefore, the type II solutions on the three hemispheres $\BIIa,\BIIb,\BIIc$ 
induce a discrete map of the circle, %: it maps the $\alpha$-limit set of any type II heteroclinic to the $\omega$-limit set of the same heteroclinic. 
 the \emph{Kasner map} $\KM:\KC \longrightarrow \KC $, given by
	\begin{align} \label{KM}
		\begin{split}
			%\KM:\KC &\longrightarrow \KC \\
			p  &\longmapsto 
			\KM(p):=\begin{cases} 
				%g_\alpha(p)\left(p+ \frac{\T_\alpha}{v} \right)- \frac{\T_\alpha}{v}
				\KM_\alpha(p)
				&\text{if  } p \in \A_\alpha(v) \\
				p &\text{if  }  p \in \S ,
			\end{cases}
		\end{split}
	\end{align}
where $\A(v):=\A_1(v) \cup \A_2(v) \cup \A_3(v)$ is the subset of $\KC$ where at least one of the maps $\KM_\alpha$ is defined, and its complement, $\S:=\KC\setminus \A(v)$, is called the \emph{stable set}.
%, as depicted in Figure \ref{FIG:KASNERMAPS} for a geometrical description of the map. 
See \autoref{figures_KM}. % and \cite{HellLappicyUggla}.
Note that iterations of $\KM$ represent a sequence of heteroclinic Bianchi type II trajectories.
        \vspace{-0.25cm}
        %
        %
	%\subsection{Properties of the Kasner map.}\label{sec:prop}
	The properties of the Kasner map strongly depend on the parameter $v\in (0,1)$, as was recently pointed out in \cite{HellLappicyUggla}, where three qualitatively different cases are considered. %For a visual impression, see \autoref{figures_KM}. 
    The well-studied case of $v=\frac 12$ is a bifurcation point, where chaotic behaviour becomes generic, see~\cite{mis69a,bkl70,Chernoff83,khaetal85,Cornish97,ugg13a,ugg13b} and references therein. 
    For $v\in (1/2,1)$, the map $\KM$ is chaotic in a Cantor set of measure zero, see \cite{HellLappicyUggla}. % by means of conjugation to the shift map.
    For $v\in(0,1/2)$, the (multivalued) map $\KM$ is chaotic, see \cite{LappicyDaniel}. Thus, $v=1/2$ corresponds to a bifurcation from non-generic to generic chaos. 
    We collect some characteristics of the different cases. 
 	
	\subsubsection*{The critical GR case: $v=1/2$.}
	
	This case was extensively treated in the literature over the past 50 years, see \cite{Mixmaster} and references therein.
	Each two neighboring tangent points $\talphabeta,\tbetaalpha$ coincide with the Taub point $\T_\gamma$, dividing $\KC$ into the three unstable arcs $\A_\alpha(v)$, each with length $\frac{2\pi}{3}$.
	The map $\KM(p)$ is thereby uniquely defined via $\KM_\alpha(p)$, and fixed on the stable set $\S=\{\T_1,\T_2,\T_3\}$.

    One can quotient out the $D_3$-equivariance and obtain a frame-invariant \emph{Kasner parameter}, $u \in (1,\infty)$, which parametrizes each sector between a $\Q$-point and a Taub point.
    %    
	%The Kasner circle $\KC$ is the inscribed circle of the triangle formed by the three "shooting points" $\frac{\Q_\alpha}{v}$.
	%With the changes of coordinates according to \eqref{6dim_coordinates} and \eqref{kasner_exponents},each of the sectors between a $\Q$-point and a Taub point is described by a permutation $(\alpha\beta\gamma)$ of $(123)$ via the condition $p_\alpha<p_\beta< p_\gamma$. The \emph{Kasner parameter} $u \in (1,\infty)$ parametrizes the sector $(\alpha\beta\gamma)$ via
	%\begin{align}\label{parametrization_via_u}
	%	p_\alpha = \frac{-u}{1+u+u^2}, \hspace{1.2cm} p_\beta = \frac{1+u}{1+u+u^2}, \hspace{1.2cm} p_\gamma = \frac{u(1+u)}{1+u+u^2}, \hspace{1cm}
	%\end{align}
	The point $\Q_\alpha$ corresponds to the parameter value $u=1$ and the Taub point, $\T_\gamma$, corresponds to $u=\infty$.
	For $u \in [1,\infty)$, the induced map by $\KM$ under the $D_3$-group action is called the \emph{BKL map}:
	% (\cite{bkl70}, \cite{Mixmaster})
	\begin{align}\label{KM_u}
		u \mapsto \begin{cases}
			u-1 &\text{ if \,} u \in [2,\infty) \\
			\frac{1}{u-1} &\text{ if \,} u \in [1,2).
		\end{cases}
	\end{align}
	To recover the corresponding orbit of $\KM$, it is necessary to specify an initial point in one of the six open sectors.
	Note that $u=\infty$ is not considered in \eqref{KM_u}, since the Taub points are fixed. 
	With the convention $0^{-1}:=\infty$, the value $u=1$ is mapped to $\infty$, which is consistent with the point $\Q_\alpha$ being mapped to $\T_\alpha$, for $\alpha=1,2,3$.
	
	A sequence of points $(u_\ell)_{\ell \in \NN}$ generated by \eqref{KM_u} can be partitioned into pieces, called \emph{eras}. Indeed, starting with a maximal value $u_{\max}$, each iteration, called \emph{epoch}, decreases according to $u_{\ell+1} =  u_\ell - 1$. An era ends at a minimal value $u_{\min} \in [1,2)$, when a new era begins with $\tilde{u}_{\max}:=1/(u_{\min}-1)$, provided $u_{\min} \neq 1$. The dynamics within an era corresponds to the oscillations between two unstable arcs.
	The larger $u_{max}$, the closer it is to a Taub point, and the larger the number of oscillations before the next era begins. %The next era corresponds to the oscillations between one of the previous arcs and the third remaining arc.
	
	%   \textcolor{red}{(also relevant for other cases (oscillations). describe in caption of figure: this era length is 5 but in general very large)}
	
	The dynamics of the map \eqref{KM_u} can be understood by representing the Kasner parameter $u$ as a continued fraction, i.e.
 \begin{equation}
     u= k_0 + \frac{1}{k_1 + \frac{1}{k_2+ \frac{1}{k_3 + ...}}} =: [k_0;k_1, k_2, k_3, ...], \shortquad k_i \in \NN.
 \end{equation}
	The asymptotic behaviour of \eqref{KM_u} naturally follows from the theory of continued fractions.
	Indeed, for $k_0\ge 2$, the map \eqref{KM_u} becomes $[k_0;k_1, k_2, k_3, ...] \mapsto [k_0-1;k_1, k_2, k_3, ...]$ within an era, until $(k_0-n)\in [1,2)$ for some $n\in\NN$.
	For an era change, where $u_{\text{min}} \in [1,2)$ and thus $k_0=1$, the map \eqref{KM_u} becomes the left-shift: $[1;k_1, k_2, k_3, ...] \mapsto [k_1; k_2, k_3, ...]$. Thus, the number $k_i$ describes the length of the $i^{th}$ era.
	To capture the qualitative behaviour it is also useful to reduce the dynamics of \eqref{KM_u} by introducing the \emph{era map}. It takes the first point of an era, $u_{\max}$ and maps it to the first point of the next era, $\tilde{u}_{\max}:=1/(u_{\min}-1)$, and is thus given by the shift $[k_0;k_1, k_2, k_3, ...]\mapsto [k_1; k_2, k_3, ...]$.
	For the inverse parameter $x=u^{-1}$, this is the \emph{Gauss map}, given by $x\mapsto \frac1x - \lfloor\frac1x\rfloor$.
%
	%The glimpse of the dynamical behaviour given here already provides some intuition, why the Kasner map exhibits chaotic behaviour.
	%There are several notions of chaoticity in the literature, see for example (ref).  According to Devaney (ref) the classical characteristics of chaotic behaviour are sensitive dependence on initial conditions, transitivity and denseness of periodic orbits.
	For a more detailed investigation of the Gauss map and its chaotic behaviour, see, e.g., \cite{gaussmap_chotic}. For a rigorous proof of chaoticity of the Kasner map by showing topological conjugacy to the so-called \emph{shift map}, see \cite{HellLappicyUggla}. 
	%A proof of topological conjugacy to the shift map for the critical case is given in \cite{Bernhard}, \cite{HellLappicyUggla}.

	%Using properties of continued fractions we obtain a characterization of the dynamics.
	Either one of the following possibilities occurs:
	\begin{itemize}
		\item[(i)] $u = [k_0;k_1, ..., k_n]$ has a finite continued fraction expansion. Consequently, after $k_0+k_1+...+k_n$ iterations of the BKL map and $n+1$ iterations of the era map the initial point ends up at a Taub point. Recall that $u\in \QQ_{\ge 1}$ if and only if its continued fraction expansion is finite. Thus, the finite heteroclinic chains are associated with a countable dense subset of $\KC$, and thereby of Lebesgue measure zero. Hence this case is non-generic.
		\item[(ii)] $u$ has an infinite continued fraction expansion. We distinguish two cases.
		\begin{itemize}
			\item[(a)] $u$ is badly approximable, i.e. the sequence of coefficients is bounded. As (i) this is not generic. A special case is when
			$u =[k_0; k_1,...,k_m,\overline{{k}_{m+1},...,{k}_{m+n}}]$ has a periodic continued fraction expansion. %This occurs if $u$ is a quadratic irrational, that is, an irrational number that solves a quadratic polynomial with integer coefficients.
			In this case $u$ yields for both the Kasner map and the era map, an orbit which is preperiodic, i.e. eventually periodic.
			%	 The orbit is periodic when $k_i=0$ for $i=0,...,m-1$, in which case the continued fraction expansion is said to be purely periodic.
			%	the continued fraction expansion of a generic $u$ is not bounded, let alone periodic (reference).
			\item[(b)]
			Generically, $u$ is a well approximable irrational number, i.e. the continued fraction expansion of $u$ is infinite and its sequence of coefficients is not bounded. %, let alone periodic or preperiodic. 
            Consequently, the orbit of $u$ comes arbitrarily close to a Taub point.
		\end{itemize}
	\end{itemize}
	
	We consider two simple examples. First, the golden ratio $u = [1;\bar{1}] = \frac{1+\sqrt{5}}{2}$ yields a fixed point, i.e. a point of period one, for both, the BKL map and the era map. For the Kasner map $\KM$ as a function on $\KC$, this represents two period three orbits related by $D_3$ equivariance, which traverse $\KC$ either clockwise or counter-clockwise, depending on the initial point. %This yields six points in the Kasner circle, which are related by rotation, yielding period three orbits of the Kasner map.
	%These orbits are  see also \autoref{fig:per3}.
	%
	Second, the silver ratio $u = [2;\bar{2}] = 1+\sqrt{2}$. The era map is again one-periodic, whereas the BKL orbit is of period two, given by $ 1+\sqrt{2} \,\to\, \sqrt{2} \to\, 1+\sqrt{2} \,\to\, \sqrt{2} \,\to\, ... \,$.
	%To understand the corresponding orbit of $\KM$, consider $p \in \A_1(v)$.
	%Within an era there are two oscillations. Thus $p$ visits a neighboring unstable arc and then lands back in $\A_1(v)$, but at a point different from $p$, since there are no orbits of period two of $\KM$ when $v=\frac 12$. Since the era map is one-periodic, this point should yield the same point in the quotient $\KC/D_3$, and is therefore the reflection of $p$ along the $\Sigma_+$-axis. The following two oscillations are between $\A_1$ and the remaining unstable arc until the point lands at $p$ again. 
        For the Kasner map $\KM$ as a function on $\KC$, this represents three orbits of period four, related by $D_3$ equivariance. We discuss the orbits of period three and four in more detail in \autoref{sec:periods}.
	
	\subsubsection*{The supercritical case: $v\in (1/2,1)$.}
 
	As $v$ increases, the three points $\frac{\Q_\alpha}{v}$ move towards $\KC$, see \autoref{figures_KM}. Thus, the size of each arc $\A_\alpha(v)$ decreases and its complement, the stable set $\S$, is of positive Lebesgue measure.
	The Kasner map $\KM$ is uniquely defined via $\KM_\alpha(p)$ for $p \in \A_\alpha(v)$ and fixed otherwise. This yields a continuous function with discontinuities of the derivative at the tangent points.
	
	The infinite heteroclinic chains are associated with all points $p \in \KC$ for which the orbit of $p$ is disjoint from $\S$.
	We define
	\begin{align}\label{Cantor_def}
		\Cantor(v) := \left\{p \in \KC : \KM^n(p) \notin \S \text{ for all } n \in \NN\right\}.
	\end{align}
	Theorems 4.1 and 4.2 in \cite{HellLappicyUggla} prove that the set $\Cantor(v)$ is a Cantor set of Lebesgue measure zero for all $ v \in (\frac12,1)$. Moreover, the Kasner circle map $\KM$ restricted to $\Cantor(v)$ is topologically conjugate to the shift map and it generates a chaotic discrete dynamical system.

 	Recall that a nonempty set is called a \emph{Cantor set} if it is compact, perfect and totally disconnected.
	Thus, in contrast to the critical case, chaos is not generic. The points associated with finite heteroclinic chains are the complement of $\Cantor(v)$ and thus of full Lebesgue measure.
	% \textcolor{red}{also Baire?!, why is this then not chaotic? }

        The Cantor set is delimited by period 2 chains, see  \cite{HellLappicyUggla}, which are bounded away\footnote{In the limit $v\to 1/2$, each of the period 2 chains degenerates to a Taub point.} from the tangential points, where the expansion rate of the Kasner map $\KM$ is minimal with value 1. Thus, we let $m(v)$ (resp. $M(v)$) denote the minimum (resp. maximum) expansion rate of $\KM$ in the Cantor set, for $v\in (1/2,1)$, %occurs at the period 2 orbit (resp. the pre-image of a period 2 point), 
see \cite[Lemma 4.4]{HellLappicyUggla}. The explicit rates are given by
\begin{subequations}
\begin{align}
m(v) &:= \min_{p\in \Cantor(v)} |D\mathcal{K}(p)|= \frac{2(1-v^2)}{1+2v^2-\sqrt{12 v^2 - 3}}>1, \label{minC}\\
M(v) &:= \max_{p\in \Cantor(v)} |D\mathcal{K}(p)|= \frac{2+v^2}{1-v^2}>3. \label{maxC}
\end{align}
\end{subequations}
         
	\subsubsection*{The subcritical case: $v\in (0,1/2)$.}%\label{sec:Kasnersub}
	
	In this case the intersection of each two arcs $\A_\alpha(v) \cap \A_\beta(v)$, which we denote by  $\A_{\alpha\beta}(v)$, is non-empty. Consequently, the stable set vanishes and for points in these overlaps the assignment \eqref{KM} is not unique, i.e. the Kasner map $\KM$ is multivalued for some points in its domain.
	Similar to the set $\Cantor(v)$ in the supercritical case, we define
	\begin{subequations}\label{setsCF}
		\begin{align}
			\tilde{\mathcal{C}}(v)&:= \{ p\in \KC \text{ $ : $ } \mathcal{K}^n(p)\notin A_{12}(v)\cup A_{13}(v)\cup A_{23}(v),\forall n \in \mathbb{N}_0 \},\\
			\tilde{\mathcal{F}}(v)&:= \{ p\in \KC \text{ $ : $ } \exists  n \in \mathbb{N}_0,  \,
			\mathcal{K}^n(p) \in  A_{12}(v)\cup A_{13}(v)\cup A_{23}(v)\}.
		\end{align}
	\end{subequations}
	The circle $\KC= \tilde{\mathcal{C}} \cup \tilde{\mathcal{F}}$ is thereby split into disjoint invariant sets, such that $\KM$ is well-defined on $\tilde{\mathcal{C}}$, but eventually multivalued in $\tilde{\mathcal{F}}$. In \cite{LappicyDaniel} it is shown that $\tilde{\mathcal{F}}$ is dense.
	%\textcolor{red}{include: Conjecture 5.2 from HL was formulated more precisely and proven [ref]}
	Moreover, the dynamical behaviour of $\KM$, in particular its chaotic nature, is described in \cite{LappicyDaniel}. Note that a new notion of chaotic dynamics for multivalued maps is necessary.
	%The classical characteristics of chaotic behaviour according to Devaney are sensitive dependence on initial conditions, transitivity and denseness of periodic orbits \cite{Devaney} . In \cite{LappicyDaniel} it is proven that the multivalued Kasner map \emph{realizes} these characteristics.

    \subsection{Shadowing the Kasner map}

        Next, we discuss the role of the induced discrete dynamics, given by the Kasner map $\KM$, in the continuous dynamical system of Bianchi type VIII and IX solutions.

		Let $x(t)$ be a Bianchi type VIII or IX solution and let $(p_k)_{k \in \NN_0}$ be a sequence of points obtained by iteration of the Kasner map, i.e. $p_{k+1} = \KM(p_k)$, for all $k \in \NN_0$.  
            We are interested in orbits of $\KM$ that never enter the stable set $\S$.
            If %the orbit $\left\{\KM^n(p)\right\}_{n \in \NN_0}$ is a subset of 
            $p_k\in A(v)$ for all $k\in\NN_0$, we call the corresponding sequence of heteroclinic solutions an \emph{infinite heteroclinic chain}, denoted by $(h_k(t))_{k\in \NN_0}$, such that each $h_k(t)$ is a Bianchi type II solution %within the subset $\BIIalpha$, for $\alpha=1,2,3$, 
        satisfying  %if the following holds. For all $k\in \NN_0$, $h_k$ is a solution of Bianchi type II satisfying 
        %\begin{align}\label{het_chain_def}
			$\lim_{t \to -\infty} h_k(t) = p_k$ and  $\lim_{t \to \infty} h_k(t) = p_{k+1}$. %\text{ for all } k \in \NN
	%\end{align}
		%such that all elements of $(h_k(t))_{k\in \NN_0}$ of the subset $\BIIalpha$ have the same sign, for $\alpha=1,2,3$.
        %
		If, in addition, the sequence is periodic with minimal period $N\in \NN$, we call $(h_k(t))_{k\in \NN_0}$ a \emph{heteroclinic cycle of length $N$}.
            %On the other hand, if there exists $N \in \NN$ such that $p_{N} \in \S$, then $p_{k} = p_N$ for all $k \ge N$, and we call the sequence of finite associated heteroclinics $(h_k(t))_{k=0}^{N-1}$ a \emph{finite heteroclinic chain}. %In this case \eqref{het_chain_def} is only required to hold for $k \le N$.

\begin{figure}[H]
		\centering
		\begin{tikzpicture}[scale=1.0]
%              \begin{scope}[yscale=1,xscale=-1]
			% points p_k
\def \dis {2.5}

\def \n {5}
\foreach \i in {0,...,\n}
{\coordinate (p\i) at (\dis*\i,0);}

\foreach \i in {1,...,4}
{\filldraw (p\i) circle (1pt); }

\filldraw (p1) circle (0pt) node[anchor=north] {\footnotesize$p_{k-1}$}; 
\filldraw (p2) circle (0pt) node[anchor=north] {\footnotesize$p_{k}$}; 
\filldraw (p3) circle (0pt) node[anchor=north] {\footnotesize$p_{k+1}$}; 
\filldraw (p4) circle (0pt) node[anchor=north] {\footnotesize$p_{k+2}$};

% heteroclinics

\def \n {3}
\foreach \i in {1,...,\n}
{\coordinate (h\i) at (1.2+\dis*\i,0.45);
	\filldraw (h\i) circle (0pt); }
\filldraw (h1) circle (0pt) node[anchor=north] {\footnotesize$h_{k-1}(t)$}; 
\filldraw (h2) circle (0pt) node[anchor=north] {\footnotesize$h_{k}(t)$}; 
\filldraw (h3) circle (0pt) node[anchor=north] {\footnotesize$h_{k+1}(t)$}; 

\foreach \i/\j in {1/2, 2/3, 3/4}
{\draw [-Latex, shorten <=0.08cm, shorten >=0.1cm, bend left = 35] (p\i) to (p\j);}
%\draw [shorten <=0.08cm, shorten >=0.08cm, bend left = 30] (p\i)-- (p\j);}
\draw [_-Latex, shorten <=1cm, shorten >=0.1cm, bend left = 35] (p0) to (p1);
\draw [-Latex, shorten <=0.08cm, shorten >=1cm, bend left = 35] (p4) to (p5);

\path let \p1=(p1) in coordinate (x0) at (\x1-0.8cm,		\y1+1.2 cm);
\path let \p1=(p1) in coordinate (x1) at (\x1,					\y1+1 cm);
\path let \p1=(p2) in coordinate (x2) at (\x1-1.4cm ,	   	\y1+0.85 cm);
\path let \p1=(p2) in coordinate (x3) at (\x1-0.6cm ,	   \y1+0.6cm);
\path let \p1=(p2) in coordinate (x4) at (\x1-0.05cm ,	   \y1+0.4cm);
\path let \p1=(p2) in coordinate (x5) at (\x1+0.65cm ,	   \y1+0.55 cm);
\path let \p1=(p2) in coordinate (x6) at (\x1+1.6cm ,	   \y1+0.58 cm);
\path let \p1=(p3) in coordinate (x7) at (\x1-0.4cm,	   \y1+0.4 cm);
\path let \p1=(p3) in coordinate (x8) at (\x1 + 0.05cm,	   \y1+0.24 cm);
\path let \p1=(p3) in coordinate (x9) at (\x1+0.6cm ,	   \y1+0.44 cm);
\path let \p1=(p3) in coordinate (x10) at (\x1+1.3cm ,	   \y1+0.53 cm);
\path let \p1=(p4) in coordinate (x11) at (\x1-0.5cm,			\y1+0.36 cm);
\path let \p1=(p4) in coordinate (x12) at (\x1-0cm,			\y1+0.14 cm);
\path let \p1=(p4) in coordinate (x13) at (\x1+0.5cm ,			\y1+0.34 cm);
\path let \p1=(p4) in coordinate (x14) at (\x1+1cm ,			\y1+0.47 cm);
\path let \p1=(p4) in coordinate (x15) at (\x1+1.3cm ,			\y1+0.48 cm);

\filldraw (3,1) circle (0pt) node[anchor=south] {\footnotesize$x(t)$}; 
%	
%\foreach \k in {0,1,2,3,4,5,6,7,8,9,10,11,12,13,14,15}{\filldraw (x\k) circle (1pt) node[anchor=south] {\tiny$x_{\k}$};}
\draw [xshift=3cm] plot [smooth, tension=0.8] coordinates {(x0) (x1) (x2) (x3) (x4) (x5) (x6) (x7) (x8) (x9) (x10) (x11) (x12) (x13) (x14) (x15)};

%\filldraw (x1) circle (1pt) node[anchor=south] {\scriptsize$x(t_{k-1})$};
\filldraw (x4) circle (1pt) node[anchor=south] {\scriptsize$x(t_k)$};
\filldraw (x8) circle (1pt) node[anchor=south] {\scriptsize$x(t_{k+1})$};
%\filldraw (x12) circle (1pt) node[anchor=south] {\footnotesize$x(t_{k+2})$};
%              \end{scope}
        \end{tikzpicture} 
		\caption{A solution $x$ \emph{shadows} a given infinite heteroclinic chain $(h_k(t))_{k\in \NN_0}$ if it approaches the chain in Hausdorff distance when $t \to \infty$.}\label{figure_shadowing}
	\end{figure}
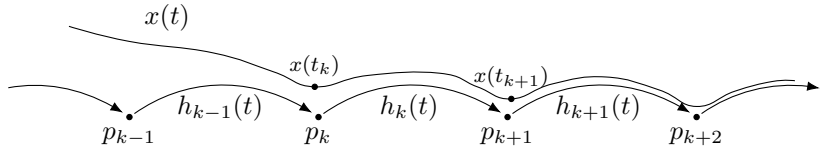

	%We are only interested in infinite heteroclinic chains, and therefore in orbits of $\KM$ which never enter the stable set $\S$.
	%Note that a heteroclinic chain is unique only up to sign changes of its elements. The sign of Bianchi type VIII and IX solutions is invariant. For a heteroclinic chain to be contained in the boundary of the type VIII and IX subset we therefore assume that all elements of the heteroclinic chains from the subset $\BIIalpha$ have the same sign, for $\alpha=1,2,3$.
	%By itself, the heteroclinic chains have no significance.
	%To make the idea of BKL and Misner precise, i.e. to connect the discrete dynamics of $\KM$ to the asymptotic behaviour of generic solutions of Bianchi type VIII and IX, we introduce the following notion.
	%Their importance is due to the work by BKL \cite{bkl70} and Misner \cite{mis69a}, who connected the discrete dynamics of the Kasner map to the asymptotic behaviour of generic solutions of Bianchi type VIII and IX. The following notion is used to make this idea precise
        %	
	%\begin{defi} \normalfont
	%	Let $x$ be a solution of Bianchi type VIII or IX and let $(h_k(t))_{k \in \NN_0}$ be an infinite heteroclinic chain with corresponding sequence $(p_k)_{k \in \NN_0}$ of equilibria. 
    We say that a solution $x(t)$ of Bianchi type VIII or IX \emph{shadows} $(h_k(t))_{k \in \NN_0}$ if there exists a strictly increasing sequence $(t_k)_{k\in \NN_0}$ such that
		\begin{equation}
			\begin{aligned}
				(i) \shortquad&\| x(t_k) - p_k \| \xrightarrow{k \to \infty} 0,\\
				(ii) \shortquad& d_H \left( \{x(t) | t_k \le t \le t_{k+1}\}, h_k(\RR) \right) \xrightarrow{k \to \infty} 0,
			\end{aligned}
		\end{equation}
		where $d_H$ denotes the Hausdorff distance between two sets. %, let
		%\begin{equation}
		%d_H(A,B) := \max \left\{ \sup_{a \in A} \,\Inf_{b \in B} \,\|a-b\|\, ,\, \, \sup_{b\in B} \,\Inf_{a \in A} \|a-b\| \right\}.
          %\end{equation}
		The set of all solutions shadowing $(h_k(t))_{k\in \NN_0}$ is called the \emph{stable set} of $(h_k(t))_{k\in \NN_0}$, see also \autoref{figure_shadowing}.
	%\end{defi}
        %
		Recall that the sign of the $N$-variables of a Bianchi type VIII or IX solution is invariant. Thus all elements of a heteroclinic chain $(h_k(t))_{k\in \NN_0}$ of the same Bianchi type II subset, $\BIIalpha$, for each fixed $\alpha=1,2,3$, must have the same sign. Indeed, the heteroclinic chain can only be shadowed if it is contained in the boundary of a specific type VIII or IX subset orthant.
	
	%Before we briefly discuss what is known about shadowing, we investigate the Kasner map $\KM$.

	We now discuss shadowing results for $v=\frac 12$.
	%The existence of solutions shadowing a given heteroclinic chain depends on the corresponding orbit of the Kasner map.
	By the characterization in \autoref{sec:kasner}, a generic orbit gets arbitrarily close to a Taub point, where the linearization degenerates. This poses a problem for shadowing the corresponding heteroclinic chain, and thereby all known results possess assumptions on the coefficients of the continued fraction expansion of the Kasner parameter. %i.e. the linear stability analysis yields no characterization of the local dynamics.

        For each element in a (generic) set satisfying a subpolynomial bound on the Kasner parameter, there is a Bianchi type IX solution following the corresponding heteroclinic chain, see \cite{ReitererTrubowitz}.
	However, the dimensionality or the characterization (e.g. a graph with certain regularity) of the stable set is not provided, see also \cite{ugg13a}.
	%It is also not answered if the asymptotic behaviour of generic solutions of type IX are represented by a heteroclinic chain, see also \cite{ugg13a}
 
	The proof in \cite{Liebscher} concerns the heteroclinic cycle of length three in Bianchi type VIII and IX, but their method is applicable to arbitrary periodic and aperiodic chains that %in the boundary of the Bianchi type VIII and IX subset that 
    keep a uniform distance away from the Taub points. This deals with the case (ii)(a) in  \autoref{sec:kasner}. The stable set of the heteroclinic chain is given as a codimension-one Lipschitz manifold.
	
	%The formal heteroclinic chains obtained by iteration of $\KM$ can be characterized according to.
	The approach in \cite{Beguin} applies Takens linearization at each point along the sequence of equilibria. This %yields a simple representation of the local vector field, but 
    assumes \emph{Sternberg's condition}, which %requires that the eigenvalues are linearly independent over $\QQ$, and %. This requirement directly 
    excludes quadratic irrationals of the Kasner parameter (such as pre-periodic sequences). More precisely, it is shown that infinite heteroclinic chains associated with closed forward invariant subsets of $\KC$ without preperiodic points admit $C^1$-stable manifolds of codimension one. Manifolds of this kind are also constructed in \cite{Buchner} using similar methods, which additionally allow certain periodic orbits. Although the set of admissible points is dense in $\KC$, it is of measure zero, as in \cite{Liebscher}.
	%The Kasner parameter however is given as a function (more precisely, the ratio) of the eigenvalues. Therefore, Takens linearization is not applicable excludes the (pre)periodic orbits, as the eigenvalues in these points are resonant.
	%A similar approach is chosen in Buchner for what?
	%In both of these cases the set of points on $\KC$ whose corresponding heteroclinic chains attract Bianchi type VIII and IX solutions is of measure zero.
	
	In \cite{Dutilleul}, the method of \cite{Liebscher} is refined, resulting in a wider class of heteroclinic chains with a codimension-one stable Lipschitz manifold in Bianchi types VIII and IX. The proof allows Kasner parameters $u$ with unbounded continued fraction expansion, as in (ii)(b), satisfying a moderate growth condition. %$k^4_{n+4} = o_{n\to \infty}\left(\sum_{i=1}^{n} k_i^5\right)$.
	%\begin{equation}\label{moderate_growth}
	%	k^4_{n+4} = o_{n\to \infty}\left(\sum_{i=1}^{n} k_i^5\right).
	%\end{equation}
    This yields a different subset of $\KC$ than imposing the growth condition of \cite{ReitererTrubowitz}, but which is also of full one-dimensional Lebesgue measure (in the Kasner circle).	
    Moreover, the union of all type IX orbits shadowing heteroclinic chains satisfying the growth condition has positive four-dimensional Lebesgue measure (in phase-space). %It is, however, not answered if the union of \emph{all} type IX orbits shadowing a heteroclinic chain is of full measure (in phase-space). 

	%For this reason, each iteration of $\eqref{KM_u}$ encodes six, respectively three iterations of the Kasner map. In the case of a $\Q$-point there are only three. Taub points are not part of the definition. Thus, to recover an orbit of the Kasner map, we need the additional information about the sector of the initial point.
	
	%By a classical argument (see [KH97]), the Kasner map is topologically conjugate to θ ↦ −2θ (on the circle R/Z) which has a well understood dynamics.
	
	%	Dutilleul: -degree -2 \\
	%	Periodic points of F are dense in K .
	%	• There exists points in K whose forward orbit under F are dense in K . The set of all such points
	%	is a Gδ dense.
	%	• For every point p of the Kasner circle, the complete backward orbit of p under F is dense in K .
	%	• The topological entropy of the Kasner map is positive (it is equal to log(2)).
	%	• F possesses an invariant measure (of infinite mass) absolutely continuous with respect to Lebesgue measure.

    \section{Heteroclinic chains: linear structure, examples and robustness}\label{sec:warm-up}

	\subsection{Linearization at the Kasner circle}\label{chap:eigenvalues}
	The domains $\A_\alpha(v)$ can be characterized via the linearization of \eqref{full:subs} along the Kasner circle $\KC$, rather than the geometric approach we choose in \autoref{sec:kasner}. Let $p \in \KC$.
	For $\alpha=1,2,3$, the linear stability in direction of $N_\alpha$, which depends on the sign of the corresponding eigenvalue, determines whether $p$ is an $\alpha$-limit or $\omega$-limit point for a $\BIIalpha$ solution.
	Moreover, knowing the magnitude of the eigenvalues at $p$ enables us to compare contraction and expansion rates along the corresponding eigendirections in a local neighborhood of $p$. %This gives more insight than just knowing if or if not there is a heteroclinic solution converging to $p$, which is necessary to prove shadowing.
	
	%The Bianchi type II solutions are each contained in a bounded invariant two dimensional hemisphere spanning in $N_\alpha$-direction with a circle of equilibria $\KC$ as its boundary. 
	%The dynamics projected to $\Sigma$-space on that hemisphere is monotone, hence solutions converge to equilibria on the boundary $\KC$. and
	Thus, we linearize \eqref{full:subs} at a point $p = \left(0,0,0,\Sigma_+, \Sigma_-\right) \in \KC$, i.e. with $\Sigma^2 = 1$. We obtain
	\begin{align*} \label{linerazation_Kasner}
		\begingroup
		\begin{pmatrix} 
			4( \Sigma_+ - v)		 &			0	&	0	&	0	&	0	       \\
			0		&		-4v - 2\Sigma_+ - 2\sqrt 3 \Sigma_-	&		0	&	0 & 0	\\
			0		&		0		&		-4v - 2\Sigma_+ + 2\sqrt 3 \Sigma_-  	 &		0	&	0	\\
			0	&	 0	& 		0  		& 	-8v \,\Sigma_+^2 				 & 		 -8v \,\Sigma_+\Sigma_-	  	\\
			0	&	 0	& 		0  		&  -8v \,\Sigma_+\Sigma_-		&		-8v \,\Sigma_-^2			      \\
		\end{pmatrix}_.
		\endgroup
	\end{align*}
	The matrix is in block diagonal form separating the $N$- and $\Sigma$-variables. Thus, the eigenvectors lie in these subspaces. In $\Sigma$-space one eigenvalue is $0$ with eigenvector $\partial_{\KC} := \left(-\Sigma_-,  \Sigma_+\right)$, tangential to $\KC$, the boundary of the invariant manifold defined by \eqref{constraint}. The second eigenvalue $-8v$ comes with eigenvector $\left(\Sigma_+,  \Sigma_-\right)$, transverse to $\KC$.
				The three eigenvalues corresponding to $N$-space can be read off the diagonal as
				\begin{subequations}
					\label{eigenvalues}
					\begin{align}
						\eva(p,v) &:= \,4(\Sigma_+ -v), \\
						\evb(p,v) &:= -4v - 2\Sigma_+ - 2\sqrt 3 \Sigma_-, \\
						\evc(p,v) &:= -4v - 2\Sigma_+ + 2\sqrt 3 \Sigma_-.
					\end{align}
				\end{subequations}
				Moreover, the three domains $\A_\alpha(v)$ of the maps constituting the Kasner map $\KM$ can be characterized via $\evi(p,v) > 0$, in accordance with \eqref{domain_KM_def},  as
				\begin{subequations}\label{eigenvalue_instability}
					\begin{align}
						\A_1(v) &=
						\left\{p \in \KC: \Sigma_+ > v\right\}   , \hspace{2.07cm}\\
						\A_2(v) &=
						\left\{p \in \KC:\Sigma_+ <-2v -\sqrt {3} \, \Sigma_-\right\}, \\
						\A_3(v) &=
						\left\{p \in \KC:\Sigma_+ <-2v + \sqrt {3}\, \Sigma_-\right\}.
					\end{align}
				\end{subequations}
				For each $v$, the zero set of each function $\evi(p,v)$ is a straight line in $\Sigma$-plane, which contains the two boundary points $\talphabeta(v),\talphagamma(v)$ of $\A_\alpha(v)$, denoted by
				\begin{align}\label{eigenvalue_zero}
					L_\alpha(v) := \left\{\Sigma\in \RR^2 : \evi(p,v)=0\right\}, \shortquad \alpha= 1,2,3.
				\end{align}
				Note that whether an eigenvalue $\evi(p,v)$ is positive or negative depends on the position of $p$ relative to $L_\alpha(v)$, and is determined by \eqref{eigenvalue_instability}.
                    Next, we present the relationship of the eigenvalues of a point in $\A_\alpha(v)$ and the lines $L_\alpha(v)$. %These three lines have a special meaning.
				\begin{lem}\label{lem_eigenvalues_distance}
					Let $ p = (\Sigma_+, \Sigma_-) \in \KC$ and $L_\alpha(v)$ as in \eqref{eigenvalue_zero}.
					%	the set of points for which the eigenvalue $\evi$ is zero. 
					The eigenvalues \eqref{eigenvalues} of the linearization \eqref{linerazation_Kasner} satisfy 
					%\begin{equation}
					$	|\evi(p,v)| = 4 \cdot \text{dist}\left(p,L_\alpha(v)\right), \quad \alpha=1,2,3, $
					%\end{equation}
					where $\text{dist}$
					%	\left(\Sigma,L_\alpha(v)\right)$
					denotes the shortest, i.e. orthogonal, distance in $\RR^2$. 
					%Morevover, $\text{dist}\left(\Sigma, L_\alpha(v)\right)$ 
				\end{lem}
				\begin{proof}
					Note that $L_\alpha(v)$ is a straight line in $\RR^2$. In particular, $L_1(v)$ is a line through $(v,0)$ parallel to the $\Sigma_-$-axis. Hence the claim is true for $\eva(p,v)$. Recall that, by $D_3$-equivariance of \eqref{full:subs}, the set $L_1(v)$ is obtained from $L_2(v)$ and $L_3(v)$ via rotation by $\frac{2 \pi}{3}$ and $\frac{4 \pi}{3}$, respectively. Let for example $R \in \text{SO}_2(\RR)$ denote the anti-clockwise rotation by $\frac{2 \pi}{3}$ that maps $L_2(v)$ to $L_1(v)$, i.e. $L_1(v) = R L_2(v)$. 
                    The same argument applies to $L_3(v)$ with a rotation by $\frac{4\pi}{3}$.
                    In particular 
                    \begin{equation}
                      \left(Rp\right)_+ = \cos\left(\frac{2\pi}{3}\right)\Sigma_+ - \sin\left(\frac{2\pi}{3}\right) \Sigma_- =  - \frac12 \Sigma_+  - \frac{\sqrt 3}{2} \Sigma_-.  
                    \end{equation}
                     Since the distance between a point and a set is invariant under orthogonal transformations,
					\begin{equation}  \text{dist}(p,L_2(v)) =  \text{dist}(Rp,L_1(v)) = \left|  - \frac12 \Sigma_+  - \frac{\sqrt 3}{2} \Sigma_- - v \right|  =\frac{|\evb(p,v)| }{4},
                        \end{equation}
					where the second equality uses that $L_1(v)$ is parallel at distance $v$ to the $\Sigma_-$-axis.
					%		\begin{figure}
						%			\centering
						%			\input{figures/proof_eigenvalues_supercritical}
						%		\end{figure}
				\end{proof}

				\subsection{Ordering of the eigenvalues}
				We discuss the consequences of Lemma \ref{lem_eigenvalues_distance} for the supercritical and subcritical cases.
				Our aim is to divide $\A(v)$, the domain of the Kasner map $\KM$, into sectors, where the sign and the ordering of the eigenvalues with respect to their absolute value is not changed.
				Lemma \ref{lem_eigenvalues_distance} allows us to obtain this ordering in a point $p \in \KC$ by measuring the distance between $p$ and the three lines $L_\alpha(v)$. This provides a purely geometrical approach. % for which it is useful to give names to the following geometric objects, see \autoref{figure_intro_geometric_notion:subs}.
								
				%\bigskip
				%\textbf{Geometric notions.}
				%\smallskip
				
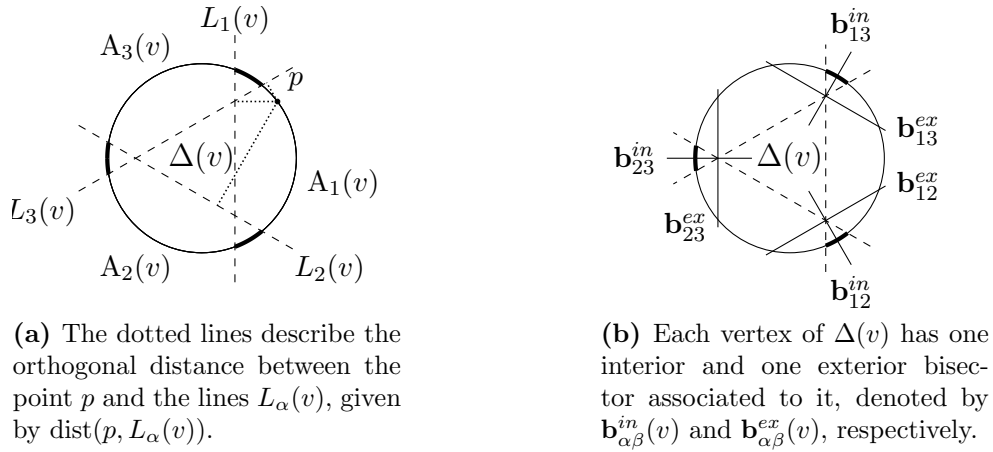
\begin{figure}[H]
					\centering
					\subfloat[The dotted lines describe the orthogonal distance between the point $p$ and the lines $L_\alpha(v)$, given by dist($p,L_\alpha(v)$).]{\begin{tikzpicture}[scale=1.25]
							\clip (-2,-1.6) rectangle (2,1.6);

	% labeling 
	%\draw (-1.9,1.3) circle(0pt) node[anchor=west] {};
	
	% coordinate system
%	\draw [-Latex] (-1.5,0)--(1.5,0);
%	\draw [-Latex] (0,-1.5)--(0,1.5);
	
	% Kasner circle
	\draw[] (1,0) arc (0:360:1cm and 1cm);
	
	% 1 unstable
	
	% lines between unstable and stable area order: 1,2,3
%	\draw[dashed, -, line width = 0.1pt] (0.5,-1.3)--(0.5,1.3);
%	\draw[dashed, -, line width = 0.1pt] (1.137,-1.23)--(-1.5,0.29);
%	\draw[dashed, -, line width = 0.1pt] (1.137,1.23)--(-1.5,-0.29);
	
	%caption v
	%\draw[line width=0.5 pt] (0.5,-0.07)--(0.5,0.07);
	%\filldraw [black] (0.5,-0.1) node[anchor=east] {$v$}; 	
	
	%Taub points
	%\filldraw [black] (-1,0) circle (0.8pt);
	%\draw  (-0.95,0.15) circle (0pt) node [anchor=east]{$\Ta$};
	%\filldraw [black] (0.5,0.866) circle (0.8pt);
	%\draw (0.5,0.866) circle (0pt)  node [anchor=south]{$\Tb$};
	%\filldraw [black] (0.5,-0.866) circle (0.8pt);
	%\draw (0.5,-0.866) circle (0pt)  node [anchor=north]{$\Tc$};

	\def \v {0.35}
\def \radius {1}

% Q points and Taub points
\foreach \s in {1,2,3}
{\coordinate (Q\s) at ({-120 * (\s - 1)}:\radius);}
\foreach \s in {1,2,3}
{\path let \p\s=(Q\s) in coordinate (T\s) at (-1*\x\s,-1*\y\s);}

% Kasner circle
\draw [] (Q1) arc (0:360:\radius cm and \radius cm);	 	

% caption circle
%\node at (0.9*\radius, \radius*0.7) [anchor = west] {$\KC$};

% caption Q1 and T1
\path let \p1=(Q1) in coordinate (capQ1) at (1.15*\x1,\y1);
\path let \p1=(T1) in coordinate (capT1) at (1.15*\x1,\y1);

%	%\filldraw  (capT1) circle (0pt) node [anchor=south]{$\T_1$};
%	%\filldraw  (T1) circle (0.5pt);
%	\filldraw  (T1) circle (0.7pt); % node [anchor=east]{\scriptsize $\T_1$};
%	\filldraw  (T2) circle (0.7pt); % node [anchor=south]{\scriptsize$\,\,\T_2$};
%	\filldraw  (T3) circle (0.7pt); % node [anchor=north]{\scriptsize$\,\,\T_3$};
%	
%	%\filldraw  (capQ1) circle (0pt) node [anchor=south]{$\Q_1$};
%	%\filldraw  (Q1) circle (0.5pt);
%	\filldraw  (Q1) circle (0.7pt); % node [anchor=west]{\scriptsize$\Q_1$};
%	\filldraw  (Q2) circle (0.7pt); % node [anchor=north]{\scriptsize$\Q_2\,\,$};
%	\filldraw  (Q3) circle (0.7pt); % node [anchor=south]{\scriptsize$\Q_3\,\,$};

%coordinate system
%\draw [->] (0,0)--(0.55,0);
%\draw [->] (0,0)--(0,0.55);

%\draw  (0.6,-0.05) circle (0pt) node [anchor=south]{\scriptsize$\Sigma_+$};
%\draw  (0,0.5) circle (0pt) node [anchor=west]{\scriptsize$\Sigma_-$};

% tangent points
\coordinate (t13) at (\v,{sqrt(1-\v^2)});
\coordinate (t12) at (\v,-{sqrt(1-\v^2)});

%rotate about -120
\path let \p1=(t13) in coordinate (t21) at (-0.5*\x1  +0.866*\y1, -0.866*\x1 -0.5*\y1);
%rotate about -120
\path let \p1=(t12) in coordinate (t23) at (-0.5*\x1  +0.866*\y1, -0.866*\x1 -0.5*\y1);

%rotate about 120
\path let \p1=(t13) in coordinate (t32) at (-0.5*\x1  -0.866*\y1, 0.866*\x1 -0.5*\y1);
%rotate about 120
\path let \p1=(t12) in coordinate (t31) at (-0.5*\x1  -0.866*\y1, 0.866*\x1 -0.5*\y1);

%	\draw [dashed] (t12)--(t13);
%	\draw [dashed] (t21)--(t23);
%	\draw [dashed] (t31)--(t32);

%allonged L-lines
\def \ppp {1.2}
\coordinate (T13) at ($(t12)!\ppp!(t13)$);
\coordinate (T12) at ($(t13)!\ppp!(t12)$);

\coordinate (T23) at ($(t21)!\ppp!(t23)$);
\coordinate (T21) at ($(t23)!\ppp!(t21)$);

\coordinate (T31) at ($(t32)!\ppp!(t31)$);
\coordinate (T32) at ($(t31)!\ppp!(t32)$);

\draw [dashed] (T12)--(T13);
\draw [dashed] (T21)--(T23);
\draw [dashed] (T31)--(T32);

% X - points : vertices of triangle DELTA
\coordinate (X1) at (-2*\v*\radius,0);
%rotate about -120
\path let \p1=(X1) in coordinate (X2) at (-0.5*\x1  +0.866*\y1, -0.866*\x1 -0.5*\y1);
%rotate about -120
\path let \p1=(X2) in coordinate (X3) at (-0.5*\x1  +0.866*\y1, -0.866*\x1 -0.5*\y1);

% caption L lines
\coordinate (Y1) at (\v,\ppp);
%rotate about -120
\path let \p1=(Y1) in coordinate (Y2) at (-0.5*\x1  +0.866*\y1, -0.866*\x1 -0.5*\y1);
%rotate about -120
\path let \p1=(Y1) in coordinate (Y3) at (-0.5*\x1  -0.866*\y1, 0.866*\x1 -0.5*\y1);

\draw (Y1) circle(0pt) node [anchor=south]{\small$L_1(v)$};
\draw (Y2) circle(0pt) node [anchor=north west]{\small$L_2(v)$};
\draw (Y3) circle(0pt) node [anchor=north east]{\small$L_3(v)$};

% triangle DELTA
%\draw [line width=1pt] (X1)--(X2)--(X3)--(X1);

% interior bisectors 
%	\def \InScale {1}
%	\foreach \j in{1,2,3} 
%	{\path let \p1=(T\j) in coordinate (Y\j) at (\InScale*\x1,\InScale*\y1);
	%		\path let \p2=(Q\j) in coordinate (Z\j) at (\InScale*\x2,\InScale*\y2);
	%		\draw (Y\j)--(Z\j);}

%exterior bisectors
\def \OutScale  {1}
\path let \p1=(X1) in coordinate (Ex23Up) at (\x1,\OutScale);
\path let \p1=(X1) in coordinate (Ex23Down) at (\x1,-\OutScale);

% rotate
\path let \p1=(Ex23Up) in coordinate (Ex13Up) at (-0.5*\x1  +0.866*\y1, -0.866*\x1 -0.5*\y1);
\path let \p1=(Ex23Down) in coordinate (Ex13Down) at (-0.5*\x1  +0.866*\y1, -0.866*\x1 -0.5*\y1);

\path let \p1=(Ex23Up) in coordinate (Ex12Up) at (-0.5*\x1  -0.866*\y1, 0.866*\x1 -0.5*\y1);
\path let \p1=(Ex23Down) in coordinate (Ex12Down) at (-0.5*\x1  -0.866*\y1, 0.866*\x1 -0.5*\y1);

%	\draw (Ex23Up)--(Ex23Down);
%	\draw (Ex13Up)--(Ex13Down);
%	\draw (Ex12Up)--(Ex12Down);

\def \scaling {1.2}
\coordinate (q1) at ($(T1)!\scaling!(Q1)$);
\coordinate (q2) at ($(T2)!\scaling!(Q2)$);
\coordinate (q3) at ($(T3)!\scaling!(Q3)$);

\coordinate (t1) at ($(Q1)!\scaling!(T1)$);
\coordinate (t2) at ($(Q2)!\scaling!(T2)$);
\coordinate (t3) at ($(Q3)!\scaling!(T3)$);

%\draw (1,0) circle(0pt) node [anchor=east]{\small$L_1(v)$};
%\draw (-0.2,-0.8) circle(0pt) node [anchor=south east]{\small$L_2(v)$};
%\draw (-0.2,0.8) circle(0pt) node [anchor=north east]{\small$L_3(v)$};
\draw (0,0) circle(0pt) node []{$\Delta(v)$};

\draw (1,0) circle(0pt) node [anchor=north west]{\small$\A_1(v)$};
\draw (-0.2,-0.9) circle(0pt) node [anchor=north east]{\small$\A_2(v)$};
\draw (-0.2,0.9) circle(0pt) node [anchor=south east]{\small$\A_3(v)$};

%	\draw (t1) -- (q1);
%	\draw (t2) -- (q2);
%	\draw (t3) -- (q3);

\foreach \s in {1,2}
{ \def \z {\s+1}
	\path let \p1=(X\s) in coordinate (X\z) at (-0.5*\x1  +0.866*\y1, -0.866*\x1 -0.5*\y1);}

%arcs 
\draw [ name path=A1] (t12) arc (-{acos(\v)}:{acos(\v)}:1);
\draw [ name path=A2] (t23) arc (-{acos(\v) -120 }:{acos(\v) -120}:1);
\draw [ name path=A3] (t31) arc (-{acos(\v) -240 }:{acos(\v) -240}:1);

\def \z {0.8}
\coordinate (P1) at (\z,{sqrt(1-\z^2)});

\path let \p1=(P1) in coordinate (P2) at (-0.5*\x1  -0.866*\y1, 0.866*\x1 -0.5*\y1);
\path let \p1=(P1) in coordinate (P3) at (-0.5*\x1  +0.866*\y1, -0.866*\x1 -0.5*\y1);

\path let \p1=(P1) in coordinate (P1B) at (\v, \y1);
\path let \p1=(P2) in coordinate (P2L) at (\v, \y1);
\path let \p1=(P3) in coordinate (P3L) at (\v, \y1);

\path let \p1=(P2L) in coordinate (P2B) at (-0.5*\x1  +0.866*\y1, -0.866*\x1 -0.5*\y1);
\path let \p1=(P3L) in coordinate (P3B) at (-0.5*\x1  -0.866*\y1, 0.866*\x1 -0.5*\y1);

\draw [semithick, densely dotted] (P1)--(P1B);
\draw [semithick, densely dotted] (P1)--(P2B);
\draw [semithick, densely dotted] (P1)--(P3B);

\filldraw  (P1) circle (0.7pt) node [anchor=south west]{\small$p$};

%overlap
%overlap
\draw [line width=1.6pt,domain={acos(\v)}:120-acos(\v)] plot ({cos(\x)}, {sin(\x)});
\draw [line width=1.6pt,domain={-acos(\v)}:-120+acos(\v)] plot ({cos(\x)}, {sin(\x)});
\draw [line width=1.6pt,domain={180}:180 + acos(\v)-60] plot ({cos(\x)}, {sin(\x)});
\draw [line width=1.6pt,domain={180}:180 +60 - acos(\v)] plot ({cos(\x)}, {sin(\x)});
						\end{tikzpicture} \label{figure_intro_geometric_notion:a}}
					\hspace{2.5cm}
					\subfloat[Each vertex of $\Delta(v)$ has one interior and one exterior bisector associated to it, denoted by $\mathbf{b}_{\alpha \beta}^{in}(v)$ and $\mathbf{b}_{\alpha \beta}^{ex}(v)$, respectively.]
					{\begin{tikzpicture}[scale=1.25]
							\clip (-2,-1.6) rectangle (2,1.6);
%\clip (0,0) circle (1cm);
%\clip (-0.1,-1.8) rectangle (1.8,1.8);

%v = 0.36
\def \v {0.38}
\def \radius {1}

% Q points and Taub points
\foreach \s in {1,2,3}
{\coordinate (Q\s) at ({-120 * (\s - 1)}:\radius);}
\foreach \s in {1,2,3}
{\path let \p\s=(Q\s) in coordinate (T\s) at (-1*\x\s,-1*\y\s);}

% Kasner circle
\draw [] (Q1) arc (0:360:\radius cm and \radius cm);	 	

% caption circle
%\node at (0.9*\radius, \radius*0.7) [anchor = west] {$\KC$};

% caption Q1 and T1
\path let \p1=(Q1) in coordinate (capQ1) at (1.15*\x1,\y1);
\path let \p1=(T1) in coordinate (capT1) at (1.15*\x1,\y1);

%\filldraw  (capT1) circle (0pt) node [anchor=south]{$\T_1$};
%\filldraw  (T1) circle (0.5pt);
%\filldraw  (T1) circle (0.7pt); % node [anchor=east]{\scriptsize $\T_1$};
%\filldraw  (T2) circle (0.7pt); % node [anchor=south]{\scriptsize$\,\,\T_2$};
%\filldraw  (T3) circle (0.7pt); % node [anchor=north]{\scriptsize$\,\,\T_3$};
%
%%\filldraw  (capQ1) circle (0pt) node [anchor=south]{$\Q_1$};
%%\filldraw  (Q1) circle (0.5pt);
%\filldraw  (Q1) circle (0.7pt); % node [anchor=west]{\scriptsize$\Q_1$};
%\filldraw  (Q2) circle (0.7pt); % node [anchor=north]{\scriptsize$\Q_2\,\,$};
%\filldraw  (Q3) circle (0.7pt); % node [anchor=south]{\scriptsize$\Q_3\,\,$};

%coordinate system
%\draw [->] (0,0)--(0.55,0);
%\draw [->] (0,0)--(0,0.55);

%\draw  (0.6,-0.05) circle (0pt) node [anchor=south]{\scriptsize$\Sigma_+$};
%\draw  (0,0.5) circle (0pt) node [anchor=west]{\scriptsize$\Sigma_-$};

% tangent points
\coordinate (t13) at (\v,{sqrt(1-\v^2)});
\coordinate (t12) at (\v,-{sqrt(1-\v^2)});

%rotate about -120
\path let \p1=(t13) in coordinate (t21) at (-0.5*\x1  +0.866*\y1, -0.866*\x1 -0.5*\y1);
%rotate about -120
\path let \p1=(t12) in coordinate (t23) at (-0.5*\x1  +0.866*\y1, -0.866*\x1 -0.5*\y1);

%rotate about 120
\path let \p1=(t13) in coordinate (t32) at (-0.5*\x1  -0.866*\y1, 0.866*\x1 -0.5*\y1);
%rotate about 120
\path let \p1=(t12) in coordinate (t31) at (-0.5*\x1  -0.866*\y1, 0.866*\x1 -0.5*\y1);

\def \ppp {1.15}
\coordinate (T13) at ($(t12)!\ppp!(t13)$);
\coordinate (T12) at ($(t13)!\ppp!(t12)$);

\coordinate (T23) at ($(t21)!\ppp!(t23)$);
\coordinate (T21) at ($(t23)!\ppp!(t21)$);

\coordinate (T31) at ($(t32)!\ppp!(t31)$);
\coordinate (T32) at ($(t31)!\ppp!(t32)$);

\draw [dashed] (T12)--(T13);
\draw [dashed] (T21)--(T23);
\draw [dashed] (T31)--(T32);

% X - points : vertices of triangle DELTA
\coordinate (X1) at (-2*\v*\radius,0);
%rotate about -120
\path let \p1=(X1) in coordinate (X2) at (-0.5*\x1  +0.866*\y1, -0.866*\x1 -0.5*\y1);
%rotate about -120
\path let \p1=(X2) in coordinate (X3) at (-0.5*\x1  +0.866*\y1, -0.866*\x1 -0.5*\y1);

% triangle DELTA
%\draw [dashed] (X1)--(X2)--(X3)--(X1);
%\draw (1,0) circle(0pt) node [anchor=east]{\small$L_1(v)$};
%\draw (-0.2,-0.8) circle(0pt) node [anchor=south east]{\small$L_2(v)$};
%\draw (-0.2,0.8) circle(0pt) node [anchor=north east]{\small$L_3(v)$};
\draw (0,0) circle(0pt) node []{$\Delta(v)$};

% interior bisectors 
\def \InScale {1}
\foreach \j in{1,2,3} 
{\path let \p1=(T\j) in coordinate (Y\j) at (\InScale*\x1,\InScale*\y1);
	\path let \p2=(Q\j) in coordinate (Z\j) at (\InScale*\x2,\InScale*\y2);}
%	\draw (Y\j)--(Z\j);}

%exterior bisectors
\def \OutScale  {0.73}
\path let \p1=(X1) in coordinate (Ex23Up) at (\x1,\OutScale);
\path let \p1=(X1) in coordinate (Ex23Down) at (\x1,-\OutScale);

% rotate
\path let \p1=(Ex23Up) in coordinate (Ex13Up) at (-0.5*\x1  +0.866*\y1, -0.866*\x1 -0.5*\y1);
\path let \p1=(Ex23Down) in coordinate (Ex13Down) at (-0.5*\x1  +0.866*\y1, -0.866*\x1 -0.5*\y1);

\path let \p1=(Ex23Up) in coordinate (Ex12Up) at (-0.5*\x1  -0.866*\y1, 0.866*\x1 -0.5*\y1);
\path let \p1=(Ex23Down) in coordinate (Ex12Down) at (-0.5*\x1  -0.866*\y1, 0.866*\x1 -0.5*\y1);

\draw (Ex23Up)--(Ex23Down) node [anchor=east] {\small $\mathbf{b}^{ex}_{23}$};
\draw (Ex13Down)--(Ex13Up) node [anchor=west] {\small $\mathbf{b}^{ex}_{13}$};
\draw (Ex12Up)--(Ex12Down) node [anchor=west] {\small $\mathbf{b}^{ex}_{12}$};

\def \scaling {0.3}
\def \scalingA {1.15}
\coordinate (q1) at ($(T1)!\scaling!(Q1)$);
\coordinate (q2) at ($(T2)!\scaling!(Q2)$);
\coordinate (q3) at ($(T3)!\scaling!(Q3)$);

\coordinate (t1) at ($(Q1)!\scalingA!(T1)$);
\coordinate (t2) at ($(Q2)!\scalingA!(T2)$);
\coordinate (t3) at ($(Q3)!\scalingA!(T3)$);

\draw (q1) -- (t1) node [anchor=east] {\small $\mathbf{b}^{in}_{23}$};
\draw (q2) -- (t2) node [anchor=south] {\small $\mathbf{b}^{in}_{13}$};
\draw (q3) -- (t3) node [anchor=north] {\small $\mathbf{b}^{in}_{12}$};

\foreach \s in {1,2}
{ \def \z {\s+1}
	\path let \p1=(X\s) in coordinate (X\z) at (-0.5*\x1  +0.866*\y1, -0.866*\x1 -0.5*\y1);}

%overlap
%overlap
\draw [line width=1.6pt,domain={acos(\v)}:120-acos(\v)] plot ({cos(\x)}, {sin(\x)});
\draw [line width=1.6pt,domain={-acos(\v)}:-120+acos(\v)] plot ({cos(\x)}, {sin(\x)});
\draw [line width=1.6pt,domain={180}:180 + acos(\v)-60] plot ({cos(\x)}, {sin(\x)});
\draw [line width=1.6pt,domain={180}:180 +60 - acos(\v)] plot ({cos(\x)}, {sin(\x)});
						\end{tikzpicture} \label{figure_intro_geometric_notion:b}}
					\caption{The unstable arc $\A_\alpha(v)$ is determined by the intersections of the (dashed) line $L_\alpha(v)$ with $\KC$, for $\alpha=1,2,3$. The three lines $L_\alpha(v)$ span the triangle $\Delta(v)$.}  \label{figure_intro_geometric_notion:subs}
				\end{figure}

				We denote by $\Delta(v)$ the equilateral triangle formed by the three lines $L_1(v), L_2(v), L_3(v)$. Each vertex of $\Delta(v)$ is the intersection of two lines $L_{\alpha}(v)$ and $L_{\beta}(v)$. Thus, two angle bisectors are associated with each vertex of $\Delta(v)$, yielding two lines whose points are at equal distance to both $L_\alpha(v)$ and $L_\beta(v)$.
				One bisector intersects the interior of $\Delta(v)$, thereby we call it an \emph{interior bisector}, denoted by $\mathbf{b}_{\alpha \beta}^{in}(v)$. Another bisector corresponds to the supplementary angle and it is perpendicular to $\mathbf{b}_{\alpha \beta}^{in}(v)$, which we call an \emph{exterior bisector}, denoted by $\mathbf{b}_{\alpha\beta}^{ex}(v)$, see \autoref{figure_intro_geometric_notion:subs}.

				%
				%\smallskip
				%
				Consider a point $p \in \mathbf{b}_{\alpha \beta}^{in}(v)$.
				By construction, when $p \in \mathbf{b}_{\alpha \beta}^{in}(v)$ is in the interior of $\Delta(v)$, both eigenvalues $\evi(p,v)$ and $\evj(p,v)$ are negative.
				As $p$ crosses the vertex of $\Delta(v)$ both lines $L_\alpha(v)$ and $L_\beta(v)$ are traversed simultaneously and both signs of the eigenvalues switch. Hence, if $p \in \mathbf{b}_{\alpha \beta}^{in}(v)$, then $\evi(p,v)=\evj(p,v)$, and if $p \in \mathbf{b}_{\alpha \beta}^{ex}(v)$, then $\evi(p,v)=-\evj(p,v)$.
				Moreover, the interior bisectors do not depend on $v$.
				Indeed, consider $\mathbf{b}_{2 3}^{in}(v)$, which contains all points $p$ such that $\evb(p,v) = \evc(p,v)$. By symmetry (or the expressions \eqref{eigenvalues} for the eigenvalues) we find that $\mathbf{b}_{2 3}^{in}(v)$ coincides with the $\Sigma_+$-axis, i.e. with the line spanned by $\T_1$ and $\Q_1$, independently of $v$.
				By $D_3$-equivariance the analogous statement is true for $\mathbf{b}_{1 2}^{in}(v)$ and $\mathbf{b}_{1 3}^{in}(v)$. Thus, for a permutation $(\alpha\beta\gamma)$ of $(123)$ we obtain
				\begin{align}\label{inbisector}
					\mathbf{b}_{\alpha \beta}^{in}(v) = \eta \T_\gamma+ (1-\eta)\Q_\gamma, \hspace{0.8cm} \eta \in \RR, \shortquad v \in (0,1).
				\end{align}
				%
				%
				%\bigskip
				%\textbf{Corollaries supercritical and subcritical case.}
				%\smallskip
				%
				Now assume that $p$ is a point of intersection of a bisector with $\KC$. According to Lemma \ref{lem_eigenvalues_distance} the two eigenvalues $\evi(p,v)$ and $\evj(p,v)$ are equal in modulus. Consequently, to either side of $p$ in $\KC$ they are no longer equal in modulus and the ordering is strict.
				Thus, by determining all intersections of the bisectors with $\KC$, we obtain a division of $\KC$ in the desired sectors.
				%In \autoref{cor_supercritical_eigenvalues} and \autoref{cor_subcritical_eigenvalues} we investigate the dependence of these sectors on $v$, i.e. we determine all points of intersection of a bisector with $\KC$.
				%
				%To determine these sectors it suffices to compute the points of intersection of $\KC$ with the bisectors $\mathbf{b}_{\alpha\beta}^{in}(v)$ and $ \mathbf{b}_{\alpha\beta}^{ex}(v)$ for $\{\alpha, \beta\} \subset \{1,2,3\}$. \\
				%because each bisector splits the plane in two halfplanes, and thus divides the circle in two parts as well.\\
				%
				\begin{lem}[Critical and supercritical case]\label{cor_supercritical_eigenvalues} 
					Let $v \in \left[\frac12, 1\right)$ and $(\alpha\beta\gamma)$ be a permutation of $(123)$. Each point $p \in \A_\alpha(v)$ has exactly three non-zero eigenvalues, one of which is positive and two are negative. The positive eigenvalue is given by $\evi(p,v)$ and it is always the smallest in absolute value.
					% The ordering among the two negative eigenvalues $\evj(p,v)$ and $\evk(p,v)$ is strict and depends on the side of $p$ with respect to $\Q_\alpha$.
					Moreover, all eigenvalues in the (open) half-arc between $\talphabeta(v)$ and $\Q_\alpha$ satisfy
					\begin{align}\label{eigenvalues_corollary}
						0<\evi<-\evj<-\evk. % \quad \text{or} \quad 0<\evi<-\evk<-\evj.
					\end{align}
				\end{lem}
				\begin{remark}
					Consider $\A^\pm_1(v):=\{p\in \A_1 : \text{sgn}(\Sigma_-)=\pm\}$. Then
					\begin{equation}\label{eigenvalue_condition_cor}
						\begin{aligned}
							&0<\eva(p,v)<-\evc(p,v)< -\evb(p,v) \quad \text{ if } \quad p \in \A^+_1(v), \\
							&0<\eva(p,v)<-\evb(p,v)< -\evc(p,v) \quad \text{ if } \quad p \in \A^-_1(v).
						\end{aligned}
					\end{equation}
					%\textcolor{blue}{add?: In our definition of the arc $\A_\alpha$ we exclude the boundary points $\talphabeta$ and $\talphagamma$, which have an eigenvalue zero in direction of $\partial_{N_\alpha}$.}
				\end{remark}
				\begin{proof}
					%For the rest of the proof we consider $p \in \A_1$. This choice is without loss of generality, since the other arcs yield the same structure with roles of the eigenvalues interchanged and are obtained by $D_3$-equivariance.
					Note that $v=\frac 12$ is a special value, since $\Delta(v)$ is the inscribed triangle to $\KC$, with set of vertices $\{\T_1, \T_2, \T_3\}$.
					%\textcolor{blue}{Thus, each Taub point has a double eigenvalue zero.}
					%	, i.e. the intersection of each two lines $L_\alpha\left(\frac 12\right)$ and $L_\beta\left(\frac 12\right)$ coincides with the Taub point $\T_\gamma$.
					The interior bisectors are perpendicular to $\KC$, due to \eqref{inbisector}. Hence the exterior bisectors are tangent. That is, 
					apart from a single contact point, the Kasner circle $\KC$ is entirely to one side of each exterior bisector, see \autoref{figure_geometric_notion:subs} (middle). Moreover, since the points of contact are Taub points, which are disjoint from $\A(v)$, none of the exterior bisectors intersects $\A(v)$. With increasing $v$, each exterior bisector moves outwards along the corresponding interior bisector and there is no intersection with $\KC$.
					%	, i.e. the ordering is the same as for $v=\frac 12$.\textcolor{red}{explain again what I'm doing?} 
                    %
				\begin{figure}[H]
					\captionsetup[subfigure]{labelformat=empty}
					\centering
					\subfloat[$v\in(0,1/2)$: The vertices of $\Delta(v)$ are inside of $\KC$. Each exterior bisector intersects $\KC$ at two points.]{
						\begin{tikzpicture}[scale=1.25]
							\clip (-1.5,-1.6) rectangle (1.6,1.6);
%\clip (-0.1,-1.8) rectangle (1.8,1.8);
%\clip (0,0) circle (1cm);

%v = 0.36
\def \v {0.35}
\def \radius {1}

% Q points and Taub points
\foreach \s in {1,2,3}
{\coordinate (Q\s) at ({-120 * (\s - 1)}:\radius);}
\foreach \s in {1,2,3}
{\path let \p\s=(Q\s) in coordinate (T\s) at (-1*\x\s,-1*\y\s);}

% Kasner circle
\draw [semithick] (Q1) arc (0:360:\radius cm and \radius cm);	 	

% caption circle
%\node at (0.9*\radius, \radius*0.7) [anchor = west] {$\KC$};

% caption Q1 and T1
\path let \p1=(Q1) in coordinate (capQ1) at (1.15*\x1,\y1);
\path let \p1=(T1) in coordinate (capT1) at (1.15*\x1,\y1);

% points 
%\filldraw  (capT1) circle (0pt) node [anchor=south]{$\T_1$};
%%\filldraw  (T1) circle (0.5pt);
%\filldraw  (T1) circle (0.7pt); % node [anchor=east]{\scriptsize $\T_1$};
%\filldraw  (T2) circle (0.7pt); % node [anchor=south]{\scriptsize$\,\,\T_2$};
%\filldraw  (T3) circle (0.7pt); % node [anchor=north]{\scriptsize$\,\,\T_3$};
%
%%\filldraw  (capQ1) circle (0pt) node [anchor=south]{$\Q_1$};
%%\filldraw  (Q1) circle (0.5pt);
%\filldraw  (Q1) circle (0.7pt); % node [anchor=west]{\scriptsize$\Q_1$};
%\filldraw  (Q2) circle (0.7pt); % node [anchor=north]{\scriptsize$\Q_2\,\,$};
%\filldraw  (Q3) circle (0.7pt); % node [anchor=south]{\scriptsize$\Q_3\,\,$};

%coordinate system
%\draw [->] (0,0)--(0.55,0);
%\draw [->] (0,0)--(0,0.55);

%\draw  (0.6,-0.05) circle (0pt) node [anchor=south]{\scriptsize$\Sigma_+$};
%\draw  (0,0.5) circle (0pt) node [anchor=west]{\scriptsize$\Sigma_-$};

% tangent points
\coordinate (t13) at (\v,{sqrt(1-\v^2)});
\coordinate (t12) at (\v,-{sqrt(1-\v^2)});

%rotate about -120
\path let \p1=(t13) in coordinate (t21) at (-0.5*\x1  +0.866*\y1, -0.866*\x1 -0.5*\y1);
%rotate about -120
\path let \p1=(t12) in coordinate (t23) at (-0.5*\x1  +0.866*\y1, -0.866*\x1 -0.5*\y1);

%rotate about 120
\path let \p1=(t13) in coordinate (t32) at (-0.5*\x1  -0.866*\y1, 0.866*\x1 -0.5*\y1);
%rotate about 120
\path let \p1=(t12) in coordinate (t31) at (-0.5*\x1  -0.866*\y1, 0.866*\x1 -0.5*\y1);

\draw [dashed] (t12)--(t13);
\draw [dashed] (t21)--(t23);
\draw [dashed] (t31)--(t32);

%intersection points
\filldraw [] (t12) circle (0.8pt);
\filldraw [] (t21) circle (0.8pt);
\filldraw [] (t13) circle (0.8pt);
\filldraw [] (t31) circle (0.8pt);
\filldraw [] (t23) circle (0.8pt);
\filldraw [] (t32) circle (0.8pt);

\filldraw [] (T1) circle (0.8pt);
\filldraw [] (T2) circle (0.8pt);
\filldraw [] (T3) circle (0.8pt);
\filldraw [] (Q1) circle (0.8pt);
\filldraw [] (Q2) circle (0.8pt);
\filldraw [] (Q3) circle (0.8pt);

% X - points : vertices of triangle DELTA
\coordinate (X1) at (-2*\v*\radius,0);
%rotate about -120
\path let \p1=(X1) in coordinate (X2) at (-0.5*\x1  +0.866*\y1, -0.866*\x1 -0.5*\y1);
%rotate about -120
\path let \p1=(X2) in coordinate (X3) at (-0.5*\x1  +0.866*\y1, -0.866*\x1 -0.5*\y1);

% triangle DELTA
%\draw [dashed] (X1)--(X2)--(X3)--(X1);

% interior bisectors 
%\def \InScale {1}
%\foreach \j in{1,2,3} 
%{\path let \p1=(T\j) in coordinate (Y\j) at (\InScale*\x1,\InScale*\y1);
% \path let \p2=(Q\j) in coordinate (Z\j) at (\InScale*\x2,\InScale*\y2);
% \draw [very thick]  (Y\j)--(Z\j);}

%exterior bisectors
\def \OutScale  {1}
\path let \p1=(X1) in coordinate (Ex23Up) at (\x1,{sqrt(1-4*\v^2)});
\path let \p1=(X1) in coordinate (Ex23Down) at (\x1,-{sqrt(1-4*\v^2)});

% rotate
\path let \p1=(Ex23Up) in coordinate (Ex13Up) at (-0.5*\x1  +0.866*\y1, -0.866*\x1 -0.5*\y1);
\path let \p1=(Ex23Down) in coordinate (Ex13Down) at (-0.5*\x1  +0.866*\y1, -0.866*\x1 -0.5*\y1);

\path let \p1=(Ex23Up) in coordinate (Ex12Up) at (-0.5*\x1  -0.866*\y1, 0.866*\x1 -0.5*\y1);
\path let \p1=(Ex23Down) in coordinate (Ex12Down) at (-0.5*\x1  -0.866*\y1, 0.866*\x1 -0.5*\y1);

\draw [] (Ex23Up)--(Ex23Down);
\draw [] (Ex13Up)--(Ex13Down);
\draw [] (Ex12Up)--(Ex12Down);

\def \scalingA {1}
\def \scalingB {1}
\coordinate (q1) at ($(T1)!\scalingA!(Q1)$);
\coordinate (q2) at ($(T2)!\scalingA!(Q2)$);
\coordinate (q3) at ($(T3)!\scalingA!(Q3)$);

\coordinate (t1) at ($(Q1)!\scalingB!(T1)$);
\coordinate (t2) at ($(Q2)!\scalingB!(T2)$);
\coordinate (t3) at ($(Q3)!\scalingB!(T3)$);

\filldraw [] (Ex13Up) circle (0.8pt);
\filldraw [] (Ex13Down) circle (0.8pt);
\filldraw [] (Ex23Up) circle (0.8pt);
\filldraw [] (Ex23Down) circle (0.8pt);
\filldraw [] (Ex12Up) circle (0.8pt);
\filldraw [] (Ex12Down) circle (0.8pt);

\draw []  (t1) -- (q1);
\draw []  (t2) -- (q2);
\draw []  (t3) -- (q3);

\foreach \s in {1,2}
{ \def \z {\s+1}
	\path let \p1=(X\s) in coordinate (X\z) at (-0.5*\x1  +0.866*\y1, -0.866*\x1 -0.5*\y1);}

%overlap
\draw [line width=1.6pt,domain={acos(\v)}:120-acos(\v)] plot ({cos(\x)}, {sin(\x)});
\draw [line width=1.6pt,domain={-acos(\v)}:-120+acos(\v)] plot ({cos(\x)}, {sin(\x)});
\draw [line width=1.6pt,domain={180}:180 + acos(\v)-60] plot ({cos(\x)}, {sin(\x)});
\draw [line width=1.6pt,domain={180}:180 +60 - acos(\v)] plot ({cos(\x)}, {sin(\x)});
						\end{tikzpicture} \label{figure_geometric_notion:subcritical}}
					\shortquad \qquad
					\subfloat[$v=1/2$: The triangle $\Delta(v)$ is inscribed to $\KC$. The exterior bisectors are tangent to $\KC$ at the Taub points.]{
						\begin{tikzpicture}[scale=1.25]
							\clip (-1.5,-1.6) rectangle (1.6,1.6);
%\clip (0,0) circle (1cm);
%\clip (-0.1,-1.8) rectangle (1.8,1.8);

%v = 0.36
\def \v {0.5}
\def \radius {1}

% Q points and Taub points
\foreach \s in {1,2,3}
{\coordinate (Q\s) at ({-120 * (\s - 1)}:\radius);}
\foreach \s in {1,2,3}
{\path let \p\s=(Q\s) in coordinate (T\s) at (-1*\x\s,-1*\y\s);}

% Kasner circle
\draw [semithick] (Q1) arc (0:360:\radius cm and \radius cm);	 	

% caption circle
%\node at (0.9*\radius, \radius*0.7) [anchor = west] {$\KC$};

% caption Q1 and T1
\path let \p1=(Q1) in coordinate (capQ1) at (1.15*\x1,\y1);
\path let \p1=(T1) in coordinate (capT1) at (1.15*\x1,\y1);

%\filldraw  (capT1) circle (0pt) node [anchor=south]{$\T_1$};
%\filldraw  (T1) circle (0.5pt);
%\filldraw  (T1) circle (0.7pt); % node [anchor=east]{\scriptsize $\T_1$};
%\filldraw  (T2) circle (0.7pt); % node [anchor=south]{\scriptsize$\,\,\T_2$};
%\filldraw  (T3) circle (0.7pt); % node [anchor=north]{\scriptsize$\,\,\T_3$};
%
%%\filldraw  (capQ1) circle (0pt) node [anchor=south]{$\Q_1$};
%%\filldraw  (Q1) circle (0.5pt);
%\filldraw  (Q1) circle (0.7pt); % node [anchor=west]{\scriptsize$\Q_1$};
%\filldraw  (Q2) circle (0.7pt); % node [anchor=north]{\scriptsize$\Q_2\,\,$};
%\filldraw  (Q3) circle (0.7pt); % node [anchor=south]{\scriptsize$\Q_3\,\,$};

%coordinate system
%\draw [->] (0,0)--(0.55,0);
%\draw [->] (0,0)--(0,0.55);

%\draw  (0.6,-0.05) circle (0pt) node [anchor=south]{\scriptsize$\Sigma_+$};
%\draw  (0,0.5) circle (0pt) node [anchor=west]{\scriptsize$\Sigma_-$};

% tangent points
\coordinate (t13) at (\v,{sqrt(1-\v^2)});
\coordinate (t12) at (\v,-{sqrt(1-\v^2)});

%rotate about -120
\path let \p1=(t13) in coordinate (t21) at (-0.5*\x1  +0.866*\y1, -0.866*\x1 -0.5*\y1);
%rotate about -120
\path let \p1=(t12) in coordinate (t23) at (-0.5*\x1  +0.866*\y1, -0.866*\x1 -0.5*\y1);

%rotate about 120
\path let \p1=(t13) in coordinate (t32) at (-0.5*\x1  -0.866*\y1, 0.866*\x1 -0.5*\y1);
%rotate about 120
\path let \p1=(t12) in coordinate (t31) at (-0.5*\x1  -0.866*\y1, 0.866*\x1 -0.5*\y1);

%\draw (t12)--(t13);
%\draw (t21)--(t23);
%\draw (t31)--(t32);

% X - points : vertices of triangle DELTA
\coordinate (X1) at (-2*\v*\radius,0);
%rotate about -120
\path let \p1=(X1) in coordinate (X2) at (-0.5*\x1  +0.866*\y1, -0.866*\x1 -0.5*\y1);
%rotate about -120
\path let \p1=(X2) in coordinate (X3) at (-0.5*\x1  +0.866*\y1, -0.866*\x1 -0.5*\y1);

% triangle DELTA
\draw [dashed] (X1)--(X2)--(X3)--(X1);

%intersection points
\filldraw [] (t12) circle (0.8pt);
\filldraw [] (t21) circle (0.8pt);
\filldraw [] (t13) circle (0.8pt);
\filldraw [] (t31) circle (0.8pt);
\filldraw [] (t23) circle (0.8pt);
\filldraw [] (t32) circle (0.8pt);

\filldraw [] (T1) circle (0.8pt);
\filldraw [] (T2) circle (0.8pt);
\filldraw [] (T3) circle (0.8pt);
\filldraw [] (Q1) circle (0.8pt);
\filldraw [] (Q2) circle (0.8pt);
\filldraw [] (Q3) circle (0.8pt);

% interior bisectors 
\def \InScale {1}
\foreach \j in{1,2,3} 
{\path let \p1=(T\j) in coordinate (Y\j) at (\InScale*\x1,\InScale*\y1);
	\path let \p2=(Q\j) in coordinate (Z\j) at (\InScale*\x2,\InScale*\y2);
	\draw (Y\j)--(Z\j);}

%exterior bisectors
\def \OutScale  {0.7}
\path let \p1=(X1) in coordinate (Ex23Up) at (\x1,\OutScale);
\path let \p1=(X1) in coordinate (Ex23Down) at (\x1,-\OutScale);

% rotate
\path let \p1=(Ex23Up) in coordinate (Ex13Up) at (-0.5*\x1  +0.866*\y1, -0.866*\x1 -0.5*\y1);
\path let \p1=(Ex23Down) in coordinate (Ex13Down) at (-0.5*\x1  +0.866*\y1, -0.866*\x1 -0.5*\y1);

\path let \p1=(Ex23Up) in coordinate (Ex12Up) at (-0.5*\x1  -0.866*\y1, 0.866*\x1 -0.5*\y1);
\path let \p1=(Ex23Down) in coordinate (Ex12Down) at (-0.5*\x1  -0.866*\y1, 0.866*\x1 -0.5*\y1);

\draw (Ex23Up)--(Ex23Down);
\draw (Ex13Up)--(Ex13Down);
\draw (Ex12Up)--(Ex12Down);

\def \scaling {1}
\coordinate (q1) at ($(T1)!\scaling!(Q1)$);
\coordinate (q2) at ($(T2)!\scaling!(Q2)$);
\coordinate (q3) at ($(T3)!\scaling!(Q3)$);

\coordinate (t1) at ($(Q1)!\scaling!(T1)$);
\coordinate (t2) at ($(Q2)!\scaling!(T2)$);
\coordinate (t3) at ($(Q3)!\scaling!(T3)$);

\draw (t1) -- (q1); % node [anchor=south] {\small $\mathbf{b}^{in}_{23}$};
\draw (q2) -- (t2); % node [anchor=south] {\small $\mathbf{b}^{in}_{13}$};
\draw (t3) -- (q3); % node [anchor=south] {\small $\mathbf{b}^{in}_{12}$};

\foreach \s in {1,2}
{ \def \z {\s+1}
	\path let \p1=(X\s) in coordinate (X\z) at (-0.5*\x1  +0.866*\y1, -0.866*\x1 -0.5*\y1);}
						\end{tikzpicture} \label{figure_geometric_notion:critical}}
					\qquad \shortquad
					\subfloat[$v\in (1/2,1)$: The vertices of $\Delta(v)$ are outside of $\KC$. The exterior bisectors do not intersect $\KC$.]{
						\begin{tikzpicture}[scale=1.25]
							\clip (-1.6,-1.6) rectangle (1.6,1.6);
%\clip (-0.1,-1.8) rectangle (1.8,1.8);

%v = 0.36
\def \v {0.76}
\def \radius {1}

% Q points and Taub points
\foreach \s in {1,2,3}
{\coordinate (Q\s) at ({-120 * (\s - 1)}:\radius);}
\foreach \s in {1,2,3}
{\path let \p\s=(Q\s) in coordinate (T\s) at (-1*\x\s,-1*\y\s);}

% Kasner circle
\draw [very thin] (Q1) arc (0:360:\radius cm and \radius cm);	 	

% caption circle
%\node at (0.9*\radius, \radius*0.7) [anchor = west] {$\KC$};

% caption Q1 and T1
\path let \p1=(Q1) in coordinate (capQ1) at (1.15*\x1,\y1);
\path let \p1=(T1) in coordinate (capT1) at (1.15*\x1,\y1);

%%\filldraw  (capT1) circle (0pt) node [anchor=south]{$\T_1$};
%%\filldraw  (T1) circle (0.5pt);
%\filldraw  (T1) circle (0.7pt); % node [anchor=east]{\scriptsize $\T_1$};
%\filldraw  (T2) circle (0.7pt); % node [anchor=south]{\scriptsize$\,\,\T_2$};
%\filldraw  (T3) circle (0.7pt); % node [anchor=north]{\scriptsize$\,\,\T_3$};
%
%%\filldraw  (capQ1) circle (0pt) node [anchor=south]{$\Q_1$};
%%\filldraw  (Q1) circle (0.5pt);
%\filldraw  (Q1) circle (0.7pt); % node [anchor=west]{\scriptsize$\Q_1$};
%\filldraw  (Q2) circle (0.7pt); % node [anchor=north]{\scriptsize$\Q_2\,\,$};
%\filldraw  (Q3) circle (0.7pt); % node [anchor=south]{\scriptsize$\Q_3\,\,$};

%coordinate system
%\draw [->] (0,0)--(0.55,0);
%\draw [->] (0,0)--(0,0.55);

%\draw  (0.6,-0.05) circle (0pt) node [anchor=south]{\scriptsize$\Sigma_+$};
%\draw  (0,0.5) circle (0pt) node [anchor=west]{\scriptsize$\Sigma_-$};

% tangent points
\coordinate (t13) at (\v,{sqrt(1-\v^2)});
\coordinate (t12) at (\v,-{sqrt(1-\v^2)});

%rotate about -120
\path let \p1=(t13) in coordinate (t21) at (-0.5*\x1  +0.866*\y1, -0.866*\x1 -0.5*\y1);
%rotate about -120
\path let \p1=(t12) in coordinate (t23) at (-0.5*\x1  +0.866*\y1, -0.866*\x1 -0.5*\y1);

%rotate about 120
\path let \p1=(t13) in coordinate (t32) at (-0.5*\x1  -0.866*\y1, 0.866*\x1 -0.5*\y1);
%rotate about 120
\path let \p1=(t12) in coordinate (t31) at (-0.5*\x1  -0.866*\y1, 0.866*\x1 -0.5*\y1);

%intersection points
\filldraw [] (t12) circle (0.8pt);
\filldraw [] (t21) circle (0.8pt);
\filldraw [] (t13) circle (0.8pt);
\filldraw [] (t31) circle (0.8pt);
\filldraw [] (t23) circle (0.8pt);
\filldraw [] (t32) circle (0.8pt);

\filldraw [] (T1) circle (0.8pt);
\filldraw [] (T2) circle (0.8pt);
\filldraw [] (T3) circle (0.8pt);
\filldraw [] (Q1) circle (0.8pt);
\filldraw [] (Q2) circle (0.8pt);
\filldraw [] (Q3) circle (0.8pt);

%arcs 
\draw [ semithick, name path=A1] (t12) arc (-{acos(\v)}:{acos(\v)}:1);
\draw [ semithick, name path=A2] (t23) arc (-{acos(\v) -120 }:{acos(\v) -120}:1);
\draw [ semithick, name path=A3] (t31) arc (-{acos(\v) -240 }:{acos(\v) -240}:1);

%\draw (t12)--(t13);
%\draw (t21)--(t23);
%\draw (t31)--(t32);

% X - points : vertices of triangle DELTA
\coordinate (X1) at (-2*\v*\radius,0);
%rotate about -120
\path let \p1=(X1) in coordinate (X2) at (-0.5*\x1  +0.866*\y1, -0.866*\x1 -0.5*\y1);
%rotate about -120
\path let \p1=(X2) in coordinate (X3) at (-0.5*\x1  +0.866*\y1, -0.866*\x1 -0.5*\y1);

% triangle DELTA
\draw [dashed] (X1)--(X2)--(X3)--(X1);

% interior bisectors 
\def \InScale {1}
\foreach \j in{1,2,3} 
{\path let \p1=(T\j) in coordinate (Y\j) at (\InScale*\x1,\InScale*\y1);
	\path let \p2=(Q\j) in coordinate (Z\j) at (\InScale*\x2,\InScale*\y2);
	\draw (Y\j)--(Z\j);}

%exterior bisectors
\def \OutScale  {0.7}
\path let \p1=(X1) in coordinate (Ex23Up) at (\x1,\OutScale);
\path let \p1=(X1) in coordinate (Ex23Down) at (\x1,-\OutScale);

% rotate
\path let \p1=(Ex23Up) in coordinate (Ex13Up) at (-0.5*\x1  +0.866*\y1, -0.866*\x1 -0.5*\y1);
\path let \p1=(Ex23Down) in coordinate (Ex13Down) at (-0.5*\x1  +0.866*\y1, -0.866*\x1 -0.5*\y1);

\path let \p1=(Ex23Up) in coordinate (Ex12Up) at (-0.5*\x1  -0.866*\y1, 0.866*\x1 -0.5*\y1);
\path let \p1=(Ex23Down) in coordinate (Ex12Down) at (-0.5*\x1  -0.866*\y1, 0.866*\x1 -0.5*\y1);

\draw (Ex23Up)--(Ex23Down);
\draw (Ex13Up)--(Ex13Down);
\draw (Ex12Up)--(Ex12Down);

\def \scaling {1.16}
\def \scalingA {1}
\coordinate (q1) at ($(T1)!\scalingA!(Q1)$);
\coordinate (q2) at ($(T2)!\scalingA!(Q2)$);
\coordinate (q3) at ($(T3)!\scalingA!(Q3)$);

\coordinate (t1) at ($(Q1)!\scaling!(T1)$);
\coordinate (t2) at ($(Q2)!\scaling!(T2)$);
\coordinate (t3) at ($(Q3)!\scaling!(T3)$);

\draw (t1) -- (q1);
\draw (t2) -- (q2);
\draw (t3) -- (q3);

\foreach \s in {1,2}
{ \def \z {\s+1}
	\path let \p1=(X\s) in coordinate (X\z) at (-0.5*\x1  +0.866*\y1, -0.866*\x1 -0.5*\y1);}
						\end{tikzpicture} \label{figure_geometric_notion:supercritical}}
					\caption{The dashed lines are $L_\alpha(v)$, for $\alpha=1,2,3$, and the solid lines are the corresponding angle bisectors, as in \autoref{figure_intro_geometric_notion:subs}. According to Lemma~\ref{lem_eigenvalues_distance}, intersections of $\KC$ with these lines split $\KC$ in sectors, where the ordering of the eigenvalues is unchanged. The ordering is determined in Lemmata~\ref{cor_supercritical_eigenvalues}\, -~\ref{cor_subcritical_eigenvalues}. %, where we name the sectors.
						%	The Taub and the $\Q$-points are without caption
					}\label{figure_geometric_notion:subs}
				\end{figure}
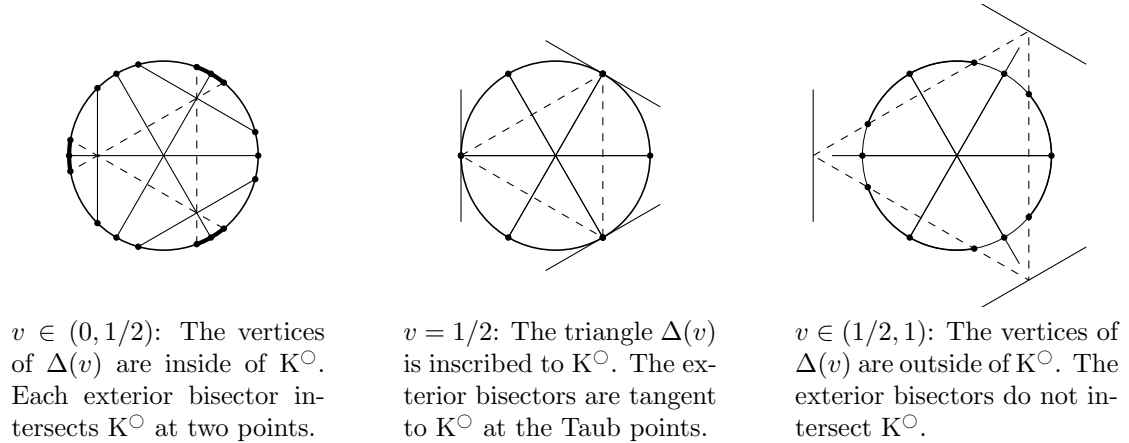
					The three interior bisectors are independent of $v$ by \eqref{inbisector}, but $\A_\alpha(v)$ depends on $v$. Initially, when $v=\frac 12$, the boundary points of $\A_\alpha(v)$ coincide with the two Taub points $\T_{\beta},\T_{\gamma}$. With increasing $v$ the boundary points monotonically move towards $\Q_\alpha$. Thus, for all $\frac 12\le v<1$, the only interior bisector intersecting $\A_\alpha(v)$ is $\mathbf{b}_{\beta \gamma}^{in}(v)$, in the point $\Q_{\alpha}$.
					
					Thus, to determine the precise ordering, it suffices to consider a single value $v \approx 1$, see \autoref{figure_geometric_notion:supercritical}. Moreover, we can choose $p \in \A_1^+(v)$. The lower half $\A_1^-(v)$ as well as the other arcs yield the same structure with roles of the eigenvalues interchanged and are obtained by $D_3$-equivariance. Clearly then $p$ is closer to $L_1(v)$ than it is to both $L_2(v)$ and $L_3(v)$. Furthermore, since $\mathbf{b}_{2 3}^{in}(v)$ is the $\Sigma_+$-axis, $p$ is closer to $L_3(v)$ than it is to $L_2(v)$, due to reflectional symmetry of $L_2(v)$ and $L_3(v)$. The distance is the same if and only if $p=\Q_1$.
				\end{proof}
				%Each two arcs $\A_\alpha(v)$ and $\A_\beta(v)$ overlap, i.e. the intersection point of the corresponding lines $L_\alpha(v)$ and $L_\beta(v)$ is inside the circle. Consequently, each exterior bisector intersects the Kasner circle in two points.
				%This also means that the Taub points have a two-dimensional unstable manifold with two equal eigenvalues.
                %
				For $v<\frac12$, however, the situation is more involved. Let $\A_{\alpha\beta}(v) = \A_\alpha(v) \cap \A_\beta(v)$ denote the overlap of two unstable arcs.
				\begin{lem}[Subcritical case]\label{cor_subcritical_eigenvalues}
					Let $v \in \left(0,\frac 12\right)$ and $(\alpha\beta\gamma)$ a permutation of $(123)$.
					%	Each point $p \in \KC$ has one trivial eigenvalue zero to the eigenvector tangential to $\KC$(boundary points).
					Each half-arc between the points $\talphabeta(v)$ and $\Q_\alpha$ is divided into disjoint open sectors where the ordering of the eigenvalues does not change. More precisely,
					%	 there are three qualitatively different cases with $v=\frac 14$ as a point of bifurcation. More precisely.
					\begin{itemize}
						\item [(i)] The overlap $\A_{\alpha\beta}(v)$ is split in two symmetric sectors where \begin{equation}\label{lemma_sub_overlap}
							0<\evi<\evj<-\evk \quad \text{or} \quad 0<\evj<\evi<-\evk.
						\end{equation}
						\item [(ii)] There is a sector $\Mid$ where \begin{equation}\label{lemma_sub_mid}
							0<-\evj<\evi<-\evk.
						\end{equation}
						\item [(iii)] \begin{itemize}
							\item [(a)] For all $v \in \left(\frac 14, \frac 12\right)$ there is a sector $\Stab$ where \begin{equation}\label{lemma_sub_stab}
								0<\evi<-\evj<-\evk.
							\end{equation}
							\item [(b)] For all $v \in \left(0,\frac 14\right)$ there is a sector $\Unstab$ where \begin{equation}\label{lemma_sub_unstab}
								0<-\evj<-\evk<\evi.
							\end{equation}
						\end{itemize}
					\end{itemize}
				\end{lem}
				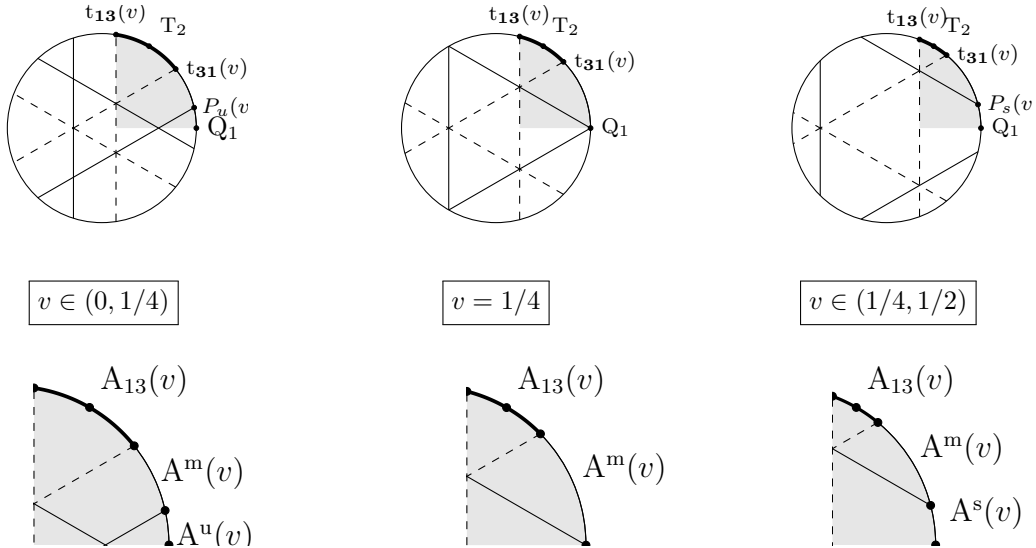
\begin{figure}[H]
					\captionsetup[subfigure]{labelformat=empty}
					\centering
					\subfloat[\boxed{\text{$v\in (0,1/4)$}}]{
						\begin{tikzpicture}[scale=1.25]
							\definecolor{light-gray}{gray}{0.9}
\clip (-1.5,-1.5) rectangle (1.54,1.3);
%\clip (-0.1,-1.8) rectangle (1.8,1.8);
%\clip (0,0) circle (1cm);

%v = 0.36
\def \v {0.15}
\def \radius {1}

% Q points and Taub points
\foreach \s in {1,2,3}
{\coordinate (Q\s) at ({-120 * (\s - 1)}:\radius);}
\foreach \s in {1,2,3}
{\path let \p\s=(Q\s) in coordinate (T\s) at (-1*\x\s,-1*\y\s);}

\path [fill = light-gray] (\v,0) -- (1,0) arc (0:{acos(\v)}:1) -- cycle;

% Kasner circle
\draw [] (Q1) arc (0:360:\radius cm and \radius cm);	 	
\draw [name path=A1] (Q1) arc (0:{acos(\v)}:1);	 	
% caption circle
%\node at (0.9*\radius, \radius*0.7) [anchor = west] {$\KC$};

% caption Q1 and T1
\path let \p1=(Q1) in coordinate (capQ1) at (1.15*\x1,\y1);
\path let \p1=(T1) in coordinate (capT1) at (1.15*\x1,\y1);

%%\filldraw  (capT1) circle (0pt) node [anchor=south]{$\T_1$};
%%\filldraw  (T1) circle (0.5pt);
%\filldraw  (T1) circle (0.7pt); % node [anchor=east]{\scriptsize $\T_1$};
\filldraw  (T2) circle (0.7pt) node [anchor=south west]{\scriptsize $\T_2$};
% [anchor=south]{\scriptsize$\,\,\T_2$};
%\filldraw  (T3) circle (0.7pt); % node [anchor=north]{\scriptsize$\,\,\T_3$};
%
%%\filldraw  (capQ1) circle (0pt) node [anchor=south]{$\Q_1$};
%%\filldraw  (Q1) circle (0.5pt);
\filldraw  (Q1) circle (0.7pt) node [anchor=west]{\footnotesize$\Q_1$};
%\filldraw  (Q2) circle (0.7pt); % node [anchor=north]{\scriptsize$\Q_2\,\,$};
%\filldraw  (Q3) circle (0.7pt); % node [anchor=south]{\scriptsize$\Q_3\,\,$};

%coordinate system
%\draw [->] (0,0)--(0.55,0);
%\draw [->] (0,0)--(0,0.55);

%\draw  (0.6,-0.05) circle (0pt) node [anchor=south]{\scriptsize$\Sigma_+$};
%\draw  (0,0.5) circle (0pt) node [anchor=west]{\scriptsize$\Sigma_-$};

% tangent points
\coordinate (t13) at (\v,{sqrt(1-\v^2)});
\coordinate (t12) at (\v,-{sqrt(1-\v^2)});

%rotate about -120
\path let \p1=(t13) in coordinate (t21) at (-0.5*\x1  +0.866*\y1, -0.866*\x1 -0.5*\y1);
%rotate about -120
\path let \p1=(t12) in coordinate (t23) at (-0.5*\x1  +0.866*\y1, -0.866*\x1 -0.5*\y1);

%rotate about 120
\path let \p1=(t13) in coordinate (t32) at (-0.5*\x1  -0.866*\y1, 0.866*\x1 -0.5*\y1);
%rotate about 120
\path let \p1=(t12) in coordinate (t31) at (-0.5*\x1  -0.866*\y1, 0.866*\x1 -0.5*\y1);

\draw [dashed] (t12)--(t13);
\draw [dashed] (t21)--(t23);
\draw [dashed] (t31)--(t32);

% X - points : vertices of triangle DELTA
\coordinate (X1) at (-2*\v*\radius,0);
%rotate about -120
\path let \p1=(X1) in coordinate (X2) at (-0.5*\x1  +0.866*\y1, -0.866*\x1 -0.5*\y1);
%rotate about -120
\path let \p1=(X2) in coordinate (X3) at (-0.5*\x1  +0.866*\y1, -0.866*\x1 -0.5*\y1);

% triangle DELTA
%\draw [line width=1pt] (X1)--(X2)--(X3)--(X1);

% interior bisectors 
\def \InScale {1}
\foreach \j in{1,2,3} 
{\path let \p1=(T\j) in coordinate (Y\j) at (\InScale*\x1,\InScale*\y1);
	\path let \p2=(Q\j) in coordinate (Z\j) at (\InScale*\x2,\InScale*\y2);}
%	\draw (Y\j)--(Z\j);}

%exterior bisectors
\def \OutScale  {1}
\path let \p1=(X1) in coordinate (Ex23Up) at (\x1,{sqrt(1-4*\v^2)});
\path let \p1=(X1) in coordinate (Ex23Down) at (\x1,-{sqrt(1-4*\v^2)});

% rotate
\path let \p1=(Ex23Up) in coordinate (Ex13Up) at (-0.5*\x1  +0.866*\y1, -0.866*\x1 -0.5*\y1);
\path let \p1=(Ex23Down) in coordinate (Ex13Down) at (-0.5*\x1  +0.866*\y1, -0.866*\x1 -0.5*\y1);

\path let \p1=(Ex23Up) in coordinate (Ex12Up) at (-0.5*\x1  -0.866*\y1, 0.866*\x1 -0.5*\y1);
\path let \p1=(Ex23Down) in coordinate (Ex12Down) at (-0.5*\x1  -0.866*\y1, 0.866*\x1 -0.5*\y1);

\draw (Ex23Up)--(Ex23Down);
\draw [name path = line3] (Ex13Up)--(Ex13Down);
\draw [name path = line2] (Ex12Up)--(Ex12Down);

\def \scaling {1}
\coordinate (q1) at ($(T1)!\scaling!(Q1)$);
\coordinate (q2) at ($(T2)!\scaling!(Q2)$);
\coordinate (q3) at ($(T3)!\scaling!(Q3)$);

\coordinate (t1) at ($(Q1)!\scaling!(T1)$);
\coordinate (t2) at ($(Q2)!\scaling!(T2)$);
\coordinate (t3) at ($(Q3)!\scaling!(T3)$);

%\draw (t1) -- (q1);
%\draw (t2) -- (q2);
%\draw (t3) -- (q3);

\foreach \s in {1,2}
{ \def \z {\s+1}
	\path let \p1=(X\s) in coordinate (X\z) at (-0.5*\x1  +0.866*\y1, -0.866*\x1 -0.5*\y1);}

\path [name intersections={of=line2 and A1 ,by=P2}];
\path [name intersections={of=line3 and A1 ,by=P3}];

%\filldraw  (P2) circle (0.8pt);

%
\filldraw  (P2) circle (0.7pt) node [anchor=west]{\scriptsize$ \hspace{-0.04cm}P_u(v)$};
%\draw  (P2) circle (0pt) node [anchor=north west]{\scriptsize$\Unstab$};
%\draw  (t31) circle (0pt) node [anchor=north west]{\scriptsize$\,\,\,\Mid$};

\filldraw  (t13) circle (0.7pt) node [anchor=south]{\scriptsize $\tac(v)$};
\filldraw  (t31) circle (0.7pt) node [anchor=west]{\scriptsize $\tca(v)$};
%overlap
\draw [line width=1.4pt,domain={acos(\v)}:120-acos(\v)] plot ({cos(\x)}, {sin(\x)});
						\end{tikzpicture} \label{figure_geometric_notion:subcritical1}}
					\hspace{1cm}
					\subfloat[\boxed{\text{$v= 1/4$}} ]{
						\begin{tikzpicture}[scale=1.25]
							\definecolor{light-gray}{gray}{0.9}
\clip (-1.5,-1.5) rectangle (1.5,1.3);
%\clip (-0.1,-1.8) rectangle (1.8,1.8);
%\clip (0,0) circle (1cm);

%v = 0.36
\def \v {0.25}
\def \radius {1}

% Q points and Taub points
\foreach \s in {1,2,3}
{\coordinate (Q\s) at ({-120 * (\s - 1)}:\radius);}
\foreach \s in {1,2,3}
{\path let \p\s=(Q\s) in coordinate (T\s) at (-1*\x\s,-1*\y\s);}

\path [fill = light-gray] (\v,0) -- (1,0) arc (0:{acos(\v)}:1) -- cycle; 

% Kasner circle
\draw [] (Q1) arc (0:360:\radius cm and \radius cm);	 	
\draw [name path=A1] (Q1) arc (0:{acos(\v)}:1);	 	

% caption circle
%\node at (0.9*\radius, \radius*0.7) [anchor = west] {$\KC$};

% caption Q1 and T1
\path let \p1=(Q1) in coordinate (capQ1) at (1.15*\x1,\y1);
\path let \p1=(T1) in coordinate (capT1) at (1.15*\x1,\y1);

%%\filldraw  (capT1) circle (0pt) node [anchor=south]{$\T_1$};
%%\filldraw  (T1) circle (0.5pt);
%\filldraw  (T1) circle (0.7pt); % node [anchor=east]{\scriptsize $\T_1$};
\filldraw  (T2) circle (0.7pt) node [anchor=south]{\scriptsize$\qquad \T_2$};
%\filldraw  (T3) circle (0.7pt); % node [anchor=north]{\scriptsize$\,\,\T_3$};
%
%%\filldraw  (capQ1) circle (0pt) node [anchor=south]{$\Q_1$};
%%\filldraw  (Q1) circle (0.5pt);
\filldraw  (Q1) circle (0.7pt) node [anchor=west]{\scriptsize$\Q_1$};
%\filldraw  (Q2) circle (0.7pt); % node [anchor=north]{\scriptsize$\Q_2\,\,$};
%\filldraw  (Q3) circle (0.7pt); % node [anchor=south]{\scriptsize$\Q_3\,\,$};

%coordinate system
%\draw [->] (0,0)--(0.55,0);
%\draw [->] (0,0)--(0,0.55);

%\draw  (0.6,-0.05) circle (0pt) node [anchor=south]{\scriptsize$\Sigma_+$};
%\draw  (0,0.5) circle (0pt) node [anchor=west]{\scriptsize$\Sigma_-$};

% tangent points
\coordinate (t13) at (\v,{sqrt(1-\v^2)});
\coordinate (t12) at (\v,-{sqrt(1-\v^2)});

%rotate about -120
\path let \p1=(t13) in coordinate (t21) at (-0.5*\x1  +0.866*\y1, -0.866*\x1 -0.5*\y1);
%rotate about -120
\path let \p1=(t12) in coordinate (t23) at (-0.5*\x1  +0.866*\y1, -0.866*\x1 -0.5*\y1);

%rotate about 120
\path let \p1=(t13) in coordinate (t32) at (-0.5*\x1  -0.866*\y1, 0.866*\x1 -0.5*\y1);
%rotate about 120
\path let \p1=(t12) in coordinate (t31) at (-0.5*\x1  -0.866*\y1, 0.866*\x1 -0.5*\y1);

\draw [dashed] (t12)--(t13);
\draw [dashed] (t21)--(t23);
\draw [dashed] (t31)--(t32);

% X - points : vertices of triangle DELTA
\coordinate (X1) at (-2*\v*\radius,0);
%rotate about -120
\path let \p1=(X1) in coordinate (X2) at (-0.5*\x1  +0.866*\y1, -0.866*\x1 -0.5*\y1);
%rotate about -120
\path let \p1=(X2) in coordinate (X3) at (-0.5*\x1  +0.866*\y1, -0.866*\x1 -0.5*\y1);

% triangle DELTA
%\draw [line width=1pt] (X1)--(X2)--(X3)--(X1);

% interior bisectors 
\def \InScale {1}
\foreach \j in{1,2,3} 
{\path let \p1=(T\j) in coordinate (Y\j) at (\InScale*\x1,\InScale*\y1);
	\path let \p2=(Q\j) in coordinate (Z\j) at (\InScale*\x2,\InScale*\y2);}
%	\draw (Y\j)--(Z\j);}

%exterior bisectors
\def \OutScale  {1}
\path let \p1=(X1) in coordinate (Ex23Up) at (\x1,{sqrt(1-4*\v^2)});
\path let \p1=(X1) in coordinate (Ex23Down) at (\x1,-{sqrt(1-4*\v^2)});

% rotate
\path let \p1=(Ex23Up) in coordinate (Ex13Up) at (-0.5*\x1  +0.866*\y1, -0.866*\x1 -0.5*\y1);
\path let \p1=(Ex23Down) in coordinate (Ex13Down) at (-0.5*\x1  +0.866*\y1, -0.866*\x1 -0.5*\y1);

\path let \p1=(Ex23Up) in coordinate (Ex12Up) at (-0.5*\x1  -0.866*\y1, 0.866*\x1 -0.5*\y1);
\path let \p1=(Ex23Down) in coordinate (Ex12Down) at (-0.5*\x1  -0.866*\y1, 0.866*\x1 -0.5*\y1);

\draw (Ex23Up)--(Ex23Down);
\draw (Ex13Up)--(Ex13Down);
\draw (Ex12Up)--(Ex12Down);

\def \scaling {1}
\coordinate (q1) at ($(T1)!\scaling!(Q1)$);
\coordinate (q2) at ($(T2)!\scaling!(Q2)$);
\coordinate (q3) at ($(T3)!\scaling!(Q3)$);

\coordinate (t1) at ($(Q1)!\scaling!(T1)$);
\coordinate (t2) at ($(Q2)!\scaling!(T2)$);
\coordinate (t3) at ($(Q3)!\scaling!(T3)$);

%\draw (t1) -- (q1);
%\draw (t2) -- (q2);
%\draw (t3) -- (q3);

\foreach \s in {1,2}
{ \def \z {\s+1}
	\path let \p1=(X\s) in coordinate (X\z) at (-0.5*\x1  +0.866*\y1, -0.866*\x1 -0.5*\y1);}

%\draw  (t31) circle (0pt) node [anchor=north west]{\scriptsize$\,\,\,\Mid$}; 

\filldraw  (t13) circle (0.7pt) node [anchor=south]{\scriptsize $\tac(v)$};
\filldraw  (t31) circle (0.7pt) node [anchor=west]{\scriptsize $\tca(v)$};
%overlap
\draw [line width=1.4pt,domain={acos(\v)}:120-acos(\v)] plot ({cos(\x)}, {sin(\x)});
						\end{tikzpicture} \label{figure_geometric_notion:subcritical2}}
					\hspace{1cm}
					\subfloat[\boxed{\text{$v\in (1/4,1/2)$}}]{
						\begin{tikzpicture}[scale=1.25]
							\definecolor{light-gray}{gray}{0.9}
\clip (-1.5,-1.5) rectangle (1.54,1.3);
%\clip (-0.1,-1.8) rectangle (1.8,1.8);
%\clip (0,0) circle (1cm);

%v = 0.36
\def \v {0.35}
\def \radius {1}

% Q points and Taub points
\foreach \s in {1,2,3}
{\coordinate (Q\s) at ({-120 * (\s - 1)}:\radius);}
\foreach \s in {1,2,3}
{\path let \p\s=(Q\s) in coordinate (T\s) at (-1*\x\s,-1*\y\s);}

\path [fill = light-gray] (\v,0) -- (1,0) arc (0:{acos(\v)}:1) -- cycle; 

% Kasner circle
\draw [] (Q1) arc (0:360:\radius cm and \radius cm);	 	
\draw [name path=A1] (Q1) arc (0:{acos(\v)}:1);	 	

% caption circle
%\node at (0.9*\radius, \radius*0.7) [anchor = west] {$\KC$};

% caption Q1 and T1
\path let \p1=(Q1) in coordinate (capQ1) at (1.15*\x1,\y1);
\path let \p1=(T1) in coordinate (capT1) at (1.15*\x1,\y1);

%%\filldraw  (capT1) circle (0pt) node [anchor=south]{$\T_1$};
%%\filldraw  (T1) circle (0.5pt);
%\filldraw  (T1) circle (0.7pt); % node [anchor=east]{\scriptsize $\T_1$};
\filldraw  (T2) circle (0.7pt) node [anchor=south ]{\scriptsize$\qquad \, \T_2$};
%\filldraw  (T3) circle (0.7pt); % node [anchor=north]{\scriptsize$\,\,\T_3$};
%
%%\filldraw  (capQ1) circle (0pt) node [anchor=south]{$\Q_1$};
%%\filldraw  (Q1) circle (0.5pt);
\filldraw  (Q1) circle (0.7pt) node [anchor=west]{\scriptsize$\Q_1$};
%\filldraw  (Q2) circle (0.7pt); % node [anchor=north]{\scriptsize$\Q_2\,\,$};
%\filldraw  (Q3) circle (0.7pt); % node [anchor=south]{\scriptsize$\Q_3\,\,$};

%coordinate system
%\draw [->] (0,0)--(0.55,0);
%\draw [->] (0,0)--(0,0.55);

%\draw  (0.6,-0.05) circle (0pt) node [anchor=south]{\scriptsize$\Sigma_+$};
%\draw  (0,0.5) circle (0pt) node [anchor=west]{\scriptsize$\Sigma_-$};

% tangent points
\coordinate (t13) at (\v,{sqrt(1-\v^2)});
\coordinate (t12) at (\v,-{sqrt(1-\v^2)});

%rotate about -120
\path let \p1=(t13) in coordinate (t21) at (-0.5*\x1  +0.866*\y1, -0.866*\x1 -0.5*\y1);
%rotate about -120
\path let \p1=(t12) in coordinate (t23) at (-0.5*\x1  +0.866*\y1, -0.866*\x1 -0.5*\y1);

%rotate about 120
\path let \p1=(t13) in coordinate (t32) at (-0.5*\x1  -0.866*\y1, 0.866*\x1 -0.5*\y1);
%rotate about 120
\path let \p1=(t12) in coordinate (t31) at (-0.5*\x1  -0.866*\y1, 0.866*\x1 -0.5*\y1);

\draw [dashed] (t12)--(t13);
\draw [dashed] (t21)--(t23);
\draw [dashed] (t31)--(t32);

% X - points : vertices of triangle DELTA
\coordinate (X1) at (-2*\v*\radius,0);
%rotate about -120
\path let \p1=(X1) in coordinate (X2) at (-0.5*\x1  +0.866*\y1, -0.866*\x1 -0.5*\y1);
%rotate about -120
\path let \p1=(X2) in coordinate (X3) at (-0.5*\x1  +0.866*\y1, -0.866*\x1 -0.5*\y1);

% triangle DELTA
%\draw [line width=1pt] (X1)--(X2)--(X3)--(X1);

% interior bisectors 
\def \InScale {1}
\foreach \j in{1,2,3} 
{\path let \p1=(T\j) in coordinate (Y\j) at (\InScale*\x1,\InScale*\y1);
	\path let \p2=(Q\j) in coordinate (Z\j) at (\InScale*\x2,\InScale*\y2);}
% \draw (Y\j)--(Z\j);}

%exterior bisectors
\def \OutScale  {1}
\path let \p1=(X1) in coordinate (Ex23Up) at (\x1,{sqrt(1-4*\v^2)});
\path let \p1=(X1) in coordinate (Ex23Down) at (\x1,-{sqrt(1-4*\v^2)});

% rotate
\path let \p1=(Ex23Up) in coordinate (Ex13Up) at (-0.5*\x1  +0.866*\y1, -0.866*\x1 -0.5*\y1);
\path let \p1=(Ex23Down) in coordinate (Ex13Down) at (-0.5*\x1  +0.866*\y1, -0.866*\x1 -0.5*\y1);

\path let \p1=(Ex23Up) in coordinate (Ex12Up) at (-0.5*\x1  -0.866*\y1, 0.866*\x1 -0.5*\y1);
\path let \p1=(Ex23Down) in coordinate (Ex12Down) at (-0.5*\x1  -0.866*\y1, 0.866*\x1 -0.5*\y1);

\draw (Ex23Up)--(Ex23Down);
\draw [name path=line3] (Ex13Up)--(Ex13Down);
\draw (Ex12Up)--(Ex12Down);

\def \scaling {1}
\coordinate (q1) at ($(T1)!\scaling!(Q1)$);
\coordinate (q2) at ($(T2)!\scaling!(Q2)$);
\coordinate (q3) at ($(T3)!\scaling!(Q3)$);

\coordinate (t1) at ($(Q1)!\scaling!(T1)$);
\coordinate (t2) at ($(Q2)!\scaling!(T2)$);
\coordinate (t3) at ($(Q3)!\scaling!(T3)$);

%\draw (t1) -- (q1);
%\draw (t2) -- (q2);
%\draw (t3) -- (q3);

\foreach \s in {1,2}
{ \def \z {\s+1}
	\path let \p1=(X\s) in coordinate (X\z) at (-0.5*\x1  +0.866*\y1, -0.866*\x1 -0.5*\y1);}

\path [name intersections={of=line3 and A1 ,by=P3}];

%\filldraw  (P3) circle (0.8pt); 
%\draw  (P3) circle (0pt) node [anchor=north east]{\scriptsize$\, P_3$};
%\path let \p1=(Ex13Up) in coordinate (Ex13UpCap) at (\x1,\y1-0.1cm);
%\draw  (Ex13UpCap) circle (0pt) node [anchor=north west]{\scriptsize $\,\Stab$};
%\draw  (t31) circle (0pt) node [anchor=north west]{\scriptsize$\quad \Mid$};

\filldraw  (t31) circle (0.7pt) node [anchor=west]{\scriptsize $\tca(v)$};
\filldraw  (t13) circle (0.7pt) node [anchor=south]{\scriptsize $\tac(v)\,$};
\filldraw  (P3) circle (0.7pt) node [anchor=west]{\scriptsize$ \hspace{-0.02cm}P_s(v)$};

%overlap
\draw [line width=1.4pt,domain={acos(\v)}:120-acos(\v)] plot ({cos(\x)}, {sin(\x)});
						\end{tikzpicture} \label{figure_geometric_notion:subcritical3}}
					
					\qquad  
					\subfloat[]{
						\begin{tikzpicture}[scale=2.1]
							\definecolor{light-gray}{gray}{0.9}
\clip (0.15,-0.001) rectangle (1.565,1.2);
%\clip (-0.1,-1.8) rectangle (1.8,1.8);
%\clip (0,0) circle (1cm);

%v = 0.36
\def \v {0.15}
\def \radius {1}

% Q points and Taub points
\foreach \s in {1,2,3}
{\coordinate (Q\s) at ({-120 * (\s - 1)}:\radius);}
\foreach \s in {1,2,3}
{\path let \p\s=(Q\s) in coordinate (T\s) at (-1*\x\s,-1*\y\s);}

\path [fill = light-gray] (\v,0) -- (1,0) arc (0:{acos(\v)}:1) -- cycle; 

% Kasner circle
\draw [] (Q1) arc (0:360:\radius cm and \radius cm);	 	
\draw [name path=A1] (Q1) arc (0:{acos(\v)}:1);	 	

% caption circle
%\node at (0.9*\radius, \radius*0.7) [anchor = west] {$\KC$};

% caption Q1 and T1
\path let \p1=(Q1) in coordinate (capQ1) at (1.15*\x1,\y1);
\path let \p1=(T1) in coordinate (capT1) at (1.15*\x1,\y1);

%%\filldraw  (capT1) circle (0pt) node [anchor=south]{$\T_1$};
%%\filldraw  (T1) circle (0.5pt);
%\filldraw  (T1) circle (0.7pt); % node [anchor=east]{\scriptsize $\T_1$};
\filldraw  (T2) circle (0.7pt); % node [anchor=south]{\scriptsize$\,\,\T_2$};
%\filldraw  (T3) circle (0.7pt); % node [anchor=north]{\scriptsize$\,\,\T_3$};
%
%%\filldraw  (capQ1) circle (0pt) node [anchor=south]{$\Q_1$};
%%\filldraw  (Q1) circle (0.5pt);
\filldraw  (Q1) circle (0.7pt); % node [anchor=west]{\scriptsize$\Q_1$};
%\filldraw  (Q2) circle (0.7pt); % node [anchor=north]{\scriptsize$\Q_2\,\,$};
%\filldraw  (Q3) circle (0.7pt); % node [anchor=south]{\scriptsize$\Q_3\,\,$};

%coordinate system
%\draw [->] (0,0)--(0.55,0);
%\draw [->] (0,0)--(0,0.55);

%\draw  (0.6,-0.05) circle (0pt) node [anchor=south]{\scriptsize$\Sigma_+$};
%\draw  (0,0.5) circle (0pt) node [anchor=west]{\scriptsize$\Sigma_-$};

% tangent points
\coordinate (t13) at (\v,{sqrt(1-\v^2)});
\coordinate (t12) at (\v,-{sqrt(1-\v^2)});

%rotate about -120
\path let \p1=(t13) in coordinate (t21) at (-0.5*\x1  +0.866*\y1, -0.866*\x1 -0.5*\y1);
%rotate about -120
\path let \p1=(t12) in coordinate (t23) at (-0.5*\x1  +0.866*\y1, -0.866*\x1 -0.5*\y1);

%rotate about 120
\path let \p1=(t13) in coordinate (t32) at (-0.5*\x1  -0.866*\y1, 0.866*\x1 -0.5*\y1);
%rotate about 120
\path let \p1=(t12) in coordinate (t31) at (-0.5*\x1  -0.866*\y1, 0.866*\x1 -0.5*\y1);

\draw [dashed] (t12)--(t13);
\draw [dashed] (t21)--(t23);
\draw [dashed] (t31)--(t32);

% X - points : vertices of triangle DELTA
\coordinate (X1) at (-2*\v*\radius,0);
%rotate about -120
\path let \p1=(X1) in coordinate (X2) at (-0.5*\x1  +0.866*\y1, -0.866*\x1 -0.5*\y1);
%rotate about -120
\path let \p1=(X2) in coordinate (X3) at (-0.5*\x1  +0.866*\y1, -0.866*\x1 -0.5*\y1);

% triangle DELTA
%\draw [line width=1pt] (X1)--(X2)--(X3)--(X1);

% interior bisectors 
\def \InScale {1}
\foreach \j in{1,2,3} 
{\path let \p1=(T\j) in coordinate (Y\j) at (\InScale*\x1,\InScale*\y1);
	\path let \p2=(Q\j) in coordinate (Z\j) at (\InScale*\x2,\InScale*\y2);}
%	\draw (Y\j)--(Z\j);}

%exterior bisectors
\def \OutScale  {1}
\path let \p1=(X1) in coordinate (Ex23Up) at (\x1,{sqrt(1-4*\v^2)});
\path let \p1=(X1) in coordinate (Ex23Down) at (\x1,-{sqrt(1-4*\v^2)});

% rotate
\path let \p1=(Ex23Up) in coordinate (Ex13Up) at (-0.5*\x1  +0.866*\y1, -0.866*\x1 -0.5*\y1);
\path let \p1=(Ex23Down) in coordinate (Ex13Down) at (-0.5*\x1  +0.866*\y1, -0.866*\x1 -0.5*\y1);

\path let \p1=(Ex23Up) in coordinate (Ex12Up) at (-0.5*\x1  -0.866*\y1, 0.866*\x1 -0.5*\y1);
\path let \p1=(Ex23Down) in coordinate (Ex12Down) at (-0.5*\x1  -0.866*\y1, 0.866*\x1 -0.5*\y1);

\draw [] (Ex23Up)--(Ex23Down);
\draw [name path = line3] (Ex13Up)--(Ex13Down);
\draw [name path = line2] (Ex12Up)--(Ex12Down);

\def \scaling {1}
\coordinate (q1) at ($(T1)!\scaling!(Q1)$);
\coordinate (q2) at ($(T2)!\scaling!(Q2)$);
\coordinate (q3) at ($(T3)!\scaling!(Q3)$);

\coordinate (t1) at ($(Q1)!\scaling!(T1)$);
\coordinate (t2) at ($(Q2)!\scaling!(T2)$);
\coordinate (t3) at ($(Q3)!\scaling!(T3)$);

%\draw (t1) -- (q1);
%\draw (t2) -- (q2);
%\draw (t3) -- (q3);

\foreach \s in {1,2}
{ \def \z {\s+1}
	\path let \p1=(X\s) in coordinate (X\z) at (-0.5*\x1  +0.866*\y1, -0.866*\x1 -0.5*\y1);}

\path [name intersections={of=line2 and A1 ,by=P2}];
\path [name intersections={of=line3 and A1 ,by=P3}];

\path let \p1=(P2) in coordinate (P2Cap) at (\x1,\y1+0.04cm);

%\draw  (P3) circle (0pt) node [anchor=north east]{\small$\, P_3$};
%\draw  (P2Cap) circle (0pt) node [anchor=east]{\small$\, P_u\,$};
\draw  (P2) circle (0pt) node [anchor=north west]{$\Unstab$};
%\draw  (Ex13Up) circle (0pt) node [anchor=south east]{$\Unstab$};
\draw  (t31) circle (0pt) node [anchor=north west]{\shortquad$\Mid$};
\draw  (T2) circle (0pt) node [anchor=south west]{$\A_{13}(v)$};

\filldraw  (P2) circle (0.7pt);
\filldraw  (t13) circle (0.7pt);
\filldraw  (t31) circle (0.7pt);

%overlap
\draw [line width=1.4pt,domain={acos(\v)}:120-acos(\v)] plot ({cos(\x)}, {sin(\x)});
					\end{tikzpicture} }
					\qquad \qquad \qquad
					\subfloat[]{
						\begin{tikzpicture}[scale=2.1]
							\definecolor{light-gray}{gray}{0.9}
\clip (0.25,0) rectangle (1.7,1.2);
%\clip (-0.1,-1.8) rectangle (1.8,1.8);
%\clip (0,0) circle (1cm);

%v = 0.36
\def \v {0.25}
\def \radius {1}

% Q points and Taub points
\foreach \s in {1,2,3}
{\coordinate (Q\s) at ({-120 * (\s - 1)}:\radius);}
\foreach \s in {1,2,3}
{\path let \p\s=(Q\s) in coordinate (T\s) at (-1*\x\s,-1*\y\s);}

\path [fill = light-gray] (\v,0) -- (1,0) arc (0:{acos(\v)}:1) -- cycle;

% Kasner circle
\draw [] (Q1) arc (0:360:\radius cm and \radius cm);	 	
\draw [name path=A1] (Q1) arc (0:{acos(\v)}:1);	 	

% caption circle
%\node at (0.9*\radius, \radius*0.7) [anchor = west] {$\KC$};

% caption Q1 and T1
\path let \p1=(Q1) in coordinate (capQ1) at (1.15*\x1,\y1);
\path let \p1=(T1) in coordinate (capT1) at (1.15*\x1,\y1);

%%\filldraw  (capT1) circle (0pt) node [anchor=south]{$\T_1$};
%%\filldraw  (T1) circle (0.5pt);
%\filldraw  (T1) circle (0.7pt); % node [anchor=east]{\scriptsize $\T_1$};
\filldraw  (T2) circle (0.7pt); % node [anchor=south]{\scriptsize$\,\,\T_2$};
%\filldraw  (T3) circle (0.7pt); % node [anchor=north]{\scriptsize$\,\,\T_3$};
%
%%\filldraw  (capQ1) circle (0pt) node [anchor=south]{$\Q_1$};
%%\filldraw  (Q1) circle (0.5pt);
\filldraw  (Q1) circle (0.7pt); % node [anchor=west]{\scriptsize$\Q_1$};
%\filldraw  (Q2) circle (0.7pt); % node [anchor=north]{\scriptsize$\Q_2\,\,$};
%\filldraw  (Q3) circle (0.7pt); % node [anchor=south]{\scriptsize$\Q_3\,\,$};

%coordinate system
%\draw [->] (0,0)--(0.55,0);
%\draw [->] (0,0)--(0,0.55);

%\draw  (0.6,-0.05) circle (0pt) node [anchor=south]{\scriptsize$\Sigma_+$};
%\draw  (0,0.5) circle (0pt) node [anchor=west]{\scriptsize$\Sigma_-$};

% tangent points
\coordinate (t13) at (\v,{sqrt(1-\v^2)});
\coordinate (t12) at (\v,-{sqrt(1-\v^2)});

%rotate about -120
\path let \p1=(t13) in coordinate (t21) at (-0.5*\x1  +0.866*\y1, -0.866*\x1 -0.5*\y1);
%rotate about -120
\path let \p1=(t12) in coordinate (t23) at (-0.5*\x1  +0.866*\y1, -0.866*\x1 -0.5*\y1);

%rotate about 120
\path let \p1=(t13) in coordinate (t32) at (-0.5*\x1  -0.866*\y1, 0.866*\x1 -0.5*\y1);
%rotate about 120
\path let \p1=(t12) in coordinate (t31) at (-0.5*\x1  -0.866*\y1, 0.866*\x1 -0.5*\y1);

\coordinate (XX) at (\v,0);

\draw [dashed] (XX)--(t13);
\draw [dashed] (t21)--(t23);
\draw [dashed] (t31)--(t32);

% X - points : vertices of triangle DELTA
\coordinate (X1) at (-2*\v*\radius,0);
%rotate about -120
\path let \p1=(X1) in coordinate (X2) at (-0.5*\x1  +0.866*\y1, -0.866*\x1 -0.5*\y1);
%rotate about -120
\path let \p1=(X2) in coordinate (X3) at (-0.5*\x1  +0.866*\y1, -0.866*\x1 -0.5*\y1);

% triangle DELTA
%\draw [line width=1pt] (X1)--(X2)--(X3)--(X1);

% interior bisectors 
\def \InScale {1}
\foreach \j in{1,2,3} 
{\path let \p1=(T\j) in coordinate (Y\j) at (\InScale*\x1,\InScale*\y1);
	\path let \p2=(Q\j) in coordinate (Z\j) at (\InScale*\x2,\InScale*\y2);
%	\draw (Y\j)--(Z\j);
}

%exterior bisectors
\def \OutScale  {1}
\path let \p1=(X1) in coordinate (Ex23Up) at (\x1,{sqrt(1-4*\v^2)});
\path let \p1=(X1) in coordinate (Ex23Down) at (\x1,-{sqrt(1-4*\v^2)});

% rotate
\path let \p1=(Ex23Up) in coordinate (Ex13Up) at (-0.5*\x1  +0.866*\y1, -0.866*\x1 -0.5*\y1);
\path let \p1=(Ex23Down) in coordinate (Ex13Down) at (-0.5*\x1  +0.866*\y1, -0.866*\x1 -0.5*\y1);

\path let \p1=(Ex23Up) in coordinate (Ex12Up) at (-0.5*\x1  -0.866*\y1, 0.866*\x1 -0.5*\y1);
\path let \p1=(Ex23Down) in coordinate (Ex12Down) at (-0.5*\x1  -0.866*\y1, 0.866*\x1 -0.5*\y1);

\draw (Ex23Up)--(Ex23Down);
\draw (Ex13Up)--(Ex13Down);
%\draw (Ex12Up)--(Ex12Down);

\def \scaling {1}
\coordinate (q1) at ($(T1)!\scaling!(Q1)$);
\coordinate (q2) at ($(T2)!\scaling!(Q2)$);
\coordinate (q3) at ($(T3)!\scaling!(Q3)$);

\coordinate (t1) at ($(Q1)!\scaling!(T1)$);
\coordinate (t2) at ($(Q2)!\scaling!(T2)$);
\coordinate (t3) at ($(Q3)!\scaling!(T3)$);

%\draw (0,0) -- (q1);
%\draw (t2) -- (0,0);
%\draw (t3) -- (q3);

\foreach \s in {1,2}
{ \def \z {\s+1}
	\path let \p1=(X\s) in coordinate (X\z) at (-0.5*\x1  +0.866*\y1, -0.866*\x1 -0.5*\y1);}

\path let \p1=(t31) in coordinate (t31Cap) at (\x1,\y1-0.05cm);

%\draw  (Ex12Down) circle (0pt) node [anchor=north east]{$\Stab$};
%\draw  (Ex13Up) circle (0pt) node [anchor=south east]{$\Unstab$};
\draw  (t31Cap) circle (0pt) node [anchor=north west]{\quad$\Mid$};
\draw  (T2) circle (0pt) node [anchor=south west]{$\A_{13}(v)$};

\path let \p1=(Q1) in coordinate (BLA) at (\x1-0.05cm,\y1);

\filldraw  (Q1) circle (0.4pt); % node [anchor=south west]{\small$P_2=P_3$};
%\filldraw  (BLA) node [anchor=south west]{\small$P_u=P_s$};

\filldraw  (t13) circle (0.7pt) ;
\filldraw  (t31) circle (0.7pt) ;

%overlap
\draw [line width=1.4pt,domain={acos(\v)}:120-acos(\v)] plot ({cos(\x)}, {sin(\x)});
					\end{tikzpicture}}
					\qquad \qquad
					\subfloat[]{
						\begin{tikzpicture}[scale=2.1]
							\definecolor{light-gray}{gray}{0.9}
\clip (0.35,-0.001) rectangle (1.6,1.2);
%\clip (-0.1,-1.8) rectangle (1.8,1.8);
%\clip (0,0) circle (1cm);

%v = 0.36
\def \v {0.35}
\def \radius {1}

% Q points and Taub points
\foreach \s in {1,2,3}
{\coordinate (Q\s) at ({-120 * (\s - 1)}:\radius);}
\foreach \s in {1,2,3}
{\path let \p\s=(Q\s) in coordinate (T\s) at (-1*\x\s,-1*\y\s);}

\path [fill = light-gray] (\v,0) -- (1,0) arc (0:{acos(\v)}:1) -- cycle; 

% Kasner circle
\draw [] (Q1) arc (0:360:\radius cm and \radius cm);
\draw [name path=A1] (Q1) arc (0:{acos(\v)}:1);	 	

% caption circle
%\node at (0.9*\radius, \radius*0.7) [anchor = west] {$\KC$};

% caption Q1 and T1
\path let \p1=(Q1) in coordinate (capQ1) at (1.15*\x1,\y1);
\path let \p1=(T1) in coordinate (capT1) at (1.15*\x1,\y1);

%%\filldraw  (capT1) circle (0pt) node [anchor=south]{$\T_1$};
%%\filldraw  (T1) circle (0.5pt);
%\filldraw  (T1) circle (0.7pt); % node [anchor=east]{\scriptsize $\T_1$};
\filldraw  (T2) circle (0.7pt); % node [anchor=south]{\scriptsize$\,\,\T_2$};
%\filldraw  (T3) circle (0.7pt); % node [anchor=north]{\scriptsize$\,\,\T_3$};
%
%%\filldraw  (capQ1) circle (0pt) node [anchor=south]{$\Q_1$};
%%\filldraw  (Q1) circle (0.5pt);
\filldraw  (Q1) circle (0.7pt); % node [anchor=west]{\scriptsize$\Q_1$};
%\filldraw  (Q2) circle (0.7pt); % node [anchor=north]{\scriptsize$\Q_2\,\,$};
%\filldraw  (Q3) circle (0.7pt); % node [anchor=south]{\scriptsize$\Q_3\,\,$};

%coordinate system
%\draw [->] (0,0)--(0.55,0);
%\draw [->] (0,0)--(0,0.55);

%\draw  (0.6,-0.05) circle (0pt) node [anchor=south]{\scriptsize$\Sigma_+$};
%\draw  (0,0.5) circle (0pt) node [anchor=west]{\scriptsize$\Sigma_-$};

% tangent points
\coordinate (t13) at (\v,{sqrt(1-\v^2)});
\coordinate (t12) at (\v,-{sqrt(1-\v^2)});

%rotate about -120
\path let \p1=(t13) in coordinate (t21) at (-0.5*\x1  +0.866*\y1, -0.866*\x1 -0.5*\y1);
%rotate about -120
\path let \p1=(t12) in coordinate (t23) at (-0.5*\x1  +0.866*\y1, -0.866*\x1 -0.5*\y1);

%rotate about 120
\path let \p1=(t13) in coordinate (t32) at (-0.5*\x1  -0.866*\y1, 0.866*\x1 -0.5*\y1);
%rotate about 120
\path let \p1=(t12) in coordinate (t31) at (-0.5*\x1  -0.866*\y1, 0.866*\x1 -0.5*\y1);

\draw [dashed] (t12)--(t13);
\draw [dashed] (t21)--(t23);
\draw [dashed] (t31)--(t32);

% X - points : vertices of triangle DELTA
\coordinate (X1) at (-2*\v*\radius,0);
%rotate about -120
\path let \p1=(X1) in coordinate (X2) at (-0.5*\x1  +0.866*\y1, -0.866*\x1 -0.5*\y1);
%rotate about -120
\path let \p1=(X2) in coordinate (X3) at (-0.5*\x1  +0.866*\y1, -0.866*\x1 -0.5*\y1);

% triangle DELTA
%\draw [line width=1pt] (X1)--(X2)--(X3)--(X1);

% interior bisectors 
\def \InScale {1}
\foreach \j in{1,2,3} 
{\path let \p1=(T\j) in coordinate (Y\j) at (\InScale*\x1,\InScale*\y1);
	\path let \p2=(Q\j) in coordinate (Z\j) at (\InScale*\x2,\InScale*\y2);}
%	\draw (Y\j)--(Z\j);}

%exterior bisectors
\def \OutScale  {1}
\path let \p1=(X1) in coordinate (Ex23Up) at (\x1,{sqrt(1-4*\v^2)});
\path let \p1=(X1) in coordinate (Ex23Down) at (\x1,-{sqrt(1-4*\v^2)});

% rotate
\path let \p1=(Ex23Up) in coordinate (Ex13Up) at (-0.5*\x1  +0.866*\y1, -0.866*\x1 -0.5*\y1);
\path let \p1=(Ex23Down) in coordinate (Ex13Down) at (-0.5*\x1  +0.866*\y1, -0.866*\x1 -0.5*\y1);

\path let \p1=(Ex23Up) in coordinate (Ex12Up) at (-0.5*\x1  -0.866*\y1, 0.866*\x1 -0.5*\y1);
\path let \p1=(Ex23Down) in coordinate (Ex12Down) at (-0.5*\x1  -0.866*\y1, 0.866*\x1 -0.5*\y1);

\draw (Ex23Up)--(Ex23Down);
\draw [name path=line3] (Ex13Up)--(Ex13Down);
\draw (Ex12Up)--(Ex12Down);

\def \scaling {1}
\coordinate (q1) at ($(T1)!\scaling!(Q1)$);
\coordinate (q2) at ($(T2)!\scaling!(Q2)$);
\coordinate (q3) at ($(T3)!\scaling!(Q3)$);

\coordinate (t1) at ($(Q1)!\scaling!(T1)$);
\coordinate (t2) at ($(Q2)!\scaling!(T2)$);
\coordinate (t3) at ($(Q3)!\scaling!(T3)$);

%\draw (t1) -- (q1);
%\draw (t2) -- (q2);
%\draw (t3) -- (q3);

\foreach \s in {1,2}
{ \def \z {\s+1}
	\path let \p1=(X\s) in coordinate (X\z) at (-0.5*\x1  +0.866*\y1, -0.866*\x1 -0.5*\y1);}

\path [name intersections={of=line3 and A1 ,by=P3}];

\filldraw  (P3) circle (0.4pt); 

%\draw  (P3) circle (0pt) node [anchor=north east]{\small$\, P_s$};
%\draw  (P2) circle (0pt) node [anchor=south east]{\small$\, P_2$};

\path let \p1=(Ex13Up) in coordinate (Ex13UpCap) at (\x1-0.05cm,\y1-0.05cm);

%\draw  (Ex12Down) circle (0pt) node [anchor=north east]{$\Stab$};
\draw  (Ex13UpCap) circle (0pt) node [anchor=west]{$\,\,\, \Stab$};
\draw  (t31) circle (0pt) node [anchor=north west]{\quad$\Mid$};
\draw  (T2) circle (0pt) node [anchor=south west]{$\A_{13}(v)$};

\filldraw  (t31) circle (0.7pt);
\filldraw  (t13) circle (0.7pt);
\filldraw  (P3) circle (0.7pt) ;
%overlap
\draw [very thick,domain={acos(\v))}:{acos(1-\v-0.01)}] plot ({cos(\x)}, {sin(\x)});
					\end{tikzpicture}}
					\caption{ %The dashed line $L_\alpha(v)$ determines the boundary points of the unstable arc $\A_\alpha(v)$, for $\alpha=1,2,3$. The points of intersection of $\KC$ with these lines and with the angle bisectors separate each arc into sectors, where the ordering of the eigenvalues is unchanged.
						The dashed lines $L_\alpha$, the exterior bisectors and the resulting points of intersection with $\KC$. The interior bisectors are omitted, as their points of intersection are independent of $v$.
						The bottom line depicts the corresponding sectors for the half-arc $\A_1^+(v)$. The overlap (bold) and the sector $\Mid$ exist for all $v\in \left(0,\frac 12\right)$. For $v\neq \frac 14$, there exists either the sector $\Stab$ or the sector $\Unstab$.}
					\label{figure_lem_sub}
				\end{figure}
     				\begin{remark}
					For each $v \in \left(0,\frac 12\right)$, these sectors $\A_{\alpha\beta}(v),\Mid,\Stab,\Unstab$ are adjacent, obtained in this ordering from $\talphabeta(v)$ to $\Q_\alpha$, i.e. from the boundary towards the inside of a half-arc, see \autoref{figure_lem_sub}. Note that $\Mid$ either precedes $\Stab$ or $\Unstab$, depending on $v\neq \frac 14$. In this notation, $\A_1^+(v)=\Stab$ in the supercritical case.
				\end{remark}
				\begin{proof}
					As in the proof of Lemma~\ref{cor_supercritical_eigenvalues}, it suffices to examine $\A_1^+(v)$, due to equivariance. We determine points of intersection of $\KC$ with the bisectors and we begin with the interior bisectors.
					The boundary points of $\A_1^+(v)$ are $\tac(v)$ and $\Q_1$. Since $\tac(v)_+ = v < \frac 12$, the Taub point $\Tb$ is contained in $\A_1^+(v)$. Thus, with \eqref{inbisector} the interior bisector $\mathbf{b}_{13}^{in}(v)$ intersects the interior of $\A_1^+(v)$. Moreover, since none of the other Taub points or $\Q$-points is contained in $\A_1^+(v)$, no other interior bisector intersects $\A_1^+(v)$.
					
					Next, we consider the exterior bisectors.
					$\mathbf{b}_{23}^{ex}(v)$ contains all points for which $\evb=-\evc$. With \eqref{eigenvalues} we see that $\mathbf{b}_{23}^{ex}(v)$ is a line parallel to the $\Sigma_-$-axis at $\Sigma_+=-2v$, and  is therefore disjoint from $\A_1^+(v)$. Consequently, only the dependence of $\mathbf{b}_{12}^{ex}(v)$ and $\mathbf{b}_{13}^{ex}(v)$ on $v$ remains to be investigated.
					The triangle $\Delta(v)$ is inside of $\KC$, and both, $\mathbf{b}_{12}^{ex}(v)$ and $\mathbf{b}_{13}^{ex}(v)$ intersect $\KC$ in two points. Each one of these intersection points is contained in $\A(v)$. We call these points of intersection $P_u(v)$ and $P_s(v)$, respectively.
					
					For $v=\frac 14$, we obtain $P_u(v)=P_s(v) = \Q_1$. Indeed, inserting $\Q_1=(1,0)$ and $v=\frac14$ in the formula for the eigenvalues \eqref{eigenvalues} shows that all eigenvalues are equal to three in modulus. Therefore, $\A_1^+(v)$ is split in three parts, see \autoref{figure_lem_sub}, middle.
					In the upper most part, between the boundary point $\tac(v)$ and $\T_2$, the eigenvalues satisfy
					\begin{equation}
    					 0 < \eva < \evc < - \evb.   
					\end{equation}
					As $\T_2$ is the intersection of $\mathbf{b}_{13}^{in}(v)$ with $\KC$, then $\eva(\T_2) = \evc(\T_2)$. Consequently, in the neighboring sector from $\T_2$ to $\tca(v)$ the two positive eigenvalues change ordering, i.e.
	               \begin{equation}
	                  0 < \evc < \eva < - \evb. 
	               \end{equation}				
					These two sectors are of the same size and their union with $\T_2$ yields the overlap $\A_{13}(v)$.
					As there are no more intersections of any bisector with $\A_1^+(v)$, there is only the sector $\Mid$ left. This sector is between $\tca(v)$ and $\Q_1$. Consequently, the eigenvalue $\evc$ is negative, i.e. there is only one unstable direction and the ordering is
					\begin{equation}
					   0 < -\evc < \eva < - \evb.   
					\end{equation} 
					Thus, $\evc$ changes sign but the ordering with respect to absolute value remains the same.

					With increasing $v$ the exterior bisectors $\mathbf{b}_{12}^{ex}(v)$ and $\mathbf{b}_{13}^{ex}(v)$ move outwards, away from the origin. Consequently, $P_u(v)$ moves down, and $P_s(v)$ moves up, to the interior of $A_1(v)$.
					The sectors in $\A_1^+(v)$ above $P_s(v)$ are the same as for $v=\frac 14$. Together with the sector $\Stab$ between $P_s(v)$ and $\Q_1$ we obtain four different sectors, see \autoref{figure_lem_sub}, right. As $P_s(v) \in \mathbf{b}_{13}^{ex}(v)$, the eigenvalues $-\evc$ and $\eva$ change their ordering. Thus, in $\Stab$ the ordering is
					\begin{equation}
					    0 < \eva < -\evc < - \evb,
					\end{equation} 
					and therefore the same as in supercritical case.
					
					Starting at $v=\frac 14$ and decreasing $v$ yields a similar situation with the role of $\mathbf{b}_{12}^{ex}(v)$ and $\mathbf{b}_{13}^{ex}(v)$ interchanged, see \autoref{figure_lem_sub}, left.
					Both bisectors move towards the origin. Consequently, $P_u(v)$ moves up and $P_{s}(v)$ moves down. As before, we obtain four different sectors, where the three sectors above $P_u(v)$ are the same as for $v=\frac 14$. But in the fourth sector $\Unstab$ the ordering among $-\evb$ and $\eva$ is changed. Therefore,
					\begin{equation}
					    0 <  -\evc < - \evb<\eva.
					\end{equation} 
				\end{proof}
				\begin{remark}
					When $v\to \frac 12$ the point $P_s(v)$ converges to the Taub point $\T_2$. All sectors except for $\Stab$ vanish as shown in Lemma~\ref{cor_supercritical_eigenvalues}.
					
					When $v\to 0$ the triangle $\Delta(v)$ vanishes and the parallel lines $\mathbf{b}_{12}^{ex}(v)$ and $L_3(v)$ coalesce. Thus, $P_u(v)$ and $\tca(v)$ coincide and the sector $\Mid$ vanishes. Therefore, $\A_1^+(v)$ consists of $\A_{13}(v)$, $\Unstab$ and their common boundary point $P_u(v) = \tca(v)$ where $\evc=0$ and $\eva = -\evb$.
				\end{remark}
				
				\subsection{Example of periodic heteroclinic chains}\label{sec:periods}
				The proof of stable manifolds for arbitrary heteroclinic chains presented in \autoref{chap:local} relies on the eigenvalue condition \eqref{eigenvalues_corollary} for all points of the orbit. It is satisfied in the supercritical case, due to Lemma~\ref{cor_supercritical_eigenvalues}, but it does not hold for all orbits in the subcritical case as shown in Lemma~\ref{cor_subcritical_eigenvalues}.
				%\textcolor{red}{In fact, most of the orbits contain a point in the overlap.(Is the set of initial points that never enter an overlap a cantor set as well?)}
				In the following, we investigate in more detail how these results affect the two simplest examples of periodic orbits of $\KM$ that exist for all $v \in (0,1)$, which are of period three and four.
				%see \autoref{fig:per3} and \autoref{fig:per4}, respectively
				These are particularly simple because points on these orbits are related by $D_3$-symmetry. In the $D_3$ orbit space, the local picture is the same for each point of the period three orbit, and we obtain two different situations in the case of period four. Thus, when considered in $\KC/D_3$ the period three orbit represents a homoclinic solution, whereas the period four orbit represents a heteroclinic cycle of length two.
				
    		      The Kasner circle is divided into parts with the same eigenvalue structure, by the sectors depending on $v$ from Lemmata~\ref{cor_supercritical_eigenvalues}\, -~\ref{cor_subcritical_eigenvalues}. Similarly, the orbits of period three and four are parametrized by $v$ and the points on these orbits move along different sectors of the circle as $v$ varies.
				For the period three orbit there is only one bifurcation, i.e. where the point changes the sector, but the period four orbit exhibits several bifurcations.
				% which is already indicated by the fact that there are two points affecting the stability of the cycle.
				%In general, each bifurcation reduces the local stability, and we describe more precisely in what sense exemplary for the period four orbit.
				
				%We start by reviewing some facts
				%The Lemmata on existence, uniqueness and shape of these orbits are stated without proof. \textcolor{red}{reference to bachelor thesis?}
				
				\subsubsection{Orbit of minimal period three.}
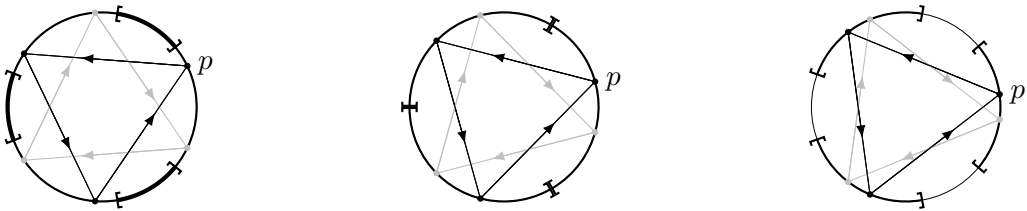
\begin{figure}[H]
					%		\captionsetup[subfigure]{labelformat=empty}
					\centering
					\begin{tikzpicture}[scale=1.25]
						\clip (-1.3,-1.3) rectangle (1.3,1.3);
%\clip (0,0) circle (1cm);
%\clip (-0.1,-1.8) rectangle (1.8,1.8);

%v = 0.378
\def \v {0.15}
\def \radius {1}

% Q points and Taub points
\foreach \s in {1,2,3}
{\coordinate (Q\s) at ({-120 * (\s - 1)}:\radius);}
\foreach \s in {1,2,3}
{\path let \p\s=(Q\s) in coordinate (T\s) at (-1*\x\s,-1*\y\s);}

\path let \p1=(Q1) in coordinate (Q1v) at (\x1/\v,\y1);

% Kasner circle
\draw [thin] (Q1) arc (0:360:\radius cm and \radius cm);	 	

% caption circle
%\node at (0.9*\radius, \radius*0.7) [anchor = west] {$\KC$};

% caption Q1 and T1
\path let \p1=(Q1) in coordinate (capQ1) at (1.15*\x1,\y1);
\path let \p1=(T1) in coordinate (capT1) at (1.15*\x1,\y1);

%coordinate system
%\draw [->] (0,0)--(0.55,0);
%\draw [->] (0,0)--(0,0.55);

%\draw  (0.6,-0.05) circle (0pt) node [anchor=south]{\scriptsize$\Sigma_+$};
%\draw  (0,0.5) circle (0pt) node [anchor=west]{\scriptsize$\Sigma_-$};

% tangent points
\coordinate (t13) at (\v,{sqrt(1-\v^2)});
\coordinate (t12) at (\v,-{sqrt(1-\v^2)});

%rotate about -120
\path let \p1=(t13) in coordinate (t21) at (-0.5*\x1  +0.866*\y1, -0.866*\x1 -0.5*\y1);
%rotate about -120
\path let \p1=(t12) in coordinate (t23) at (-0.5*\x1  +0.866*\y1, -0.866*\x1 -0.5*\y1);

%rotate about 120
\path let \p1=(t13) in coordinate (t32) at (-0.5*\x1  -0.866*\y1, 0.866*\x1 -0.5*\y1);
%rotate about 120
\path let \p1=(t12) in coordinate (t31) at (-0.5*\x1  -0.866*\y1, 0.866*\x1 -0.5*\y1);

% X - points : vertices of triangle DELTA
\coordinate (X1) at (-2*\v*\radius,0);
%rotate about -120
\path let \p1=(X1) in coordinate (X2) at (-0.5*\x1  +0.866*\y1, -0.866*\x1 -0.5*\y1);
%rotate about -120
\path let \p1=(X2) in coordinate (X3) at (-0.5*\x1  +0.866*\y1, -0.866*\x1 -0.5*\y1);

% triangle DELTA
%\draw [line width=1pt] (X1)--(X2)--(X3)--(X1);

% interior bisectors 
\def \InScale {1}
\foreach \j in{1,2,3} 
{\path let \p1=(T\j) in coordinate (Y\j) at (\InScale*\x1,\InScale*\y1);
	\path let \p2=(Q\j) in coordinate (Z\j) at (\InScale*\x2,\InScale*\y2);}
%\draw (Y\j)--(Z\j);}

%exterior bisectors
\def \OutScale  {1}
\path let \p1=(X1) in coordinate (Ex23Up) at (\x1,\OutScale);
\path let \p1=(X1) in coordinate (Ex23Down) at (\x1,-\OutScale);

% rotate
\path let \p1=(Ex23Up) in coordinate (Ex13Up) at (-0.5*\x1  +0.866*\y1, -0.866*\x1 -0.5*\y1);
\path let \p1=(Ex23Down) in coordinate (Ex13Down) at (-0.5*\x1  +0.866*\y1, -0.866*\x1 -0.5*\y1);

\path let \p1=(Ex23Up) in coordinate (Ex12Up) at (-0.5*\x1  -0.866*\y1, 0.866*\x1 -0.5*\y1);
\path let \p1=(Ex23Down) in coordinate (Ex12Down) at (-0.5*\x1  -0.866*\y1, 0.866*\x1 -0.5*\y1);

%\draw (Ex23Up)--(Ex23Down);
%\draw (Ex13Up)--(Ex13Down);
%\draw (Ex12Up)--(Ex12Down);

\def \scaling {1.3}
\coordinate (q1) at ($(T1)!\scaling!(Q1)$);
\coordinate (q2) at ($(T2)!\scaling!(Q2)$);
\coordinate (q3) at ($(T3)!\scaling!(Q3)$);

\coordinate (t1) at ($(Q1)!\scaling!(T1)$);
\coordinate (t2) at ($(Q2)!\scaling!(T2)$);
\coordinate (t3) at ($(Q3)!\scaling!(T3)$);

\foreach \s in {1,2}
{ \def \z {\s+1}
\path let \p1=(X\s) in coordinate (X\z) at (-0.5*\x1  +0.866*\y1, -0.866*\x1 -0.5*\y1);}

% T/Q points
%%\filldraw  (capT1) circle (0pt) node [anchor=south]{$\T_1$};
%%\filldraw  (T1) circle (0.5pt);
%\filldraw  (T1) circle (0.7pt); % node [anchor=east]{\scriptsize $\T_1$};
%\filldraw  (T2) circle (0.7pt); % node [anchor=south]{\scriptsize$\,\,\T_2$};
%\filldraw  (T3) circle (0.7pt); % node [anchor=north]{\scriptsize$\,\,\T_3$};
%
%%\filldraw  (capQ1) circle (0pt) node [anchor=south]{$\Q_1$};
%%\filldraw  (Q1) circle (0.5pt);
%\filldraw  (Q1) circle (0.7pt); % node [anchor=west]{\scriptsize$\Q_1$};
%\filldraw  (Q2) circle (0.7pt); % node [anchor=north]{\scriptsize$\Q_2\,\,$};
%\filldraw  (Q3) circle (0.7pt); % node [anchor=south]{\scriptsize$\Q_3\,\,$};

% tangents
%\draw (t12)--(t13);
%\draw (t21)--(t23);
%\draw (t31)--(t32);

%arcs 
\draw [Bracket-Bracket, thick, name path=A1] (t12) arc (-{acos(\v)}:{acos(\v)}:1);
\draw [Bracket-Bracket, thick, name path=A2] (t23) arc (-{acos(\v) -120 }:{acos(\v) -120}:1);
\draw [Bracket-Bracket, thick, name path=A3] (t31) arc (-{acos(\v) -240 }:{acos(\v) -240}:1);

%\draw (t1) -- (q1);
%\draw (t2) -- (q2);
%\draw (t3) -- (q3);

%points of period 3
\coordinate (p1) at ({0.25*(sqrt(12-3*\v^2) +\v)}, {sqrt(1-(1/16)*(sqrt(12-3*\v^2)+\v)^2)});
\path let \p1=(p1) in coordinate (p3) at (-0.5*\x1  -0.866*\y1, 0.866*\x1 -0.5*\y1);
\path let \p1=(p1) in coordinate (p2) at (-0.5*\x1  +0.866*\y1, -0.866*\x1 -0.5*\y1);

\foreach \j in {1,2,3}
{\filldraw (p\j) circle (0.8pt);}

\draw [] (p1)--(p3)--(p2)--(p1);

\draw [-Latex,shorten >=0.75cm] (p1)--(p3);
\draw [-Latex,shorten >=0.75cm] (p3)--(p2);
\draw [-Latex,shorten >=0.75cm] (p2)--(p1);

\filldraw (p1) circle (0pt) node[anchor=west]{\small$p$};

% second orbit
\path let \p1=(p1) in coordinate (q1) at (\x1, -\y1);
\path let \p1=(p2) in coordinate (q2) at (\x1, -\y1);
\path let \p1=(p3) in coordinate (q3) at (\x1, -\y1);

\draw [lightgray] (q1)--(q3)--(q2)--(q1);
\draw [lightgray,-Latex,shorten >=0.75cm] (q1)--(q3);
\draw [lightgray,-Latex,shorten >=0.75cm] (q3)--(q2);
\draw [lightgray,-Latex,shorten >=0.75cm] (q2)--(q1);

\foreach \j in {1,2,3}
{\filldraw[lightgray] (q\j) circle (0.8pt);}

%redraw 1st orbit to appear on top
\draw [] (p1)--(p3)--(p2)--(p1);

%overlap
\draw [line width=1.6pt,domain={acos(\v)}:120-acos(\v)] plot ({cos(\x)}, {sin(\x)});
\draw [line width=1.6pt,domain={-acos(\v)}:-120+acos(\v)] plot ({cos(\x)}, {sin(\x)});
\draw [line width=1.6pt,domain={180}:180 + acos(\v)-60] plot ({cos(\x)}, {sin(\x)});
\draw [line width=1.6pt,domain={180}:180 +60 - acos(\v)] plot ({cos(\x)}, {sin(\x)});
					\end{tikzpicture}\qquad\qquad\quad
					\begin{tikzpicture}[scale=1.25]
						\clip (-1.3,-1.3) rectangle (1.3,1.3);

\def \v {0.5}
\def \radius {1}

% Q points and Taub points
\foreach \s in {1,2,3}
{\coordinate (Q\s) at ({-120 * (\s - 1)}:\radius);}
\foreach \s in {1,2,3}
{\path let \p\s=(Q\s) in coordinate (T\s) at (-1*\x\s,-1*\y\s);}

% Q_1/v
\path let \p1=(Q1) in coordinate (Q1v) at (\x1/\v,\y1);

% Kasner circle
\draw [thin] (Q1) arc (0:360:\radius cm and \radius cm);	 	

% caption circle
%\node at (0.9*\radius, \radius*0.7) [anchor = west] {$\KC$};

% caption Q1 and T1
\path let \p1=(Q1) in coordinate (capQ1) at (1.15*\x1,\y1);
\path let \p1=(T1) in coordinate (capT1) at (1.15*\x1,\y1);

% tangent points
\coordinate (t13) at (\v,{sqrt(1-\v^2)});
\coordinate (t12) at (\v,-{sqrt(1-\v^2)});

%rotate about -120
\path let \p1=(t13) in coordinate (t21) at (-0.5*\x1  +0.866*\y1, -0.866*\x1 -0.5*\y1);
\path let \p1=(t12) in coordinate (t23) at (-0.5*\x1  +0.866*\y1, -0.866*\x1 -0.5*\y1);

%rotate about 120
\path let \p1=(t13) in coordinate (t32) at (-0.5*\x1  -0.866*\y1, 0.866*\x1 -0.5*\y1);
\path let \p1=(t12) in coordinate (t31) at (-0.5*\x1  -0.866*\y1, 0.866*\x1 -0.5*\y1);

% X - points : vertices of triangle DELTA
\coordinate (X1) at (-2*\v*\radius,0);
%rotate about -120
\path let \p1=(X1) in coordinate (X2) at (-0.5*\x1  +0.866*\y1, -0.866*\x1 -0.5*\y1);
%rotate about -120
\path let \p1=(X2) in coordinate (X3) at (-0.5*\x1  +0.866*\y1, -0.866*\x1 -0.5*\y1);

% triangle DELTA
%\draw [line width=1pt] (X1)--(X2)--(X3)--(X1);

% interior bisectors 
\def \InScale {1}
\foreach \j in{1,2,3} 
{\path let \p1=(T\j) in coordinate (Y\j) at (\InScale*\x1,\InScale*\y1);
	\path let \p2=(Q\j) in coordinate (Z\j) at (\InScale*\x2,\InScale*\y2);}
%\draw (Y\j)--(Z\j);}

%exterior bisectors
\def \OutScale  {1}
\path let \p1=(X1) in coordinate (Ex23Up) at (\x1,\OutScale);
\path let \p1=(X1) in coordinate (Ex23Down) at (\x1,-\OutScale);

% rotate
\path let \p1=(Ex23Up) in coordinate (Ex13Up) at (-0.5*\x1  +0.866*\y1, -0.866*\x1 -0.5*\y1);
\path let \p1=(Ex23Down) in coordinate (Ex13Down) at (-0.5*\x1  +0.866*\y1, -0.866*\x1 -0.5*\y1);

\path let \p1=(Ex23Up) in coordinate (Ex12Up) at (-0.5*\x1  -0.866*\y1, 0.866*\x1 -0.5*\y1);
\path let \p1=(Ex23Down) in coordinate (Ex12Down) at (-0.5*\x1  -0.866*\y1, 0.866*\x1 -0.5*\y1);

%\draw (Ex23Up)--(Ex23Down);
%\draw (Ex13Up)--(Ex13Down);
%\draw (Ex12Up)--(Ex12Down);

\def \scaling {1.3}
\coordinate (q1) at ($(T1)!\scaling!(Q1)$);
\coordinate (q2) at ($(T2)!\scaling!(Q2)$);
\coordinate (q3) at ($(T3)!\scaling!(Q3)$);

\coordinate (t1) at ($(Q1)!\scaling!(T1)$);
\coordinate (t2) at ($(Q2)!\scaling!(T2)$);
\coordinate (t3) at ($(Q3)!\scaling!(T3)$);

\foreach \s in {1,2}
{ \def \z {\s+1}
\path let \p1=(X\s) in coordinate (X\z) at (-0.5*\x1  +0.866*\y1, -0.866*\x1 -0.5*\y1);}

 % T/Q points
%\filldraw  (capT1) circle (0pt) node [anchor=south]{$\T_1$};
%\filldraw  (T1) circle (0.5pt);
%\filldraw  (T1) circle (0.7pt); % node [anchor=east]{\scriptsize $\T_1$};
%\filldraw  (T2) circle (0.7pt); % node [anchor=south]{\scriptsize$\,\,\T_2$};
%\filldraw  (T3) circle (0.7pt); % node [anchor=north]{\scriptsize$\,\,\T_3$};
%
%\filldraw  (capQ1) circle (0pt) node [anchor=south]{$\Q_1$};
%\filldraw  (Q1) circle (0.5pt);
%\filldraw  (Q1) circle (0.7pt); % node [anchor=west]{\scriptsize$\Q_1$};
%\filldraw  (Q2) circle (0.7pt); % node [anchor=north]{\scriptsize$\Q_2\,\,$};
%\filldraw  (Q3) circle (0.7pt); % node [anchor=south]{\scriptsize$\Q_3\,\,$};

% tangents
%\draw (t12)--(t13);
%\draw (t21)--(t23);
%\draw (t31)--(t32);

%arcs 
\draw [Bracket-Bracket, thick, name path=A1] (t12) arc (-{acos(\v)}:{acos(\v)}:1);
\draw [Bracket-Bracket, thick, name path=A2] (t23) arc (-{acos(\v) -120 }:{acos(\v) -120}:1);
\draw [Bracket-Bracket, thick, name path=A3] (t31) arc (-{acos(\v) -240 }:{acos(\v) -240}:1);

%points of period 3, p1 using coordinates; p2 and p3 by rotation
\coordinate (p1) at ({0.25*(sqrt(12-3*\v^2) +\v)}, {sqrt(1-(1/16)*(sqrt(12-3*\v^2)+\v)^2)});
\path let \p1=(p1) in coordinate (p3) at (-0.5*\x1  -0.866*\y1, 0.866*\x1 -0.5*\y1);
\path let \p1=(p1) in coordinate (p2) at (-0.5*\x1  +0.866*\y1, -0.866*\x1 -0.5*\y1);

\foreach \j in {1,2,3}
{\filldraw (p\j) circle (0.8pt);}

% projected lines
\draw [] (p1)--(p3)--(p2)--(p1);

% arrows indicating the direction of the orbit
\draw [-Latex,shorten >=0.75cm] (p1)--(p3);
\draw [-Latex,shorten >=0.75cm] (p3)--(p2);
\draw [-Latex,shorten >=0.75cm] (p2)--(p1);

\filldraw (p1) circle (0pt) node[anchor=west]{\small$p$};

% second orbit -> reflect
\path let \p1=(p1) in coordinate (q1) at (\x1, -\y1);
\path let \p1=(p2) in coordinate (q2) at (\x1, -\y1);
\path let \p1=(p3) in coordinate (q3) at (\x1, -\y1);

\draw [lightgray] (q1)--(q3)--(q2)--(q1);
\draw [lightgray,-Latex,shorten >=0.75cm] (q1)--(q3);
\draw [lightgray,-Latex,shorten >=0.75cm] (q3)--(q2);
\draw [lightgray,-Latex,shorten >=0.75cm] (q2)--(q1);

\foreach \j in {1,2,3}
{\filldraw[lightgray] (q\j) circle (0.8pt);}

% redraw first orbit to be on top
\draw [] (p1)--(p3)--(p2)--(p1);
					\end{tikzpicture}\qquad\qquad\quad
					\begin{tikzpicture}[scale=1.25]
						\clip (-1.3,-1.3) rectangle (1.3,1.3);
%\clip (0,0) circle (1cm);
%\clip (-0.1,-1.8) rectangle (1.8,1.8);

%v = 0.36
\def \v {0.76}
\def \radius {1}

% Q points and Taub points
\foreach \s in {1,2,3}
{\coordinate (Q\s) at ({-120 * (\s - 1)}:\radius);}
\foreach \s in {1,2,3}
{\path let \p\s=(Q\s) in coordinate (T\s) at (-1*\x\s,-1*\y\s);}

\path let \p1=(Q1) in coordinate (Q1v) at (\x1/\v,\y1);

% Kasner circle
\draw [very thin] (Q1) arc (0:360:\radius cm and \radius cm);	 	

% caption circle
%\node at (0.9*\radius, \radius*0.7) [anchor = west] {$\KC$};

% caption Q1 and T1
\path let \p1=(Q1) in coordinate (capQ1) at (1.15*\x1,\y1);
\path let \p1=(T1) in coordinate (capT1) at (1.15*\x1,\y1);

%coordinate system
%\draw [->] (0,0)--(0.55,0);
%\draw [->] (0,0)--(0,0.55);

%\draw  (0.6,-0.05) circle (0pt) node [anchor=south]{\scriptsize$\Sigma_+$};
%\draw  (0,0.5) circle (0pt) node [anchor=west]{\scriptsize$\Sigma_-$};

% tangent points
\coordinate (t13) at (\v,{sqrt(1-\v^2)});
\coordinate (t12) at (\v,-{sqrt(1-\v^2)});

%rotate about -120
\path let \p1=(t13) in coordinate (t21) at (-0.5*\x1  +0.866*\y1, -0.866*\x1 -0.5*\y1);
%rotate about -120
\path let \p1=(t12) in coordinate (t23) at (-0.5*\x1  +0.866*\y1, -0.866*\x1 -0.5*\y1);

%rotate about 120
\path let \p1=(t13) in coordinate (t32) at (-0.5*\x1  -0.866*\y1, 0.866*\x1 -0.5*\y1);
%rotate about 120
\path let \p1=(t12) in coordinate (t31) at (-0.5*\x1  -0.866*\y1, 0.866*\x1 -0.5*\y1);

% X - points : vertices of triangle DELTA
\coordinate (X1) at (-2*\v*\radius,0);
%rotate about -120
\path let \p1=(X1) in coordinate (X2) at (-0.5*\x1  +0.866*\y1, -0.866*\x1 -0.5*\y1);
%rotate about -120
\path let \p1=(X2) in coordinate (X3) at (-0.5*\x1  +0.866*\y1, -0.866*\x1 -0.5*\y1);

% triangle DELTA
%\draw [line width=1pt] (X1)--(X2)--(X3)--(X1);

% interior bisectors 
\def \InScale {1}
\foreach \j in{1,2,3} 
{\path let \p1=(T\j) in coordinate (Y\j) at (\InScale*\x1,\InScale*\y1);
	\path let \p2=(Q\j) in coordinate (Z\j) at (\InScale*\x2,\InScale*\y2);}
%\draw (Y\j)--(Z\j);}

%exterior bisectors
\def \OutScale  {1}
\path let \p1=(X1) in coordinate (Ex23Up) at (\x1,\OutScale);
\path let \p1=(X1) in coordinate (Ex23Down) at (\x1,-\OutScale);

% rotate
\path let \p1=(Ex23Up) in coordinate (Ex13Up) at (-0.5*\x1  +0.866*\y1, -0.866*\x1 -0.5*\y1);
\path let \p1=(Ex23Down) in coordinate (Ex13Down) at (-0.5*\x1  +0.866*\y1, -0.866*\x1 -0.5*\y1);

\path let \p1=(Ex23Up) in coordinate (Ex12Up) at (-0.5*\x1  -0.866*\y1, 0.866*\x1 -0.5*\y1);
\path let \p1=(Ex23Down) in coordinate (Ex12Down) at (-0.5*\x1  -0.866*\y1, 0.866*\x1 -0.5*\y1);

%\draw (Ex23Up)--(Ex23Down);
%\draw (Ex13Up)--(Ex13Down);
%\draw (Ex12Up)--(Ex12Down);

\def \scaling {1.3}
\coordinate (q1) at ($(T1)!\scaling!(Q1)$);
\coordinate (q2) at ($(T2)!\scaling!(Q2)$);
\coordinate (q3) at ($(T3)!\scaling!(Q3)$);

\coordinate (t1) at ($(Q1)!\scaling!(T1)$);
\coordinate (t2) at ($(Q2)!\scaling!(T2)$);
\coordinate (t3) at ($(Q3)!\scaling!(T3)$);

\foreach \s in {1,2}
{ \def \z {\s+1}
\path let \p1=(X\s) in coordinate (X\z) at (-0.5*\x1  +0.866*\y1, -0.866*\x1 -0.5*\y1);}

% T/Q points
%%\filldraw  (capT1) circle (0pt) node [anchor=south]{$\T_1$};
%%\filldraw  (T1) circle (0.5pt);
%\filldraw  (T1) circle (0.7pt); % node [anchor=east]{\scriptsize $\T_1$};
%\filldraw  (T2) circle (0.7pt); % node [anchor=south]{\scriptsize$\,\,\T_2$};
%\filldraw  (T3) circle (0.7pt); % node [anchor=north]{\scriptsize$\,\,\T_3$};
%
%%\filldraw  (capQ1) circle (0pt) node [anchor=south]{$\Q_1$};
%%\filldraw  (Q1) circle (0.5pt);
%\filldraw  (Q1) circle (0.7pt); % node [anchor=west]{\scriptsize$\Q_1$};
%\filldraw  (Q2) circle (0.7pt); % node [anchor=north]{\scriptsize$\Q_2\,\,$};
%\filldraw  (Q3) circle (0.7pt); % node [anchor=south]{\scriptsize$\Q_3\,\,$};

% tangents
%\draw (t12)--(t13);
%\draw (t21)--(t23);
%\draw (t31)--(t32);

%arcs 
\draw [Bracket-Bracket, thick, name path=A1] (t12) arc (-{acos(\v)}:{acos(\v)}:1);
\draw [Bracket-Bracket, thick, name path=A2] (t23) arc (-{acos(\v) -120 }:{acos(\v) -120}:1);
\draw [Bracket-Bracket, thick, name path=A3] (t31) arc (-{acos(\v) -240 }:{acos(\v) -240}:1);

%\draw (t1) -- (q1);
%\draw (t2) -- (q2);
%\draw (t3) -- (q3);

%points of period 3
\coordinate (p1) at ({0.25*(sqrt(12-3*\v^2) +\v)}, {sqrt(1-(1/16)*(sqrt(12-3*\v^2)+\v)^2)});

\path let \p1=(p1) in coordinate (p3) at (-0.5*\x1  -0.866*\y1, 0.866*\x1 -0.5*\y1);

\path let \p1=(p1) in coordinate (p2) at (-0.5*\x1  +0.866*\y1, -0.866*\x1 -0.5*\y1);

\foreach \j in {1,2,3}
{\filldraw (p\j) circle (0.8pt);}

\filldraw (p1) circle (0pt) node[anchor=west]{\small$p$};

\draw [] (p1)--(p3)--(p2)--(p1);
\draw [-Latex,shorten >=0.75cm] (p1)--(p3);
\draw [-Latex,shorten >=0.75cm] (p3)--(p2);
\draw [-Latex,shorten >=0.75cm] (p2)--(p1);

% second orbit
\path let \p1=(p1) in coordinate (q1) at (\x1, -\y1);
\path let \p1=(p2) in coordinate (q2) at (\x1, -\y1);
\path let \p1=(p3) in coordinate (q3) at (\x1, -\y1);

\draw [lightgray] (q1)--(q3)--(q2)--(q1);
\draw [lightgray,-Latex,shorten >=0.75cm] (q1)--(q3);
\draw [lightgray,-Latex,shorten >=0.75cm] (q3)--(q2);
\draw [lightgray,-Latex,shorten >=0.75cm] (q2)--(q1);

\foreach \j in {1,2,3}
{\filldraw[lightgray] (q\j) circle (0.8pt);}

%redraw 1st triangle
\draw [] (p1)--(p3)--(p2)--(p1);
					\end{tikzpicture} \caption{The orbits rotate with increasing $v$, see \cite{HellLappicyUggla}. %Points in each orbit are related by rotation of $\frac{2\pi}{3}$, whereas the orbits themselves are reflectional symmetric. 
                    All six points are symmetrically related to $p\in \A_1(v)$, under the $D_3$ group orbit. } \label{period3_orbit_supercritical}
				\end{figure}

				\begin{lem}
					For $v \in (0,1)$ there exist exactly two orbits of minimal period three. The points on each orbit are related by rotation by $\frac{2\pi}{3}$, and the two orbits are related by reflectional symmetry across the $\T_\alpha\Q_\alpha$-axis, for $\alpha=1,2,3$.
					%	such that the projected heteroclinic cycle forms an equilateral triangle.
					%Moreover, the incoming heteroclinic solution always enters via the strong stable direction.
				\end{lem}
				%
                    %
				%\textcolor{red}{no need for a proof because I give the points explicitely?!}
				We consider the orbit which traverses anti-clockwise, i.e.  $\A_1(v)\to \A_3(v) \to \A_2(v)\to \A_1(v)$, see \autoref{period3_orbit_supercritical}. %The points of this orbit are related by rotation about $\frac {2\pi}{3}$, just as the arcs itself. 
                Due to symmetry, all points of this orbit possess the same local structure, and thus we consider $p \in \A_1(v)$.
				The coordinates of $p$ can be found by symmetry, i.e. solving the equation $\KM_1(p) = R p$, where $R$ denotes the rotation by $\frac{2\pi}{3}$. We obtain that
				\begin{align}\label{coordinates_period3}
					\Sigma_+(v) = \frac{1}{4} \left(\sqrt{12-3v^2} + v\right), \quad \Sigma_-(v)= \sqrt{1-\frac{1}{16} \left(\sqrt{12-3v^2} + v\right)^2}.
				\end{align}
                    %\textcolor{red}{this is not in agreement with claes paper, eqn 34. there is a negative sign in front of the square root there.}
				%(where we used $\Sigma_- = \sqrt{1-\Sigma_+^2}$.)
				%
    The corresponding eigenvalues are depicted in \autoref{period3_eigenvalues}.
				The order in which the heteroclinic chains traverses the circle determines the incoming and outgoing directions in $p \in \A_1^+(v)$, due to \eqref{coordinates_period3}.
                For instance, $\A_1(v) \to \A_3(v)$ implies that the outgoing heteroclinic solution always leaves along $\partial_{N_1}$, and $\A_2(v) \to \A_1(v)$ implies that the incoming heteroclinic solution of this cycle is along $\partial_{N_2}$.
				%
				%\begin{figure}
				%	\centering
				%	\includegraphics[scale=0.45]{figures/period3_evalues.png}\caption{figure similar to this using tikz}
				%\end{figure}
				%
				%
				%Recall that $\partial_{N_1}$ is the unstable direction and due to order in which the heteroclinic cycle traverses the circle, the incoming heteroclinic is tangent to $\partial_{N_2}$, independent of $v$.
				%(Somewhere: always if a point is in the upper hemisphere and uses $\KM_1$).
				%
				%From Corollary \ref{cor_supercritical_eigenvalues} we already know that no change in the ordering among the eigenvalues can occur for $v\ge \frac12$.
			  According to Lemma~\ref{cor_supercritical_eigenvalues} the eigenvalues satisfy
				\begin{equation}\label{period3_ordering}
					0<\eva(p,v) < -\evc(p,v)<-\evb(p,v)
				\end{equation}
				for all $v \ge \frac 12$. Thus, the heteroclinic cycle enters non-generically, along the strong stable direction and leaves along the only unstable direction, which is the weakest with respect to absolute value. The ordering \eqref{period3_ordering} is preserved until 
                \begin{equation}
                    v = \frac{1}{\sqrt{7}} \approx 0.378,
                \end{equation} in which case $\eva(p,v) = -\evc(p,v)$. For $v < \frac{1}{\sqrt{7}}$, the point $p$ transitions from the sector $\Stab$ to $\Mid$ in the notation of Lemma~\ref{cor_subcritical_eigenvalues}, and we have that
                \begin{equation}
                    0< -\evc(p,v) < \eva(p,v)< -\evb(p,v).
                \end{equation}
                %
				%i.e. the expansion rate in unstable direction $\partial_{N_1}$ dominates the lateral contration rate, coming from the weak stable direction $\partial_{N_3}$. The contraction rate in the incoming direction $\partial_{N_2}$ is still the strongest. \textcolor{red}{How related to sectors?}     
                Even though the existence of the stable manifold of the period three cycle in \autoref{chap:shadowing} assumes that $v>\frac{1}{2}$, the proof works almost verbatim for $v>\frac {1}{\sqrt{7}}$. It is only when $v\le\frac {1}{\sqrt{7}}$ that the assumption \eqref{eigenvalue_condition} is not fulfilled. In this case, however, we may adapt \cite{Liebscher_magnetic}. %, which we briefly discuss in \autoref{chap:conclusion}.
				%Note that $p$ never enters the overlap $\A_{13}$.
				%
                    %
				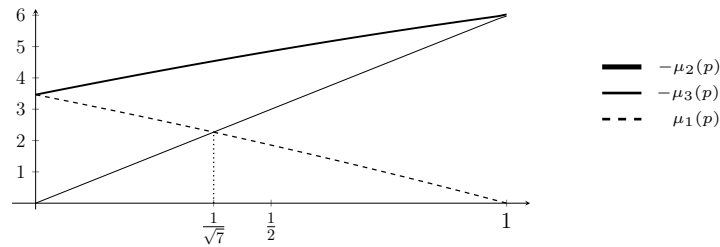
\begin{figure}[H]\centering
					\begin{tikzpicture}[scale=0.6]
						%arccos(0.8) = 0.6435
% first term  -2 tan^(-1)(((1 + v) tan(arccos(0.25)/2))/(-1 + v)) = 2.85665
\clip [] (-1.5,-0.8) rectangle (16,5);
\pgfplotsset{
	compat=newest,
	every axis plot/.append style={no marks,thick},
	every axis/.style={
		axis lines=middle,
		width=13cm,
		height=6cm,
	}
}

\begin{axis}
	[axis lines = center,
	domain=0,
	xmin= -0.05,
	xmax = 1.05,
	ymin = -0.2,
	ymax = 6.2,
	samples=100,
	trig format plots=rad,
	xtick={0,0.378,0.5,1},
	xticklabels={\huge $0$,\large$\frac{1}{\sqrt{7}}$,\large$\frac 12$, \large$1$},
	ytick={0,1,2,3,4,5,6},
	yticklabels={$0$,$1$,$2$, $3$,$4$,$5$, $6$}
	]
	%ytick = {0.6435 , 1.45089, 2.73789, 3.54529, 4.83229},
	%yticklabels={$\delta$, $\tca$, $\tcb$, $\tbc$, $\tba$}

%	Sigma1 = ((sqrt(3)*sqrt(4-x^2)+x)/4);
%	Sigma2 = sqrt(1-((sqrt(3)*sqrt(4-x^2)+x)/4)^2);
%	nu1(i) = 4*((sqrt(3)*sqrt(4-x^2)+x)/4-x);
%   nu2(i) = -4*x - 2*Sigma1 - 2*sqrt(3)*Sigma2;
%   nu3(i) = -4*x - 2*Sigma1 + 2*sqrt(3)*Sigma2;

	\addplot [dashed, domain=0:1, smooth] {4*((sqrt(3)*sqrt(4-x^2)+x)/4-x)};
	\addplot [very thick, domain=0:1, smooth] {abs(-4*x - 2*((sqrt(3)*sqrt(4-x^2)+x)/4) - 2*sqrt(3)*sqrt(1-((sqrt(3)*sqrt(4-x^2)+x)/4)^2))};
	\addplot [thin, domain=0:1, smooth] {abs(-4*x - 2*((sqrt(3)*sqrt(4-x^2)+x)/4) + 2*sqrt(3)*sqrt(1-((sqrt(3)*sqrt(4-x^2)+x)/4)^2))};

%	\addplot [ultra thick] coordinates {(0,0)(0.11,0)};
%	\addplot [ultra thick] coordinates {(0.22,0)(0.33,0)};
%	\addplot [ultra thick] coordinates {(0.66,0)(0.77,0)};
%	\addplot [ultra thick] coordinates {(0.88,0)(1,0)};
	
	%\coordinate (A) at (0.8,0.5);
	%\coordinate (B) at (0.8,1);
	%\coordinate (C) at (0.8,1.5);

	\coordinate (A1) at (0.7,2);
	\coordinate (A2) at (0.75,2);    

        \coordinate (B1) at (0.7,1.5);
	\coordinate (B2) at (0.75,1.5);    

        \coordinate (C1) at (0.7,1);
	\coordinate (C2) at (0.75,1);    

%	\addplot [domain=0.9:0.9] {0.1};
	%\addplot [domain=0:0.22, dashed] {0.6};
	\addplot [dotted] coordinates {(0.378,0)(0.378,2.3)};
%	\addplot [dotted] coordinates {(0.22,0)(0.22,0.66)};
%	\addplot [dotted] coordinates {(0.77,0)(0.77,0.33)};
%	\addplot [dotted] coordinates {(0.88,0)(0.88,0.66)};
\end{axis}
%\node []  at (A) {\tiny$\eva(p)$};
%\node []  at (B) {\tiny$-\evc(p)$};
%\node []  at (C) {\tiny$-\evb(p)$};

\draw [line width = 1.9pt] (A1)--(A2) node [anchor=west] {\tiny$\,-\evb(p)$};
\draw [line width = 1pt] (B1)--(B2) node [anchor=west] {\tiny$\,-\evc(p)$};
\draw [dashed, line width = 1pt] (C1)--(C2) node [anchor=west]  {\tiny$\quad\eva(p)$};
%\draw [line width = 0.9pt] (A3)--(B3) node [anchor=west] {\tiny$\, \shortquad|\evc(p)|$};

%\draw[dotted] (2.47,0.5) -- (2.47,5.2);
%\draw[dotted] (6.3,0.5) -- (6.3,5.2);

%\draw[red] (1.528,5.461) -- (1.528,2.85);
%\draw[red] (3.08,0.234) -- (3.07,2.85);

%    \addplot [domain=0.6:1, smooth, thick] { (1-x)/x };
%%%  \draw[->] (0,0) -- (0.6435+0.1, 0) node[right] {};
%%%  \draw[->] (0,-0.6435-0.1) -- (0, 0.6435+0.1) node[above] {$\,$};
%\draw[domain=0:0.5, smooth, variable=\x] plot ({\x}, {(0.5-\x)/ (0.5+\x)});

%%%  \draw[domain=0:0.6435, smooth, variable=\x] plot ({\x}, {\x});

% x axis beschriftung
%\draw[dotted] (0.54,0) -- (0.54,1);
%\draw[] (0.54,0.02) -- (0.54,-0.02);
%\draw[] (0.76,0.02) -- (0.76,-0.02);
%\draw[<-](0.54,-0.05) arc (0:-10:1.5cm and 1cm) node [anchor = north] {\tiny $\KI^{-1}(b) \, $};
%\draw[<-] (0.76,-0.05) arc (180:190:1.5cm and 1cm) node [anchor = north] {\tiny $ \,\KI^{-1}(a)$};
%\draw[] (0.5,0.02) -- (0.5,-0.02) node [anchor = north] {$\Q$};

%%%  \draw[] (0.6435,0.02) -- (0.6435,-0.02) node [anchor = north] {$\delta$};
%%%  \draw[] (0,0.02) -- (0,-0.02) node [anchor = north] {$0$};

%y-achsenabschnitt
%\draw[dashed] (0.54,1) -- (0,1);
%\draw[] (-0.02,1) -- (0.02,1) node [anchor = east] {$\delta \,$};
%\draw[] (-0.02,0) -- (0.02,0) node [anchor = east] {$-\delta \, $};
					\end{tikzpicture} 
					\caption{The eigenvalues in $p \in \A_1(v)$. The incoming heteroclinic direction is along the direction corresponding to the non-leading stable eigenvalue $-\evb(p)$ (bold). At $v=\frac{1}{\sqrt{7}}$ the ordering of the unique unstable eigenvalue $\eva(p)$ (dashed) and the leading stable eigenvalue $-\evc(p)$ switches. This indicates that as $v$ decreases, $p$ leaves the set $\Stab$ and enters the set $\Mid$ at $v=\frac{1}{\sqrt{7}}$.}\label{period3_eigenvalues}
				\end{figure}
				
				%\begin{figure}
				%	\centering
				%	\begin{tikzpicture}[scale=0.9]
					%		\input{figures/period3_eigenvalues}
					%	\end{tikzpicture}\caption{The eigenvalues in the point of period three $p \in \A_1$. The incoming heteroclinic direction is along the direction corresponding to the non-leading stable eigenvalue $\evb(p)$ (bold). At $v=\frac{1}{\sqrt{7}}$ the ordering of the unique unstable eigenvalue $\eva(p)$ (dashed) and the leading stable eigenvalue $\evc(p)$ switches. This indicates a transition of $p$ from the set $\Mid$ to the set $\Stab$ at $v=\frac 17$.} \label{period3_eigenvalues}
				%\end{figure}

				%\textcolor{blue}{Does it fail for all other chains?}
				%\textcolor{red}{non-euclidean metric}
				
				%Different metric in insections, require a non-Euclidean metric in the insection. "Unfortunately the metric used to obtain the contraction in the proof of the main theorem of this paper is singular on the invariant subspaces. We have no result on the way in which the manifolds constructed approach these boundaries."\\
				
				\subsubsection{Orbit of minimal period four.}\label{sec:per4}
\begin{figure}[H]
					\captionsetup[subfigure]{labelformat=empty}
					\centering
					\subfloat[$0<v<\frac12$]{\begin{tikzpicture}[scale=1.25]
							\definecolor{light-gray}{gray}{0.55}

\clip (-1.2,-1.2) rectangle (1.2,1.2);
%\clip (0,0) circle (1cm);
%\clip (-0.1,-1.8) rectangle (1.8,1.8);

%v = 0.36
\def \v {0.15}
\def \radius {1}

% Q points and Taub points
\foreach \s in {1,2,3}
{\coordinate (Q\s) at ({-120 * (\s - 1)}:\radius);}
\foreach \s in {1,2,3}
{\path let \p\s=(Q\s) in coordinate (T\s) at (-1*\x\s,-1*\y\s);}

\path let \p1=(Q1) in coordinate (Q1v) at (\x1/\v,\y1);

% Kasner circle
\draw [thin] (Q1) arc (0:360:\radius cm and \radius cm);	 	

% caption circle
%\node at (0.9*\radius, \radius*0.7) [anchor = west] {$\KC$};

% caption Q1 and T1
\path let \p1=(Q1) in coordinate (capQ1) at (1.15*\x1,\y1);
\path let \p1=(T1) in coordinate (capT1) at (1.15*\x1,\y1);

%coordinate system
%\draw [->] (0,0)--(0.55,0);
%\draw [->] (0,0)--(0,0.55);

%\draw  (0.6,-0.05) circle (0pt) node [anchor=south]{\scriptsize$\Sigma_+$};
%\draw  (0,0.5) circle (0pt) node [anchor=west]{\scriptsize$\Sigma_-$};

% tangent points
\coordinate (t13) at (\v,{sqrt(1-\v^2)});
\coordinate (t12) at (\v,-{sqrt(1-\v^2)});

%rotate about -120
\path let \p1=(t13) in coordinate (t21) at (-0.5*\x1  +0.866*\y1, -0.866*\x1 -0.5*\y1);
%rotate about -120
\path let \p1=(t12) in coordinate (t23) at (-0.5*\x1  +0.866*\y1, -0.866*\x1 -0.5*\y1);

%rotate about 120
\path let \p1=(t13) in coordinate (t32) at (-0.5*\x1  -0.866*\y1, 0.866*\x1 -0.5*\y1);
%rotate about 120
\path let \p1=(t12) in coordinate (t31) at (-0.5*\x1  -0.866*\y1, 0.866*\x1 -0.5*\y1);

% X - points : vertices of triangle DELTA
\coordinate (X1) at (-2*\v*\radius,0);
%rotate about -120
\path let \p1=(X1) in coordinate (X2) at (-0.5*\x1  +0.866*\y1, -0.866*\x1 -0.5*\y1);
%rotate about -120
\path let \p1=(X2) in coordinate (X3) at (-0.5*\x1  +0.866*\y1, -0.866*\x1 -0.5*\y1);

% triangle DELTA
%\draw [line width=1pt] (X1)--(X2)--(X3)--(X1);

% interior bisectors 
\def \InScale {1}
\foreach \j in{1,2,3} 
{\path let \p1=(T\j) in coordinate (Y\j) at (\InScale*\x1,\InScale*\y1);
	\path let \p2=(Q\j) in coordinate (Z\j) at (\InScale*\x2,\InScale*\y2);}
%\draw (Y\j)--(Z\j);}

%exterior bisectors
\def \OutScale  {1}
\path let \p1=(X1) in coordinate (Ex23Up) at (\x1,\OutScale);
\path let \p1=(X1) in coordinate (Ex23Down) at (\x1,-\OutScale);

% rotate
\path let \p1=(Ex23Up) in coordinate (Ex13Up) at (-0.5*\x1  +0.866*\y1, -0.866*\x1 -0.5*\y1);
\path let \p1=(Ex23Down) in coordinate (Ex13Down) at (-0.5*\x1  +0.866*\y1, -0.866*\x1 -0.5*\y1);

\path let \p1=(Ex23Up) in coordinate (Ex12Up) at (-0.5*\x1  -0.866*\y1, 0.866*\x1 -0.5*\y1);
\path let \p1=(Ex23Down) in coordinate (Ex12Down) at (-0.5*\x1  -0.866*\y1, 0.866*\x1 -0.5*\y1);

%\draw (Ex23Up)--(Ex23Down);
%\draw (Ex13Up)--(Ex13Down);
%\draw (Ex12Up)--(Ex12Down);

\def \scaling {1.3}
\coordinate (q1) at ($(T1)!\scaling!(Q1)$);
\coordinate (q2) at ($(T2)!\scaling!(Q2)$);
\coordinate (q3) at ($(T3)!\scaling!(Q3)$);

\coordinate (t1) at ($(Q1)!\scaling!(T1)$);
\coordinate (t2) at ($(Q2)!\scaling!(T2)$);
\coordinate (t3) at ($(Q3)!\scaling!(T3)$);

\foreach \s in {1,2}
{ \def \z {\s+1}
\path let \p1=(X\s) in coordinate (X\z) at (-0.5*\x1  +0.866*\y1, -0.866*\x1 -0.5*\y1);}

% T/Q points
%%\filldraw  (capT1) circle (0pt) node [anchor=south]{$\T_1$};
%%\filldraw  (T1) circle (0.5pt);
%\filldraw  (T1) circle (0.7pt); % node [anchor=east]{\scriptsize $\T_1$};
%\filldraw  (T2) circle (0.7pt); % node [anchor=south]{\scriptsize$\,\,\T_2$};
%\filldraw  (T3) circle (0.7pt); % node [anchor=north]{\scriptsize$\,\,\T_3$};
%
%%\filldraw  (capQ1) circle (0pt) node [anchor=south]{$\Q_1$};
%%\filldraw  (Q1) circle (0.5pt);
%\filldraw  (Q1) circle (0.7pt); % node [anchor=west]{\scriptsize$\Q_1$};
%\filldraw  (Q2) circle (0.7pt); % node [anchor=north]{\scriptsize$\Q_2\,\,$};
%\filldraw  (Q3) circle (0.7pt); % node [anchor=south]{\scriptsize$\Q_3\,\,$};

% tangents
%\draw (t12)--(t13);
%\draw (t21)--(t23);
%\draw (t31)--(t32);

%arcs 
\draw [Bracket-Bracket, thick, name path=A1] (t12) arc (-{acos(\v)}:{acos(\v)}:1);
\draw [Bracket-Bracket, thick, name path=A2] (t23) arc (-{acos(\v) -120 }:{acos(\v) -120}:1);
\draw [Bracket-Bracket, thick, name path=A3] (t31) arc (-{acos(\v) -240 }:{acos(\v) -240}:1);

%\draw (t1) -- (q1);
%\draw (t2) -- (q2);
%\draw (t3) -- (q3);

\def \M {{-1/(2*\v) +   ((2*\v^2+1)* (\v^2+2+sqrt(4*\v^4+\v^2+4)*\v)  ) / (4*\v*(\v^4+\v^2+1))            }}

\def \N {   {    sqrt(1-(-1/(2*\v) +   ((2*\v^2+1)* (\v^2+2+sqrt(4*\v^4+\v^2+4)*\v)  ) / (4*\v*(\v^4+\v^2+1)))^2 )     }    }

\coordinate (p1) at (\M,\N);
\filldraw  (p1) circle (0.7pt);

%  intersection of arc A3 with the allonged line from Q1/v
\coordinate (PP) at ($(Q1v)!10!(p1)$);
\path [name path = line] (Q1v)--(PP);
\path [name intersections={of=line and A3 ,by={p1,p2}}];
%\path [name intersections={of=line and A3 ,by=p2}];

\path let \p1=(p1) in coordinate (p3) at (\x1,-\y1);
\path let \p1=(p2) in coordinate (p4) at (\x1,-\y1);

\foreach \j in {1,2,3,4}
{\filldraw (p\j) circle (0.8pt);}

\draw (p1)--(p2)--(p3)--(p4)--(p1);
\draw [-Latex,shorten >=1cm] (p1)--(p2);
\draw [-Latex,shorten >=1.6cm] (p2)--(p3);
\draw [-Latex,shorten >=1cm] (p3)--(p4);
\draw [-Latex,shorten >=1.6cm] (p4)--(p1);

%overlap
\draw [line width=1.6pt,domain={acos(\v)}:120-acos(\v)] plot ({cos(\x)}, {sin(\x)});
\draw [line width=1.6pt,domain={-acos(\v)}:-120+acos(\v)] plot ({cos(\x)}, {sin(\x)});
\draw [line width=1.6pt,domain={180}:180 + acos(\v)-60] plot ({cos(\x)}, {sin(\x)});
\draw [line width=1.6pt,domain={180}:180 +60 - acos(\v)] plot ({cos(\x)}, {sin(\x)});

%points
\filldraw [] (p1) circle (1.4pt);% node [anchor=south west]{$\,p$};
\filldraw [] (p3) circle (1.4pt);
\filldraw [light-gray] (p2) circle (1.4pt);
\filldraw [light-gray] (p4) circle (1.4pt);

\filldraw (p4) circle (0pt);% node [anchor=north east]{$q$};
						\end{tikzpicture}\label{period4_orbit_subcritical}} \qquad\qquad\quad
					\subfloat[$v = \frac 12$]{\begin{tikzpicture}[scale=1.25]
							\definecolor{light-gray}{gray}{0.55}
\clip (-1.2,-1.2) rectangle (1.2,1.2);
%\clip (0,0) circle (1cm);
%\clip (-0.1,-1.8) rectangle (1.8,1.8);

%v = 0.36
\def \v {0.5}
\def \radius {1}

% Q points and Taub points
\foreach \s in {1,2,3}
{\coordinate (Q\s) at ({-120 * (\s - 1)}:\radius);}
\foreach \s in {1,2,3}
{\path let \p\s=(Q\s) in coordinate (T\s) at (-1*\x\s,-1*\y\s);}

\path let \p1=(Q1) in coordinate (Q1v) at (\x1/\v,\y1);

% Kasner circle
\draw [thin] (Q1) arc (0:360:\radius cm and \radius cm);	 	

% caption circle
%\node at (0.9*\radius, \radius*0.7) [anchor = west] {$\KC$};

% caption Q1 and T1
\path let \p1=(Q1) in coordinate (capQ1) at (1.15*\x1,\y1);
\path let \p1=(T1) in coordinate (capT1) at (1.15*\x1,\y1);

%coordinate system
%\draw [->] (0,0)--(0.55,0);
%\draw [->] (0,0)--(0,0.55);

%\draw  (0.6,-0.05) circle (0pt) node [anchor=south]{\scriptsize$\Sigma_+$};
%\draw  (0,0.5) circle (0pt) node [anchor=west]{\scriptsize$\Sigma_-$};

% tangent points
\coordinate (t13) at (\v,{sqrt(1-\v^2)});
\coordinate (t12) at (\v,-{sqrt(1-\v^2)});

%rotate about -120
\path let \p1=(t13) in coordinate (t21) at (-0.5*\x1  +0.866*\y1, -0.866*\x1 -0.5*\y1);
%rotate about -120
\path let \p1=(t12) in coordinate (t23) at (-0.5*\x1  +0.866*\y1, -0.866*\x1 -0.5*\y1);

%rotate about 120
\path let \p1=(t13) in coordinate (t32) at (-0.5*\x1  -0.866*\y1, 0.866*\x1 -0.5*\y1);
%rotate about 120
\path let \p1=(t12) in coordinate (t31) at (-0.5*\x1  -0.866*\y1, 0.866*\x1 -0.5*\y1);

% X - points : vertices of triangle DELTA
\coordinate (X1) at (-2*\v*\radius,0);
%rotate about -120
\path let \p1=(X1) in coordinate (X2) at (-0.5*\x1  +0.866*\y1, -0.866*\x1 -0.5*\y1);
%rotate about -120
\path let \p1=(X2) in coordinate (X3) at (-0.5*\x1  +0.866*\y1, -0.866*\x1 -0.5*\y1);

% triangle DELTA
%\draw [line width=1pt] (X1)--(X2)--(X3)--(X1);

% interior bisectors 
\def \InScale {1}
\foreach \j in{1,2,3} 
{\path let \p1=(T\j) in coordinate (Y\j) at (\InScale*\x1,\InScale*\y1);
	\path let \p2=(Q\j) in coordinate (Z\j) at (\InScale*\x2,\InScale*\y2);}
%\draw (Y\j)--(Z\j);}

%exterior bisectors
\def \OutScale  {1}
\path let \p1=(X1) in coordinate (Ex23Up) at (\x1,\OutScale);
\path let \p1=(X1) in coordinate (Ex23Down) at (\x1,-\OutScale);

% rotate
\path let \p1=(Ex23Up) in coordinate (Ex13Up) at (-0.5*\x1  +0.866*\y1, -0.866*\x1 -0.5*\y1);
\path let \p1=(Ex23Down) in coordinate (Ex13Down) at (-0.5*\x1  +0.866*\y1, -0.866*\x1 -0.5*\y1);

\path let \p1=(Ex23Up) in coordinate (Ex12Up) at (-0.5*\x1  -0.866*\y1, 0.866*\x1 -0.5*\y1);
\path let \p1=(Ex23Down) in coordinate (Ex12Down) at (-0.5*\x1  -0.866*\y1, 0.866*\x1 -0.5*\y1);

%\draw (Ex23Up)--(Ex23Down);
%\draw (Ex13Up)--(Ex13Down);
%\draw (Ex12Up)--(Ex12Down);

\def \scaling {1.3}
\coordinate (q1) at ($(T1)!\scaling!(Q1)$);
\coordinate (q2) at ($(T2)!\scaling!(Q2)$);
\coordinate (q3) at ($(T3)!\scaling!(Q3)$);

\coordinate (t1) at ($(Q1)!\scaling!(T1)$);
\coordinate (t2) at ($(Q2)!\scaling!(T2)$);
\coordinate (t3) at ($(Q3)!\scaling!(T3)$);

\foreach \s in {1,2}
{ \def \z {\s+1}
\path let \p1=(X\s) in coordinate (X\z) at (-0.5*\x1  +0.866*\y1, -0.866*\x1 -0.5*\y1);}

% T/Q points
%%\filldraw  (capT1) circle (0pt) node [anchor=south]{$\T_1$};
%%\filldraw  (T1) circle (0.5pt);
%\filldraw  (T1) circle (0.7pt); % node [anchor=east]{\scriptsize $\T_1$};
%\filldraw  (T2) circle (0.7pt); % node [anchor=south]{\scriptsize$\,\,\T_2$};
%\filldraw  (T3) circle (0.7pt); % node [anchor=north]{\scriptsize$\,\,\T_3$};
%
%%\filldraw  (capQ1) circle (0pt) node [anchor=south]{$\Q_1$};
%%\filldraw  (Q1) circle (0.5pt);
%\filldraw  (Q1) circle (0.7pt); % node [anchor=west]{\scriptsize$\Q_1$};
%\filldraw  (Q2) circle (0.7pt); % node [anchor=north]{\scriptsize$\Q_2\,\,$};
%\filldraw  (Q3) circle (0.7pt); % node [anchor=south]{\scriptsize$\Q_3\,\,$};

% tangents
%\draw (t12)--(t13);
%\draw (t21)--(t23);
%\draw (t31)--(t32);

%arcs 
\draw [Bracket-Bracket, thick, name path=A1] (t12) arc (-{acos(\v)}:{acos(\v)}:1);
\draw [Bracket-Bracket, thick, name path=A2] (t23) arc (-{acos(\v) -120 }:{acos(\v) -120}:1);
\draw [Bracket-Bracket, thick, name path=A3] (t31) arc (-{acos(\v) -240 }:{acos(\v) -240}:1);

%\draw (t1) -- (q1);
%\draw (t2) -- (q2);
%\draw (t3) -- (q3);

\def \M {{-1/(2*\v) +   ((2*\v^2+1)* (\v^2+2+sqrt(4*\v^4+\v^2+4)*\v)  ) / (4*\v*(\v^4+\v^2+1))            }}

\def \N {   {    sqrt(1-(-1/(2*\v) +   ((2*\v^2+1)* (\v^2+2+sqrt(4*\v^4+\v^2+4)*\v)  ) / (4*\v*(\v^4+\v^2+1)))^2 )     }    }

\coordinate (p1) at (\M,\N);
\filldraw  (p1) circle (0.7pt);

%  intersection of arc A3 with the allonged line from Q1/v
\coordinate (PP) at ($(Q1v)!10!(p1)$);
\path [name path = line] (Q1v)--(PP);
%\path [name intersections={of=line and A3 ,by={p1,p2}}];
\path [name intersections={of=line and A3 ,by=p2}];

\path let \p1=(p1) in coordinate (p3) at (\x1,-\y1);
\path let \p1=(p2) in coordinate (p4) at (\x1,-\y1);

\foreach \j in {1,2,3,4}
{\filldraw (p\j) circle (0.8pt);}

\draw (p1)--(p2)--(p3)--(p4)--(p1);
\draw [-Latex,shorten >=1cm] (p1)--(p2);
\draw [-Latex,shorten >=1.6cm] (p2)--(p3);
\draw [-Latex,shorten >=1cm] (p3)--(p4);
\draw [-Latex,shorten >=1.6cm] (p4)--(p1);

%points
\filldraw [] (p1) circle (1.4pt);% node [anchor=south west]{$\,p$};
\filldraw [] (p3) circle (1.4pt);
\filldraw [light-gray] (p2) circle (1.4pt);
\filldraw [light-gray] (p4) circle (1.4pt);

\filldraw (p4) circle (0pt);% node [anchor=north east]{$q$};
						\end{tikzpicture}\label{period4_orbit_critical}}\qquad\qquad\quad
					\subfloat[$\frac12<v<1$]{\begin{tikzpicture}[scale=1.25]
							\definecolor{light-gray}{gray}{0.55}

\clip (-1.2,-1.2) rectangle (1.2,1.2);
%\clip (0,0) circle (1cm);
%\clip (-0.1,-1.8) rectangle (1.8,1.8);

%v = 0.36
\def \v {0.75}
\def \radius {1}

% Q points and Taub points
\foreach \s in {1,2,3}
{\coordinate (Q\s) at ({-120 * (\s - 1)}:\radius);}
\foreach \s in {1,2,3}
{\path let \p\s=(Q\s) in coordinate (T\s) at (-1*\x\s,-1*\y\s);}

\path let \p1=(Q1) in coordinate (Q1v) at (\x1/\v,\y1);

% Kasner circle
\draw [thin] (Q1) arc (0:360:\radius cm and \radius cm);	 	

% caption circle
%\node at (0.9*\radius, \radius*0.7) [anchor = west] {$\KC$};

% caption Q1 and T1
\path let \p1=(Q1) in coordinate (capQ1) at (1.15*\x1,\y1);
\path let \p1=(T1) in coordinate (capT1) at (1.15*\x1,\y1);

%coordinate system
%\draw [->] (0,0)--(0.55,0);
%\draw [->] (0,0)--(0,0.55);

%\draw  (0.6,-0.05) circle (0pt) node [anchor=south]{\scriptsize$\Sigma_+$};
%\draw  (0,0.5) circle (0pt) node [anchor=west]{\scriptsize$\Sigma_-$};

% tangent points
\coordinate (t13) at (\v,{sqrt(1-\v^2)});
\coordinate (t12) at (\v,-{sqrt(1-\v^2)});

%rotate about -120
\path let \p1=(t13) in coordinate (t21) at (-0.5*\x1  +0.866*\y1, -0.866*\x1 -0.5*\y1);
%rotate about -120
\path let \p1=(t12) in coordinate (t23) at (-0.5*\x1  +0.866*\y1, -0.866*\x1 -0.5*\y1);

%rotate about 120
\path let \p1=(t13) in coordinate (t32) at (-0.5*\x1  -0.866*\y1, 0.866*\x1 -0.5*\y1);
%rotate about 120
\path let \p1=(t12) in coordinate (t31) at (-0.5*\x1  -0.866*\y1, 0.866*\x1 -0.5*\y1);

% X - points : vertices of triangle DELTA
\coordinate (X1) at (-2*\v*\radius,0);
%rotate about -120
\path let \p1=(X1) in coordinate (X2) at (-0.5*\x1  +0.866*\y1, -0.866*\x1 -0.5*\y1);
%rotate about -120
\path let \p1=(X2) in coordinate (X3) at (-0.5*\x1  +0.866*\y1, -0.866*\x1 -0.5*\y1);

% triangle DELTA
%\draw [line width=1pt] (X1)--(X2)--(X3)--(X1);

% interior bisectors 
\def \InScale {1}
\foreach \j in{1,2,3} 
{\path let \p1=(T\j) in coordinate (Y\j) at (\InScale*\x1,\InScale*\y1);
	\path let \p2=(Q\j) in coordinate (Z\j) at (\InScale*\x2,\InScale*\y2);}
%\draw (Y\j)--(Z\j);}

%exterior bisectors
\def \OutScale  {1}
\path let \p1=(X1) in coordinate (Ex23Up) at (\x1,\OutScale);
\path let \p1=(X1) in coordinate (Ex23Down) at (\x1,-\OutScale);

% rotate
\path let \p1=(Ex23Up) in coordinate (Ex13Up) at (-0.5*\x1  +0.866*\y1, -0.866*\x1 -0.5*\y1);
\path let \p1=(Ex23Down) in coordinate (Ex13Down) at (-0.5*\x1  +0.866*\y1, -0.866*\x1 -0.5*\y1);

\path let \p1=(Ex23Up) in coordinate (Ex12Up) at (-0.5*\x1  -0.866*\y1, 0.866*\x1 -0.5*\y1);
\path let \p1=(Ex23Down) in coordinate (Ex12Down) at (-0.5*\x1  -0.866*\y1, 0.866*\x1 -0.5*\y1);

%\draw (Ex23Up)--(Ex23Down);
%\draw (Ex13Up)--(Ex13Down);
%\draw (Ex12Up)--(Ex12Down);

\def \scaling {1.3}
\coordinate (q1) at ($(T1)!\scaling!(Q1)$);
\coordinate (q2) at ($(T2)!\scaling!(Q2)$);
\coordinate (q3) at ($(T3)!\scaling!(Q3)$);

\coordinate (t1) at ($(Q1)!\scaling!(T1)$);
\coordinate (t2) at ($(Q2)!\scaling!(T2)$);
\coordinate (t3) at ($(Q3)!\scaling!(T3)$);

\foreach \s in {1,2}
{ \def \z {\s+1}
\path let \p1=(X\s) in coordinate (X\z) at (-0.5*\x1  +0.866*\y1, -0.866*\x1 -0.5*\y1);}

% T/Q points
%%\filldraw  (capT1) circle (0pt) node [anchor=south]{$\T_1$};
%%\filldraw  (T1) circle (0.5pt);
%\filldraw  (T1) circle (0.7pt); % node [anchor=east]{\scriptsize $\T_1$};
%\filldraw  (T2) circle (0.7pt); % node [anchor=south]{\scriptsize$\,\,\T_2$};
%\filldraw  (T3) circle (0.7pt); % node [anchor=north]{\scriptsize$\,\,\T_3$};
%
%%\filldraw  (capQ1) circle (0pt) node [anchor=south]{$\Q_1$};
%%\filldraw  (Q1) circle (0.5pt);
%\filldraw  (Q1) circle (0.7pt); % node [anchor=west]{\scriptsize$\Q_1$};
%\filldraw  (Q2) circle (0.7pt); % node [anchor=north]{\scriptsize$\Q_2\,\,$};
%\filldraw  (Q3) circle (0.7pt); % node [anchor=south]{\scriptsize$\Q_3\,\,$};

% tangents
%\draw (t12)--(t13);
%\draw (t21)--(t23);
%\draw (t31)--(t32);

%arcs 
\draw [Bracket-Bracket, thick, name path=A1] (t12) arc (-{acos(\v)}:{acos(\v)}:1);
\draw [Bracket-Bracket, thick, name path=A2] (t23) arc (-{acos(\v) -120 }:{acos(\v) -120}:1);
\draw [Bracket-Bracket, thick, name path=A3] (t31) arc (-{acos(\v) -240 }:{acos(\v) -240}:1);

%\draw (t1) -- (q1);
%\draw (t2) -- (q2);
%\draw (t3) -- (q3);

\def \M {{-1/(2*\v) +   ((2*\v^2+1)* (\v^2+2+sqrt(4*\v^4+\v^2+4)*\v)  ) / (4*\v*(\v^4+\v^2+1))            }}

\def \N {   {    sqrt(1-(-1/(2*\v) +   ((2*\v^2+1)* (\v^2+2+sqrt(4*\v^4+\v^2+4)*\v)  ) / (4*\v*(\v^4+\v^2+1)))^2 )     }    }

\coordinate (p1) at (\M,\N);
\filldraw  (p1) circle (0.7pt);

%  intersection of arc A3 with the allonged line from Q1/v
\coordinate (PP) at ($(Q1v)!10!(p1)$);
\path [name path = line] (Q1v)--(PP);
%\path [name intersections={of=line and A3 ,by={p1,p2}}];
\path [name intersections={of=line and A3 ,by=p2}];

\path let \p1=(p1) in coordinate (p3) at (\x1,-\y1);
\path let \p1=(p2) in coordinate (p4) at (\x1,-\y1);

\foreach \j in {1,2,3,4}
{\filldraw (p\j) circle (0.8pt);}

\draw (p1)--(p2)--(p3)--(p4)--(p1);
\draw [-Latex,shorten >=1cm] (p1)--(p2);
\draw [-Latex,shorten >=1.6cm] (p2)--(p3);
\draw [-Latex,shorten >=1cm] (p3)--(p4);
\draw [-Latex,shorten >=1.6cm] (p4)--(p1);

%points
\filldraw [] (p1) circle (1.4pt);% node [anchor=south west]{$\,p$};
\filldraw [] (p3) circle (1.4pt);
\filldraw [light-gray] (p2) circle (1.4pt);
\filldraw [light-gray] (p4) circle (1.4pt);

\filldraw (p4) circle (0pt);% node [anchor=north east]{$q$};
						\end{tikzpicture}\label{period4_orbit_supercritical}} \caption{An orbit of period four for all $v\in (0,1)$. Points with the same color are obtained by reflection across the $\Sigma_+$-axis and possess the same eigenvalue $\eva$, whereas the other two eigenvalues are interchanged.
						As $v\in (0,1)$ increases, the (black) points monotonically rotate towards $\Q_1$, whereas the (gray) points do not rotate monotonically. There are two symmetric chains obtained by rotations of $2\pi/3$ and $4\pi/3$.
						%Also, this chain converges to an unique chain of period four as $v\to 0$, whereas it collapses to two different chains of period two as $v\to 1$. 
					} \label{fig:per4}
				\end{figure}
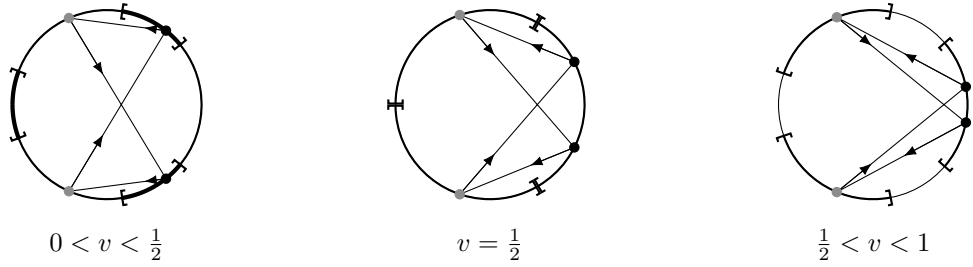

				\begin{lem}
					For $v \in (0,1)$ the Kasner map $\KM$ has exactly three orbits of minimal period four.
					%	On each of these orbits every second point is given by reflection along the $T_\alpha \Q_\alpha$-axis.
					%	The points on each orbit are reflectional symmetric with respect to exactly one the $T_\alpha \Q_\alpha$-axis.
					Each of these orbits is reflectional symmetric with respect to one of the $\T_\alpha\Q_\alpha$-axes.
					Consequently, the three orbits are obtained from each other by rotation by $\frac{2\pi}{3}$.
				\end{lem}
    				Each orbit of period four has two points in one arc and one point in each of the other two arcs, see \autoref{fig:per4}. We consider the orbit with two points in $\A_1(v)$.
				Thus, the orbit visits the arcs in the order $\A_1(v) \to \A_3(v) \to \A_1(v) \to \A_2(v)  \to \A_1(v)$. The two points in $\A_1(v)$ are symmetric with respect to the $\Sigma_+$-axis. Because $\KM$ is symmetric, the two other points are also symmetric.
				In contrast to the period three orbit, not every point is related by $D_3$-equivariance, but every second point on the period four orbit is symmetric. Thus, we investigate the eigenvalue structure of two consecutive points on the orbit.
				
				One of these two points, we call it $p$, represents the two points in $\A_1(v)$ and can be chosen from $\A_1^+(v)$. The other point represents the points in $\A_2(v)$ and $\A_3(v)$. We choose the point in $\A_2(v)$ and call it $q$.
				Coordinates of $p$ are obtained using the symmetry, i.e., by solving $\KM_3 \circ \KM_1 (p) = \left(\Sigma_+, -\Sigma_-\right)$. Similarly for $q$. The coordinates for $p$ and $q$ are, respectively,
                \begin{subequations}
				\begin{align} %\label{period4_first}
					\Sigma_+(p) &= -\frac{1}{2v} + 
					\frac{(2v^2+1)(v^2+2 + v\sqrt{4v^4+v^2+4})}{4v(v^4+v^2+1)}, \quad \Sigma_-(p) = \sqrt{1-\Sigma_+^2(p)},\\
     				\Sigma_+(q) &= -\frac{1}{2v} + 
					\frac{(2v^2+1)(v^2+2 - v\sqrt{4v^4+v^2+4})}{4v(v^4+v^2+1)}, \quad \Sigma_- (q)= -\sqrt{1-\Sigma_+^2(q)}.
				\end{align}   
                \end{subequations}
                %
				%
				%\textcolor{blue}{Rotating $q$ yields a representative in $\A_1^+(v)$. We choose this point to illustrate the corresponding sectors from Corollary \eqref{cor_subcritical_eigenvalues}, which are in $\A_1^+(v)$.}
                    %
                    \vspace{-0.3cm}
				%
				%Note that it is known that an orbit of minimal period four exists in the supercritical case, due to the symbolic dynamics construction in \cite{HL}. However, for the subcritical case a proof ... \\
				%Moreover, the above method yields orbits of any even minimal period, $2n$, by choosing the initial angle $\varphi_-\in \A_1$ such that $\KI^n(\varphi_-)=-\varphi_-$ for a particular sequences of maps and then repeating said sequence of maps, interchanging the role of $\KM_2$ and $\KM_3$, to obtain $\KI^{2n}$.
				%All these orbits of even period are symmetric with respect to reflection in $\Sigma_-\mapsto -\Sigma_-$. See for example the period six orbit in Figure
				%
				At $p$, the outgoing direction is $\partial_{N_1}$ and
				the incoming direction is $\partial_{N_2}$, and the lateral direction is $\partial_{N_3}$.
				%Consequently, unlike for the period three orbit, the incoming direction is along the weak stable direction.
				At $q$, the outgoing, incoming and lateral directions are respectively $\partial_{N_2},\partial_{N_1},\partial_{N_3}$, 
                see \autoref{period4_to_cycle_figure:b}. These directions are independent of $v$, but the ordering of the eigenvalues is not.
                Similar to the heteroclinic cycle of period three, the ordering of the eigenvalues in $p$ and $q$ is unchanged when $v \ge \frac 12$, see Lemma~\ref{cor_supercritical_eigenvalues}.
				For $v<\frac 12$, there are four bifurcations corresponding to the cases presented in Lemma~\ref{cor_subcritical_eigenvalues}. See \autoref{figure_period4_bif} for an overview. 
				%\textcolor{red}{are they really tangent?}
				%
				%\textcolor{red}{second points maps to left neighbor, not to the right as before}
				%
				%
				%\begin{figure}
				%	\centering
				%	\begin{tikzpicture}[scale=0.9]
					%		\input{figures/period4_eigenvalues}
					%	\end{tikzpicture}
				%	\caption{The eigenvalues in $p$ are in black, and the eigenvalues in $q$ are in gray. The eigenvalues corresponding to the incoming heteroclinic direction are bold, and the eigenvalues corresponding to outgoing heteroclinic direction are dashed.} \label{figure_period4_bif}
				%\end{figure}
				%		
                    \vspace{-0.5cm}
				\begin{figure}[H]
					\centering
				\subfloat[][The induced heteroclinic cycle between the two consecutive (black and gray) points, denoted by $p$ and $q$. The incoming and outgoing directions are independent of $v$. The local stability, however, depends on $v$.]	
                {\label{period4_to_cycle_figure:b}
                \begin{tikzpicture}[scale=1.1]
				\definecolor{light-gray}{gray}{0.7}
\coordinate (LOrigin) at (-2,0);
\coordinate (ROrigin) at (2,0);

%\draw (-3,0) circle (0pt);
%\draw (3,0) circle (0pt);

\filldraw [] (LOrigin) circle (1.4pt);
\filldraw [light-gray] (ROrigin) circle (1.4pt);

\def \lateral {0.6}
\def \up {0.8}

\path let \p1=(LOrigin) in coordinate (LUp) at (\x1,\y1+\up cm);
\path let \p1=(ROrigin) in coordinate (RUp) at (\x1,\y1+\up cm);

% lateral lines
\path let \p1=(LOrigin) in coordinate (LLateralB) at (\x1-\lateral cm,\y1-\lateral cm);
\path let \p1=(LOrigin) in coordinate (LLateralT) at (\x1+\lateral cm,\y1+\lateral cm);

\path let \p1=(ROrigin) in coordinate (RLateralB) at (\x1+\lateral cm,\y1-\lateral cm);
\path let \p1=(ROrigin) in coordinate (RLateralT) at (\x1-\lateral cm,\y1+\lateral cm);

%dotted
\draw [dotted] (LOrigin)--(ROrigin);

\draw  (-0.5,0)--(LOrigin);
\draw  (ROrigin)--(0.5,0);
\draw [-Latex]  (-0.7,0)--(-1.5,0);
\draw [-Latex]  (ROrigin)--(1.3,0);

\draw (LOrigin)--(LUp);
\draw (ROrigin)--(RUp);
\draw (LLateralB)--(LLateralT);
\draw (RLateralB)--(RLateralT);

%dotted
\draw [dotted] (LUp) to[ bend left=90] (RUp);

\draw [-_,shorten >=3.2cm] (LUp) to [bend left=90] (RUp);
\draw [_-,shorten <=3.2cm] (LUp) to [bend left=90] (RUp);

%left and right arrow
\draw [-Latex,shorten >=4.5cm] (LUp) to [bend left=90] (RUp);
\draw [-Latex,dotted, shorten >=0.4cm] (LUp) to [bend left=90] (RUp);

\coordinate (A1) at (-2.2,-0.4);
\coordinate (B1) at (-1.8,0.4);
\draw [] (A1)--(B1) node[ anchor=south] {\scriptsize$\KC$};
\def \T {10}
\foreach \i in {0,...,\T}
{\coordinate (A1\i) at ($(A1)!\i/\T!(B1)$);
	%	\path let \p1=(A) in coordinate (A\i) at (\i/6*\x1, 2.5/6*\i*\y1); 
	\filldraw [] (A1\i) circle (0.6pt);}

\coordinate (A1) at (2.1,-0.4);
\coordinate (B1) at (1.9,0.4);
\draw [] (A1)--(B1) node[ anchor=south] {\scriptsize$\KC$};
\def \T {10}
\foreach \i in {0,...,\T}
{\coordinate (A1\i) at ($(A1)!\i/\T!(B1)$);
	%	\path let \p1=(A) in coordinate (A\i) at (\i/6*\x1, 2.5/6*\i*\y1); 
	\filldraw [] (A1\i) circle (0.6pt);}

% arrows

%%Left
%%lateral
%\draw [-Latex, shorten >=0.3cm] (LLateralT)--(LOrigin);
%\draw [-Latex, shorten >=0.5cm] (LLateralT)--(LOrigin);
%
%\draw [-Latex, shorten >=0.3cm] (LLateralB)--(LOrigin);
%\draw [-Latex, shorten >=0.5cm] (LLateralB)--(LOrigin);
%
%
%% incoming
%\draw [-Latex, shorten >=1cm] (ROrigin)--(LOrigin);
%\draw [-Latex, shorten >=0.7cm] (ROrigin)--(LOrigin);
%\draw [-Latex, shorten >=0.4cm] (ROrigin)--(LOrigin);
%
%%outgoing
%\draw [-Latex, shorten >=0.2cm] (LOrigin)--(LUp);
%
%%Right
%%lateral
%\draw [-Latex, shorten >=0.3cm] (RLateralT)--(ROrigin);
%\draw [-Latex, shorten >=0.5cm] (RLateralT)--(ROrigin);
%\draw [-Latex, shorten >=0.7cm] (RLateralT)--(ROrigin);
%
%\draw [-Latex, shorten >=0.3cm] (RLateralB)--(ROrigin);
%\draw [-Latex, shorten >=0.5cm] (RLateralB)--(ROrigin);
%\draw [-Latex, shorten >=0.7cm] (RLateralB)--(ROrigin);
%
%% incoming
%\draw [-Latex, shorten >=0.3cm] (RUp)--(ROrigin);
%\draw [-Latex, shorten >=0.6cm] (RUp)--(ROrigin);
%
%%outgoing
%\draw [-Latex, shorten >=4cm] (ROrigin)--(LOrigin);

% captions
% Left
\draw (LLateralB) node [anchor=west] {\scriptsize $\, N_3$};
\draw (LUp) node [anchor=east] {\scriptsize $ N_1$};
\draw (-1,-0.05) node [anchor=south] {\scriptsize  $N_2$};
\draw (LOrigin) node [anchor=east] {\scriptsize $p$};

% Right
\draw (RLateralB) node [anchor=east] {\scriptsize $ N_3\,$};
\draw (RUp) node [anchor=west] {\scriptsize $ N_1$};
\draw (1,-0.05) node [anchor=south] {\scriptsize  $N_2$};
\draw (ROrigin) node [anchor=west] {\scriptsize $q$};

%\coordinate (A) at (-0.6,-1.5);
%\coordinate (B) at (0.6,1.5);
%\draw [gray] (A)--(B) node[ anchor=west] {$\xc$};
%\def \T {16}
%\foreach \i in {0,...,\T}
%{\coordinate (A\i) at ($(A)!\i/\T!(B)$);
%	%	\path let \p1=(A) in coordinate (A\i) at (\i/6*\x1, 2.5/6*\i*\y1); 
%	\filldraw [gray] (A\i) circle (1pt);}

%  font sizes
% \small
% \footnotesize
% \scriptsize
% \tiny

\filldraw [light-gray] (ROrigin) circle (1.4pt);
			\end{tikzpicture}}
                \qquad
                \subfloat[][ %The eigenvalues in $p$ are in black, and the eigenvalues in $q$ are in gray.
					The eigenvalues corresponding to the incoming heteroclinic direction are bold, the eigenvalues corresponding to the outgoing heteroclinic direction are dashed. At $v=v_3\approx 0.2532$ the eigenvalue $\evc(p,v)$ changes its sign, which implies that $p$ leaves the overlap, as $v$ increases.] 
                {\label{figure_period4_bif} 
			\begin{tikzpicture}[scale=0.5]
				%arccos(0.8) = 0.6435
% first term  -2 tan^(-1)(((1 + v) tan(arccos(0.25)/2))/(-1 + v)) = 2.85665
%\clip [draw] (1.5,1.5) circle (4.5cm);
\definecolor{light-gray}{gray}{0.8}
\pgfplotsset{
	compat=newest,
	every axis plot/.append style={no marks},
	every axis/.style={
		axis lines=middle,
		width=14cm,
		height=6.5cm,
	}
}

\begin{axis}
	[axis lines = center,
	domain=0,
	xmin= -0.05,
	xmax = 1.05,
	ymin = -0.2,
	ymax = 6.2,
	samples=100,
	trig format plots=rad,
	xtick={0,0.191,0.252,0.325, 0.439,0.5,1},
	xticklabels={ $0$,$v_4$,$v_3$, $v_2$, $v_1$,$\frac 12$, 1},
	ytick={0,1,2,3,4,5,6},
	yticklabels={$0$,$1$,$2$, $3$,$4$,$5$, $6$}
	]

	\def \M {{  (-1/(2*x) +   ((2*x^2+1)* (x^2+2+sqrt(4*x^4+x^2+4)*x)  ) / (4*x*(x^4+x^2+1))  )          }}
	
	\def \N {   {    (sqrt(1-(-1/(2*x) +   ((2*x^2+1)* (x^2+2+sqrt(4*x^4+x^2+4)*x)  ) / (4*x*(x^4+x^2+1)))^2))     }    }

	%	Sigma1 = ((sqrt(3)*sqrt(4-x^2)+x)/4);
	%	Sigma2 = sqrt(1-((sqrt(3)*sqrt(4-x^2)+x)/4)^2);
	%	nu1(i) = 4*(Sigma1_1-v);
	%   nu2(i) = -4*x - 2*Sigma1 - 2*sqrt(3)*Sigma2;
	%   nu3(i) = -4*x - 2*Sigma1 + 2*sqrt(3)*Sigma2;

	\addplot [line width = 0.9pt, dashed, domain=-0.001:1, smooth] {abs(4*(-1/(2*x) +   ((2*x^2+1)* (x^2+2+sqrt(4*x^4+x^2+4)*x)  ) / (4*x*(x^4+x^2+1))   -x)};
	\addplot [line width = 1.6pt,domain=-0.001:1, smooth] {abs(-4*x - 2*(-1/(2*x) +   ((2*x^2+1)* (x^2+2+sqrt(4*x^4+x^2+4)*x)  ) / (4*x*(x^4+x^2+1))  ) - 2*sqrt(3)*(sqrt(1-(-1/(2*x) +   ((2*x^2+1)* (x^2+2+sqrt(4*x^4+x^2+4)*x)  ) / (4*x*(x^4+x^2+1)))^2)))  };
	\addplot [line width = 0.8pt, domain=-0.001:1, smooth] {abs(-4*x - 2*(-1/(2*x) +   ((2*x^2+1)* (x^2+2+sqrt(4*x^4+x^2+4)*x)  ) / (4*x*(x^4+x^2+1))  ) + 2*sqrt(3)*(sqrt(1-(-1/(2*x) +   ((2*x^2+1)* (x^2+2+sqrt(4*x^4+x^2+4)*x)  ) / (4*x*(x^4+x^2+1)))^2)))  };

	\addplot[line width = 1.6pt, light-gray, domain=-0.001:1, smooth] {abs(4*(-1/(2*x) +   ((2*x^2+1)* (x^2+2-sqrt(4*x^4+x^2+4)*x)  ) / (4*x*(x^4+x^2+1))   -x)};
	\addplot [line width = 0.8pt,  light-gray,domain=-0.001:1, smooth] {abs(-4*x - 2*(-1/(2*x) +   ((2*x^2+1)* (x^2+2-sqrt(4*x^4+x^2+4)*x)  ) / (4*x*(x^4+x^2+1))  ) - 2*sqrt(3)*(sqrt(1-(-1/(2*x) +   ((2*x^2+1)* (x^2+2-sqrt(4*x^4+x^2+4)*x)  ) / (4*x*(x^4+x^2+1)))^2)))  };
	\addplot [line width = 1pt, dashed, light-gray,domain=-0.001:1, smooth] {abs(-4*x - 2*(-1/(2*x) +   ((2*x^2+1)* (x^2+2-sqrt(4*x^4+x^2+4)*x)  ) / (4*x*(x^4+x^2+1))  ) + 2*sqrt(3)*(sqrt(1-(-1/(2*x) +   ((2*x^2+1)* (x^2+2-sqrt(4*x^4+x^2+4)*x)  ) / (4*x*(x^4+x^2+1)))^2)))  };

	%	\addplot [ultra thick] coordinates {(0,0)(0.11,0)};
	%	\addplot [ultra thick] coordinates {(0.22,0)(0.33,0)};
	%	\addplot [ultra thick] coordinates {(0.66,0)(0.77,0)};
	%	\addplot [ultra thick] coordinates {(0.88,0)(1,0)};

%\addplot [loosely dotted] coordinates {(0.191,0)(0.191,3.215)};
%%\addplot [dotted] coordinates {(0.252,0)(0.252,6)};
%\addplot [dotted] coordinates {(0.325,0)(0.325,2.67)};
%\addplot [dotted] coordinates {(0.439,0)(0.439,1.64)};
	
%	\coordinate (A) at (0.75,0.16);
%	\coordinate (B) at (0.75,3.92);
%	\coordinate (C) at (0.75,4.78);

\end{axis}

\definecolor{light-gray}{gray}{0.65}
\coordinate (A1) at (12.5,4);
\coordinate (A2) at (12.5,3.5);
\coordinate (A3) at (12.5,3);
\coordinate (A4) at (12.5,1.8);
\coordinate (A5) at (12.5,1.3);
\coordinate (A6) at (12.5,0.8);

\foreach \i in {1,2,3,4,5,6}
{\path let \p1=(A\i) in coordinate (B\i) at (\x1+0.8cm, \y1); }

\draw [line width = 1.7pt] (A1)--(B1) node [anchor=west] {\tiny$\,-\evb(p)$};
\draw [dashed, line width = 1pt] (A2)--(B2) node [anchor=west]  {\tiny$\quad\eva(p)$};
\draw [line width = 0.9pt] (A3)--(B3) node [anchor=west] {\tiny$\, \shortquad|\evc(p)|$};

\draw [light-gray, line width = 1.9pt] (A4)--(B4); % node [anchor=west] {\small$\,-\eva(q)$};
\draw [light-gray, line width = 1.2pt, dashed] (A5)--(B5); % node [anchor=west] {\small$\,\evb(q)$};
\draw [light-gray, line width = 1.2pt] (A6)--(B6); % node [anchor=west] {\small$\,-\evc(q)$};

\draw [] (B4) node [anchor=west] {\tiny$\,-\eva(q)$};
\draw [] (B5) node [anchor=west] {\tiny$\quad\evb(q)$};
\draw [] (B6) node [anchor=west] {\tiny$\,-\evc(q)$};
			\end{tikzpicture}}\caption{The orbit of period four yields an orbit of period two in the orbit space $\KC/D_3$. %Thus, it represents a heteroclinic cycle of length two.
			}
                \label{period4_to_cycle_figure:full}
			\end{figure}
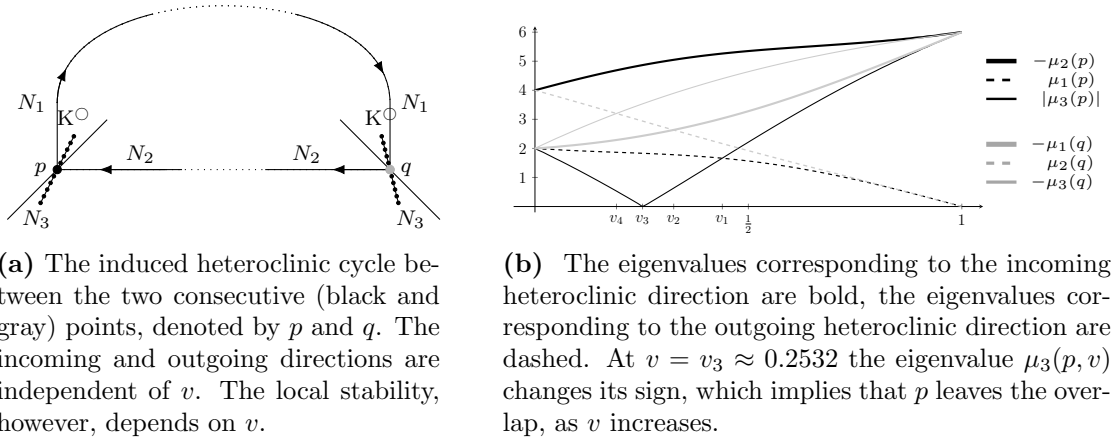 

				We discuss these bifurcations in decreasing order with $v$. They occur alternately at the two points $p$ and $q$, i.e. with each bifurcation exactly one local passage changes its dynamics.
				The bifurcation values of $v$ are given in terms of rather complicated expressions, obtained by equating the respective eigenvalues \eqref{eigenvalues}; thus we only give their numerical approximations.% in all four cases.%, and exemplarily for $v_3$ we give the exact formula.
				%Unfortunately, Galois theory claims that there is no general formula to solve a polynomial equation in terms of \textbf{radicals for such high degree polynomials}}

			%\bigskip
			%\textbf{case 1}
			%\smallskip
			
			The first change occurs at $p$, when the eigenvalues corresponding to the outgoing and the lateral direction switch order: the expansion in outgoing direction $\partial_{N_1}$ becomes stronger than the contraction in lateral direction $\partial_{N_3}$. In the notation of Lemma~\ref{cor_subcritical_eigenvalues}, $p$ enters the sector $\Mid$, which occurs when $\eva(p,v) = -\evc(p,v)$ at an approximate value given by
			\begin{equation}
				v_1 \approx 0.4387.
			\end{equation}
			The second change occurs at $q$, when the eigenvalues corresponding to the outgoing direction $\partial_{N_2}$ and the incoming direction $\partial_{N_1}$ switch order. Thus $q$ enters the sector $\Mid$ as well, i.e. $-\eva(q,v) = \evb(q,v)$ at approximate value
                \begin{equation}
				v_2 \approx 0.3246.
			\end{equation}
			Note, however, that in $p$ the incoming heteroclinic solution enters along the non-principal direction, whereas in $q$ the incoming heteroclinic solution enters along the principal direction.
			%
			%$v_*\in (0,1)$. Indeed, if one linearizes the ODE \eqref{full:subs} at some point of period four chain, then one of the eigenvalue changes sign at
   
			The third change occurs at $p$, when the eigenvalue $\evc(p,v)$ changes sign and thereby changes the dimension of the stable and unstable directions.
			%i.e. this eigenvalue is negative for $v\in (v_3,1)$ and positive for $v\in (0,v_3)$. 
			%$v_3$ is given by a root of the following polynomial,
			Two points of the period four orbit enter the overlap $\A_{\alpha\beta}(v)$, which occurs when $\evc(p,v) = 0$ at approximate value        
            %This occurs when $\evc(p,v) = 0$ at approximate value %of $v$ which satisfies
			%\begin{equation}\label{poly:v_3}
			%	v\left(v\left(4v^3+ \sqrt{4v^4+v^2+4}+7v\right)+7\right) = \sqrt{4v^4+v^2+4}.
			%\end{equation}
			%Unfortunately, Galois theory claims that there is no general formula to solve a polynomial equation in terms of radicals for such high degree polynomials.
			%Unfortunately, Galois theory \textcolor{red}{?} claims that there is no general formula to solve a polynomial equation in terms of radicals for such high degree polynomials, and thus
			%Numerically, we obtain
			\begin{equation} %\label{v_3}
				v_3\approx 0.2532,
			\end{equation}
			%Note that \eqref{poly:v_*} is a polynomial of degree 10. Numerically we obtain eight complex and two real roots: besides $v_* $ given by \eqref{v_*}, the other real root is approximately $-0.288597$, which is not in $(0,1)$.
			%
            which implies that this chain is in the set $\tilde{\mathcal{C}}$ for $v>v_3$ and in $\tilde{\mathcal{F}}$ for $v<v_3$, defined in \eqref{setsCF}.
        %
			%This means that all points of the period four orbit lie outside the overlap $\A_{\alpha\beta}(v)$ for $v\in (v_3,1)$ and two points of the period four lie inside the overlap region for $v\in (0,v_3)$. Accordingly, the period four chain is in the set $\tilde{C}$ for $v>v_3$ and in $\tilde{F}$ for $v<v_3$, which are defined in \eqref{setsCF}.
			%For $v<v_3$, this chain has two points inside the (bold) overlap region, but all points are outside these overlaps for $v>v_3$, and thus $v=v_3$ indicates the transition of the chain of period four between the sets in \eqref{setsCF}. 
			
			The last change occurs at $q$, when the expansion rate of the outgoing direction $\partial_{N_2}$ is larger than both stable contraction rates. Thus, $q$ transitions from $\Mid$ to $\Unstab$, i.e. when $\evb(q,v) = -\evc(q,v) $, at approximate value 
                \begin{equation}
                v_4 \approx 0.1908.    
                \end{equation}
			Therefore, the ordering among the stable eigenvalues at $q$ is not affected by the parameter. The incoming direction is always along the principal (weak stable) direction. The unstable eigenvalue increases with decreasing $v$, from being smallest to largest in absolute value.
			At $p$, the ordering among the incoming and the outgoing direction is not affected. For larger $v$, the lateral contraction is stronger than the expansion along the outgoing direction. With decreasing $v$, this ordering changes and finally the lateral component becomes unstable.

   	\subsection{Robustness of GR heteroclinic chains}\label{sec:periodic}

	It is well known that the critical GR case, $v = 1/2$, exhibits periodic heteroclinic chains of Bianchi type II solutions for any minimal period $n \ge 3$, see~\cite{HellLappicyUggla}.
	In Lemma~\ref{proposition_deformation}, we show that all of these orbits persist for $v$ sufficiently close to $1/2$. Thus, the bifurcation at $v=1/2$ does not affect any of these periodic chains, but it may affect the aperiodic chains. % such as the chains that end at one of the Taub points.
	%Moreover, Proposition \ref{proposition_unique} assures uniqueness of periodic chains associated with a given sequence of maps.% $\{a_i\}_{i=1}^n$ it is sufficient to show existence. 
	
	Recall that the Kasner map $\KM$ is well-defined for $v\in [1/2,1)$, but it is multivalued for $v\in(0,1/2)$.
	Thus a point might have several images in the subcritical case, and we have to clarify what is meant by periodicity. We say a point $p \in \KC$ \emph{realizes a periodic orbit} of minimal period $n\geq 2$, if $\KM^n(p) := \KM_{a_{n}} \circ ... \circ \KM_{a_1}(p) = p$ for some sequence of maps described by the finite sequence $\left({a_i}\right)_{1\le i \le n}$.
	Furthermore we enumerate the points in the periodic orbit in the order of appearance. Without loss of generality, due to equivariance, the sequence can start with $a_1=1$ such that $p=p_1 \in A_1$ and $p_{a_i} \in A_{a_i}$ for each $\left({a_i}\right)_{1\le i \le n}$. 
	We denote by $\varphi_i$ the corresponding angle in polar coordinates of each point $p_{a_i} \in \KC$.
	\begin{lem}\label{proposition_deformation}
		For every $n\in \NN, n \ge 3,$ there exists $\varepsilon>0$ such that a periodic orbit of minimal period $n$ exists for any $v \in (1/2 -\varepsilon, 1/2+\varepsilon)$.
	\end{lem}
	\begin{proof}
		Fix $n \ge 3$. For $v=1/2$ there exists a point $p^\star \in \KC$ (with corresponding angle $\varphi^\star$) which is a periodic orbit of minimal period $n$ for some sequence of maps $\{a_i\}_{i=1}^n$, i.e. $\KI^n(\varphi^\star) := \KI_{a_n} \circ ... \circ \KI_{a_1}(\varphi^\star) =\varphi^\star$. 
		Denote the points in the periodic orbit by their angle variable, according to $\{\varphi_i\}_{i=1}^n$.
		Since $\KI$ depends both on the angle $\varphi_1\in \KC$ and parameter $v\in (0,1)$, finding a point that realizes a periodic orbit with angle $\varphi$ is equivalent to solving
		\begin{equation}\label{IFT}
			\KI^n(\varphi) - \varphi =: F(v,\varphi) = 0
		\end{equation}
		In particular, $F(1/2,\varphi^\star)=0$ for any periodic orbit starting at $\varphi^\star$ of minimal period $n$.
		By chain rule and noticing that each derivative for $\psi_1 \in N(\varphi_1) $ is described after \eqref{KM1}, where $N(\varphi_1)$ is a sufficiently small neighborhood of $\varphi_1$, we obtain
		\begin{align}\label{derivative_chainrule}
			D\KI^n(\psi_1) = D\left(\KI_{a_n} \circ ... \circ \KI_{a_1}\right) (\psi_1) =\prod_{i=1}^{n} D\KI_{a_i} (\psi_i) = (-1)^n \prod_{i=1}^{n} g_{a_i} (\psi_i).
		\end{align}
		Hence, $\left|D\KI^n (\psi_1) \right| > 1$,
		%\begin{equation}\label{derivative_strict_inequality}
		%	\left|D\KI^n (\psi_1) \right| = \left|\prod_{i=1}^{n} g_{a_i} (\psi_i)\right| > 1,
		%\end{equation}
		since each $|g_{a_i}(\psi_i)| > 1$. Therefore 
		%Furthermore we use \eqref{derivative_strict_inequality} in order to obtain that
		\begin{equation}\label{IFT_derivative}
			\frac{\partial F}{\partial \varphi} = \frac{\partial\KI^n}{\partial \varphi} -1 = (-1)^n \prod_{i=1}^{n} g_{a_i} (\varphi_i)-1 \neq 0.
		\end{equation}
		The implicit function theorem implies that there exists $\varepsilon > 0$ such that $F(v,\varphi(v)) = 0$ for $v \in (1/2 -\varepsilon, 1/2+\varepsilon)$ and consequently $\varphi(v)$ is a point that realizes a periodic orbit of minimal period $n$ with sequence of maps $\{a_i\}_{i=1}^n$.
	\end{proof}
	In particular, Lemma~\ref{proposition_deformation} shows that all periodic orbits in GR persist under small perturbations in Ho{\v{r}}ava models. Therefore, all periodic chains (except the period 2) in the Cantor set for $v>1/2$ converge to a periodic chain at $v=1/2$, due to  \cite[Appendix C]{HellLappicyUggla}, and thereby %survive across the bifurcation at $v=1/2$ and 
	remain existing for $v<1/2$.
	We believe a similar result should hold for perturbations of aperiodic chains that are bounded away from the Taub points, when $v=1/2$; however, it is not clear what happens with the chains that end up at the Taub points, and the chains that get arbitrarily close to the Taub points, through the bifurcation point $v=1/2$. 
        In particular, the question of persistence of all heteroclinic chains in the Cantor set, except the period 2 orbit, remains open as the bifurcation value $v=1/2$ is crossed.
        %In turn, this implies that the full structure of the Cantor set persists across the bifurcation, and the bifurcation at $v=1/2$ only changes the structure of the chains that end at one of the Taub points.
    %
        Even though we do not prove that the full structure of the Cantor set persists across $v=1/2$, it is proven in \cite[Appendix C]{HellLappicyUggla} that all the heteroclinic chains associated with the Cantor set persist in the limiting case, as $v\to 1/2$, except for the period 2 chain that converges to the Taub point.

   \section{Stable manifolds of heteroclinic chains} \label{chap:shadowing}
						
			In this chapter we investigate the dynamics of system \eqref{full:subs} along an infinite heteroclinic chain $(h_k(t))_{k \in \NN_0}$ given by a sequence of equilibria $(p_k)_{k\in \NN_0}$ on the Kasner circle $\KC$, and obtained by iteration of the Kasner map $\KM$.
			Our consideration is restricted to the supercritical case $v \in \left(\frac 12, 1\right)$. The infinite heteroclinic chains in this case are associated with the Cantor set $\Cantor(v)$, as described in \eqref{Cantor_def}.
			We prove the existence of three-dimensional stable manifolds for all heteroclinic chains in $\Cantor(v)$. The main result is as follows.

				\begin{thm}\label{manifold_theorem}
					Let $v\in (\frac 12 ,1)$. Every infinite heteroclinic chain $(h_k)_{k \in \NN_0}$ of Bianchi type II associated with the Cantor set $\Cantor(v)$ has a codimension-one (three-dimensional) Lipschitz stable manifold of Bianchi type $\mathrm{VIII}$ and $\mathrm{IX}$ solutions depending continuously on $v\in (\frac 12 ,1)$. 
                    Moreover, the union of all stable manifolds associated with the Cantor set amounts to a continuous foliation, where the leaves are the three-dimensional Lipschitz stable manifolds that depend continuously on $p\in \mathcal{C}(v)$. We denote it by $\mathcal{W}^s(\mathcal{C}(v)):=\cup_{p\in \mathcal{C}(v)} W^s((\KM^k(p))_{k \in \NN_0})$. 
				\end{thm}
			Our proof of stable manifolds of heteroclinic chains follows the critical case $v = \frac 12$ of general relativity in \cite{Liebscher}.
			Most features of system \eqref{full:subs} are not affected by the parameter $v$, such as existence of the Kasner circle of equilibria and an invariant constraint with a hierarchy of invariant subsets.
			An important assumption relates to the eigenvalues of the linearization in each point along the sequence $(p_k)_{k\in\NN_0}$. The ordering among the eigenvalues is not always favorable in the subcritical case, but it is in the supercritical case; see the Lemmata~\ref{cor_supercritical_eigenvalues} -~\ref{cor_subcritical_eigenvalues}.
			
			The proof in \cite{Liebscher} fails for most heteroclinic chains in general relativity, since a generic orbit comes arbitrarily close to a Taub point, where the eigenvalue assumption is violated, % as it possess only one nonzero eigenvalue, 
            and therefore hyperbolicity becomes arbitrarily weak.
			Thus, the proof is only applicable to chains which keep a uniform distance to the Taub points.
			The refined method in \cite{Dutilleul} allows orbits of $\KM$ which come arbitrarily close to the Taub point, but ``not too fast".
			For fixed $v>\frac 12$, however, the Cantor set $\Cantor$
			%, where all points corresponding to infinite heteroclinic chains are located,
			is bounded away from points with zero eigenvalues. Thus, the method of \cite{Liebscher} is applicable to \emph{all} heteroclinic chains in the supercritical case. We adapt their proof to the case of a parameter dependent vector field. The proof is split in two parts, a local and a global analysis. We provide a sketch of the proof.

                \subsubsection*{Outline local analysis.}
                
			%\bigskip
			%\vspace{-0.06cm} \textbf{Outline local analysis.}
			%\smallskip
			
\begin{figure}[H]
				\centering
				\subfloat[We investigate the local neighborhood of a point $p=p_k \in \Cantor(v)$ along an infinite heteroclinic chain $(h_k)_{k\in \NN_0}$.]	{\begin{tikzpicture}[scale=1.25]
						\clip (-1.6,-1.5) rectangle (1.6,1.4);
%\clip (0,0) circle (1cm);
%\clip (-0.1,-1.8) rectangle (1.8,1.8);

%v = 0.36
\def \v {0.7}
\def \radius {1}

% Q points and Taub points
\foreach \s in {1,2,3}
{\coordinate (Q\s) at ({-120 * (\s - 1)}:\radius);}
\foreach \s in {1,2,3}
{\path let \p\s=(Q\s) in coordinate (T\s) at (-1*\x\s,-1*\y\s);}

\path let \p1=(Q1) in coordinate (Q1v) at (\x1/\v,\y1);
\path let \p1=(Q2) in coordinate (Q2v) at (\x1/\v,\y1/\v);
\path let \p1=(Q3) in coordinate (Q3v) at (\x1/\v,\y1/\v);

% Kasner circle
\draw [thin] (Q1) arc (0:360:\radius cm and \radius cm);	 	

% caption circle
%\node at (0.9*\radius, \radius*0.7) [anchor = west] {$\KC$};

% caption Q1 and T1
\path let \p1=(Q1) in coordinate (capQ1) at (1.15*\x1,\y1);
\path let \p1=(T1) in coordinate (capT1) at (1.15*\x1,\y1);

% tangent points
\coordinate (t13) at (\v,{sqrt(1-\v^2)});
\coordinate (t12) at (\v,-{sqrt(1-\v^2)});

%rotate about -120
\path let \p1=(t13) in coordinate (t21) at (-0.5*\x1  +0.866*\y1, -0.866*\x1 -0.5*\y1);
%rotate about -120
\path let \p1=(t12) in coordinate (t23) at (-0.5*\x1  +0.866*\y1, -0.866*\x1 -0.5*\y1);

%rotate about 120
\path let \p1=(t13) in coordinate (t32) at (-0.5*\x1  -0.866*\y1, 0.866*\x1 -0.5*\y1);
%rotate about 120
\path let \p1=(t12) in coordinate (t31) at (-0.5*\x1  -0.866*\y1, 0.866*\x1 -0.5*\y1);

%arcs 
\draw [Bracket-Bracket, thick, name path=A1] (t12) arc (-{acos(\v)}:{acos(\v)}:1);
\draw [Bracket-Bracket, thick, name path=A2] (t23) arc (-{acos(\v) -120 }:{acos(\v) -120}:1);
\draw [Bracket-Bracket, thick, name path=A3] (t31) arc (-{acos(\v) -240 }:{acos(\v) -240}:1);

\def \M {0.9243}

\def \N { -0.2641 }

\coordinate (p0) at (\M,\N);
\path [name path = line0] (Q1v)--(p0);
\path [name intersections={of=line0 and A1 ,by=p1}];

%  intersection of arc A3 with the allonged line from Q1/v
\coordinate (PP) at ($(Q1v)!10!(p1)$);
\path [name path = line1] (Q1v)--(PP);
\path [name intersections={of=line1 and A2 ,by=p2}];

\coordinate (QQ) at ($(Q2v)!10!(p2)$);
\path [name path = line2] (Q2v)--(QQ);
\path [name intersections={of=line2 and A1 ,by=p3}];

\coordinate (RR) at ($(Q1v)!10!(p3)$);
\path [name path = line3] (Q1v)--(RR);
\path [name intersections={of=line3 and A3 ,by=p4}];

\coordinate (SS) at ($(Q3v)!10!(p4)$);
\path [name path = line4] (Q3v)--(SS);
\path [name intersections={of=line4 and A2 ,by=p5}];

\coordinate (TT) at ($(Q2v)!10!(p5)$);
\path [name path = line5] (Q2v)--(TT);
\path [name intersections={of=line5 and A3 ,by=p6}];

%\coordinate (UU) at ($(Q3v)!10!(p6)$);
%\path [name path = line6] (Q3v)--(UU);
%\path [name intersections={of=line6 and A1 ,by=p7}];

%\path let \p1=(p1) in coordinate (p3) at (\x1,-\y1);
%\path let \p1=(p2) in coordinate (p4) at (\x1,-\y1);

%\foreach \j in {1,2,3,4,5,6,7}
%{\filldraw (p\j) circle (0.8pt) node [anchor = west]{$p\j$};}
%\filldraw (Q1v) circle (0.8pt);
%\filldraw (Q2v) circle (0.8pt);
%\filldraw (Q3v) circle (0.8pt);

\filldraw (p3) circle (0.8pt);
\draw (p3)--(p4);
\coordinate (mid) at ($(p2)!0.55!(p3)$);
\coordinate (X) at ($(p2)!0.3!(p3)$);
\coordinate (Y) at ($(p4)!0.85!(p5)$);
\draw [dashed] (X)--(p3);
\draw [dashed] (p4)--(Y);

\draw [] (mid)--(p3);
\draw [-Latex,shorten >=0.8cm] (mid)--(p3);

%\draw [-Latex,shorten >=1cm] (p2)--(p3);
\draw [-Latex,shorten >=1cm] (p3)--(p4);
\draw [-Latex,shorten >=1.2cm] (p4)--(p5);
%\draw [-Latex,shorten >=1cm] (p5)--(p6);
%\draw [dashed,-Latex, shorten >=1.5cm] (p6)--(p7);
%\draw [dashed,shorten >=1cm] (p6)--(p7);
%\draw [-Latex,shorten >=1.6cm] (p2)--(p3);
%\draw [-Latex,shorten >=1cm] (p3)--(p4);
%\draw [-Latex,shorten >=1.6cm] (p4)--(p1);

\path let \p1=(p3) in coordinate (recBL) at (\x1 -0.2cm, \y1-0.2cm);
\path let \p1=(p3) in coordinate (recTR) at (\x1 +0.2cm, \y1+0.2cm);
\draw (recBL) rectangle (recTR);

\filldraw  (p3) circle (0pt) node [anchor=west]{\small$p$};
					\end{tikzpicture}\label{figure_local_passage:1}} 
                \qquad \qquad
                    \subfloat[The map $\Psiloc: \tinsection\subseteq \insection \to \outsection$, where $\insection,\outsection$ are 3d sections transverse to the incoming and outgoing heteroclinics. The ordering of the eigenvalues in $p$ is represented by the number of arrows.]
                {\begin{tikzpicture}[scale=0.6]
                          \begin{scope}[yscale=1,xscale=-1]
						\input{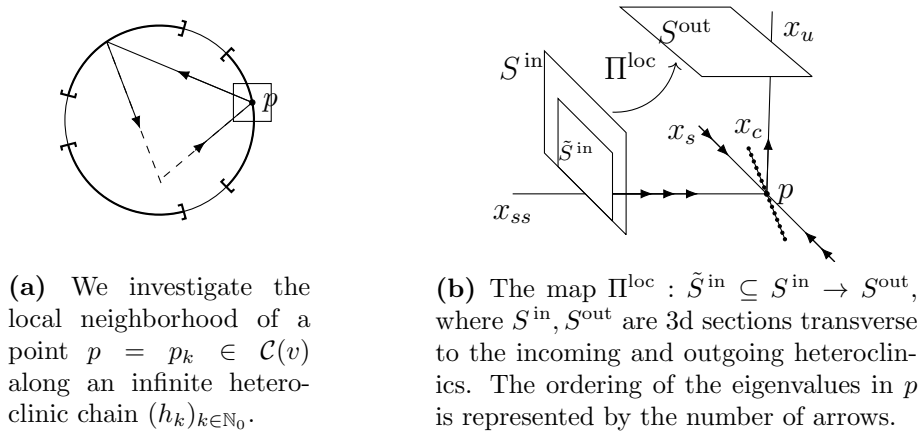}
                          \end{scope}
					\end{tikzpicture} \label{figure_local_passage:2}}
				\caption{The local passage.}\label{figure_local_passage}
			\end{figure}

			Firstly, we examine the vector field near a point $p=p_k$ of said sequence of equilibria. After rectifying the local invariant manifolds, rescaling the coordinates and time in a neighborhood of $p$, we obtain a system of equations %\eqref{general_equation_rescaled1}
			with a particularly simple expression for the unstable component.
			To capture properties of the local vector field, we introduce a local map $\Psiloc$ between the sections $\insection$ and $\outsection$, which are transverse to the incoming and outgoing heteroclinic solution, respectively.
			See \autoref{figure_local_passage}.
			We derive estimates for each component of a solution $\x$ that 
			travels between these sections. These are summarized in Proposition~\ref{proposition}. 
			The main goal of the local analysis is Theorem~\ref{local_theorem}, where 
			we prove Lipschitz continuity of the local map $\Psiloc$.
			More precisely, we show that by choosing the sections close enough to $p$, and by restricting $\Psiloc$ to a smaller subset $\tinsection \subset \insection$, we achieve an arbitrarily strong contraction of the stable components and an arbitrarily small drift in center direction.
   
                \subsubsection*{Outline global analysis.}
                
			%\smallskip
			%\textbf{Outline global analysis.}
			%\smallskip
			
			Secondly, we analyse the dynamics along a heteroclinic solution $h_k(t)$, which induces a transition map $\Psiglob$. Then we combine these results with the local estimates. 
			This yields properties of a map $\Psi_k$ between the insection $\insection_{k}$ at $p_k$ and the insection $\insection_{k+1}$ at $p_{k+1}=\KM(p_k)$. The map
			$\Psi_k$ is the concatenation of the local map $\Psiloc_k$ with the transition map $\Psiglob_k:\outsection_{k} \to \insection_{k+1}$, see \autoref{sequence_of_sections}.
			The transition map is sufficiently smooth due to regularity of the vector field.
			It expands the component tangential to the Kasner circle, as the dynamics close to the Kasner caps is governed by the Kasner map, which is expanding. The regularity of $\Psiglob_k$ yields a bounded expansion of the transverse components.
			%Analysing the concatenation of the local map with the global excursion requires some work.
			%The local contraction of the stable components can be made arbitrarily strong by choosing $\insection_{k}$ sufficiently close to $p_k$. At the same time, the local drift in the center direction is made arbitrily small too.
			Combining the local estimates with the properties of $\Psiglob$ yields a hyperbolic map
			$\Psi_k = \Psiglob_k \circ \Psiloc_k$, which expands the center component and contracts in transverse direction. This manifests in the cone transformation properties, which we prove in Lemma~\ref{cones_theorem}. The hyperbolicity of $\Psi_k$ allows us to apply a graph transform argument via the inverse maps $\Psi_k^{-1}$ to obtain three-dimensional stable manifolds for all infinite heteroclinic chains, and is done in Theorem~\ref{manifold_theorem}.
                \begin{figure}[H]\centering
				\begin{tikzpicture}[scale=0.6]
                      \begin{scope}[yscale=1,xscale=-1]
					\input{figures/sequence_of_sections}
                      \end{scope}
				\end{tikzpicture}
				\caption{To describe the dynamics along the infinite heteroclinic chain $(h_k(t))_{k \in \NN_0}$ we consider cross sections $\insection_k$ and $\outsection_{k}$ transverse to the incoming and outgoing heteroclinic solutions in each point $p_k$ and analyse the sequence of maps $\Psi_k$, each of which is composed of the local map $\Psiloc_k: \tinsection_{k} \to \outsection_{k}$ and the transition map $\Psiglob_k: \toutsection_{k} \to \insection_{k+1}$. Note that to have a well-defined map, we need to restrict the domains $\tinsection_{k}$ and $\toutsection_{k}$ appropriately, such that $\Psiloc_k(\tinsection_{k}) \subseteq \toutsection_{k}$ and $\Psiglob_k(\toutsection_{k}) \subseteq \tinsection_{k+1}$.
				}\label{sequence_of_sections}
			\end{figure}

			\subsection{Local Passage}\label{chap:local}
			We examine the local dynamics near an arbitrary equilibrium $p=p_k \in \Cantor(v)$ of the sequence of equilibria $(p_k)_{k\in \NN_0}$. To that end,
			we investigate a general setup that includes all features of system \eqref{full:subs} and can be obtained by a suitable change of coordinates.
			%\textcolor{red}{Maybe write the theorem here already and put the vector field manipulations in step 1 of proof?
				%What is the relation to Bianchi model? charts... in Dutilleul}
			\subsubsection{The setting}\label{subsec:setting}
			Consider the family of analytic vector fields
			\begin{equation}\label{general_equation}
				x^\prime = f(x,v),
				%\qquad x = \left(\xu,\xs,\xss,x_c\right)^T \in \RR^4,
				\qquad x \in \RR^4, \qquad \frac 12 <v<1.
			\end{equation}
			The dimension reduction of the phase space from five to four dimensions is due to the invariance of the Gauss constraint \eqref{constraint}. The parameter $v$ is the same as in the original system \eqref{full:subs}. Furthermore, the assumption
			\begin{equation}\label{line_of_equilibria}
				f\left(\left(0,0,0,x_c\right)^T,v\right) = 0, \qquad x_c \in \RR, \qquad \frac 12 <v<1
			\end{equation}
			yields a line of equilibria, which corresponds to the Kasner circle $\KC$ in the original system. Thus, $x_c$ parametrizes this line along which we assume the linearization to be of the form
			\begin{equation}\label{linearization_general}
				\frac{d}{dx}f\left(\left(0,0,0,x_c\right)^T, v\right) = \left(\begin{array}{cccc}
					\uev(\xc,v) & 0 & 0 & 0 \\
					0 & -\sev(\xc,v) & 0 & 0 \\
					0 & 0 & - \ssev(\xc,v) & 0 \\
					0 & 0 & 0 & 0
				\end{array}\right) =: A(\xc,v)  %\qquad \forall x_c \in \RR
			\end{equation}
			with
			\begin{equation}\label{eigenvalue_condition}
				0 < \uev(\xc,v) < \sev(\xc,v) < \ssev(\xc,v) \qquad \forall  x_c \in \RR, \qquad \frac 12 <v<1,
			\end{equation}
			as in %which is consistent with
			%with the linearization \eqref{linerazation_Kasner} and
			Lemma~\ref{cor_supercritical_eigenvalues}. Thus, in addition to the one-dimensional center manifold, each equilibrium has a one-dimensional unstable manifold and a two-dimensional stable manifold with strict ordering among the eigenvalues, which makes the local coordinates $\x = \left(\xu,\xs,\xss,x_c\right)^T $ unambiguous. 
			The center component $\xc$ corresponds to the variables $\Sigma_{\pm}$, and the other three components correspond to each of the variables $N_1,N_2,N_3$, compare \eqref{eigenvalue_condition} with \eqref{eigenvalues_corollary}.
			
			Throughout this analysis, the point of interest is the origin. The eigenvalues $\sev(0,v), \uev(0,v)$ are called \emph{principal} or \emph{leading}.
			%because typical solutions in the (un)stable manifold converge to the origin along the corresponding eigenvector for $t \to \pm \infty$.
			Their difference is called the \emph{spectral gap} and it is given by
			\begin{equation}\label{spectral_gap}
				\sg:= \sev(0,v)- \uev(0,v) > 0.
			\end{equation}
			%
			%\smallskip 
			%
			In the original coordinates, any $p \in \A_1^+(v)$ can be written as $p = (\Sigma_+, \sqrt{1-\Sigma_+^2})$ with $\Sigma_+>v$. According to \eqref{eigenvalues} and \eqref{eigenvalue_condition_cor}, the spectral gap for $p \in \A_1^+(v)$ is 
			\begin{equation}
			    \sg=-\evc(\Sigma_+,v) - \eva(\Sigma_+,v) = 8v -2 \left(\Sigma_+ + \sqrt{3(1-\Sigma_+^2)}\right).
			\end{equation}
			Taking derivative with respect to $\Sigma_+$ yields
			\begin{equation}
			    \frac{\partial\lambda}{\partial{\Sigma_+}} %= -2 \left(1 - \frac {6\Sigma_+}{2\left(3(1-\Sigma_+^2)\right)^{1/2} } \right) 
                    = 2 \left(\frac {\sqrt{3}\Sigma_+}{\sqrt{1-\Sigma_+^2}} - 1\right) > 0,
			\end{equation}
                since $\Sigma_+>v>\frac 12,$
			%In the original coordinates the spectral gap vanishes in the Taub points, independently of $v$ and
			i.e. the spectral gap increases monotonically towards the midpoint $\Q_1$.
			Consequently, the evaluation in the point $p$ of period two that is contained in $\A_1$, which is given by $\Sigma_+ = \frac{1+ 2\sqrt{3(v^2-1/4)}}{4v}$ and $\Sigma_-=  \sqrt{1-\Sigma_+^2}$ as in~\cite[Equation (51)]{HellLappicyUggla}, yields a lower bound of the spectral gap, as it constitutes the outer delimitation of the Cantor set $\Cantor(v)$, whereas the spectral gap at $\Q_1$ yields an upper bound:
            \begin{equation}
                \sg \in \left[ \lambda_{\textrm{per}2}(v) , \lambda_{\Q_1}(v) \right], 
            \end{equation}
            where $\lambda_{\textrm{per}2}(v):= \frac{2(4v^2-1)}{v}$ and $\lambda_{\Q_1}(v):= 2(4v-1).$%. Indeed, we obtain that
			% \begin{equation}
			%     \lambda_{\textrm{per}2}(v) %= 8v - \frac{1 + \sqrt{-3 + 12 v^2}}{2v} - 2 \sqrt{3} \sqrt{1 - \frac{\left(1 + \sqrt{-3 + 12 v^2}\right)^2}{16 v^2}}. 
   %                  %= 8v - \frac{1}{2v}\left[1 + \sqrt{12 v^2-3} + \sqrt{12v^2 + 6 - 6\sqrt{12 v^2-3}} \right] .
   %                  %:= 8v - \frac{1}{2v}\left[1 + 2\sqrt{3\left(v^2-\frac{1}{4}\right)} + \sqrt{12v^2 + 6 - 2\sqrt{3\left(v^2-\frac{1}{4}\right)}} \right] 
   %                  := \frac{2(4v^2-1)}{v} \qquad \text{ and } \qquad \lambda_{\Q_1}(v):= 2(4v-1).
			% \end{equation}
                Note that there are chains in $\Cantor(v)$, which get arbitrarily close to the period 2 chain, whose spectral gap tends to zero as $v \to 1/2$; e.g. the word $\omega = 123121231212123..$.

			\smallskip
			\textbf{Rectifying local invariant manifolds.} 
			\smallskip
			
			Equation $\eqref{general_equation}$ fully written out becomes
			\begin{equation}\label{general_equation_components:subs} 
				\begin{alignedat}{6}
					&\xu^\prime && \,=\, && \uev(\xc, v) \xu &&\shortquad+ \shortquad  &&P_u \, g(\x,v),\\
					&\xsss^\prime && \,=\,-\,&&\sssev(\xc,v) \xsss && \shortquad+ \shortquad &&P_{s,ss}\,g(\x,v),\\
					&x_c^\prime && \,=\, && && &&P_c \, g(\x,v),
				\end{alignedat}
			\end{equation}
			where $P_\star \,g$ denotes the projection of the nonlinear terms $g(\x,v)=f(\x,v)-A(\xc,v)$ to the respective coordinate axis.
			Moreover, we group variables according to their linear stability, i.e. we abbreviate
			\begin{align}
				\xsss := \left(\begin{array}{c}\xs \\ \xss
				\end{array}\right), \quad \sssev(\xc,v):=\left(\begin{matrix}
					\sev(\xc,v)&  0 \\ 0 & \ssev(\xc,v)
				\end{matrix}\right).
			\end{align}
			By \cite[Theorem 5.8]{Shilnikov} there exists, for every $r \in \NN$, a $C^{r}$ change of coordinates such that the vector field \eqref{general_equation_components:subs} is reduced to
			\begin{equation}\label{general_equation_components_flattened} 
				\begin{alignedat}{6}
					&\xu^\prime && \,=\,  && \uev(\xc,v) \xu \shortquad &&+ \shortquad  &&g_u(\x,v) \xu, &&\\
					&\xsss^\prime && \,=\, -&&\sssev(\xc,v) \xsss \shortquad &&+ \shortquad &&g_{s,ss}(x,v)\xsss, \\
					&x_c^\prime && \,=\, && && && G^0_c(x_c,v)x_c+ G^1_c(x,v)\xu + G^2_c(x,v) \xsss,
				\end{alignedat}
			\end{equation}
			in an open neighborhood $U$ of the origin, whose size depends on $r$. Such change of coordinates is invertible and regular, see \cite[Proposition 3.2]{Dutilleul}.
			%
			%\begin{align}
			%\xu^\prime &= \uevx \xu +  g_u(x) \xu \label{general_equation_components_flattened:unstable} \\
			%\xsss^\prime &= -\sevx \xs +  g_{s}^1(x)\xs + g_{s}^2(x)\xss \label{general_equation_components_flattened:weakstable} \\
			%\xsss^\prime &= -\sssev(\x,v) \xss +  g_{ss}^1(x)\xs + g_{ss}^2(x)\xss \label{general_equation_components_flattened:strongstable} \\
			%x_c^\prime &=  g^0_c(x_c)x_c + g^1_c(x)\xu + g^2_c(x) \xsss  \label{general_equation_components_flattened:center}
			%\end{align}
			The functions $g_u,g_{s,ss}$ in system \eqref{general_equation_components_flattened} are $C^{r}$ and vanish on the line of equilibria. The function $g_u$ is a scalar, whereas $g_{s,ss}$ is a matrix given by
            \begin{equation}                
            g_{s,ss} =\left(\begin{matrix}
				g_s^1 &  g_s^2 \\ g_{ss}^1 & g_{ss}^2
			\end{matrix}\right).
            \end{equation}
               %is a matrix.
			%(In Liebscher: $g_u,g_s$ vanish along line of eq.).
			%$G^0_c \in C^{r-1}$ with $G^0_c(0)=0$. Note however that
			The line of equilibria \eqref{line_of_equilibria} implies that $G_c^0$ vanishes identically.
			Furthermore, $G_c^1,G_c^2$ are $C^{r}$, $G_c^1$ vanishes for $\xs = \xss = 0$ and $G_c^2$ vanishes for $\xu=0$.
			% % %
			%and \eqref{general_equation_components:subs} is transformed to
			%\begin{subequations}\label{general_equation_components_flattened:subs} 
			%	\begin{align}
				%	\xu^\prime &= \uevx \xu +  g_u(x) \xu \label{general_equation_components_flattened:unstable} \\
				%	\xsss^\prime &= -\sssev(\x,v) \xsss +  g_{s,ss}(x)\xsss \label{general_equation_components_flattened:stable} \\
				%	x_c^\prime &=  g^0_c(x_c)x_c + g^1_c(x)\xu + g^2_c(x) \xsss  \label{general_equation_components_flattened:center}
				%	\end{align}
			%\end{subequations}
			%with $C^{?-1}$ functions 
			%$g_u, g_{s,ss}=\left(\begin{matrix}
				%g_s^1 & g_s^2 \\ g_{ss}^1 & g_{ss}^2
				%\end{matrix}\right)$ satisfying ... \\
				% % %5
				%Furthermore $g_c^1(x) = 0$ if $\xsss = 0$ and $g_c^2(x) = 0$ if $x_{u} = 0$.
				In other words, $x_c^\prime$ is nonzero if $\xu$ does not vanish and at least one of the terms $\xs$ and $\xss$ does not vanish. Consequently, the lowest term in the Taylor expansion have both the unstable and a stable component, i.e. the center component in
				\eqref{general_equation_components_flattened} can be written as
				\begin{equation} %\label{center_component}
					x_c^\prime =  \xu g^1_c(x,v) \xs + \xu g^2_c(x,v) \xss = \xu g_c(x,v) \xsss,
				\end{equation}
				with a $C^{r-1}$ matrix function
				$g_c := \left(\begin{matrix}
					g_c^1 &  g_c^2
				\end{matrix}\right).$
				% $g^1_c, g_c^2$ \textcolor{red}{satisfying what? }\\
				%Moreover, the nonlinear terms are Lipschitz continuous:
				%\begin{align}
				%\| g_{u}(\tx) - g_{u}(\x) \|, \quad \| g_{s,ss}(\tx) - g_{s,ss}(\x) \|, \quad \| g_{c}(\tx) - g_{c}(\x) \| \le C \|\tx-\x\|
				%\end{align}
				%The linear terms as well as
				%the nonlinear terms contributing to the stable and the center component can be summarized in matrices $$ \quad g_c := \left(\begin{matrix}
					%g_c^1 &  g_c^2
					%\end{matrix}\right),$$ and
					Thus,  \eqref{general_equation_components_flattened} becomes
					\begin{equation}	\label{general_equation_compact} 
						\begin{alignedat}{5}
							&\xu^\prime && \,=\, &&  \uev(\xc,v) \xu && \shortquad + \shortquad  && g_u(x,v) \xu, \\
							&\xsss^\prime && \,=\,  - \, &&\sssev(\xc,v) \xsss &&\shortquad + \shortquad &&g_{s,ss}(\x,v)\xsss,  \\
							&\xc^\prime && \,=\, && && && \xu g_c(\x,v) \xsss. 
						\end{alignedat}
					\end{equation}
					The local invariant manifolds are now rectified and they are given by the linear spaces
					\begin{align}\label{manifolds}
						\begin{split}
							W^{s,ss}(\xc) &:= \left\{\xc\text{ fixed}, \xu=0\right\} \\%\subseteq \insection, \\
							W^{u} (\xc) &:= \left\{\xc\text{ fixed}, \xsss= 0\right\}. % \subseteq \insection.
						\end{split}
					\end{align}
                        %\textcolor{red}{WHY CONTAINED IN $S^{in}$? i think it shouldnt be...}
					Note that $\xc$ parametrizes the Kasner circle $\KC$.
					The two-dimensional stable manifold $W^{s,ss}(\xc)$ is spanned by the weak stable manifold in direction of $\partial_{\xs}$ and the strong stable manifold in direction of $\partial_{\xss}$; this manifold intersects $ \insection$. Solutions on these axes coincide with the two incoming Bianchi type II solutions in $\xc$, whereas the one-dimensional unstable manifold $W^{u}(\xc)$ corresponds to the outgoing heteroclinic solution; this manifold intersects $ \outsection$. In particular, $W^{s,ss}(0)$ is spanned by two heteroclinic solutions converging to $p$, one of which corresponds to the heteroclinic solution encoded by the Kasner map $\KM$.
					
					Throughout this paper we use the $\ell^1$-norm, i.e. 
					\begin{align}\label{l1norm}
						\|x\| := |\xu| +   |\xs| + |\xss| + |\xc|.
					\end{align}
					Note that $\| \xsss \| =  |\xs| + |\xss|$.
					We make use of the following bounds of the nonlinear terms, which follow from their regularity. Therefore, for the neighborhood $U$, there exists a constant $M=M(U, v)>0$ such that
					\begin{align}\label{g_bound}
						\|g(x,v)\| \le M \quad \text{and} \quad \|g(\tx,v)- g(x,v) \| \le M \|\tx-\x \|,
					\end{align}
					where $g$ represents any of the components of the higher order terms in \eqref{general_equation_compact}.
					
					\smallskip
					\textbf{Rescaling.} 
					\smallskip
					
					There exists $\varepsilon_0 > 0$ such that the local neighborhood $U$ contains the cube $B_{2\eps} := (-2\varepsilon,2\varepsilon)^4$
					%of sidelength $4 \varepsilon$ centered at the origin
					for all $0<\varepsilon\le \varepsilon_0$.
					In our analysis we treat $\varepsilon$ as a parameter, which we choose small enough to obtain certain contraction properties.
					%
					%$4\eps$ to have some buffer. explain how and when choose $\eps$.
					The bounds \eqref{g_bound} are valid in $U$ and therefore independent of the exact choice of $\varepsilon$.
					%Since we want to study the local passage in a small neighborhood of the origin, 
					Following \cite{Liebscher}, we rescale the coordinates with the factor $\frac1\varepsilon$ to make the dependence on $\varepsilon$ explicit, i.e. $(\xu, \xsss, \xc) = \x \mapsto \frac{1}{\eps}{\x}$, which yields 
					% \begin{equation}
					% 	(\eps \yu)^\prime = \uev(\eps \yc,v) \eps \yu \shortquad+ \shortquad g_u(\eps \y,v) \eps \yu.
					% \end{equation}
					% Note that $g_u(x ,v)$ is quadratic in $x$, therefore 
     %                    \begin{equation}
     %                        g_u(\eps y ,v) = \eps^2 \cdot (...) 
     %                    \end{equation}
     % 					%
	%				% Analogous computations for the other components and 
    %                renaming yields a system
					\begin{equation}\label{general_equation_rescaled} 
						\begin{alignedat}{5}
							&\xu^\prime && \,=\, && \uev(\xc,v) \xu \shortquad &&+ \shortquad  && \varepsilon g_u(x,v) \xu, \\
							&\xsss^\prime && \,=\, - \, &&\sssev(\xc,v) \xsss \shortquad &&+ \shortquad &&  \varepsilon g_{s,ss}(\x,v)\xsss, \\
							& \xc^\prime && \,=\, && && &&\varepsilon  \xu g_c(\x,v) \xsss.
						\end{alignedat}
					\end{equation}
     %
     %\textcolor{red}{shouldn't we have $\epsilon^2$ in the 3rd equation}
     %
					Note that in the new coordinates the cube $B_{2\eps}$ becomes $B := (-2,2)^4$.
					The nonlinear functions $g_u, g_{s,ss}, g_c$ are in general different as in \eqref{general_equation_compact} but satisfy analogous estimates as \eqref{g_bound}, which justifies the abuse of notation.
					Moreover, the eigenvalue functions $\uev, \sssev$ in \eqref{general_equation_rescaled} are rescaled versions of those in \eqref{general_equation_compact} as well.
					%That is, the range of each eigenvalue in the new coordinates within a radius of one corresponds to a radius of $\varepsilon$ in the old coordinates.
					Thus, differentiability and ordering according to \eqref{eigenvalue_condition} are preserved.
					The parameter $\varepsilon$ thereby determines how close the dynamics of vector field \eqref{general_equation_rescaled} is to that of a linear vector field. That is, we can make the vector field ``as linear as we like" by considering small $\varepsilon$.
					%In \autoref{proposition} we determine an upper bound for the rescaling parameter $\varepsilon$.

                        %\newpage
     
					\smallskip
					\textbf{Local sections.} 
					\smallskip
					
					To capture the dynamics in the cube $B$ we introduce local sections of codimension one
					\begin{equation} \label{sections}
						\begin{alignedat}{1}
							\insection_{ss} &:= \left\{\xin \in \B \, | \, \xssin=1, \quad 0\le\xuin<1,\quad   |\xsin|<1, \quad  |\xcin| < 1 \right\}, \\
							\outsection_{ss} &:= \left\{\xout \in \B \, | \, \xuout = 1, \quad 0\le \xssout<1,\quad  |\xsout|<1, \quad |\xcout|<1 \right\}.
						\end{alignedat}
					\end{equation}
					with a map $\Psiloc_{ss}: \insection_{ss} \to \outsection_{ss}$, which is determined by the flow of \eqref{general_equation_rescaled} and described in more detail in \autoref{sec:local_map}, where we show that this map is well-defined, after restricting $\Psiloc_{ss}$ to a subset of $ \insection_{ss}$.
					%In the unscaled coordinates, these sections are located at of size $\varepsilon$.
					
					The sections $\insection_{s}, \outsection_{s}$ with the map $\Psiloc_{s}$ are defined analogously with the role of $\xs$ and $\xss$ interchanged, i.e. with the in-section perpendicular to the weak stable manifold $W^s(0)$.
					%In the original coordinates (ref) these sections are transverse to the flow/vector field because \textcolor{red}{why?}.
					We consider two different sets of sections because of the two incoming heteroclinic solutions at $p$. Depending on which we aim to describe we choose the corresponding sections and local map.
					%This choice is given the Kasner map, as it encodes a single incoming heteroclinic solution that coincides with either of the stable directions.
					However, we only consider the direction perpendicular to the $\xss$-direction in our analysis. That is, we assume that the incoming heteroclinic solution of interest enters via the strong stable (non-principal) direction.
					This choice is without loss of generality, since all subsequent estimates only use minimum contraction rates, see \eqref{weak_stable_contraction}, and are thus valid for points in the section $\insection_{s}$ as well. It slightly simplifies the notation, especially for the global analysis (the ``free" variable is $\xs$ instead of $\xss$). Consequently, we omit the subscript for sections and map, i.e. we write $\Psiloc: \insection\to \outsection$.
					%This choice of the incoming direction is not generic (from an abstract dyn sys viewpoint, name: principal eigenvalues\\
					%	but here it is?\\
					
					Note the sign restriction of the different components in \eqref{sections}.
					The sections are located at positive values for both $\xu$ and $\xss$. As the sign of the $N$-variables in the original vector field is invariant, this determines the sign of the solution for all times.
					To allow all heteroclinic chains in the boundary of the Bianchi type VIII subset, this is too restrictive. Indeed, the incoming and heteroclinics outgoing need not have the same sign. There is, however, no restriction on the sign of $\xs$. This facilitates the generalization of shadowing for Bianchi type VIII solutions, which we address in Theorem~\ref{manifold_theorem}.
					
					\smallskip
					\textbf{Addendum: The three-dimensional hyperbolic saddle.}
					\smallskip
					
					To motivate the subsequent modifications of vector field \eqref{general_equation_rescaled}, in particular the statement of Theorem~\ref{local_theorem}, and to get a geometric intuition we examine a simplified model, whose local dynamics is easily characterized.
					Consider the following decoupled linear system of equations for $y = (\yu,\ys,\yss)$ in $\RR^3$:
					\begin{equation}\label{simplified_model}
						\begin{alignedat}{3}
							\yu^\prime &= &\uev &\yu, \\
							\ys^\prime &= -&\sev &\ys, \\
							\yss^\prime &= -&\ssev &\yss,
						\end{alignedat}
					\end{equation}
					with $0<\uev < \sev < \ssev$.
					Thus, the origin is a hyperbolic equilibrium with a two-dimensional stable manifold $\{\yu = 0\}$ and one-dimensional unstable manifold $\{\ysss = 0\}$, but without the additional center direction as in \eqref{general_equation_rescaled}.
					Analogously to \eqref{sections}, we consider sections of codimension-one perpendicular to the unstable direction and the strong stable (non-leading) direction. In this case
					\begin{equation}
						\begin{aligned}
							\insection &= \left\{\yin \,\,: \yssin \,= 1, \shortquad \, 0\le\yuin\le1, \shortquad\, |\ysin|\le 1 \right\} \\
							\outsection &= \left\{\yout : \yuout = 1, \shortquad 0\le\yssout\le1, \shortquad |\ysout|\le 1 \right\}	.
						\end{aligned}
					\end{equation}
					
					A solution of \eqref{simplified_model} with initial data $\yin \in\insection$ at $t_0=0$ such that $\yuin\neq 0$ takes the form
					\begin{align}\label{simplified_model_solution}
						\begin{split}
							\yu(t) &= \ex{\,\uev t} \yuin, \\
							\ys(t) &= \ex{-\sev t} \ysin, \\
							\yss(t) &= \ex{-\ssev t} . %\yssin.
						\end{split}
					\end{align}
					Solving $\yu(t) = 1$ to obtain the time when $\y$ intersects the plane $\{\yu=1\}$, yields 
                        \begin{equation}
                         \tout = \frac{1}{\uev} \log \frac{1}{\yuin}  .
                        \end{equation}
                         Thus, the closer $\yin$ is to the stable manifold $\{\yu=0\}$, the longer the solution $y(t)$ takes to leave the local neighborhood.
					Due to exponential contraction of $\ys$ and $\yss$, the solution $\y$ intersects $\outsection$ in a point $\yout = y(\tout) = (\yuout = 1,\ysout, \yssout)$ with coordinates given by plugging in $\tout$ in \eqref{simplified_model_solution}: 
					\begin{equation}\label{coordinates_simplified}
						\ysout = %\ex{-\sev \tout} \ysin= 
						\left({\yuin}\right)^{\beta_{s}}\ysin, \qquad
						%	&&= {\yuin}^{\beta_s}\ysin\\
						\yssout = %\ex{-\ssev \tout} = 
						\left({\yuin}\right)^{\beta_{ss}},
						% &&= {\yuin}^{\beta_{ss}}.
					\end{equation}
					%Thus, $\yout:=\y(\tout) $ has coordinates $({\yuin}^{\beta_s}\ysin,{\yuin}^{\beta_{ss}})$
					where by assumption $\beta_{ss} > \beta_{s} > 1$ with the saddle quantities $\beta_{ss} := \frac{\ssev}{\uev}$ and $\beta_{s} := \frac{\sev}{\uev}$. Note that $\beta_{ss}, \beta_{s}>1$ implies that  $\|\ysssout\| < \|\ysssin\|$.
					
					From \eqref{coordinates_simplified} we can already see that by choosing an upper bound $0<\delta< 1$ for $\yuin$ we can make the contraction of $\ysss$ between the sections arbitrarily strong.
					%For the original four-dimensional vector field, this restriction is necessary to make sure that all solutions intersect $\outsection$.
					%In the analysis of (ref) we consider a smaller subset of the insection (why?) We need to do this later.
					To understand the local dynamics, we investigate how lines parallel to the lower boundary line $\{\yssin=1, \yuin=0\}$ at distance $0\le \ell \le\delta$ are transformed, see \autoref{3d_saddle_figure}.
					%Let $0<\delta<1$ and consider the box with points satisfying $0<\yuin, |\ysin|<\delta$.
					%  in this box.
					%The lower boundary line $\{\yuin = 0, -\delta\le \ysin \le \delta\}$ converges to the origin under the flow and has no image under the locmap but we can map it to 0 what makes the local map continuous, as the following consideratios shows.
					We define, for each $\ell \in [0,\delta]$:
					\begin{equation}				    
					\insection_\ell:=\{\yuin = \ell, \shortquad\, |\ysin| \le \delta, \shortquad\, \yssin = 1\}.
                        \end{equation}
					% wefix $\yuin =  c$ and let $-\delta\le \ysin \le \delta$.
					For $\ell\neq 0$ and $\yin \in \insection_\ell$ we obtain with \eqref{coordinates_simplified},
					\begin{equation}
					 \ysout = {\ell}^{\beta_s}\ysin \in 
					(- \ell^{\beta_s} \delta, \ell^{\beta_s} \delta), \qquad \yssout = {\ell}^{\beta_{ss}}.   
					\end{equation}
					%
					% ((-{\delta}^{\beta_s + 1},{\delta}^{\beta_{ss}}), ({\delta}^{\beta_s + 1}\ysin,{\delta}^{\beta_{ss}})  )$
					In particular, $\yssout$ is fixed at ${\ell}^{\beta_{ss}}< \ell = \yuin$, and therefore the image of $\insection_\ell$ is a line parallel to $\{\yuout=1, \yssout=0\}$.
					Moreover, the length of $\insection_\ell$ is scaled by a factor $\ell^{\beta_{s}}$.
					Consequently, for $\ell \to 0$, the line $\insection_\ell$ gets squeezed to the point $\ysssout = 0$, the intersection of the unstable manifold with $\outsection$.
					%
					%The line $\{0<\yuin < \delta, \ysin = \pm \delta\}$ is mapped to
					%
					% ($(1,{0}^{-\frac{\sev}{\uev}}\pm \delta,{0}^{-\frac{\ssev}{\uev}}),
					%(1,{\delta}^{-\frac{\sev}{\uev}}(\pm \delta),{\delta}^{-\frac{\ssev}{\uev}})$) = 
					% ($(1,0,0),
					%(1,\pm{\delta}^{-\frac{\sev}{\uev}+1} ,{\delta}^{-\frac{\ssev}{\uev}})$)
					Thus, although solutions with $\yuin = 0$ do not leave the local neighborhood, it is natural to define the image of the line $\insection_0$ to be $\ysssout = 0$, see \autoref{3d_saddle_figure}.
                    Note that the hyperbolic dynamics of \eqref{general_equation_rescaled} is qualitatively similar to \eqref{simplified_model} and \autoref{3d_saddle_figure} up to higher terms and a small drift in the center direction.
					\begin{figure}[H]
						\centering
						\begin{tikzpicture}[scale=0.75]
                            \begin{scope}[yscale=1,xscale=-1]
                                \input{figures/3d_saddle}
                            \end{scope}
						\end{tikzpicture}\caption{The three-dimensional saddle. The linear flow sends parallel lines $\insection_\ell$ in $\insection$ to parallel lines of smaller length in $\outsection$. Note that $\insection_0$ is mapped to $\ysssout = 0$. Moreover, $\tout = \infty$ for $\insection_0$ and $\tout = 0$ for $\insection_1$, and $\tout$ is decreasing with respect to $\ell$.} \label{3d_saddle_figure}
					\end{figure}
					
					%The situation (ref) and a wide range of other situations is described in (ref shilnikov)
										
					\subsubsection{Local estimates}
					%For the original four-dimensional vector field, this restriction is necessary to make sure that all solutions intersect $\outsection$.
					We continue modifying vector field \eqref{general_equation_rescaled}. We want to obtain a situation similar to the three-dimensional saddle for the stable and unstable components. Moreover, the expression for the center component in \eqref{general_equation_rescaled} is of order $\varepsilon$. Therefore, by considering a neighborhood close to the line of equilibria, the influence of the center component is weakened.
					We proceed component-wise with the analysis, and we summarize the main results below.

							\begin{prop} \label{proposition}
								Let $0<\alpha<1$. Then there exists $\talpha \in (\alpha,1)$ according to \eqref{relation_alpha_talpha} and $\varepsilon_0>0$ such that for a fixed choice of the rescaling parameter $0<\varepsilon \le \varepsilon_0$ in \eqref{general_equation_rescaled1} the following holds.
								A solution $\x = (\xu,\xsss, \xc) $ with initial data $\xin \in \insection$ at initial time $t_0$ satisfies
                 \begin{subequations}\label{lemma_local_estimates:subs}
									\begin{align}
										\xu(t) &=\, \ex{\uev(0,v) (t-t_0)} \xuin, \label{lemma_local_estimates:unstable}  \\
										\|\xsss(t) \| &\le\, \ex{-\talpha \sev(0,v)(t-t_0)} \|\xsssin\|, \label{lemma_local_estimates:stable} \\
										| \xc(t) - \xcin | & \le \,\varepsilon \frac{2M}{\alpha \sg} \xuin \left(1 - \ex{-\alpha\sg (t-t_0)}\right).\label{lemma_local_estimates:center} 
									\end{align}
									for $t\ge t_0$ as long as the solution stays in $B$.
								\end{subequations}
						\end{prop}
     
					\smallskip
					\textbf{The unstable component $\xu^\prime$.}
					\smallskip
					
					By the following consideration we achieve that in \eqref{general_equation_rescaled} the growth in unstable direction is independent of the other components.
					Since $\uev(\xc,v)$ depends differentiably on $x_c$, and as $g_u$ is bounded in the cube $B_{2\eps}$, we can choose an upper bound $\varepsilon_0 > 0$ for the rescaling parameter in \eqref{general_equation_rescaled} such that $\uev(0,v) \approx {\uev(\xc,v) + \varepsilon g_u(x,v)}$ for all $0<\varepsilon \le \varepsilon_0$. Thus
					\begin{align}
						\theta(x,v, \varepsilon) := \frac{\uev(0,v)}{{\uev(\xc,v) + \varepsilon g_u(x,v)}} >  0
					\end{align}
					%which contains the $\insection, \outsection$.
					in $\B$. Furthermore, $\theta$ is differentiable and thus serves as an Euler multiplier, i.e.
					the vector field $f$ in $\eqref{general_equation_rescaled}$ and $\theta f$ have the same trajectories in $U$.
		 	\begin{equation}\label{general_equation_rescaled0} 
						\begin{alignedat}{4}
							& \xu^\prime && \,=\, && \uev(0,v) \xu, &&  \\
							& \xsss^\prime && \,=\, -\,&&\tsssev(\xc,v) \xsss \shortquad + \shortquad  &&\varepsilon \tilde{g}_{s,ss}(\x,v)\xsss, \\
							&\xc^\prime &&\,=\, && &&\varepsilon  \xu \tilde{g}_c(\x,v) \xsss.
						\end{alignedat}
					\end{equation}
                         As a result, the growth in the unstable direction is linear and decoupled.
						Note that here $\theta \approx 1$ for sufficiently small $\eps>0$, and $\tsssev(\xc,v) = \theta(x,v,\eps) \cdot \sssev(\xc,v)$, which is a nonlinear function. Linearizing at $(0,0,0,\xc) \in \KC$, we obtain a linear eigenvalue sufficiently close to $\sssev$, and the extra terms are adjoined to $\tilde{g}_{s,ss}$.
						The equations for the other components are perturbed, but the bounds \eqref{g_bound} as well as the ordering among the eigenvalues \eqref{eigenvalue_condition} are preserved. We rename the resulting functions for sufficiently small $\eps$ and stick to our notation: 
					\begin{equation}\label{general_equation_rescaled1} 
						\begin{alignedat}{4}
							& \xu^\prime && \,=\, && \uev(0,v) \xu, &&  \\
							& \xsss^\prime && \,=\, -\,&&\sssev(\xc,v) \xsss \shortquad + \shortquad  &&\varepsilon g_{s,ss}(\x,v)\xsss, \\
							&\xc^\prime &&\,=\, && &&\varepsilon  \xu g_c(\x,v) \xsss,
						\end{alignedat}
					\end{equation}
					which is henceforth the vector field of our consideration.

					\smallskip
					\textbf{The stable components $\xsss^\prime$.}
					\smallskip

					%Since the stable directions together span the stable manifold it is useful to consider the norm estimates for the combined vector $\xsss := \left(\xs, \xss\right)^T$.
					%\begin{subequations}\label{estimate_stable_derivative}
					%	\begin{align}
						%\|\xsss\|^\prime &=
						%\xs^\prime + \xss^\prime \\&= -\sevx \xs -\ssevx \xss + \varepsilon \left[ g^1_{s}(x)\xs + g^2_{s}(x)\xss + g^2_{ss}(x)\xs  + g^2_{ss}(x)\xss \right]
						%\\ & \le -\sevx \xs -\sevx \xss + \varepsilon \left[ \|g^1_{s}\| \xs + \|g^2_{s}\| \xss + \|g^2_{ss}\|\xs  + \|g^2_{ss}\|\xss \right]
						%\\ & \le -\sevx (\xs + \xss) + 2\varepsilon  \|g\| (\xs + \xss) 
						%\\ & = -\sevx \|\xsss\| + 2\varepsilon  \|g\| \|\xsss\| \\&= \left(-\sevx+ 2 \eps M  \right)\|\xsss\|
						%\end{align}
						%\end{subequations}
						Using the bounds on the nonlinear terms \eqref{g_bound} and the ordering of the eigenvalues in $U$, $-\ssev < -\sev$, we obtain
						\begin{equation}\label{weak_stable_contraction}
							\begin{aligned}
								|\xs|^\prime  &\le -\sev(\xc,v) |\xs| + \eps M (|\xs| + \xss),\\
								\xss^\prime  &\le -\sev(\xc,v)\xss + \eps M (|\xs| + \xss).
							\end{aligned}
						\end{equation}
						Accordingly,
						\begin{equation}\label{preliminary_estimate_xsss}
							\|\xsss \|^\prime = |\xs|^\prime + \xss^\prime \le \left(-\sev(\xc,v)+ 2 \eps M  \right)\|\xsss\|.
						\end{equation}

						%For each $0<\alpha<1$ there exists $\varepsilon$ such that  $\|\xsss\| \le \exp(- \alpha \sev)\cdot \|\xsssin\|$ for all $\xc$ since 
						We want to simplify the last expression by getting rid of  the constant term $2\eps M$ and, as for the unstable component, the dependence on $\xc$. However, we cannot achieve an explicit expression that leaves $\xu^\prime$ untouched.
						Since our goal is to prove a contraction of the stable components, a strong decay is desirable. Therefore, we give a bound on the decay rate.
						More precisely, for given $0\ll\alpha < 1$,
						we want an estimate of the form \vspace{-0.1cm}
						\begin{equation}\label{alpha_bound}
							-\sev(\xc,v)+ 2 \eps M \le -\alpha \sev(0,v).
						\end{equation} 
                            %
                            %\textcolor{red}{Can't i choose simply $-\sev(\xc,v)+ 2 \eps M \le - \sev(per2,v).$}
                            This is possible by choosing $\varepsilon$ small enough. Indeed,
						by differentiability of $\sev(\xc,v)$ with respect to $\xc$ we can choose an upper bound $ \varepsilon_0 > 0$ such that
						\begin{subequations}
                            \begin{align}
                                0&\le 2 \eps M\le \frac{(1-\alpha)}{2} \sev(0,v)\\
                                0&\le 2  |\sev(\xc,v)-\sev(0,v)| \le \frac{(1-\alpha)}{2} \sev(0,v)
                            \end{align}
                            \end{subequations}
						for $|\xc|<2$, which corresponds to a range of $2\varepsilon$ in the unscaled coordinates, and for all choices of $\varepsilon \le \eps_0$. Then
						%\begin{align}
						%\sevx \ge \alpha \sev + 2 \eps M
						%\end{align}
						%Indeed:
						\begin{equation}
							|\sev(\xc,v)-\sev(0,v)| + 2 \eps M \le (1-\alpha)\sev(0,v),  \end{equation}
						or equivalently
						\begin{equation}
							\alpha \sev(0,v)  + 2 \eps M  \le \sev(0,v) - |\sev(\xc,v)-\sev(0,v)|.
						\end{equation}
						The term on the right is bounded from above by $\sev(\xc,v)$ using the triangle inequality, and therefore \eqref{alpha_bound} holds.
						%\begin{align}
						% - \sev(\xc,v)+ 2 \eps M \le -\alpha \sev(0,v).
						%\end{align}
						The choice of the parameter $\alpha$ thus affects the choice of the scaling parameter $\eps$ and thereby determines the size of the cube around the origin.
						Hence, \eqref{preliminary_estimate_xsss} becomes
						\begin{equation}\label{estimate_stable_derivative2}
							\|\xsss \|^\prime  \le -\alpha \sev(0,v)  \|\xsss\|.
						\end{equation}
						%For the solution $\x$ with initial condition $x^\text{in} = (\xuin, \xsin, \xssin=1, \xcin) \in \insection$ and initial time $t_0$ this yields the estimate
						%\begin{align}\label{estimate_stable}
						%\|\xsss(t)\| \le \ex{- \alpha \sev(0,v) (t-t_0)} \|\xsssin\|.
						%\end{align}
						
						\smallskip
						\textbf{The center component $\xc^\prime$.}
						\smallskip
						
						By the fundamental theorem of calculus, for $t\ge t_0$, and plugging in \eqref{general_equation_rescaled1}
						\begin{equation}\label{center}
							%\varepsilon \int_{t_0}^{t}  |\xu (\tau) g_c(x(\tau),v) \xsss (\tau)| \,d\tau 
							|\xc(t)-\xc(t_0)| = \bigg| \int_{t_0}^{t} \xc^\prime(\tau) d\tau \bigg| \le  \eps M \int_{t_0}^{t}  \xu (\tau)   \| \xsss (\tau)\| \, d\tau \qquad \text{ in } B.
						\end{equation}
						Let $0 < \alpha <1$ as above. We can choose a possibly smaller $\varepsilon$-bound to obtain the estimate \eqref{alpha_bound} for an arbitrary $\talpha \in (\alpha,1)$. Thus, we assume that \eqref{alpha_bound} holds for such $\talpha$, and explain how to choose $\talpha$ in dependence of $\alpha$. Before we do that, in order to estimate the integrand, we take the derivative, plugging in \eqref{general_equation_rescaled1} and \eqref{estimate_stable_derivative2} with $\talpha$:
						% that makes the local vector field look even linear %\eqref{estimate_stable_derivative2}
						\begin{equation} %\label{estimate_integrand_xc}
							\begin{aligned}
								\left( \xu  \, \| \xsss \| \right)^\prime &= 
								\xu ^\prime  \| \xsss \| + \xu   \| \xsss \|^\prime \\&\le 
								\uev(0,v) \, \xu   \|\xsss\| - \xu  \talpha \sev(0,v) \| \xsss \| \\&=
								-(\talpha \sev(0,v)-\uev(0,v))  \xu  \|\xsss\| 
								%\\&\le  -{\alpha} \sg  \xu  \|\xsss\|
							\end{aligned}
						\end{equation}
						Recall that $\sev(0,v)-\uev(0,v) = \sg$ denotes the spectral gap. 
						To make its role more explicit
						%We explain how to choose $\talpha$ in dependence of $\alpha$ in order to make the role of $\sg$ more explicit.
						%\begin{lem}\label{lemma_sg_alpha}
						%	Let $0<\alpha<1$ and assume that $\sg=\sev(0,v)- \uev(0,v)>0$. Then there exists $\talpha \in [\alpha, 1)$ such that
						%\begin{align}\label{spectral_gap_lemma}
						%		{\alpha} \sg\le \talpha \sev(0,v)-\uev(0,v). 
						%\end{align}
						%	\end{lem} \begin{proof}
						%	Therefore, $$\alpha +  (1-\alpha) \frac{\uev(0,v)}{\sev(0,v)}< \alpha +  (1-\alpha) = 1.$$
						we choose $\talpha$ such that
						\begin{equation}
						    \alpha +  (1-\alpha) \frac{\uev(0,v)}{\sev(0,v)} \le \tilde \alpha < 1.
						\end{equation}
						This is possible because $\frac{\uev(0,v)}{\sev(0,v)} < 1$ by assumption.
						Accordingly, \begin{align}\label{relation_alpha_talpha}
							\alpha\sg =  \alpha(\sev(0,v)-\uev(0,v))  \le \tilde \alpha \sev(0,v) - \uev(0,v).
						\end{align}
						%\end{proof}
						Consequently, the derivative of the integrand in \eqref{center} is bounded according to
						\begin{equation} 
							\begin{aligned}
								\left( \xu  \, \| \xsss \| \right)^\prime
								\le  -{\alpha} \sg  \xu  \|\xsss\|,
							\end{aligned}
						\end{equation}
						which yields the bound
						\begin{align}
							\xu(\tau)   \| \xsss(\tau) \|  \le e^{-\alpha \sg (\tau-t_0)} \, \xuin   \| \xsssin \|.
						\end{align}
						Plugging in \eqref{center},
						\begin{equation}
							\begin{aligned}
								|\xc(t)-\xc(t_0)| &\le
								\eps M \xuin \cdot \| \xsssin \|  \int_{t_0}^{t} e^{-\alpha\sg (\tau-t_0)} d\tau
								%\\ &\le \frac{2 \eps M |\xuin | }{-\alpha\sg } \cdot e^{-\alpha\sg (\tau-t_0)} \Bigg|_{t_1}^{t}
								\\&\le  \varepsilon \frac{2M   }{\alpha\sg } \xuin \cdot \left(1 - \ex{-\alpha\sg (t-t_0)}\right)
							\end{aligned}
						\end{equation}
						using $\|\xsssin\| = 1+ |\xsin| \le 2$.

						%\begin{remark}
						%\textcolor{blue}{By choosing the parameter $\alpha$ close to 1...}
						%\end{remark}
						%\begin{remark}
						%	For some estimates it suffices to use weaker bounds than \eqref{lemma_local_estimates:subs}. By definition \eqref{sections} of the insection $\|\xsssin\| = 1+ |\xuin| \le 2$. Moreover, with $\sg > 0$ we obtain
						%\begin{subequations}
						%		\begin{align}
							%	\|\xsss(t) \| &\le  2 \ex{-\alpha \sev(0,v)(t-t_0)}\le 2 \\
							%	| \xc(t) - \xcin |  & \le \frac{2 \eps M}{\alpha \sg} \xuin .
							%	\end{align}
						%\end{subequations}
						%\end{remark}
						
						\subsubsection{The local map $\Psiloc$}\label{sec:local_map}
						
						With Proposition~\ref{proposition} we are able to define the local map $\Psiloc$ between the sections \eqref{sections}.
						We must, however, reduce the size of the domain to make sure that $\Psiloc$ is well-defined.
						%Let $\Phi_t$ denote the flow of the local vector field \eqref{general_equation_rescaled1}  and
						Consider a solution $\x$ of \eqref{general_equation_rescaled1} with initial condition $\xin=\left(\xuin, \xsin, \xssin=1, \xcin\right) \in \insection$ at initial time $t_0=0$ with $\xuin\neq 0$.
						In complete analogy to the three-dimensional saddle \eqref{simplified_model}, the simple expression \eqref{lemma_local_estimates:unstable} for $\xu^\prime$ enables us to compute the time $\tout$ when $\x$ intersects $\{\xu = 1\}$ via \begin{align}\label{tout}
							1 = \xu(\tout) = \ex{\uev \tout} \xuin \iff  \tout= \frac{1}{\uev(0,v)} \log \frac{1}{\xuin},
						\end{align}
						Inserting $t=\tout$ in the estimate \eqref{lemma_local_estimates:stable} yields a contraction of the stable component, provided $\talpha > \alpha > \frac{\uev(0,v)}{\sev(0,v)}$:
						%
						%		Assume $ \frac{\uev(0,v)}{\sev(0,v)}<\alpha<1$ and let $\tout$ as in (ref). Then there exist $\varepsilon_0>0$ and constants $\beta, L_c>0$ such that
						%
						%		\begin{subequations}\label{cor_prop:subs}
							%				\begin{align}
								%				\| \xsss(\tout) \| & \le {\left(\xuin\right)}^\beta  \color{red}{( |\xuin| + |\xsin| )},\label{cor_prop:stable} \\
								%				| \xc(\tout) - \xcin | &\le  \varepsilon L_c \xuin \label{cor_prop:center}
								%			\end{align}
							%		\end{subequations}
						%
						%
						%		for a fixed choice of the rescaling parameter $0<\varepsilon\le \varepsilon_0$ in \eqref{general_equation_rescaled1}. \\
						%		Moreover, the local map $\Psiloc$ as defined in \eqref{definition_localmap} is well-defined after restricting the domain to a (nondegenerate) subset $\tinsection$ depends on $\eps$.
						%
						%
						%	With initial time $t_0=0$ in $\insection$ and $t=\tout$ as in \eqref{tout}, \autoref{proposition} yields
						\begin{align}\label{stable_est}
							\|\xsss(\tout)\| &\le 
							\ex{-\alpha \sev(0,v)\tout} \|\xsssin\| =
							\ex{-\frac{\alpha \sev(0,v)}{\uev(0,v)} \log\frac{1}{\xuin}} \|\xsssin\|  =
							{\left(\xuin\right)}^{\frac{\alpha \sev(0,v)}{\uev(0,v)}}\|\xsssin\|.
						\end{align}
						%and $\beta:= \frac{\talpha \sev(0,v)}{\uev(0,v)} >1$
						%we therby get a local contraction of the stable components near the line of equilibria.
						By choosing solutions close to the stable manifold, i.e. with $\xuin$ small, the contraction becomes arbitrarily strong.
						%(It remains to show that there exists a proper subset $\tinsection \subseteq \insection$ such that a solution with initial condition $\xin \in \tinsection$ intersects $\outsection$.)
						%
						%Indeed, when $t=\tout$ the unstable component equals one. Moreover, both stable components $\xs$ and $\xss$ decrease exponentially, and are initially at most one.
						%The center estimate provides a growth bound which can be made arbitrarily small by choosing a sufficiently small upper bound $\varepsilon_0$ for the parameter $\varepsilon$.
						Moreover, \eqref{lemma_local_estimates:center} yields the time-independent estimate
						\begin{align}\label{center_est}
							| \xc(\tout) - \xcin | \le \frac{2 \eps M}{\alpha \sg} \xuin 
						\end{align}
						%The choice of $\varepsilon$ fixes the position of the local neighborhood and thereby of the local sections. A sufficiently small upper bound $\varepsilon_0$ for the parameter $\varepsilon$ does not increase $L_c$ %=2M\alpha^{-1}\sg^{-1}$
						Note that even though the coefficient $ \frac{2 M}{\alpha \sg} $ depends on the local vector field, the right-hand side can be made arbitrarily small for some sufficiently small $\varepsilon \le \varepsilon_0$.
						Thus, for a suitable choice of $\varepsilon_0$ there exists $0<\delta<1$ such that the restriction of initial conditions to
						\begin{align} \label{section_delta_local}
							\tinsection &:= \left\{\xin \in B \, | \, \xssin=1, \shortquad 0\le\xuin\le{\delta},\shortquad   |\xsin|\le{\delta}, \shortquad  |\xcin| \le {\delta} \right\}
						\end{align}
						guarantees that $|\xc(\tout)|$ does not exceed one, i.e. the solution $\x$ intersects $\outsection$.
						
						% $\Psiloc\big|_{\tinsection}$ .
						
						% \textcolor{red}{following is the limit of the above procedure, thus probably continuous}
						
						%\smallskip
						
						An initial condition $\xin$ with $\xuin=0$ represents the heteroclinic solutions converging to $(0,0,0,\xcin)$ due to the rectified invariant manifolds \eqref{manifolds}. Thus, these solutions do not leave the local neighborhood, and hence they do not intersect $\outsection$. There exists exactly one heteroclinic emerging from $(0,0,0,\xcin)$, intersecting $\outsection$ in $(\xuout,\xsout,\xssout, \xcout) = (1,0,0,\xcin)$.
						%	\textcolor{red}{Explain why define image then as this point? to have compact domain?}
						% To capture Kasner map? To have a "complete" definition?}
					Therefore we define the image of $\xin$ to be this point. This yields continuity of $\Psiloc$ because the expressions \eqref{stable_est} and \eqref{center_est} converge to zero in the limit $\xuin \to 0$.
					Note that this is independent of $\xsin$, i.e. the whole $\xs$-axis in $\tinsection$ is contracted to a single point.
					%
					%write something that close to the incoming heteroclinics, i.e. for small $\xuin$ the contraction of stable and drift of center component are also small.
					%
					%\textcolor{blue}{The heteroclinic solution of our interest given by the Kasner map converges to the origin in $U$ and is therefore represented by the additional assumption $\xcin=0$.}
					%
					%
					We summarize.
					\begin{cor}\label{corollary_proposition}
						Let %\textcolor{red}{$\alpha \in \left(\frac{\uev(0,v)}{\sev(0,v)}, 1\right)$ and}
						$\tinsection$ as in \eqref{general_equation_rescaled1}. Consider a solution $\x$ of \eqref{lemma_local_estimates:subs} with $\x(0) = \xin \in \tinsection$. There exists $\varepsilon_0>0$ such that for a fixed choice of the rescaling parameter $0<\varepsilon \le \varepsilon_0$, the following map is well-defined and continuous:
						\begin{align}\label{definition_localmap}
							\begin{split}
								\Psiloc : \tinsection &\to \outsection \\
								%\left(\xuin, \xsin, \xssin =1, \xcin \right) =
								\xin &\mapsto
								%\Psiloc(\xin) =:
								\xout 
								%= \left(\xuout =1, \xsout, \xssout, \xcout \right)
								= \begin{cases}
									\x(\tout)  & \text{ if } \xuin \neq 0 \\ \left(1,0,0,\xcin \right) & \text{ if } \xuin =0 \,.
								\end{cases}
							\end{split}
						\end{align}
						Moreover, given $\alpha \in \left(\frac{\uev(0,v)}{\sev(0,v)}, 1\right)$, then we have the following\footnote{For points in $\tinsection$, where $0\leq \xuin\leq \delta < 1$, we have $\|\xsssout\| \le {\left(\xuin\right)}^{1+\beta}\|\xsssin\| \le {\left(\xuin\right)}^{\beta}\|\xsssin\|$. } for $\beta:=\frac{\alpha \sev(0,v)}{\uev(0,v)}-1>0$: %, which is uniform in $\varepsilon$, such that
						\begin{align}\label{psiloc_estimates}
							\|\xsssout\| &\le
							{\left(\xuin\right)}^{\beta}\|\xsssin\| \\ 
							| \xcout - \xcin | &\le \varepsilon \frac{2  M}{\alpha \sg} \xuin.
						\end{align}
					\end{cor}
				%
					%$\Psiloc$ is not yet well-defined, as there may be solutions starting in $\insection$ that do not intersect $\outsection$ due to a drift in center direction. We adress this issue in Corollary \ref{corollary_proposition}, where we reduce the size of $\insection$ appropriately.
					The derivation of $\tout$ assumes the initial time $t_0$ to be the same for all solutions with initial condition in $\insection$. This is necessary for $\Psiloc$ to be a well-defined function. In the following Theorem \ref{local_theorem} we prove Lipschitz continuity of $\Psiloc$, where it is necessary to compare the solution $\x$ to another solution $\tx$ for which we pick a different initial time $-\ttzero\le 0$ which is chosen such that $\tx$ intersects $\outsection$ at $\tout=\tout(\xin)$.
				\begin{thm}\label{local_theorem}
					Let $\tinsection \subseteq \insection$ as in \eqref{section_delta_local} such that $\Psiloc: \tinsection \to \outsection $ is well-defined. Consider two solutions $\tx,\x$ of \eqref{general_equation_rescaled1} with initial conditions $\txin,\xin \in \tinsection$.
					Then there exist $\varepsilon_0>0$ such that for a fixed choice of the rescaling parameter $0<\varepsilon<\varepsilon_0$ the local map $\Psiloc$ is Lipschitz continuous and satisfies
					\begin{align}
						\left|\left(\txcout-\xcout\right) - \left(\txcin-\xcin\right) \right| &\le \varepsilon L_c \| \txin - \xin \|, \label{center_theorem} \\
						\| \txsssout- \xsssout  \| & \le L_s {\left(\max\{ \xuin,\txuin\}\right)}^{\beta} \| \txin - \xin \|. \label{stable_theorem}
					\end{align}
					The constants $L_c,L_s, \beta>0$ are uniform in $\varepsilon$. 
				\end{thm}
                %    
					%\textcolor{red}{mention $v$ in the theorem somewhere?}
					\begin{figure}
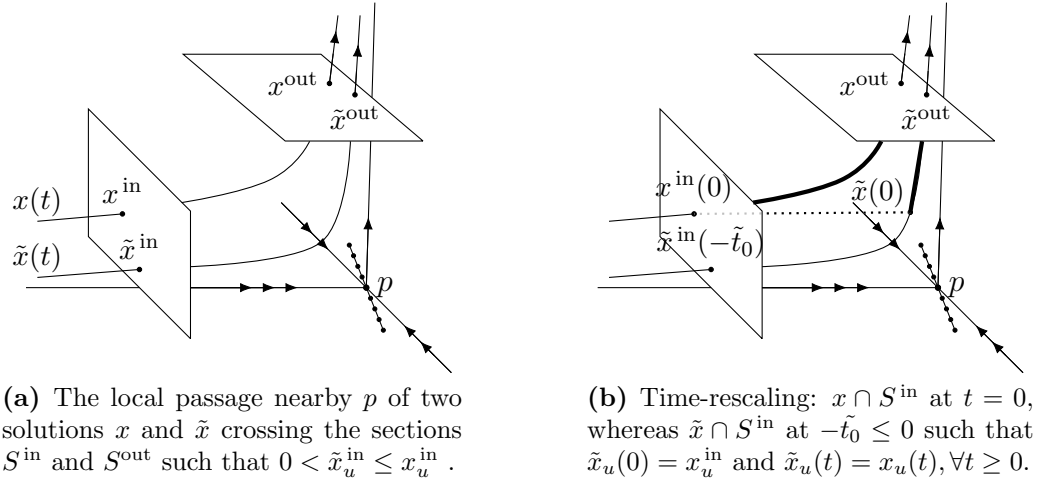

						\centering
						%	\subfloat[We consider two codimension-one cross sections $\insection$ and $\outsection$ perpendicular to the incoming and outcoming heteroclinic solution, respectively, with a local map $\Psiloc$.]
						%	{\begin{tikzpicture}[scale=0.8]
								%			\input{figures/local_passage}
								%		\end{tikzpicture} } %\label{figure_local_passage:1}}
					%	\qquad
					\subfloat[The local passage nearby $p$ of two solutions $\x$ and $\tx$ crossing the sections  $\insection$ and  $\outsection$  such that $0<\txuin \le \xuin$ . ]
					{\begin{tikzpicture}[scale=0.75]
                            \begin{scope}[yscale=1,xscale=-1]
							\input{figures/local_passage_orbits}
                            \end{scope}
						\end{tikzpicture} \label{figure_local_passage:3a}}
					\qquad\qquad
					\subfloat[Time-rescaling: $\x\cap \insection$ at $t=0$, whereas $\tx \cap \insection$ at $-\ttzero\le 0$ such that $\txu(0) = \xuin$ and $\txu(t)=\xu(t), \forall t\ge 0$.]
					{\begin{tikzpicture}[scale=0.75]
                             \begin{scope}[yscale=1,xscale=-1]
							\input{figures/local_passage_orbits2}
						\end{scope}
						\end{tikzpicture} \label{figure_local_passage:3b}}
					\caption{The comparison of the local passage of two different solutions.} \label{figure_local_passage:3}
				\end{figure}
				
				%
				%\begin{remark}
				%	Thus, the local map $\Psiloc$ is Lipschitz continuous. By choosing $\varepsilon$ sufficiently small the drift in center direction is arbitrarily small. Moreover, for a sufficiently small upper bound on $\xuin$ the contraction of the stable components is arbitrarily strong.
				%\end{remark}
                    %
					\begin{remark}
						Lipschitz continuity follows from \eqref{center_theorem} and \eqref{stable_theorem}:
						To bound $\left| \txcout-\xcout \right|$ we add and subtract $\left(\txcin-\xcin\right)$ and use the triangle inequality together with \eqref{center_theorem} to obtain
						\begin{align} %\label{center_theorem_2}
							\left| \txcout-\xcout \right|
							%&= \left| \txcout + \txcin - \txcin + \xcin - \xcin -  \xcout \right| \\
							& \le \left|\left(\txcout-\xcout\right) - \left(\txcin-\xcin\right) \right| + \left|\txcin - \xcin \right|
							\\ &\le (\varepsilon L_c + 1) \| \txin - \xin \|,
						\end{align}
						where we use that $\left| \txcin - \xcin \right| \le \| \txin - \xin \|$.
						Thus, to bound the full norm, we use \eqref{stable_theorem} to obtain
						\begin{align}\label{Lipschitz}
							\|\txout - \xout \| &\le \left(\varepsilon L_c + 1+ L_s \left(\max\{ \xuin,\txuin\}\right)^\beta \right) \| \txin - \xin \|\\
                            &\le \left(\varepsilon L_c + 1+ L_s {\delta}^\beta \right) \| \txin - \xin \| =: \tilde{L}\| \txin - \xin \|.
						\end{align}
				\end{remark}
				
				\emph{Outline of Proof.}
				The proof is organized in four main steps.
				Firstly, we treat the boundary case, when one of the two solutions lies in the stable manifold of the line of equilibria.
				Then we choose a different initial time $-\ttzero\le 0$ for the solution $\tx$ to obtain $\txu(t)=\xu(t)$ for $t\ge 0$. In the third step we derive estimates for $\tx$ for $t\le 0$. In the last, most elaborate step we derive estimates for $t\ge 0$ involving both solutions. This step is divided in several sub-steps, where cross estimates of stable and center component are resolved little by little and finally the Gr{\"o}nwall lemma yields the desired bounds.
				
				\begin{proof} 
					\textbf{Step 0: Ingredients.}
					Choose $\alpha \in (0,1)$ such that
                    \begin{equation}\label{alphaCHOICE}
                        \alpha>\frac{\uev(0,v)}{\sev(0,v)}, %\rho:= \alpha \sev(0,v) - \uev(0,v) >0.
                    \end{equation}
                    and thus $\alpha \sev(0,v) - \uev(0,v)\in (0, \alpha\lambda(v))$, since $\uev(0,v) > \alpha \uev(0,v)$. 
                    Proposition~\ref{proposition} provides $\talpha\in (\alpha,1)$ and an upper bound $\varepsilon_0>0$ on the parameter $0<\varepsilon < \varepsilon_0$ in \eqref{general_equation_rescaled1} such that 
					\begin{align}
						-\sev(\xc,v)+ 2 \eps M &\le -\alpha \sev(0,v), \label{estimate_with_alpha}\\
						- (\talpha \sev(0,v)-\uev(0,v)) &\le -{\alpha} \sg. \label{estimate_with_talpha}
					\end{align}
					Similar to \eqref{estimate_with_alpha}, the differentiability of $\sssev(\xc,v)$ with respect to $\xc$ enables us to choose $\varepsilon_0>0$ such that
					\begin{align}\label{eigenvalue_bounds}
						\talpha \sev(0,v) \le \sev(\xc, v) \le \ssev(\xc, v) \le \frac{\ssev(0,v) }{\talpha} \qquad \text{ in } \B ,
					\end{align}
					%We need this ordering in step 3. 
                    i.e., we can bound the eigenvalues depending on $\xc$ by the eigenvalues at $\xc = 0$.
					%Throughout this proof we consider two points $\txin, \xin \in \insection$. The corresponding solutions are $\tx$ and $x$, respectively. $\x$ has initial time $0$ and the initial time for $\tx$ is determined in Step 2.
					
					\smallskip
					\textbf{Step 1:}  \emph{Without loss of generality $0 < \txuin \le \xuin$.}
					\smallskip
					
					Indeed, if $\txuin= 0$ the distance in the in-section is 
                    \begin{equation}\label{norm_boundary}
						\|\txin - \xin\| = \xuin + |\txsin-\xsin| + |\txcin-\xcin|,
					\end{equation} 
                    and the resulting points in the out-sections are 
                    \begin{equation}
					    \txout = (1,0,0,\txcin), \quad \xout=(1,\xsout, \xssout, \xcout).
					\end{equation}
					With Corollary \ref{corollary_proposition}:
					\begin{equation}
						|(\txcout-\xcout) - (\txcin-\xcin)| = |\xcout - \xcin| \le \varepsilon \frac{2  M}{\alpha \sg} \xuin .
                    \end{equation}
					For the transverse component note that $\|\xsssin\| \le 2$, i.e.
					\begin{equation}
						\|\txsssout-\xsssout\| = \|\xsssout\| \le 2 {\left(\xuin\right)}^{\beta} \xuin .
					\end{equation}
					Both claims now follow from the fact that $\xuin \le \|\txin-\xin\|$, due to \eqref{norm_boundary}. 
					
					\smallskip
					\textbf{Step 2: } \emph{Initial time $-\ttzero\le 0$ for $\tx$, such that $\txu(t)=\xu(t)$ for all $t\ge 0$.}
					\smallskip
					
					The growth in unstable direction of vector field \eqref{general_equation_rescaled1} is independent of the other components. That is, if $\txu$ and $\xu$ coincide at any particular time, say $t=0$, they are equal for all times
					%we can pick different initial times for the two solutions such that $\xu(t)~=~\txu(t) $
					for all $t \ge 0$.
					By our choice of the initial time $t_0=0$ for $\x$ we have $\xu(0) = \xuin$. Since $\txuin \le \xuin$, we choose a negative initial time $-\ttzero \le 0$. The explicit expression $\txu(t) = \ex{\uev(0,v) (t+ \ttzero)} \txin$ enables us to compute how long it takes for $\tx$ to have the same unstable component as $\x$ has initially:
					% , i.e. we pick the initial time $-\ttzero \le 0$ of $\tx(t) = \ex{\uev (t+ \ttzero)} \txin $ according to
					\begin{align}\label{initial_time_for_second_solution}
						\xuin = \txu(0)  \iff \ttzero = \frac{1}{\uev} \log \frac{\xuin}{\txuin}
					\end{align}
					In particular, both solutions hit $\outsection$ at the same time $\tout$, which simplifies our calculations, see \autoref{figure_local_passage:3}.
					
					\smallskip
					\textbf{Step 3: } \textbf{Initial estimate for $\tx$.} 
                    \emph{We prove that on the initial segment $\left[-\ttzero,0\right]$, until the unstable components of $\tx$ and $\x$ coincide (see bold curve in \autoref{figure_local_passage:3}), the following holds:
						\begin{align}
							\left|\txc(0)- \txcin\right| & \le \varepsilon \tilde L_c \frac{\txuin}{\xuin} \left|\txuin - \xuin\right|,  \label{initial_segment_center} \\
							\left|\txsss(0)- \txsssin\right| & \le \tilde L_s \frac{1}{\xuin} \left|\txuin - \xuin\right|. \label{initial_segment_stable} 
						\end{align}
						for some constants $\tilde L_c, \tilde L_s>0$.}
					
					\smallskip
					\textbf{Estimate of the center component \eqref{initial_segment_center}.}
					\smallskip
					
					The center estimate \eqref{lemma_local_estimates:center} from Proposition~\ref{proposition} with $t_0 = -\ttzero$ and $t=0$ gives
					\begin{align}
						|\txc(0)-\txcin| &\le
						\varepsilon \txuin \frac{2M  }{\alpha\sg  }  \left( 1 - e^{-\alpha \sg \ttzero}  \right).
					\end{align} 
					Plugging in expression  \eqref{initial_time_for_second_solution} for $\ttzero$ from step 2 yields
					\begin{align}\label{step3_center}
						\begin{split}
							|\txc(0)-\txcin| &\le  \varepsilon  \txuin  \frac{2M}{{\alpha} \sg}   \left( 1- \ex{ -\frac{\alpha \sg}{\uev(0,v)} \log \frac{\xuin}{\txuin} } \right) \\
							&=  \varepsilon \txuin  \frac{2M}{{\alpha} \sg}   \left( 1- \left(\frac{\txuin}{\xuin}\right)^{\frac{{\alpha} \sg}{\uev(0,v)}}\right)
						\end{split}
					\end{align}

					Since $\xi:=\frac{\txuin}{\xuin} \in (0,1]$ and $\kappa:={\frac{{\alpha} \sg}{\uev(0,v)}}>0$, we can employ the estimate\footnote
					{\emph{Proof.} 
						%Let $0<\xi<1$ and $\kappa >0$. Then $1 -\xi^\kappa \le \max\{1,\kappa\} (1-\xi)$. Proof:
						We fix $\kappa$ and distinguish two cases. When $0<\kappa\le1$, then $\xi^\kappa \ge \xi$ for all $\xi \in (0,1]$ and thus $1-\xi^\kappa \le 1-\xi.$
						For $\kappa >1$ define $H_\kappa(\xi):=1-\xi^\kappa -\kappa(1-\xi).$ Then $H_\kappa \in C^1([0,1])$ with $H_\kappa(0) = 1-\kappa < 0$ and $H_\kappa(1) = 0$. Moreover, $H_\kappa^\prime(\xi) = \kappa(1-\xi^{\kappa -1} ) > 0$ implies that $H_\kappa$ is strictly increasing in $(0,1)$. Consequently, $H_\kappa$ is nonpositive in $(0,1]$, i.e. $1-\xi^\kappa \le \kappa(1-\xi)$}
					\begin{align}\label{general_estimate}
						1 -\xi^\kappa \le \max\{1,\kappa\} (1-\xi).
					\end{align}
					Plugging in \eqref{general_estimate} into \eqref{step3_center}, we obtain
					\begin{align}
						|\txc(0)-\txcin| \le  \varepsilon \txuin  \frac{2M}{{\alpha} \sg}  \cdot \max\left\{1,\frac{\alpha\sg}{\uev(0,v)} \right\} \left( 1- \frac{\txuin}{\xuin}\right),
					\end{align}
					and thus, with $\tilde{L}_c:=\max\left\{\frac{2M}{{\alpha} \sg},\frac{2M}{\uev(0,v)} \right\}$ and $0<\txuin\le\xuin\le1$
					\begin{align}
						|\txc(0)-\txcin| \le \varepsilon \tilde{L}_c \txuin \left( 1- \frac{\txuin}{\xuin}\right) = \varepsilon \tilde{L}_c \frac{\txuin}{\xuin} \left|\txuin - \xuin\right|.
					\end{align}
					
                        \smallskip
					\textbf{Estimate of the stable component \eqref{initial_segment_stable}.}
					\smallskip
					
					Recall that the stable components in \eqref{general_equation_rescaled1} are given by
					\begin{equation}
						\txsss^\prime= -\sssev(\txc,v) \txsss+  h(t).
						%\varepsilon g_{s,ss}(\tx,v)\txsss.
					\end{equation}
					%Using \eqref{eigenvalue_bounds}, we get the following differential inequalitites
					%\textcolor{red}{1 or 2: ? 
						%\begin{align}
						%\txsss^\prime \in \left[ - \frac{\ssev(0,v)}{\talpha}  , -\talpha \sev(0,v)  \right] \xsss + h(t) ,
						%\end{align}
						%\begin{align}
						%- \frac{\ssev(0,v)}{\talpha}\xsss + h(t)	\le \txsss^\prime \le  -\talpha \sev(0,v)\xsss + h(t) ,
						%\end{align}
						%}
					where 
					\begin{equation}
						h(t) = \varepsilon g_{s,ss}(\tx(t),v)\,\txsss (t).
					\end{equation}
					The linear parts in the equation for the stable components decouple, i.e. $\sssev$ is a diagonal matrix. According to the variation-of-constants formula
					\begin{equation}
						\txsss(0) = \ex{\int_{-\ttzero}^{0} -\sssev(\txc(\tau),\tau) d\tau} \txsssin + 
						\int_{-\ttzero}^{0} h(s)\ex{\int_{s}^{0}- \sssev(\txc(\tau),\tau) d\tau } ds.
					\end{equation}
					Thus, 
					\begin{align}
						\| \txsss(0) - \txsssin \| &\le \biggl\| \left(\mathbf{1}- \ex{\int_{-\ttzero}^{0} -\sssev(\txc(\tau),\tau) d\tau} \right)  \txsssin \biggr\| + \biggl\|
						\int_{-\ttzero}^{0} h(s)\ex{\int_{s}^{0}- \sssev(\txc(\tau),\tau) d\tau } ds \biggr\| \\ &=: X_1 +X_2
					\end{align}
					We derive upper bounds for $X_1$ and $X_2$. For both quantities we employ \eqref{eigenvalue_bounds}, which yields upper/lower bounds on the stable eigenvalues in $\B$.
					%Thus, the entries of the diagonal matrix $\ex{\int_{s}^{0}- a(\tau) d\tau }$ are bounded from above by $\ex{\int_{s}^{0}- \alpha \sev(0,v) d\tau } = \ex{\alpha \sev(0,v) s}$.
					The bound $\|h\|_\infty \le 2 \eps M$ implies that
					\begin{equation}
						X_2 \le
						%2 \eps M \left( \int_{-\ttzero}^{0}  \ex{\alpha \sev(0,v) s }ds \right)&= 
						\frac{2 \eps M}{\talpha \sev(0,v)} \left( 1- \ex{ - \talpha \sev(0,v) \ttzero } \right) = 
						\frac{2 \eps M}{\talpha \sev(0,v)} \left(1-\left(\frac{\txuin}{\xuin}\right)^{\frac{\talpha\sev(0,v)}{ \uev(0,v)}}\right) \le \eps \tilde{L}_2^s \left(1-\frac{\txuin}{\xuin} \right)
					\end{equation}
					where $\tilde{L}_2^s:=\frac{2 M}{\talpha \sev(0,v)} \max\left\{1, \frac{\talpha\sev(0,v)}{ \uev(0,v)} \right\} $ is chosen according to the estimate \eqref{general_estimate}.
					
					\smallskip
					
					%Analogously, the entries of the diagonal matrix $\mathbf{1} - e^{\int_{-\ttzero}^{0} -a(\tau) d\tau}$ are bounded from above by $1-\ex{- \frac{\ssev(0,v)}{\alpha} \ttzero }$.
					The upper bound $\|\txsss(t)\|\le 2$ yields
					\begin{align}
						X_1 \le 2 \left(1-\ex{- \frac{\ssev(0,v)}{\talpha} \ttzero}\right) = 2\left( 1- \left(\frac{\txuin}{\xuin}\right)^{\frac{\ssev(0,v)}{\talpha \uev(0,v)}}\right) \le \tilde{L}_1^s \left(1- \frac{\txuin}{\xuin}\right)
					\end{align}
					for $\tilde{L}_1^s:=2\frac{\ssev(0,v)}{\talpha \uev(0,v)}$, where we again use the estimate \eqref{general_estimate} and $\ssev(0,v) > \talpha \uev(0,v)$.
					
					\smallskip
					
					Finally, since $\varepsilon<1$, 
                    \begin{align}
						\|\txsss(0)-\txsssin\| \le \left(\tilde{L}_1^s + \eps \tilde{L}_2^s\right) \left(1-\frac{\txuin}{\xuin} \right) \le \tilde{L}_s \frac{1}{\xuin} |\xuin-\txuin|
					\end{align}
					where $\tilde{L}_s := \tilde{L}_1^s + \tilde{L}_2^s $ and we use that $0<\txuin \le \xuin$ by step $1$. Note that $\tilde{L}_s$ is uniform in $\varepsilon$. 
					
					\smallskip 
					\textbf{Step 4: estimates for $t \ge 0$.}
					\begin{remark} \label{remark_constant}
						Throughout this step we treat $M$ as a universal constant. That is, we ignore constant factors of $M$ for sake of readability.
					\end{remark}
					The term to estimate for the center component is
					\begin{align}\label{first_estimate_center_drift}
						\begin{split}
							\left|\left(\txcout-\xcout\right) - \left(\txcin-\xcin\right) \right| & = 
							\left|\left(\txcout-\xcout\right) - \left(\txcin-\xcin\right) +\txc(0)-\txc(0) \right|
							%\\& \le 
							%\left|(\txcout-\xcout) +(\xc(0) - \txc(0)) | +|\txc(0) - \txcin \right|
							\\& \le 
							\left|\int_0^{\tout}(\txc(\tau) - \xc(\tau) )^\prime  d\tau\right| +|\txc(0) - \txcin |
							\\& \le \int_0^{\tout}\left|(\txc(\tau) - \xc(\tau) )^\prime \right|  d\tau   + \varepsilon \tilde{L}_c |\txuin-\xuin|
							%\\&\le \varepsilon L_c \| \txin - \xin \|
						\end{split}
					\end{align}
					where we use the estimate \eqref{initial_segment_center} for $\txc$ on the initial segment $\left[-\ttzero,0\right]$ from step 3. 
                    
					For $t\ge 0$ let
					\begin{equation}
						d_c(t) := \int_{0}^{t} \left|\left(\txc(\tau)-\xc(\tau)\right)^\prime\right| d\tau . \label{dc_definition} 
					\end{equation}
						%	\\d_s(t) &:= \|\txsss(t)-\xsss(t)\| \label{ds_definition}
						Thus, it remains to find a suitable upper bound for $d_c(\tout)$ to prove the desired center estimate \eqref{center_theorem}.
							For the difference of the stable components we abbreviate by the following term, which is necessary to bound in order to prove \eqref{stable_theorem}:
					\begin{equation}
                        d_s(t) := \|\txsss(t)-\xsss(t)\|. \label{ds_definition}
					\end{equation}
					%
					%\smallskip
					%
					In the following we derive mutually dependent estimates related to the quantities $d_s(t)$ and $d_c(t)$, and a $d_s(t)$-independent estimate of $d_c(t)$. Gr\"{o}nwall's lemma provides the desired upper bound for $d_c$. Plugged in the estimate for $d_s$ this yields the claim for the stable component.
					We begin with the term $|\txc(t)-\xc(t)|$, since it  appears in the following computations.
					
					\smallskip
					\textbf{Upper bound for $|\txc-\xc|$ in dependence of $d_c$ and on the initial distance.}
					\smallskip
					
					Analogously to \eqref{first_estimate_center_drift}, the fundamental theorem of calculus and triangle inequality yield
					\begin{align}
						\left| \txc(t) - \xc(t) \right| &\le  \left| \int_{0}^{t}  (\txc(\tau) - \xc(\tau) )^\prime d\tau \right| + |\txc(0)- \xc(0)| \\&\le d_c(t)+ |\txc(0)- \txcin| +|\txcin - \xcin|.
						% \\ &\le d_c(t) + |\txc(0)- \txcin| +|\txcin - \xcin|
					\end{align}
					Together with the estimate for $\txc$ on the initial segment $\left[-\ttzero,0\right]$ from step 3, \eqref{initial_segment_center}:
					\begin{equation}
						\left| \txc(t) - \xc(t) \right| \le d_c(t)  + \varepsilon\tilde{L}_c \frac{\txuin}{\xuin} \left|\txuin - \xuin\right| + \left|\txcin - \xcin\right|.
					\end{equation}
					We can add $\left|\txsin - \xsin\right|$ to the right-hand side to obtain the full norm $\| \txin - \xin \|$, yielding	
					\begin{equation}\label{xc,txc_estimate} 
						\left| \txc(t) - \xc(t) \right|	 \le d_c(t) + \| \txin - \xin\|
					\end{equation}
					for a sufficiently small upper bound $\varepsilon_0$ on $\varepsilon$. Note that $\txuin \le \xuin$ by step 1.
     
					\smallskip
					\textbf{Preliminary bound for $d_c^\prime$ in dependence of $d_s$.}
					\smallskip
					
					The vector field \eqref{general_equation_rescaled1} yields
					%	\begin{subequations}\label{dc_derivative1.1:subs}
						\begin{align}
							d_c^\prime = \left|\txc^\prime - \xc^\prime\right| &=\varepsilon  \left| \txu  g_c(\tx,v) \txsss -  \xu g_c(\x,v) \xsss  \right|
							\\&= \varepsilon \xu  \left|  g_c(\tx,v) \txsss -  g_c(\x,v) \xsss  \right|, \end{align}
						%	\end{subequations}
					due to the choice of initial times \eqref{initial_time_for_second_solution}, which entails that $\txu(t)=\xu(t)$ for $t\ge 0$, see \autoref{figure_local_passage:3b}. Triangle inequality and the bounds $\eqref{g_bound}$ on $g_c$ imply that
					\begin{align}\label{dc_derivative1.2}
						\begin{split}
							d_c^\prime  &\le  \varepsilon \xu   \big(\left|  g_c(\tx,v) \txsss - g_c(\tx,v)\xsss \right| +  \left|  g_c(\tx,v)\xsss -    g_c(\x,v) \xsss  \right|\big)
							\\&\le  \eps M \xu   \Big(\| \txsss - \xsss \| +  \| \tx -   \x \| \, \|\xsss \|\Big),
						\end{split}
					\end{align}
					by the mean value theorem.
					The first term in the parentheses is $d_s$ and the second term contains $\|\tx- \x\| = |\txc - \xc| + d_s$, since $\txu(t) = \xu(t)$ for $t\ge 0$. Plugging this in \eqref{dc_derivative1.2} yields
					\begin{align}\label{dc_derivative1.3}
						\begin{split}
							d_c^\prime &
							\le  \eps M \xu   \Big((1+\|\xsss\|)d_s+  | \txc -   \xc | \, \|\xsss \|\Big)\\&
							\le  \eps M \xu   \Big(\left(1 + 2 \ex{-\alpha\sev(0,v)t}\right)d_s +  2 \ex{-\talpha\sev(0,v)t} | \txc -   \xc | \Big) ,
						\end{split}
					\end{align}
					where we use Proposition~\ref{proposition} to bound $\|\xsssin\| \le 2$. Note that for the prefactor of $d_s$, we can bound the exponential term by a constant. Hence %With $M$ as the universal constant we obtain the estimate
					\begin{align}\label{dc_derivative1.4}
						d_c^\prime(t) \le  \eps M \xu(t)  \Big(d_s(t) + \ex{-\talpha\sev(0,v)t}   | \txc(t) -   \xc(t)| \Big).
					\end{align}
					%Moreover, the last step uses remark \ref{remark_constant}, i.e. the constant $M$ is replaced by a multiple of itself.
					%\begin{align}
					%\ex{- \alpha \sev t}\|\xsssin\| \le 2, C\to 3C \hspace{0.5cm} & =  \eps M \xu \cdot  \Big[d_s +  | \txc -   \xc | \, \|\xsss \| \Big]
					%\end{align}
					%\textcolor{red}{why keep $\|\xsss\| |\txc- \xc|$ but not$\|\xsss\| |\txsss- \xsss|$? }
					%
					Plugging in the expression for the unstable component, i.e. $\xu=\ex{\uev(0,v)t}\xuin$, the estimate \eqref{xc,txc_estimate} for $|\txc-\xc|$, and noticing that $\talpha>\alpha$, we obtain
					\begin{equation}\label{dc_derivative2}
							d_c^\prime(t) \le
							%\eps M \ex{\uev(0,v) t} \xuin \cdot \left(  d_s(t)+  | \txc(t) -   \xc(t) | \cdot\ex{- \alpha \sev(0,v) t } \|\xsssin\|\right) \\ &\le 
							%\eps M \ex{\uev(0,v) t} \xuin  d_s(t)  + \eps M \ex{-( \alpha \sev(0,v)- \uev(0,v)) t} \xuin \cdot   | \txc(t) -   \xc(t) |  
							%\\ &\le 
							\eps M \ex{\uev(0,v) t} \xuin d_s(t)  + \eps M \ex{-( \alpha \sev(0,v)- \uev(0,v)) t} \xuin    \left(d_c(t) + \| \txin - \xin\|\right).
					\end{equation}
					%where we use \eqref{estimate_with_talpha} and \eqref{alphaCHOICE}: $- (\talpha \sev(0,v)-\uev(0,v)) \le -{\alpha} \sg \le - (\alpha \sev(0,v)- \uev(0,v))$. %, due to our choice of $\talpha$.
					%	\begin{align}
						%		d_c^\prime & \le  \eps M \xu \cdot  \Big(d_s +  | \txc -   \xc | \, \|\xsss \| + | \txsss -   \xsss | \, \|\xsss \|\Big)
						%						\\ & \le  \eps M \xu \cdot  \Big(d_s +  | \txc -   \xc | \, \|\xsss \| +| \txsss -   \xsss | \, \ex{- \alpha \sev t } \cdot \|\xsssin\|\Big)	
						%	\end{align}
					%	\begin{align}
						%	\ex{- \alpha \sev t}\|\xsssin\| \le 2, C\to 3C \hspace{0.5cm} & =  \eps M \xu \cdot  \Big[d_s +  | \txc -   \xc | \, \|\xsss \| \Big]
						%	\end{align}
					%	\end{subequations}
				%	\textcolor{red}{why keep $\|\xsss\| |\txc- \xc|$ but not$\|\xsss\| |\txsss- \xsss|$? }
				%	Together with the expression \eqref{general_equation_rescaled1:unstable} for $\xu$, the estimate \eqref{estimate_stable} for $\xsss$ and  we obtain (\textcolor{red}{put these previous things in a lemma})
				%\begin{subequations}
				%		\begin{align}\label{dc_derivative2}
					%	d_c^\prime(t) &\le \eps M \ex{\uev(0,v) t} \xuin \cdot \left(  d_s(t)+  | \txc(t) -   \xc(t) | \cdot\ex{- \alpha \sev(0,v) t } \|\xsssin\|\right) \\ &\le 
					%\eps M \ex{\uev(0,v) t} \xuin  d_s(t)  + 2 \eps M \ex{-( \alpha \sev(0,v)- \uev(0,v)) t} \xuin \cdot   | \txc(t) -   \xc(t) |  
					%	\\ &\le 
					%	\eps M \ex{\uev(0,v) t} \xuin d_s(t)  + 2 \eps M \ex{-( \alpha \sev(0,v)- \uev(0,v)) t} \xuin \cdot   \left(d_c(t) + L_1\| \txin - \xin\|\right).
					%	\end{align}
				%\end{subequations}
				
				\smallskip
				\textbf{Preliminary bound for $d_s$ in dependence of $d_c$.}
				\smallskip
				
				The initial value $d_s(0)$ is bounded using estimate \eqref{initial_segment_stable} for $\txsss$ from step $3$:
				\begin{equation}
					\begin{aligned}
						d_s(0) & =
						\|\txsss(0)-\xsssin\| \le
						\|\txsss(0)- \txsssin \| +  \| \txsssin - \xsssin\|\\ & \hspace{3.338cm}\le 
						\tilde{L}_s \frac{1}{\xuin} \left|\txuin - \xuin\right| +  \| \txsssin - \xsssin\|.
					\end{aligned}
				\end{equation}
				Since $0<\xuin\le1$ we can bound the last term above, at the expense of making $\tilde{L}_s$ at least $1$, in case it is not already:
				\begin{equation}\label{ds_initial}
					\begin{split}
						d_s(0) \le  \tilde{L}_s \frac{1}{\xuin} \|\txin - \xin\|.
					\end{split}
				\end{equation}
				%\textcolor{red}{Yes, this is a stupid estimate but still sufficient to (do what?)}
				To estimate the derivative $d_s^\prime$, due to \eqref{general_equation_rescaled1}, consider
				\begin{equation}
					\txsss^\prime - \xsss^\prime = - \,\sssev(\tx,v) \txsss + \sssev(\x,v) \xsss + \varepsilon (g_{s,ss}(\tx)\txsss    -  g_{s,ss}(\x)\xsss).
				\end{equation}
				We add and subtract $\sssev(\tx, v) \xsss$ and $\eps g_{s,ss}(\tx) \xsss$:
				\begin{equation}
					\begin{aligned}
						\txsss^\prime - \xsss^\prime &= - \,\sssev(\tx,v) (\txsss - \xsss) + (\sssev(\x,v)-\sssev(\tx,v)) \xsss \\ & \quad \, + \varepsilon  \left(g_{s,ss}(\tx)(\txsss-\xsss) + (g_{s,ss}(\tx)    -  g_{s,ss}(\x))\xsss\right),		\end{aligned}
				\end{equation}
				i.e. 
				\begin{equation} \label{ds_derivative2}
					\txsss^\prime - \xsss^\prime = - \,\sssev(\tx,v) (\txsss - \xsss) + R,
				\end{equation}
				with remainder $R = R(\tx, \x,v) $ satisfying
				\begin{align}
					\|R\| & \le  \| \left(\sssev(\txc,v) - \sssev(\xc,v)\right) \xsss\| \,+ \\ &\qquad \varepsilon  \|  (g_{s,ss}(\tx,v)(\txsss-\xsss) + (g_{s,ss}(\tx,v)    -  g_{s,ss}(\x,v))\xsss \|  \\
					& \le  \| \sssev(\txc,v) - \sssev(\xc,v) \|  \| \xsss\| + \eps M  \left( \| \txsss-\xsss\| + \|\tx - \x \| \|\xsss \| \right),
				\end{align}
				due to boundedness and Lipschitz continuity of the nonlinear terms \eqref{g_bound}. Moreover, the derivative of $\sssev(\xc,v)$ with respect to $\xc$ is bounded in the rescaled cube $\B$ as well. Thus, by mean value theorem and definition of $d_s$                          
				\begin{equation}
					\| R\|\le 
					\eps M  \left( |\txc -\xc| \|\xsss \| +d_s + \|\tx - \x \| \| \xsss \| \right).
				\end{equation}
				Similar to \eqref{dc_derivative1.2}, \eqref{dc_derivative1.3},  \eqref{dc_derivative1.4} we substitute $\|\tx - x\| = d_s + |\txc- \xc|$
				%since $\txu(t) = \xu(t)$ for $t\ge 0$,
				and we employ two different bounds for $\|\xsss\|$:
				\begin{equation} 
					\begin{aligned}
						\|R \| & \le 
						\eps M  \left( 2 |\txc -\xc| \|\xsss \| +d_s ( 1+\|\xsss\|) \right) \\ &\le
						\eps M  \left( 4 \ex{-\talpha \sev(0,v)t}  |\txc -\xc| +3 d_s \right) .
					\end{aligned}
				\end{equation}
				Thus, considering a universal constant $M$, we obtain
				\begin{equation}\label{estimate_R}
					\|R\|  \le
					\eps M  \left(  \ex{-\talpha \sev(0,v)t}  |\txc -\xc| + d_s \right).
				\end{equation}
				%The last step again uses the choice of $\varepsilon$ in dependence of $\alpha$ and also remark (ref), i.e. 
                    %
				Consequently, the remainder $R$ can be made arbitrarily small by a suitable choice of $\varepsilon$. Considering \eqref{ds_derivative2}, this means that the difference vector  $\left(\txsss-\xsss \right)\in \RR^2$ converges to zero, exponentially at least with rate $-\sev(0,v)$, up to an $\varepsilon $-bounded error.
				Accordingly, the norm of the difference vector $d_s$, converges to the zero according to 
				\begin{equation}
					d_s^\prime(t) = \|\txsss(t) - \xsss(t) \|^\prime \le - \sev(\xc,v) d_s(t) + \|R\|.
				\end{equation}
				With \eqref{estimate_R},
				\begin{align}
					d_s^\prime(t) &\le  (-\sev(\xc,v) + \eps M) d_s(t) + 
					\eps M  \ex{-\talpha \sev(0,v)t} |\txc(t) -\xc(t)|    \\&\le  -\alpha\sev(0,v) d_s(t) + 
					\eps M \ex{-\talpha \sev(0,v)t} |\txc(t) -\xc(t)|,
				\end{align}
				where the last inequality uses \eqref{estimate_with_alpha}. 
                Inserting \eqref{xc,txc_estimate}, the estimate of $|\txc-\xc|$, yields
				\begin{equation}
					\label{ds_derivative_final}
					d_s^\prime(t) \le  -\alpha \sev(0,v) d_s(t) + 
					\eps M  \ex{-\talpha\sev(0,v) t} \left( d_c(t) + \|\txin -\xin\|   \right).
				\end{equation}
				%
                %\smallskip
				%We use the derivative to estimate $d_s$ via the following consideration.
				Multiplying $d_s$ by the integrating factor $\ex{\alpha \sev(0,v) t }$, the fundamental theorem of calculus yields
                \vspace{-0.5cm}
				\begin{equation}\label{exp*ds}
					\ex{\alpha \sev(0,v) t } d_s(t) =  d_s(0) + \int_0^t \left(\ex{\alpha \sev(0,v) \tau }  d_s(\tau)\right)^\prime \, d\tau
				\end{equation}
				The integrand satisfies
				\begin{align}
					\left(\ex{\alpha \sev(0,v) \tau }  d_s(\tau)\right)^\prime &=
					\alpha \sev(0,v) \ex{\alpha \sev(0,v) \tau }  d_s(\tau) + \ex{\alpha \sev(0,v) \tau } d_s^\prime (\tau) 
					%\\&  \le \ex{\talpha \sev(0,v) \tau }  \bigg[\alpha \sev(0,v) d_s(\tau) -\alpha \sev(0,v) d_s(\tau)  \\ & \hspace{1cm} + \eps M  \ex{-\talpha\sev(0,v) \tau} \left( d_c(\tau) + \|\txin -\xin\|   \right)\bigg]
					\\&\le \eps M \ex{-(\talpha-\alpha) \sev(0,v) \tau } \left( d_c(\tau) + \|\txin -\xin\|   \right)
					\\&\le \eps M \left( d_c(\tau) + \|\txin -\xin\|   \right),
				\end{align}
				where we plugged in \eqref{ds_derivative_final}, noticing that $\talpha>\alpha$ and $\tau\geq 0$. 
				Inserting in \eqref{exp*ds} and multiplying with $\ex{-\alpha \sev(0,v) t}$ yields
				\begin{align}\label{ds_estimate1}
					d_s(t) & \le
					% \ex{ - \alpha \sev(0,v) t } \left(d_s(0) + 2\eps M \int_0^t   \left( d_c(\tau) + L_1 \|\txcin -\xcin\| \right)  d\tau  \right) \\ &=
					\ex{ - \alpha \sev(0,v) t } \left(d_s(0) + \eps M t  \|\txin -\xin\| +\eps M  \int_0^t  d_c(\tau)  \, d\tau  \ \right).
				\end{align} 
                
				\textbf{Bound for $d_s$ independent of $d_c$.}
				\normalfont
				\smallskip
				
				We want to eliminate the integral term in the last estimate \eqref{ds_estimate1}. The function $d_c$ is continuous with $d_c(0)=0$. Therefore,  \begin{align}
					d_c(t)\le \|\txin -\xin\|
				\end{align} at least for small $t\ge 0$, i.e.
				\begin{align}\label{definition_T}
					T:=\sup \, \{{t\ge 0} : d_c(\tau)\le \|\txin -\xin\| \text{ for all } 0\le \tau\le t \}\,>\,0.
				\end{align}
				Clearly, $T\ge \tout$ would imply that we can replace $d_c$ permanently. However, this is not a priori evident. Hence, we consider $T<\tout$ a possibility, i.e. we restrict $0\le t< T$ for the following estimates.
				%We use the obtained estimates to prove that indeed $T\ge \tout$.
				With this assumption, \eqref{ds_estimate1} becomes
				\begin{equation}\label{ds_estimate2} d_s(t)  \le
					% \ex{ - \alpha \sev(0,v) t } \left(d_s(0) + 2\eps M \int_0^t   \left( d_c(\tau) + L_1 \|\txcin -\xcin\| \right)  d\tau  \right) \\ &=
					\ex{ - \alpha \sev(0,v) t } \left(d_s(0) + \eps M t  \|\txin -\xin\| \right), \qquad \text{ for } t \in [0,T).
					%\\&\le\ex{ - \talpha \sev(0,v) t } \left(d_s(0) + (\eps M t+1)  \|\txin -\xin\| \right).
				\end{equation}
				%\subsubsection{cross estimates with $d_c$ and $d_s$.}\\
				
				\textbf{Final estimate for $d_c^\prime$.}
				\smallskip
				
				Recall the preliminary estimate \eqref{dc_derivative2}:
				\begin{equation}\label{dc_derivative2MID}
					d_c^\prime(t) \le \eps M \ex{\uev(0,v) t} \xuin d_s(t)  + \eps M \ex{- (\alpha \sev(0,v)- \uev(0,v)) t} \xuin   \left(d_c(t) + \| \txin - \xin\|\right).
				\end{equation}
				Note that we can bound the first term like the second term by an exponential decay with rate $\alpha \sev(0,v) - \uev(0,v)$. Indeed, we plug in the estimate \eqref{ds_estimate2} for $d_s$, which amounts to
				\begin{equation}
					\ex{\uev(0,v)t} d_s(t)
					\le
					\ex{-(\alpha \sev(0,v)- \uev(0,v)) t}  \left(d_s(0) + \eps M t  \|\txin -\xin\|\right)
				\end{equation}
				for $t \in [0,T)$. %Therefore,
				%\begin{equation}
				%	\ex{\uev(0,v)t} d_s(t)	\le 	\ex{-\alpha \sg t}  \left(d_s(0) + \eps M t  \|\txin -\xin\|\right),
				%\end{equation}
				%due to \eqref{estimate_with_talpha}.
				Now, the estimate $\eqref{dc_derivative2MID}$ becomes
				\begin{equation}\label{dc_derivative3}
					d_c^\prime(t)
					%&\le  \eps M \ex{-\alpha \sg t} \xuin  \left(d_s(0) + \eps M t  \|\txin -\xin\| + d_c(t) + \| \txin - \xin\| \right) \\
					\le
					\eps M \ex{- (\alpha \sev(0,v)- \uev(0,v)) t} \xuin  \left(d_c(t) + d_s(0) + \left(\eps t M +1 \right) \| \txin - \xin\| \right)
				\end{equation}
				for $t \in [0,T)$.
				
				\smallskip
				\textbf{Gr\"{o}nwall's lemma for $d_c$.}
				\smallskip
				
				The estimate \eqref{dc_derivative3} rewrites as
				\begin{equation}
					d_c(t)^\prime \le A(t) \,d_c(t) + b(t)
				\end{equation}
				with
				\begin{equation}
					\begin{aligned}
						A(t) &:=  \eps M \ex{-( \alpha \sev(0,v)- \uev(0,v)) t} \xuin\\
						b(t) &:=  \eps M \ex{-( \alpha \sev(0,v)- \uev(0,v)) t} \xuin  \cdot \left( d_s(0) + \left(\eps t M +1 \right) \| \txin - \xin\| \right).
					\end{aligned}
				\end{equation}
				With $d_c(0)=0$ the fundamental theorem of calculus yields
				\vspace{-0.1cm}
				\begin{equation}
					d_c(t) \le B(t) + \int_0^t A(\tau) d_c(\tau) \, d\tau,
				\end{equation}
				where $B(t) := \int_0^t b(\tau) d\tau$. Gr\"{o}nwall's lemma then yields a right side which is independent of $d_c$ and whose terms remain to be bounded:
				\begin{align}\label{Gronwall}
					d_c(t) \le B(t) + \int_0^t A(\tau) B(\tau)\ex{\int_\tau^tA(s)ds} d\tau.
				\end{align}
				%\smallskip
				Since $\xuin\le 1$,
				\begin{equation}\label{gronwall1}
					A(t) \le \eps M \ex{-( \alpha \sev(0,v)- \uev(0,v)) t}
				\end{equation}
				and thus, for $0\le \tau \le t$,
				\begin{align}\label{gronwall2}
					\int_\tau^tA(s) \,ds &\le  \frac{\eps M}{( \alpha \sev(0,v)- \uev(0,v)) }   \left(\ex{-( \alpha \sev(0,v)- \uev(0,v))\tau} - \ex{-( \alpha \sev(0,v)- \uev(0,v)) t}\right) \\ 
                    &\le \frac{\eps M}{ \alpha \sev(0,v)- \uev(0,v)}.
				\end{align}
				
				By definition,
				\begin{equation}\label{gronwall3}
					B(t) = \eps M \xuin  \int_0^t\ex{-( \alpha \sev(0,v)- \uev(0,v)) \tau}   \left(  d_s(0)  + \left(\eps \tau M +1 \right) \|\txin-\xin\|\right)  d\tau.
				\end{equation}
				
				Plugging in \eqref{ds_initial} in \eqref{gronwall3} this rewrites as
				\begin{equation}\label{gronwall4}
					B(t) \le \eps M \xuin \left( \left(\tilde{L}_s \frac{1}{\xuin}  + 1 \right) \int_0^t\ex{-( \alpha \sev(0,v)- \uev(0,v)) \tau}    d\tau
					+ \eps M \int_0^t \tau \ex{-( \alpha \sev(0,v)- \uev(0,v)) \tau}  d\tau\right) \|\txin - \xin\| .
				\end{equation}
				The first integral term satisfies
				\begin{align}\label{gronwall5}
					%\left(L_s \frac{1}{\xuin}  + L_2 \right) 
					\int_0^t\ex{-( \alpha \sev(0,v)- \uev(0,v)) \tau}    d\tau &= \frac{1-\ex{-( \alpha \sev(0,v)- \uev(0,v)) t} }{ \alpha \sev(0,v)- \uev(0,v)} \\
					%\left(L_s \frac{1}{\xuin}  + L_2 \right)
					&\le \frac{1}{ \alpha \sev(0,v)- \uev(0,v)},
				\end{align}
				and the second
				\begin{align}\label{gronwall6}
					\int_0^t \tau \ex{-( \alpha \sev(0,v)- \uev(0,v)) \tau}  d\tau &= \frac{1-\ex{-( \alpha \sev(0,v)- \uev(0,v)) t}\left(1+( \alpha \sev(0,v)- \uev(0,v)) t \right)}{( \alpha \sev(0,v)- \uev(0,v))^2} \\
                    &\le \frac{1}{( \alpha \sev(0,v)- \uev(0,v))^2}.
				\end{align}
				Thus, plugging \eqref{gronwall5}, \eqref{gronwall6} in \eqref{gronwall4}:
				\begin{equation}
					B(t) \le \eps M \xuin \left(  \left(\tilde{L}_s \frac{1}{\xuin}  + 1 \right) \frac{1}{( \alpha \sev(0,v)- \uev(0,v))} +  \frac{\eps M}{( \alpha \sev(0,v)- \uev(0,v))^2}  \right) \|\txin - \xin\| .
				\end{equation}
				Since $0\le \xuin \le 1$ and $\eps<1$,
				\begin{align}\label{gronwall7}
					\begin{split}
						B(t) &\le \eps M  \left( \frac{\tilde{L}_s   + 1}{( \alpha \sev(0,v)- \uev(0,v))} + \frac{ M}{( \alpha \sev(0,v)- \uev(0,v))^2}  \right) \|\txin - \xin\|  \\&=:
						\varepsilon L \|\txin - \xin\|,
					\end{split}
				\end{align}
				which is a time-independent estimate. Inserting the bounds \eqref{gronwall1}, \eqref{gronwall2}, \eqref{gronwall7} in the Gr\"{o}nwall inequality \eqref{Gronwall} yields
				\begin{equation}
					d_c(t) 
					%&\le B(t) + \int_0^t A(\tau) B(\tau) \ex{\int_\tau^tA(s)ds} d\tau \\&\le
					%\varepsilon L \|\txin - \xin\|  + \varepsilon L \|\txin - \xin\|  \ex{\frac{2 \eps M}{\alpha \sg}} \int_0^t A(\tau) d\tau \\&\le
					%\varepsilon L \|\txin - \xin\| \left(1 +  2 \eps M \ex{\frac{2 \eps M}{\alpha \sg}} \int_0^t \ex{-\alpha \sg \tau} d\tau\right)
					%\\&\le
					%\varepsilon L \|\txin - \xin\| \left(1 + \int_0^t A(\tau) \ex{\int_\tau^tA(s) \, ds}d\tau\right)\\&
					\le
					\varepsilon L \|\txin - \xin\| \left(1 +  \frac{ \eps M }{( \alpha \sev(0,v)- \uev(0,v))} \ex{\frac{\eps M}{( \alpha \sev(0,v)- \uev(0,v))}}\right)
					%\\&\le
					%\varepsilon L \|\txin - \xin\|.
				\end{equation}
				With $L_c:=L\left(1 +  \frac{ M }{( \alpha \sev(0,v)- \uev(0,v))} \ex{\frac{ M}{( \alpha \sev(0,v)- \uev(0,v))}}\right)$, noticing $\varepsilon < 1$, we obtain the time-independent estimate
				\begin{align}\label{Gronwall2}
					d_c(t) \le \varepsilon L_c \|\txin - \xin\|.
				\end{align}
				Recall that this estimate is only valid for $t \le T$, whereby $T$ is defined in \eqref{definition_T}. However, our estimate is independent of $t$.
				In particular, $$d_c(T) \le \varepsilon L_c \|\txin - \xin\| < \|\txin - \xin\| $$ for sufficiently small $\varepsilon_0$ and all $0<\varepsilon \le\varepsilon_0$, since $L_c$ is uniform in $\eps$. 
				As $d_c$ is continuous for $0\le t \le \tout$, this is a contradiction to the maximality of $T$. Consequently, \eqref{Gronwall2} is also valid for $t=\tout$. This upper bound is exactly what we needed. Inserting in \eqref{first_estimate_center_drift}
				yields the desired bound for the drift in center direction.
				
				\smallskip
				
				For the claim of the stable components recall estimate \eqref{ds_estimate1}:
				\begin{align}
					d_s(t) \le
					\ex{ - \alpha \sev(0,v) t } \left(d_s(0) + \eps M t  \|\txin -\xin\| +\eps M  \int_0^t  d_c(\tau)  \, d\tau  \ \right).
				\end{align}
				We plug in the newly obtained bound on $d_c$ and \eqref{ds_initial}, the estimate of the initial value $d_s(0)$:
				\begin{align} 
					d_s(t) & \le
					%\ex{ - \alpha \sev(0,v) t } \left(d_s(0) + \eps M L_1 t  \|\txin -\xin\| +\eps^2  C L  \|\txin - \xin\| \right) \\& \le
					\ex{ - \alpha \sev(0,v) t } \left(\tilde{L}_s \frac{1}{\xuin} \|\txin - \xin\|+ \eps M t  \|\txin -\xin\| +\eps^2  M  L_c t  \|\txin - \xin\| \right)
					\\
					& =
					\ex{ - \alpha \sev(0,v) t } \|\txin - \xin\| \left(\tilde{L}_s \frac{1}{\xuin} + \eps M t   +\eps^2  M L_c t  \right).
				\end{align}
				With $\tout= \frac{1}{\uev(0,v)} \log \frac{1}{\xuin}$ and the fact that $0 < \log\frac{1}{\xuin} < \frac{1}{\xuin}$,
				%$\log\frac{1}{\xuin}$ is bounded from above by  $\frac{1}{\xuin}$,
				\begin{align} %\label{ds_estimate_almost_final}
					d_s(\tout) & \le
					%\ex{ - \alpha \sev(0,v) \tout } \|\txin - \xin\| \left(L_s \frac{1}{\xuin} + \eps M L_1 \tout   +\eps^2  C L  \right) \\ &\le
					{\left(\xuin\right)}^{\frac{\alpha \sev(0,v)}{\uev(0,v)} }    \|\txin - \xin\| \left(\tilde{L}_s \frac{1}{\xuin} + \frac{\eps M }{\uev(0,v)} \log \frac{1}{\xuin}   + \frac{\eps^2  M L_c}{\uev(0,v)} \log \frac{1}{\xuin} \right) \\& \le 
					{\left(\xuin\right)}^{\frac{\alpha \sev(0,v)}{\uev(0,v)} - 1}   \|\txin - \xin\| \left(\tilde{L}_s+ \frac{\eps M}{\uev(0,v)}  +\frac{\eps^2  M L_c}{\uev(0,v)}   \right).
				\end{align}
				Note that $\beta := \frac{\alpha \sev(0,v)}{\uev(0,v)} - 1 > 0$ due to our choice of $\alpha$ in \eqref{alphaCHOICE}. Consequently, since $\varepsilon<1$, with $L_s:=\tilde{L}_s+ \frac{ M}{\uev(0,v)}  +\frac{  M L_c}{\uev(0,v)}$ we obtain
				\begin{align}
					d_s(\tout) & \le L_s {\left(\xuin\right)}^{\beta}  \|\txin - \xin\|.
				\end{align}
			\end{proof}
			
			\subsection{Global Passage and Concatenation}
			
			For each point $p_k$ of the sequence $(p_k)_{k\in \NN_0}$ representing the infinite heteroclinic chain $(h_k(t))_{k\in \NN_0}$ there is a local neighborhood $U_k$ with rectified invariant manifolds containing the box $\B_k=\B_k(\varepsilon)$.
			Within $\B_k$ there are local sections $\tinsection_{k}$ and $ \outsection_{k}$ transverse to the incoming and outgoing heteroclinic solutions with the local map $\Psiloc_k$.
			
			In this section we derive maps $\Psi_k : \insection_k \to \insection_{k+1}$ to describe the dynamics close to the heteroclinic chain. Each of these maps is the concatenation of the local map $\Psiloc_k: \tinsection_k \to \outsection_k$ from \autoref{chap:local} and a transition map $\Psiglob_k: \outsection_k \to \insection_{k+1}$ that connects the two local neighborhoods $U_k$ and $U_{k+1}$, see \autoref{sequence_of_sections}.
			
			While properties of the local map follow from the ordering\footnote{For $v > 1/2$, where the infinite heteroclinic chains are associated with the Cantor set $\Cantor(v)$ in \eqref{Cantor_def}. For fixed $v$, the compact set $\Cantor(v)$ is bounded away from the boundary points of $\A(v)$, where the positive eigenvalue becomes zero. Thus, in each point $p_k$ the three non-zero eigenvalues, as well as the spectral gap, are uniformly bounded away from zero. A lower bound is given by the evaluation in a point of period two. Consequently,
			% (together with the uniform boundedness of the higher order terms)
			there exists a minimal size of the local neighborhood, and thus of $\varepsilon$, where the analysis of \autoref{chap:local} is valid. Therefore, all estimates are uniform in $k$.
			} of the nonzero eigenvalues of the linearization in $p_k$, the behaviour of $\Psiglob_k$ is related to the Kasner map $\KM$. The transition map $\Psiglob_k$ describes the evolution of $B_k$ along the heteroclinic solution $h_k(t)$, i.e. close to one of the Kasner caps, the hemispheres where the heteroclinic solutions reside. The dynamics on the Kasner caps in turn is described by the Kasner map $\KM$, which expands $\KC$. In the local coordinates the Kasner circle is represented by the $\xc$-axis, and thus $\Psiglob_k$ expands the center component.
			Moreover, since the vector field of our consideration is analytic, and the local coordinate change is $\mathcal{C}^r$ for sufficiently large $r$, the transition map $\Psiglob_k$ is sufficiently smooth, which ensures bounded expansion of all components. In particular, any expansion in transverse direction forced by the transition map $\Psiglob_k$ is dominated by the contraction of the local map $\Psiloc_k$ in transverse direction, while the influence of the local passage on the center component is negligible, provided $B_k$ is chosen small enough.
			%\textcolor{red}{In X we have shown that the choosing $U_k$ small enough yields arbitrarily strong contraction of what? and at the same time, makes the drift in center direction arbitrarily small.}
			Accordingly, the concatenation of $\Psiloc_k$ and $\Psiglob_k$ yields a hyperbolic map $\Psi_k$.

			%To prove hyperbolicity of a single map $\Psi_k$ we study transformations of the cones defined in \eqref{cones_definition}

			The expansion of the center component prevents the heteroclinic chain $(h_k(t))_{k\in \NN_0}$ from attracting sets of full Lebesgue measure. But there are lower dimensional sets asymptotically approaching $(h_k(t))_{k\in \NN_0}$. More precisely, our goal is to construct stable manifolds of codimension one for both periodic and aperiodic heteroclinic chains. That is, we construct a three-dimensional set of solutions in the four-dimensional phase space shadowing the given heteroclinic chain. This three-dimensional set in turn is given in terms of a two-dimensional set of initial conditions within $\tinsection_{k}$. It is the local stable set of the origin $0 \in \tinsection_{k}$ under iteration of the hyperbolic map $\Psi_k$, and obtained by a graph transform. 
			The proof presented in \cite{Liebscher} concerns the case of the heteroclinic cycle of length three, which becomes a homoclinic solution in the $D_3$ orbit space. Thus, there is one local map and one transition map and their concatenation is a proper return map. Hyperbolicity then yields the stable set of the origin. The same idea applies directly to periodic heteroclinic chains, i.e. when there are finitely many maps $\Psi_k$ concatenated, since the concatenation of hyperbolic maps is hyperbolic.
			However, the proof for aperiodic chains requires a slightly more abstract setting, where we follow \cite{Liebscher_magnetic} to treat both periodic and aperiodic infinite heteroclinic chains in a unified way. Nonetheless, the main part of our analysis concerns properties of a single map $\Psi_k$, due to the uniformity of bounds in $k$. We therefore omit the index $k$, whenever we discuss a single map. 
			
			\subsubsection{The transition map $\Psiglob$} \label{subsection_global}
			\smallskip
			
			\textbf{Definition of $\Psiglob$ on the boundary subset $\xsssout=0$.}
			\smallskip

			% Note that points $(\xsout, \xssout, \xcout)$ where exactly one stable component is nonzero do not have a preimage under the local map.
			The point $\xout = (0, 0, \xcout) \in \outsection$
			%, the projection of $\xout$ onto $\BII$
			is the intersection of the locally rectified unstable manifold $W^u(\xcout)$ with $\outsection$, due to \eqref{manifolds}.
			It represents the heteroclinic solution with $\alpha$-limit point $ \left(0,0,0,\xcout \right) \in U_k$ and $\omega$-limit point $\left( 0,0,0,\KM(\xcout) \right) \in U_{k+1}$ on the Kasner circle in the respective local coordinates.
			The image of $\xout$ under the transition map is thereby%%iven by
			\begin{equation}\label{global_map_on_singular_set}
				\Psiglob\left(0, 0, \xcout \right) = \left(0, 0, \KM(\xcout) \right).
				% \in \insection_{k+1}.
			\end{equation} 
			Note that the Kasner map $\KM$ is not given by \eqref{KM} anymore, due to changes of coordinates.
			In particular, $\xc=0$ represents the point $(0,0,0) \in \outsection_{}$, which encodes the heteroclinic orbit joining $p_k$ and $p_{k+1}$, and the local neighborhoods $B_k$ and $B_{k+1}$ are centered around these points.
			However, we abuse notation and use the same symbol $\KM$ as before.

			Note that $m(v):=\min_{p\in \Cantor(v)}|\mathrm{D}\KM|>1$, since the (compact) Cantor set $\Cantor(v)$ is bounded away from the tangent points, see \eqref{minC}.
            Thus, by continuity $\mathrm{D}\KM$, for any $\eta>0$, there exists $\varepsilon_0$ such that for all $\varepsilon\in(0,\varepsilon_0)$ and $\xcout\txcout\in \outsection$, we have that%we define $\tilde{K}_c:= \min_{p\in \Cantor(v)} |D\mathcal{K}(p)|>1$ such that
			\begin{align}\label{KM_expansion}
				|\KM(\txcout)- \KM(\xcout)| \ge (m(v)-\eta) |\txcout- \xcout|.
			\end{align}
			This expansion, however, may prevent $\left(0, 0, \KM(\xcout) \right)$ from being contained in $\insection_{k+1}$.
			
			%However, the continuous dependence on initial conditions of the flow yields a subset $\toutsection_{k} \subseteq \outsection_k$ of points intersecting $\insection_{k+1}$ under the flow.
			
			\textbf{Well-definedness
            of $\Psiglob$ on a subset $\toutsection_{} \subseteq \outsection_{}$.}
			\smallskip
			
			For $\Psiglob$ to be a well-defined map, i.e. for the image of $\Psiglob$ to be contained in the next in-section, we must reduce the size of the domain appropriately. As $\Psiglob$ is given by the evolution of a sufficiently smooth vector field, it is a diffeomorphism onto its image. Thus, all components of $\Psiglob$ as well as its derivatives are bounded by a common constant $\Cglob>0$.
			Depending on this bound we find a subset $\toutsection \subseteq \outsection$ containing $\xout = (0,0,0)$,
			where $\Psiglob$ is well-defined. Of course, this affects the domain of well-definedness of $\Psi$. We address this issue on the next subsection.
			
			%\textcolor{red}{Maybe comment/reference to Tom, he bounds global time.}

			\smallskip
			\textbf{Estimate for $\Psiglob$.}
			\smallskip
			
			Let $\xout=\left(\xsout,\xssout,\xcout\right) \in \outsection$.
			With Taylor's theorem and \eqref{global_map_on_singular_set} %\eqref{global_map_on_singular_set}, %($f(x_0 +h) = f(x_0) + f^\prime(x_0)h+h.o.t, x_0=(0,0,\xc), h=(\xs,\xss,0)$)
			\begin{align}\label{taylor}
				\Psiglob(\xout) = (0,0, \KM(\xcout)) + \tPsiglob(\xout) \xsssout
			\end{align} with a sufficiently smooth matrix-valued function $\tPsiglob$.
			Hence the components of $\tPsiglob$ as well as the components of the derivative $\mathrm{D}\tPsiglob$ are bounded with respect to sup-norm by a common constant $M^{\text{glob}}>0$. 
Note that one can compare how close $\Psi^{\text{glob}}$ is to the Kasner map $\KM$: for sufficiently small $\xsssout$, the image under $\Psi^{\text{glob}}(\xout)$ and the Kasner map $\KM(\xcout)$ are close to each other.
 %  
			%\textcolor{red}{(bounded) expansion of $\KM$.
				%Somehow derive a formula that shows that the map can be made well-defined.}
			%Applying the upper bound $\Cglob$ to the difference of their higher order terms yields
		
            Next, we consider a second point $\txout \in \toutsection$ and we show that if it is sufficiently close to $\xout$, then their images under $\Psi^{\text{glob}}$ and the Kasner map remain close:			%\begin{subequations}\label{glob_hot_estimate:subs}
			\begin{equation}\label{global_drift}
				\begin{aligned}
					\|\Psiglob(\txout)  - &\Psiglob(\xout) - \left(0,0,\KM(\txcout)- \KM(\xcout)\right) \| \\
					= &\,\| \, \tPsiglob(\txout) \txsssout - \tPsiglob(\xout) \xsssout \| 
					%	\\ &=  \|\tPsiglob(\txout) \left(\txsssout- \xsssout\right) + \left(\tPsiglob(\txout) - \tPsiglob(\xout)\right) \xsssout \| 
					\\  \le &\,\|\,\tPsiglob(\txout) \left(\txsssout- \xsssout\right)\| + \|\left(\tPsiglob(\txout) - \tPsiglob(\xout)\right) \xsssout\|  \\ 
					\le  &\,\Cglob \left(  \| \txsssout- \xsssout \| + \| \txout - \xout \| \| \xsssout\| \right), 
				\end{aligned}
			\end{equation}
			%\end{subequations}
                %
			where the last inequality uses the mean value theorem. 
			%Note that for $\txout$ and $\xout$ sufficiently close, the right-hand side in \eqref{global_drift} is arbitrarily small.

			%With the previous estimate we can bound the drift of arbitrariy points from the outsection to the images of the respective projected points to mixmaster attractor.The distance of two image points compared to their initial distance is bounded.
			%	\begin{align} \label{global_drift}
				%\begin{split}
				%	&\|\left(\Psiglob(\txout) - \Psiglob(\xout)\right) - \left(\Psiglob(0,0,\txcout) - \Psiglob(0,0,\xcout)\right) \| \\
				%	&=\|\Psiglob(\txout) - \Psiglob(\xout) - \left(0,0,\KM(\txcout)- \KM(\xcout)\right) \| \\ &\le \,  \Cglob \left(  \| \txsssout- \xsssout \| + \| \txout - \xout \| \| \xsssout\| \right).
				%\end{split}
				%	\end{align}
			
			%\textcolor{red}{summarize $\Psiglob$ in a lemma?}
			
			\subsubsection{The map $\Psi = \Psiglob \circ \Psiloc$}
			We now derive properties for the return map $ \Psi_k:\tinsection_{k} \to \insection_{k+1}$. Note that $\Psi = \Psi_k$ is Lipschitz continuous, as it is the concatenation of the Lipschitz map $\Psiloc$ with the smooth map $\Psiglob$.
			%	 Firstly, we describe $\Psi$ on the set $\{\xuin=0\}$ and we discuss well-definedness on a suitable subset of $\insection_{k}$.
			
			\smallskip
			\textbf{Description of $\Psi$ on the boundary subset $\xuin=0$.}
			\smallskip
			
			In \eqref{definition_localmap} we define the local map for $\xin = (\xuin=0, \xsin, \xcin)$ via \begin{equation}
				\Psiloc(\xin) = \left(\xsout, \xssout, \xcout \right)= \left(0, 0, \xcin \right) \in \outsection,
			\end{equation} that is, the $x_c$-component is not changed locally.
			An arbitrary point $(0,0,\xcout) \in \outsection$ of this form is mapped to the point $(0,0,\KM(\xcout))$ in the next insection, due to \eqref{global_map_on_singular_set}. Thus, on the boundary set, $\{\xuin=0\} \subseteq \tinsection_{k}$, the map $\Psi = \Psiglob \circ \Psiloc$ is given by
			\begin{align}\label{return_map_on_singluar_set}
				\Psi\left(0, \xsin, \xcin \right) = \left(0, 0, \KM(\xcin) \right).
			\end{align}

            \newpage
            
   			\smallskip
			\textbf{Well-definedness of $\Psi$ on a subset of the in-section.}
			\smallskip
			
			In Corollary~\ref{corollary_proposition} we restrict $\Psiloc$ to a cubical subset $\tinsection$ of size $\delta$ to take the local drift in center direction into account, and thereby ensure that all solutions intersect $\outsection$.
			In \autoref{subsection_global}, however, we restrict the transition map $\Psiglob$ to a subset $\toutsection \subseteq \outsection$.
			Consequently, we must decrease the size of the domain of $\Psiloc$ further to make sure that $\Psi=\Psiglob \circ \Psiloc$ is well-defined.
			%\textcolor{red}{But here the map goes to the larger next insection, so iterating eventually kicks the points out of the insections...?}
			%Note that the restriction is after the choice of rescaling parameter $\varepsilon$.
			By Lipschitz continuity of the local map, we can choose a possibly smaller value for $\delta$, such that the preimage of $\toutsection$ under $\Psiloc$ contains the cube
			\begin{align} \label{section_delta}
				\tinsection &:= \left\{\xin \in \B \, | \, \xssin=1, \shortquad 0\le\xuin\le \delta,\shortquad   |\xsin|\le \delta, \shortquad  |\xcin| \le \delta \right\},
			\end{align}
			where $\Psi$ yields a well-defined map. Recall that $B$ denotes the cube around the origin with side length $4$ in the local coordinates after rescaling, i.e. of side length $4 \varepsilon$ before rescaling.
			
			%\textcolor{red}{Comment on order of fixing the parameter $\varepsilon,\delta$.}
			% Now we must also take into account the global expansion. We choose this subset in (ref), where we also summarize the order of choosing the parameters.

			\smallskip
					\textbf{Estimates for $\Psi$.}
					\smallskip

                    We combine the estimate \eqref{global_drift} of $\Psiglob$ with the estimates of the local map $\Psiloc$ from Theorem~\ref{local_theorem} to obtain estimates for center and transverse component of $\Psi = \Psiglob \circ \Psiloc$. %These estimates are	used in the remaining four steps to prove the cone conditions \eqref{cone_conditions:full}.
                    First, note that $\Psi(\xin)$ is close to $(0,0,\KM(\Psiloc (\xin)_c))$ for sufficiently small $\xout_{s,ss}=\Psiloc(\xin)_{s,ss}$, which amounts to sufficiently small $\xin_{u}$, due to \eqref{taylor} and the local estimate \eqref{psiloc_estimates}.
                   
					To get an estimate for $\Psi = \Psiglob \circ \Psiloc$ that compares different points, we consider $\xin \in \tinsection$ and $\xout= \Psiloc(\xin)$. Then the global estimate \eqref{global_drift} for $\Psiglob$ becomes
					\begin{align}\label{estimate_return0}
						\begin{split}
							\|\Psi(\txin) - \Psi(\xin) &- (0,0,\KM(\Psiloc(\txin)_c)- \KM(\Psiloc(\xin)_c)) \|\\
							& \le \Cglob \left(  \| \txsssout- \xsssout \| + \| \txout - \xout \| \| \xsssout\| \right).
						\end{split}
					\end{align}
					Corollary~\ref{corollary_proposition} yields $\|\xsssout\| \le{\left(\xuin\right)}^{\beta} \|\xsssin\| \le 2{\left(\xuin\right)}^{\beta}$, 
					which is used to bound the last term in \eqref{estimate_return0}.
					Moreover, since $\max\{ \xuin,\txuin\}\le\delta$, the local estimate \eqref{stable_theorem} in Theorem~\ref{local_theorem} and Lipschitz continuity \eqref{Lipschitz} imply
					\begin{align}\label{estimate_return}
						\begin{split}%
							\|\Psi(\txin) - \Psi(\xin) &- (0,0,\KM(\Psiloc(\txin)_c)- \KM(\Psiloc(\xin)_c)) \|\\
							& \le \Cglob \left(  L_s \delta^{\beta} \| \txin - \xin \| + \tilde{L} \| \txin - \xin \| 2 \delta^{\beta} \right)
							\\& = \Cglob \left(  L_s + 2\tilde{L} \right) \delta^{\beta} \| \txin - \xin \|
							\\&=: \Cret \delta^{\beta} \|\txin-\xin\|.
						\end{split}
					\end{align}
     The reverse triangle inequality in \eqref{estimate_return} yields that %$\Psi$ and $\KM$ are close to each other for fixed $\xin_u$ and $\txin$ sufficiently close to $\xin$, i.e.
     				\begin{equation}\label{estimate_return2}
							\|\Psi(\txin) - \Psi(\xin)\| \le |\KM(\Psiloc(\txin)_c)- \KM(\Psiloc(\xin)_c) |+ \Cret \delta^{\beta} \|\txin-\xin\|.
					\end{equation}
					Note that the estimate \eqref{estimate_return} contains information about both the center and transverse directions. We will use them separately. Indeed, since the norm in the first line of $\eqref{estimate_return}$ is the sum of the center and transverse components, we respectively obtain a bound for each of these components, separately:
					\begin{align}\label{estimate_return_center:subs}
						%&\left|\left(\Psi(\txin) - \Psi(\xin) - (0,0,\KM(\Psiloc(\txin)_c)- \KM(\Psiloc(\xin)_c))\right)_c \right| \\
						\left|\Psi(\txin)_c - \Psi(\xin)_c - \left(\KM(\Psiloc(\txin)_c)- \KM(\Psiloc(\xin)_c)\right) \right| &\le \Cret \delta^{\beta} \|\txin-\xin\|, \\ %\label{estimate_return_center:last}  \\
						\label{estimate_return_hyperbolic}
						%&\left\|\Big(\Psi(\txin) - \Psi(\xin) - (0,0,\KM(\Psiloc(\txin)_c)- \KM(\Psiloc(\xin)_c))\Big)_{u,s} \right\|  \\
						\left\|\Psi(\txin)_{u,s} - \Psi(\xin)_{u,s} \right\| &\le \Cret \delta^{\beta} \|\txin-\xin\|.
					\end{align} 

            \newpage 
            
			\smallskip
			\textbf{Hyperbolicity of $\Psi$: cone properties}
			\smallskip
   
			Next, in Lemma~\ref{cones_theorem}, we guarantee that $\Psi$ is hyperbolic via transformation properties of certain cones. This is the main technical property which will be used to prove the main Theorem~\ref{manifold_theorem} in the upcoming Section, which yields a two-dimensional stable set of the hyperbolic fixed point $(0,0,0)$ of the return map $\Psi$. 
			
			For any $\sigma>0$ and $\xin \in \tinsection$, where $\tinsection$ is given by \eqref{section_delta}, we respectively define
				the \emph{center $\sigma$-cone} and the \emph{transverse $\sigma$-cone} at $\xin$ as
				\begin{subequations}\label{cones_definition}
					\begin{align}
						\Cc{\sigma}{\xin}  &:= \left\{ \txin \in \tinsection  : \|\txusin - \xusin \| \le \sigma \left| \txcin -\xcin\right|  \right\},\\
                            \Cus{\sigma}{\xin} & := \left\{ \txin \in \tinsection  :  \left| \txcin -\xcin\right| \le \sigma \|\txusin - \xusin \|  \right\}.
					\end{align}
				\end{subequations}
			%			
                %\textcolor{red}{PL: it seems that with the new proof, we have that FOR ANY $\sigma \in (0,1)$, there are $K_c>1$ and $K_{u,s}>1$ such that.}   
                %
			\begin{lem}\label{cones_theorem}
				Let $v\in (1/2,1)$. For any $\sigma \in (0,1)$ and sufficiently small $\epsilon_0,\delta>0$, according to $\tinsection$ in \eqref{section_delta}, 
				there are constants $K_c , K_{u,s}>1$, such that the following conditions hold: %hold for all $\xin \in \tinsection$.
				\begin{subequations}\label{cone_conditions:full} 
					\begin{enumerate}[label = \normalfont \textbf{(\roman*)}]
						\item \normalfont \textbf{Forward invariance of the center cone:}
						\begin{equation}\label{forward_invariance_center_condition}
							\Psi\left(\Cc{\sigma}{\xin}\right) \cap \tinsection  \subset \{\Psi(\xin)\}  \,\cup \, {\normalfont \text{int }} \Cc{\sigma}{\Psi(\xin)}. 
						\end{equation}
						\item \normalfont \textbf{Forward expansion in center direction:}	For all $\txin \in \Cc{\sigma}{\xin}$
						\begin{equation}\label{forward_expansion_center_condition}
							K_c \left|\txcin -\xcin\right| \leq \left|\Psi(\txin)_c - \Psi(\xin)_c\right|.
						\end{equation}
						\item \normalfont \textbf{Backward invariance of the transverse cone: }
						\begin{equation}\label{backwards_invariance_transverse_condition}
							\Psi^{-1}\left(\Cus{\sigma}{\Psi(\xin)} \right) \cap \tinsection \subset  \{\xin\}  \cup  {\normalfont \text{int }} \Cus{\sigma}{\xin}.
						\end{equation}
						\item \normalfont \textbf{Forward contraction in transverse direction:} For all  $\Psi(\txin) \in \Cus{\sigma}{\Psi(\xin)}$,
						\begin{equation}\label{forward_contraction_transverse_condition}
							\|\Psi(\txin)_{u,s} -\Psi(\xin)_{u,s}\| \le \frac{1}{K_{u,s}}\|\txusin - \xusin\|.
						\end{equation} 
					\end{enumerate}
				\end{subequations}
\vspace{-1cm}
				\begin{figure}[H]
					\centering
					\subfloat[The center $\sigma$-cone $\Cc{\sigma}{\x}$ (in gray) is \\ forward invariant and expanding.]{\begin{tikzpicture}[scale=0.35]
							% left-right shift
\def \s {11}

% length of coordinate system
\def \ly {5}
\def \lx {7}

\def \mx {0.7}
\def \my {0.7}

\coordinate (origin1) at (0,0);
\coordinate (origin2) at (\s,0);

\foreach \i in{1,2}
{\path let \p1=(origin\i) in coordinate (xneg\i) at (\x1-\mx cm ,\y1);
	\path let \p1=(origin\i) in coordinate (xpos\i) at (\x1+\lx cm,  \y1);
	
	\path let \p1=(origin\i) in coordinate (yneg\i) at (\x1,\y1-\my cm);
	\path let \p1=(origin\i) in coordinate (ypos\i) at (\x1,\y1+\ly cm);
	
\draw (xneg\i)--(xpos\i) node [anchor=west] {\small $\xus$};
\draw (yneg\i)--(ypos\i) node [anchor=south] {\small $\xc$};
}

% cone centers 
\coordinate (center1) at (4.5,1.5);
\path let \p1=(center1) in coordinate (center2) at (\x1- 2.2cm + \s cm ,\y1+1.2cm);

% cone opening
\def \sigma {0.6}

% cone length
\def \conelength {2}

\definecolor{light-gray}{gray}{0.8}

\foreach \i in{1,2}
{\path let \p1=(center\i) in coordinate (coneBL\i) at (\x1-\conelength*\sigma cm ,\y1- \conelength cm);
	\path let \p1=(center\i) in coordinate (coneTL\i) at (\x1-\conelength*\sigma cm ,\y1+ \conelength cm);
	
	\path let \p1=(center\i) in coordinate (coneBR\i) at (\x1+\conelength*\sigma cm ,\y1- \conelength cm);
	\path let \p1=(center\i) in coordinate (coneTR\i) at (\x1+\conelength*\sigma cm ,\y1+ \conelength cm);
	
	\path [fill = light-gray] (coneBL1)--(coneBR1)--(center1)--(coneBL1);
	\path [fill = light-gray] (coneTR1)--(coneTL1)--(center1)--(coneTR1);
	
	\draw (coneBL\i)--(coneTR\i);
	\draw (coneTL\i)--(coneBR\i);
}

%caption cone centers
\filldraw (center1) circle (1.5pt) ;
\filldraw (center2) circle (1.5pt); 
\path let \p1=(center1) in coordinate (center1cap) at (\x1, \y1+0 cm);
\path let \p1=(center2) in coordinate (center2cap) at (\x1, \y1+0 cm);
\filldraw (center1cap) circle (0pt) node [anchor=east] {\small $\x$};
\filldraw (center2cap) circle (0pt) node [anchor=west] {\small $\Psi(\x)$};

%\path let \p1=(center1) in coordinate (coneBL1) at (\x1-\conelength cm ,\y1- \conelength* \sigma cm);
\draw (coneBL1)--(coneTR1);
\draw (coneTL1)--(coneBR1);

\def \conelengthim {2.6}
\def \sigmaim {0.3}

\foreach \i in{1,2}
{\path let \p1=(center\i) in coordinate (coneBL\i) at (\x1-\conelengthim*\sigmaim cm ,\y1- \conelengthim cm);
	\path let \p1=(center\i) in coordinate (coneTL\i) at (\x1-\conelengthim*\sigmaim cm ,\y1+ \conelengthim cm);
	
	\path let \p1=(center\i) in coordinate (coneBR\i) at (\x1+\conelengthim*\sigmaim cm ,\y1- \conelengthim cm);
	\path let \p1=(center\i) in coordinate (coneTR\i) at (\x1+\conelengthim*\sigmaim cm ,\y1+ \conelengthim cm);
}
	\path [fill = light-gray] (coneBL2)--(coneBR2)--(center2)--(coneBL2);
	\path [fill = light-gray] (coneTR2)--(coneTL2)--(center2)--(coneTR2);
	
	\draw (coneBL2)--(coneTR2);
	\draw (coneTL2)--(coneBR2);

\filldraw (\s-1.8,3.2) circle (0pt) node [anchor=south] {\small $\Psi$};

\draw [line width = 0.3pt, -Latex] (\s-3,3) to[bend left=30] (\s-0.6,3);
						\end{tikzpicture}}
					\hspace{0.5cm}
					\subfloat[The transverse $\sigma$-cone $\Cus{\sigma}{\Psi(\x)}$ (in gray) is backward invariant and expanding.]
					{\begin{tikzpicture}[scale=0.35]
							% left-right shift
\def \s {11}

% length of coordinate system
\def \ly {5}
\def \lx {7}

\def \mx {0.7}
\def \my {0.7}

\coordinate (origin1) at (0,0);
\coordinate (origin2) at (\s,0);

\foreach \i in{1,2}
{\path let \p1=(origin\i) in coordinate (xneg\i) at (\x1-\mx cm ,\y1);
\path let \p1=(origin\i) in coordinate (xpos\i) at (\x1+\lx cm,  \y1);

\path let \p1=(origin\i) in coordinate (yneg\i) at (\x1,\y1-\my cm);
\path let \p1=(origin\i) in coordinate (ypos\i) at (\x1,\y1+\ly cm);

%coordinate axes
\draw (xneg\i)--(xpos\i) node [anchor=west] {\small $\xus$};
\draw (yneg\i)--(ypos\i) node [anchor=south] {\small $\xc$};
}

% cone centers 
\coordinate (center1) at (4.5,1.5);
\path let \p1=(center1) in coordinate (center2) at (\x1- 2.2cm + \s cm ,\y1+1.2cm);

% cone opening
\def \sigma {0.6}

% cone length
\def \conelength {2.1}
\definecolor{light-gray}{gray}{0.8}

\foreach \i in{1,2}
{\path let \p1=(center\i) in coordinate (coneBL\i) at (\x1-\conelength cm ,\y1- \conelength* \sigma cm);
\path let \p1=(center\i) in coordinate (coneTL\i) at (\x1-\conelength cm ,\y1+ \conelength* \sigma cm);

\path let \p1=(center\i) in coordinate (coneBR\i) at (\x1+\conelength cm ,\y1- \conelength* \sigma cm);
\path let \p1=(center\i) in coordinate (coneTR\i) at (\x1+\conelength cm ,\y1+ \conelength* \sigma cm);

}
	\path [fill = light-gray] (coneBL2)--(coneTL2)--(center2)--(coneBL2);
	\path [fill = light-gray] (coneBR2)--(coneTR2)--(center2)--(coneBR2);
	
	\draw (coneBL1)--(coneTR1);
	\draw (coneTL1)--(coneBR1);
	
	\draw (coneBL2)--(coneTR2);
	\draw (coneTL2)--(coneBR2);

\path let \p1=(center1) in coordinate (coneBL1) at (\x1-\conelength cm ,\y1- \conelength* \sigma cm);

\def \conelengthim {2.9}
\def \sigmaim {0.35}

\foreach \i in{1,2}
{\path let \p1=(center\i) in coordinate (coneBL\i) at (\x1-\conelengthim cm ,\y1- \conelength* \sigmaim cm);
	\path let \p1=(center\i) in coordinate (coneTL\i) at (\x1-\conelengthim cm ,\y1+ \conelength* \sigmaim cm);
	
	\path let \p1=(center\i) in coordinate (coneBR\i) at (\x1+\conelengthim cm ,\y1- \conelength* \sigmaim cm);
	\path let \p1=(center\i) in coordinate (coneTR\i) at (\x1+\conelengthim cm ,\y1+ \conelength* \sigmaim cm);
	
}
\path [fill = light-gray] (coneBL1)--(coneTL1)--(center1)--(coneBL1);
\path [fill = light-gray] (coneBR1)--(coneTR1)--(center1)--(coneBR1);

\draw (coneBL1)--(coneTR1);
\draw (coneTL1)--(coneBR1);

%\draw (coneBL2)--(coneTR2);
%\draw (coneTL2)--(coneBR2);

\filldraw (center1) circle (1.5pt) ;
\filldraw (center2) circle (1.5pt);

%caption
\path let \p1=(center1) in coordinate (center1cap) at (\x1, \y1+0.2 cm);
\path let \p1=(center2) in coordinate (center2cap) at (\x1, \y1+0.2 cm);

\filldraw (center1cap) circle (0pt) node [anchor=south] {\small $\x$};
\filldraw (center2cap) circle (0pt) node [anchor=south] {\small $\Psi(\x)$};

%arrow and map
\filldraw (\s-1.6,3.2) circle (0pt) node [anchor=south] {\small $\Psi^{-1}$};
\draw [line width = 0.3pt, Latex-] (\s-3,3) to[bend left=30] (\s-0.6,3);

%\filldraw (\s+4,2.8) circle (0pt) node [anchor=west] {\small $\Cus{\sigma}{\Psi(\x)}$};
						\end{tikzpicture} }
					\caption{The properties of cones in Lemma~\ref{cones_theorem} under the maps $\Psi$ and $\Psi^{-1}$.}  \label{fig:cones}
				\end{figure}
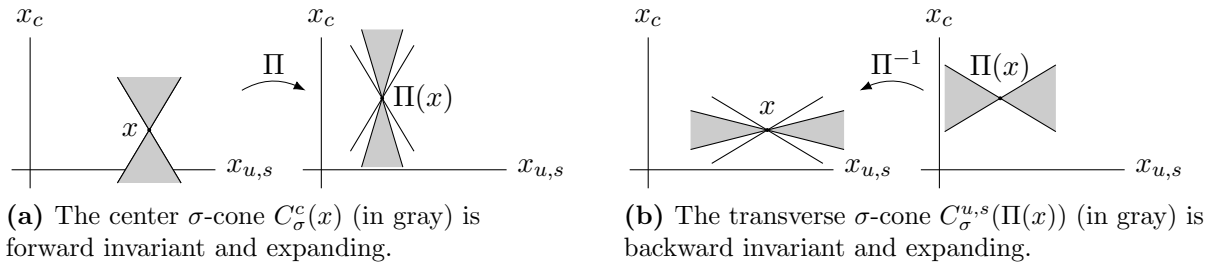
   \end{lem}
			\begin{proof}
           The proof is organized in four steps, one for each of the claims. The idea of these points is depicted in \autoref{fig:cones}. The main ingredients are the estimates for center and transverse components of $\Psi = \Psiglob \circ \Psiloc$ given by \eqref{estimate_return}.
				%	\textcolor{red}{all estimates are uniform in $k$!}
				%Theorem \ref{local_theorem} provides $\varepsilon_0>0$ such that for any choice of $\varepsilon \le \varepsilon_0$ the estimates
				%\begin{align}\label{local_theorem_step0}
				%\begin{split}
				%\left|\left(\txcout-\xcout\right) - \left(\txcin-\xcin\right) \right| &\le \varepsilon L_c \| \txin - \xin \| \\
				%\| \txsssout- \xsssout  \| & \le L_s{\left(\xuin\right)}^{\beta} \| \txin - \xin \|
				%\end{split}
				%\end{align}
				%hold, uniformly in $k$, and for constants $L_c,L_s,\beta>0$, which are independent of $\eps$.
				
				%	Triangle inequality applied to the center component in \eqref{local_theorem_step0} yields \begin{equation} %\label{center_theorem_2}
					%|\txcout-\xcout|\le (\varepsilon L_c + 1) \| \txin - \xin \|.
					%\end{equation}
					%Thus
					%\begin{equation}\label{Lipschitz}
					%\|\txout - \xout \| \le \left(\varepsilon L_c + 1+ L_s {\left(\xuin\right)}^\beta \right) \| \txin - \xin \| := \tilde{L}\| \txin - \xin \|,
					%\end{equation}
					%which makes the Lipschitz continuity of $\Psiloc$ explicit.
					%
					
					%Since we we make use of estimate \eqref{global_drift} in the first step we rewrite it here:
					%\begin{align}
					%\begin{split}
					%&\|\Psiglob(\txout) - \Psiglob(\xout) - \left(0,0,\KM(\txcout)- \KM(\xcout)\right) \| \\ \le \, &  \Cglob \left(  \| \txsssout- \xsssout \| + \| \txout - \xout \| \| \xsssout\| \right),
					%\end{split}
					%\end{align}

					\smallskip
					\textbf{Step 1: Forward expansion of the center cone $\Cc{\sigma}{\xin}$.}
					\smallskip
					
					With the reverse triangle inequality in the local estimate \eqref{center_theorem} we find the lower bound
					\begin{equation}\label{estimate_local_center}
						|\txcout - \xcout | \ge |\txcin - \xcin| - \eps L_c \| \txin - \xin \| .
					\end{equation}
					Analogously, the reverse triangle inequality applied to \eqref{estimate_return_center:subs} yields
					\begin{align}
						\left| \Psi(\txin)_c - \Psi(\xin)_c \right|  \ge \left|\KM(\txcout)- \KM(\xcout)\right| - \Cret \delta^{\beta} \|\txin-\xin\|.
					\end{align}
					Employing the minimal expansion rate $m(v)>1$ of $\KM$ in  \eqref{KM_expansion} and estimate \eqref{estimate_local_center}:
					\begin{equation}\label{return_center_expansion}
						\begin{aligned}
							\left| \Psi(\txin)_c - \Psi(\xin)_c \right|
							%&\ge \left|\KM(\txcout)- \KM(\xcout) \right| - \Cret \delta^{\beta} \|\txin-\xin\| \\
							&\ge  (m(v) -\eta) \left| \txcout- \xcout \right| - \Cret \delta^{\beta} \|\txin-\xin\| \\
							&\ge  (m(v) -\eta) |\txcin - \xcin| - \big(\eps L_c (m(v)-\eta)  + \Cret\delta^{\beta}   \big)\| \txin - \xin \| .
						\end{aligned}
					\end{equation}
					%Let $\sigma..$. Also let $\vartheta >0$. We use $\vartheta$ as opening of the cone, too. However, the following estimates are more general and we specify $\vartheta$ afterwards/set in relation with $\sigma$. 
					Note that the estimates \eqref{estimate_local_center} and \eqref{return_center_expansion} are valid for all points in $\tinsection$.
					Let $\txin \in\Cc{\sigma}{\xin}$, which by definition implies that $\|\txusin -\xusin \| \le \sigma |\txcin - \xcin|$,
			%\begin{equation}\label{center_cone_condition}
				%		\|\txusin -\xusin \| \le \sigma |\txcin - \xcin|
				%	\end{equation}
					and consequently \begin{equation}\label{norm_sigma_center}
						\| \txin - \xin\| \le (1+\sigma) |\txcin -\xcin|.
					\end{equation}
					Using this relation, \eqref{return_center_expansion} yields an estimate for the center component only:
                        \begin{equation}                         
                        \begin{aligned}
						\left| \Psi(\txin)_c - \Psi(\xin)_c \right| &\ge  \left( m(v) -\eta - \big(\eps L_c (m(v)-\eta)  + \Cret{\delta}^{\beta}   \big) (1+\sigma) \right)  |\txcin -\xcin|\\
                        &=:K_c |\txcin -\xcin|.\label{estimate_return_center_final}
					\end{aligned}                                \end{equation}
					To finish the proof of the expansion in center direction, we need to guarantee that $K_c>1$. 
					Indeed, we can decrease $\varepsilon$ because all local estimates are uniform in $\eps$, i.e. a smaller $\eps$ does not increase $L_c$ and $m(v)>1$. Moreover,
					after fixing $\eps$, and thereby $\insection$, $\Psi$ and $\Cret$, the subset $\tinsection$ in \eqref{section_delta} is determined by the choice of a sufficiently small $\delta$, and thereby constraining $\xuin \le \delta$.
					Thus, there exist sufficiently small $\eps,\delta,\eta$ such that $\big(\eps L_c (m(v)-\eta)  + \Cret{\delta}^{\beta}   \big) (1+\sigma)  $ is arbitrarily small for any $\sigma \in (0,1)$. 
                    Hence $K_c\in (1,m(v)-\eta)$ and $K_c$ can be chosen arbitrarily close to $m(v)$. Note, however, that this proof fails for some heteroclinic chains in the case $v=1/2$, since the chain of period 2 converges to the Taub points, where $m(v)\to 1$ and thereby the admissible $\varepsilon,\delta\to 0$.
%       
 %    
					%\begin{align}\label{expansion_constant1}
				%		\tilde{K_c}- \big(\eps L_c \tilde{K_c}  + \Cret\delta^{\beta}   \big) (1+\sigma)  \ge K_c,
				%	\end{align}
					%and we finally obtain
		           %
             %\begin{align}\label{estimate_return_center_final}
					%	\left| \Psi(\txin)_c - \Psi(\xin)_c \right| &\ge  K_c |\txcin -\xcin|,
					%\end{align}
					%which proves the expansion in center direction. 
					
					%In \eqref{KM_expansion} we introduced the expansion constant $\tilde{K}_c>1$ of the Kasner map $\KM$. 
     
					\smallskip
					\textbf{Step 2: Forward invariance of the center cone $\Cc{\sigma}{\xin}$.}
					\smallskip
					
					For forward invariance of $\Cc{\sigma}{\xin}$, we must show that if %applying $\Psi$ preserves the inequality \eqref{center_cone_condition}, i.e. we must show that whenever
					\begin{equation}\label{step3.1}
						\| \txusin-\xusin \| \le \sigma | \txcin-\xcin |, 
					\end{equation}
					then
					\begin{equation}\label{step3.2}
						\|\Psi(\txin)_{u,s}-\Psi(\xin)_{u,s}\| \le \sigma \left| \Psi(\txin)_c-\Psi(\xin)_c \right|.
					\end{equation}
					Combining the estimates \eqref{estimate_return_hyperbolic} and \eqref{norm_sigma_center} together with \eqref{estimate_return_center_final} yield
					\begin{equation}
						\begin{aligned}
							\| \Psi(\txin)_{u,s}-\Psi(\xin)_{u,s} \|  & \le
							\Cret \delta^{\beta} (1+\sigma) |\txcin - \xcin | \\ & \le  \frac{\Cret \delta^{\beta} (1+\sigma) }{K_c}\left| \Psi(\txin)_c - \Psi(\xin)_c \right|.
							%\\ & \le  \sigma \left| \Psi(\txin)_c - \Psi(\xin)_c \right|
						\end{aligned}
					\end{equation}
					As in \eqref{estimate_return_center_final}, we can decrease the size of $\tinsection_k$ (if $\xuin$ is not small enough yet) by choosing small $\delta > 0$ to achieve the following inequality that yields the desired claim: \begin{equation}\label{expansion_constant2}
						\frac{\Cret \delta^{\beta} (1+\sigma) }{K_c}  < \sigma.
					\end{equation}
					%Thus, 
					%\begin{equation}
					%\| \Psi(\txin)_{u,s}-\Psi(\xin)_{u,s} \| \le  \sigma \left| \Psi(\txin)_c - \Psi(\xin)_c \right|,
					%\end{equation}
					%which yields the claim.
					%
					%	As for the inequality \eqref{expansion_constant1} this does not affect previous estimates. \textcolor{red}{They only get better when the admissible points are restricted to a part closer to the heteroclinic chain.}
					%
					
					\smallskip
					\textbf{Step 3: Backward invariance of the transverse cone $\Cus{\sigma}{\xin}$.}
					\smallskip
					
					We must show that for any $\Psi(\txin) \in \Cus{\sigma}{\Psi(\xin)}$, i.e. 
					\begin{equation}\label{step4.1}
						\left|\Psi(\txin)_c-\Psi(\xin)_c\right| \le \sigma \left\| \Psi(\txin)_{u,s}  - \Psi(\xin)_{u,s}\right\|,
					\end{equation}
					it holds that 
					\begin{equation}\label{step4.2}
						|\txcin-\xcin| \le \sigma \| \txusin -\xusin \|.
					\end{equation}
					By contraposition, this is equivalent to showing that whenever
					%\begin{align}
					% |\txcin-\xcin| > \sigma \| \txusin -\xusin \| \implies |\Psi(\txin)_c-\Psi(\xin)_c| > \sigma \| \Psi(\txin)_{u,s}  - \Psi(\xin)_{u,s}\|,
					%\end{align}
					%which in turn is equivalent to
					\begin{equation}
						\| \txusin -\xusin \| < \sigma^{-1} |\txcin-\xcin|,\end{equation} then \begin{equation} \| \Psi(\txin)_{u,s}  - \Psi(\xin)_{u,s}\| <  \sigma^{-1}  |\Psi(\txin)_c-\Psi(\xin)_c|.
					\end{equation}
					Note that this statement is of the same form as for the expansion of the center cone in Step 2, with $\sigma$ replaced by $\sigma^{-1}$ in \eqref{step3.1}-\eqref{step3.2}. We want to obtain \eqref{expansion_constant2} with $\sigma$ replaced by $\sigma^{-1}$. This is achieved due to \eqref{expansion_constant2}, since for $\sigma \in (0,1)$ we have that $\sigma/(1+\sigma) \le \sigma^{-1}/(1+\sigma^{-1}) $.
					%\begin{equation}
					%	\frac{\sigma}{1+\sigma} \le \frac{\sigma^{-1}}{1+\sigma^{-1}} 
					%\end{equation}
					%
					%Estimate \eqref{expansion_constant2}, which yields invariance and forward expansion of the center cone, can be adapted to hold true for $\sigma^{-1}$. Therefore, a possibly smaller bound $\delta$ is chosen, which does not affect the result for $\sigma < \sigma^{-1}$.
					%
					%\textcolor{red}{works for $\sigma^{-1}$ as well, but must check everything carefully. also not yet clear to me when $\sigma$ is chosen.}
					
					\smallskip
					\textbf{Step 4: Forward contraction of the transverse cone $\Cus{\sigma}{\xin}$}
					\smallskip
					
					We must show that there exists a constant $K_{u,s}>1$ such that
					\begin{equation}
						\| \Psi(\txin)_{u,s}  - \Psi(\xin)_{u,s}\| \le \frac{1}{K_{u,s}} \|\txusin-\xusin\|, 
					\end{equation} 
					provided  \begin{equation} \label{step5.1}
						|\Psi(\txin)_c-\Psi(\xin)_c| \le \sigma \| \Psi(\txin)_{u,s}  - \Psi(\xin)_{u,s}\|.
					\end{equation}
					By backward invariance \eqref{step4.1} and \eqref{step4.2} we know that \eqref{step5.1} implies
					\begin{equation}
						|\txcin-\xcin| \le \sigma \| \txusin -\xusin \|,
					\end{equation} i.e., 
					\begin{equation}\label{step5.2}
						\| \txin - \xin  \| \le (1+\sigma) \|\txusin -\xusin \|.
					\end{equation}
					The bound for the transverse component  \eqref{estimate_return_hyperbolic} together with \eqref{step5.2} implies that
					\begin{equation}
						\begin{aligned}
							\left\|\Psi(\txin)_{u,s} - \Psi(\xin)_{u,s} \right\|  &\le \Cret \delta^{\beta} \|\txin-\xin\| \\& \le \Cret \delta^{\beta} (1+\sigma) \|\txusin -\xusin \| .
						\end{aligned}
					\end{equation}
					We can choose $\delta>0$ sufficiently small such that
					%For a suitable choice of the upper bound $\delta$ for $\xuin$ we achieve 
				\begin{equation}
				\frac{1}{K_{u,s}}:=\Cret {\delta}^{\beta} (1+\sigma) < 1,    
				\end{equation}	
                    and have thereby the desired forward contraction.
                    \end{proof}
                    %
				%
				%\bigskip
				%
				%\textcolor{red}{write this here?}
				%
				%Recall that the goal of our analysis is to obtain a stable manifold of dimension three for the given heteroclinic chain $(h_k)_{k \in \NN}$.
				%We give this manifold by a two-dimensional set of initial conditions within $\tinsection_{k}$.
				%This set of initial conditions is the stable manifold of the origin in $\tinsection$ under the map $\Psi$. It is given as the graph of a Lipschitz continuous function $\lipstar$ expressing $\xcin$ in terms of $\xus^\text{in}=\left(\xuin,\xsin\right)$. We obtain $\lipstar$ as the fixed point of a graph transform $G$ on the space of Lipschitz functions. This graph transform is defined via the inverse map $\Psi^{-1}$. The expansion and contraction properties of the cones from theorem \ref{cones_theorem} make sure that the graph transform $G$ is a contraction on the space of Lipschitz functions with sup-norm. Banach's fixed point theorem yields said unique fixed point $\lipstar$. We make this idea precise in the following
				%
				%
    %\begin{theorem}\label{theorem_stable_set_of_return_map}
				%Let \textcolor{red}{what?}.
				%The stable manifold of $(0,0,0) \in ?$ is \textcolor{red}{tangent to the $\xus$-plane and } given as the graph of a Lipschitz continuous function \textcolor{red}{which Lipschitz constant?}
				%\end{theorem}
%
                \subsubsection{Proof of Theorem \ref{manifold_theorem}}\label{subsub:pfthm}
				
				We provide an outline of the proof, which is organized in 10 steps.
				A set of initial conditions yielding the stable set of $(h_k(t))_{k\in \NN_0}$ is given by the stable set of $\xin = 0$ under iteration of the return maps $\Psi_k$. The stable set of $\xin = 0$ in turn is given as the graph of a Lipschitz continuous function $\xcin = \xcin(\xuin,\xsin)$.
				In step 1 we introduce the complete metric space of Lipschitz continuous functions $X$ in which we aim to find said graph. More precisely, we consider the space of sequences of Lipschitz functions $X^{\NN_0}$. This setting assumes that the heteroclinic chain resides in the Bianchi type IX boundary set, i.e. the local variables are positive. In step 2 we define a graph transform $G$ on $X^{\NN_0}$, which is a sequence of graph transformations $G_k$ on $X$.
				Steps 3 to 5 prove that $G_k$ is well-defined. Step 6 shows that $G_k$ is a contraction on $X$ and thus has a unique fixed point $\lipstar_k$. In Steps 7 and 8, we argue why choosing $\graph(\lipstar_k)$ as set of initial conditions yields a three-dimensional stable set of $(h_k(t))_{k \in \NN_0}$. Step 9 provides continuity arguments. In the last step 10, we explain the case of Bianchi type VIII.

				\begin{proof}
					\textbf{Step 0: Ingredients.}
					\smallskip
					
					We denote by $\left(p_k\right)_{k\in \NN_0}$ the sequence of equilibria on the Kasner circle $\KC$ associated with $(h_k(t))_{k\in \NN_0}$.
					The cone conditions \eqref{cone_conditions:full} require certain contraction and expansion rates of the components of the map $\Psi_k$. These are valid only in a smaller subset $\tinsection_{k}$, determined by the parameters $\varepsilon$ and $\delta$. More precisely, $\varepsilon$ determines the size and position of the section $\insection_{k}$ in the unscaled coordinates (and thereby fixes the map $\Psi_k$), whereas $\delta$ restricts the domain of $\Psiloc$ according to \eqref{section_delta}. Lemma~\ref{cones_theorem} provides an upper bound for these parameters and the cone opening $0<\sigma<1$. These parameters do not depend on $k$, due to the uniform bound of $\Cantor$ away from the boundary of the unstable arcs.
					The center cone is expanding with expansion rate $K_c > (1-\sigma^2)^{-1}$. We make use of this relation in step 6.

					%The parameter $0<\delta<1$ in \eqref{section_delta} restricts the size of admissible points in the insection $\insection_k$ for the return map $\Psi$ to be well-defined on a subset $\tinsection_{k}$. \autoref{cones_theorem} provides $0<\sigma<1$ such that the cones defined in \eqref{cones_definition} satisfy the conditions \eqref{cone_conditions:full}.These cone conditions however require certain contraction and expansion rates of the components of the map $\Psi_k$, which are valid only in an possibly smaller subset of $\tinsection_{k}$, that is, a choice of a smaller $\delta$.

					By Lemma~\ref{cones_theorem}, we eliminate the intermediate outsection $\outsection_k$. Thus, we only consider the maps $\Psi_k$ between the insections in this proof and therefore omit the superscript 'in'. However, to introduce the setting and to indicate at which $p_k$ the section is located, we replace it with the index $k$, which we again omit when we discuss properties of a single map.
					
					% \textcolor{red}{write somewhere that the graphs are desired to be invariant.}
					
					\smallskip
					\textbf{Step 1: Space of sequences of Lipschitz functions.}
					\smallskip

					Within a section $\tinsection_{k}$ we consider Lipschitz continuous functions  %\textcolor{red}{is this correct???? isnt it a codim 1 subset? shouldnt it be from $R3$ to $R$? below we have a 2dim set... No, it is OK. This is the graph for the discrete dynamics; we still have to flow this to get the continuous stable set}
					\begin{align}\label{lipschitz_def1}
						\begin{split}
							\hspace{3cm}\lip_k: [0,\delta]^2 &\longrightarrow [-\delta, \delta] \hspace{3cm} \\ 
							\hspace{3cm}
							\xus^k \hspace{0.25cm} &\longmapsto \lip(\xus^k) = \xc^k  \hspace{3cm}
						\end{split}
					\end{align}
					%\textcolor{red}{Again, note that $\delta$ is independent of $k$.}
					with Lipschitz constant Lip$(\lip)\le \sigma$. 
                    Throughout this proof we employ the relation to the transversal cones defined in \eqref{cones_definition}. For instance, 
					%\begin{remark}
						$\lip_k$ is Lipschitz continuous with Lipschitz constant $\text{Lip}(\lip_k)\le\sigma$ if and only if  $(\txus^k, \lip(\txus^k)) \in \Cus{\sigma}{\xus^k,\lip(\xus^k)}$ for all $\xus^k,\txus^k \in [0,\delta]^2$.
					%\end{remark}
					%\begin{remark}
					%	$\lip$ is Lipschitz continuous with Lipschitz constant $\text{Lip}(\lip)\le \sigma$ if and only if $\graph(\lip) \subseteq \Cus{\sigma}{\xus,\lip(\xus)}$ for all $\xus \in [0,\delta]^2.$
					%\end{remark}
					
					We restrict our consideration to the case of functions satisfying
					\begin{equation}\label{lipschitz_def2}
						\lip(0,\xs^k) = 0 \text{ for all } \xs^k \in [0,\delta].
					\end{equation}
					This is justified as follows. Assume for simplicity that $h$ is a homoclinic solution as, for example, the heteroclinic cycle of length three considered in the $D_3$ orbit space. Thus, we can omit the index $k$ and the return map is $\Psi:\tinsection \to \insection$.
					Now suppose we have the desired $\Psi$-invariant set given as the graph of a Lipschitz function $\lipstar$.
					%That is, graph($\lip$) contains the image of graph($\lip$) under $\Psi$.
					Consider a point $\x =(0, \xs, \lipstar(0,\xs))$ on that graph. What should $\xc = \lip(0,\xs)$ be for $\x$ to be mapped to graph($\lipstar$) again?
					The point $\x$ is in the stable manifold of the Kasner circle and therefore mapped to $(0, 0, \KM(\xc))$ under $\Psi$, due to \eqref{return_map_on_singluar_set}.
					For this point in turn to be contained in graph($\lipstar$) it must also map to the graph again. Its image is $(0, 0, \KM^2(\xc))$ and thus $\xc$ should be a fixed point of $\KM$.
					By expansion of $\KM$, there is exactly one fixed point with $\xus=0$, which is $(0,0,0)$, the intersection of $h$ with $\tinsection$.
					
					Let
					\begin{align}
						X_{}:=\left\{\zeta:[0,\delta]^2 \to [-\delta, \delta] \,\, | \,\, \text{Lip}(\lip) \le \sigma, \lip(0,\xs) = 0\right\}
					\end{align}
					denote the space of all functions with the above features.\footnote{The space $X_{}$ equipped with sup-norm $\|\cdot \|_\infty$ is a complete metric space. Indeed, $X$ is a closed subspace of 
						the Banach space  $ \,\mathcal{C}\big([0,\delta^2], \mathbb{R}\big)$ of (bounded) continuous functions.
						%Moreover, by the Arzelà-Ascoli theorem, $X_{}$ is compact, and therefore it is complete. %\textcolor{red}{PL: this seems a weird phrasing of things... You conclude relative compactness by AA? This is weird... which version of A-A are you using? Also, $X$ is compact? Really? The functions $x^n$ are not there, because the lipschitz constant is not uniform? }
					%\end{remark}
                        This property is crucial to apply a contraction mapping principle.}
     
     To assign to each $p_k$ such a graph we follow \cite{Liebscher_magnetic} and consider the product space
					\begin{align}\label{def_sequence_of_lipschitz_functions}
						X^{\NN_0}:= \left\{\left( \lip_n \right)_{n \in \NN_0} : \lip_n \in X_{} \right\}
					\end{align}
					of countably infinite sequences of Lipschitz functions. Since the metric on each $X$ is bounded, we can equip $X^{\NN_0}$ with the metric induced by the norm
					\begin{align} \label{norm_productspace}
						\|(\lip_k)_{k \in \NN_0}\| := \sup_{k \in \NN_0} \|\lip_k\|_{\infty},
					\end{align}
					which makes $X^{\NN_0}$ a complete metric space \cite[Theorem 43.5]{munkres}.
					%
					%
					%Analogously to \textcolor{red}{what?} the stable set of $0^\NN$ is obtained as the fixed point of a suitable contraction on $X^{\NN_0}$. The contraction $G = (G_k)_{k \in \NN}$ in this case is a sequence of contractions, each of which acts on $X_{}$, and the fixed point $\lipstar = (\lipstar_k)_{k \in \NN}$ is a sequence of $\Psi_k$-invariant Lipschitz functions.
					%
					
					\smallskip
					\textbf{Step 2: Definition of the graph transform $G$. }
					\smallskip
					
					The next step is to define a graph transform on $X^{\NN_0}$, which is a common approach to construct the stable manifold of a hyperbolic fixed point, introduced by Hadamard \cite{hadamard}, see also \cite{katok_hasselblatt}.
					The general idea is as follows. Consider a function $\psi:Y_1 \to Y_2$ and a map $F:Y_1 \times Y_2 \to Y_1 \times Y_2$ which is contracting in direction of $Y_1$ and expanding in direction of $Y_2$. Under certain conditions, the preimage $F^{-1}(\text{graph}(\psi))$ is again the graph of a function, $G\psi:Y_1 \to Y_2$. In this case we call $G\psi$ the \emph{(preimage) graph transform} of $\psi$.
					Note that the inverse map $F^{-1}$ expands in $Y_1$-direction and contracts in $Y_2$-direction, that is, the action of $F^{-1}$ ``squeezes" the graph of $\psi$. Therefore, the preimage graph transform is a contraction mapping on the space of graphs with properly chosen metric. The fixed point, which is a function $Y_1\to Y_2$, yields an $F$-invariant graph, which is the stable manifold, see \autoref{fig:graph_transform_motivation}.
					%The aim is to obtain an $F$-invariant graph, which is then the stable set \textcolor{blue}{of the origin}.
					
                    \begin{figure}[H]
						\centering
						\begin{tikzpicture}[scale=1.5]
							% length of coordinate system
\def \ly {1}
\def \lx {2.3}

\def \mx {2.3}
\def \my {1}

\coordinate (origin1) at (0,0);

\foreach \i in{1}
{\path let \p1=(origin\i) in coordinate (xneg\i) at (\x1-\mx cm ,\y1);
	\path let \p1=(origin\i) in coordinate (xpos\i) at (\x1+\lx cm,  \y1);
	
	\path let \p1=(origin\i) in coordinate (yneg\i) at (\x1,\y1-\my cm);
	\path let \p1=(origin\i) in coordinate (ypos\i) at (\x1,\y1+\ly cm);
	
	%coordinate axes
	%	\draw (xneg\i)--(xpos\i) node [anchor=north] {\small $\xus$};
	%	\draw (yneg\i)--(ypos\i) node [anchor=north east] {\small $\xc$};
}

\definecolor{light-gray}{gray}{0.65}

\draw (xneg1)--(xpos1) node [anchor=north] {\small $Y_1$};
\draw (yneg1)--(ypos1) node [anchor=north east] {\small $Y_2$};

\draw[-Stealth] (-1.2,0.2)--(-1.6,0.2);
\draw[-Stealth] (1.2,0.2)--(1.6,0.2);

\draw[-Stealth] (0.6,1)--(0.6,0.6);
\draw[-Stealth] (0.6,-1)--(0.6,-0.6);

\draw[-Stealth] (-0.6,1)--(-0.6,0.6);
\draw[-Stealth] (-0.6,-1)--(-0.6,-0.6);

\draw [semithick, xshift=0cm, name path=line1] plot [smooth, tension=0.8] coordinates { (-1.5,-0.6) (-1.1,-0.4) (-0.8,-0.36) (-0.3,-0.24) (0,0) (0.3,0.3) (0.8,0.4) (1.5,0.8)} node [anchor=south ]{$\psi$};
\draw [light-gray, semithick, xshift=0cm, name path=line1] plot [smooth, tension=0.7] coordinates { (-2,-0.3) (-1.4,-0.2) (-1,-0.25) (-0.4,-0.1) (0,0) (0.5,0.23) (1.4,0.3) (2,0.4)} ; %node [anchor=south ]{$G\psi$};
\draw[light-gray] (2,0.4) node [anchor=south ]{$G\psi$};
						\end{tikzpicture} \caption{The graph of a function $\psi:Y_1\to Y_2$ is contracted vertically and expanded horizontally under $F^{-1}$. The image is the graph of a function $G\psi: Y_1\to Y_2$, called the preimage graph transform.}\label{fig:graph_transform_motivation}
					\end{figure}
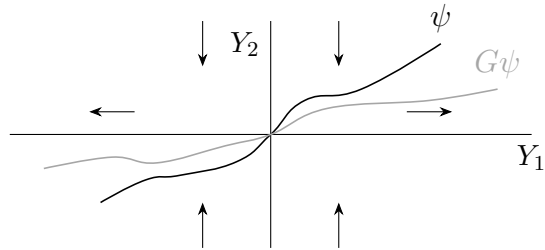

					\smallskip

					In our setting, consider a heteroclinic chain $(h_k(t))_{k \in \NN_0}$ that comes with a sequence of maps $\Psi_k: \tinsection_{k} \to \insection_{k+1}$, and its inverses $\Psi_k^{-1}: \Psi_k(\tinsection_{k}) \to \tinsection_{k}$.
					Associated with $(\tinsection_k)_{k\in \NN_0}$ is the product space $X^{\NN_0}$.
					%, whose elements are sequences of Lipschitz functions $\lip = (\lip_k)_{k\in \NN}$.
					That is, within each insection $\tinsection_{k}$ there is a space of Lipschitz continuous functions $X_k$, where $X_k=X$ for all $k\in \NN_0$.
					The graph transform $G$ acts on $X^{\NN_0}$, i.e. it maps sequences of (graphs of) Lipschitz functions to sequences of (graphs of) Lipschitz functions via the sequence of inverse return maps $(\Psi_k^{-1})_{k \in \NN_0}$. Consequently, $G = (G_k)_{k\in \NN_0}$ is a sequence of graph transformations $G_k:X_{k+1} \to X_k{}$.
					In contrast to the situation in the motivation, we consider only graphs of functions defined in a smaller subset $\tinsection_{k+1}\subset \insection_{k+1}$, 
     see \autoref{fig:graphtransforms}.
					%because we need the stronger estimates for the cone transformation properties from \autoref{cones_theorem}.
					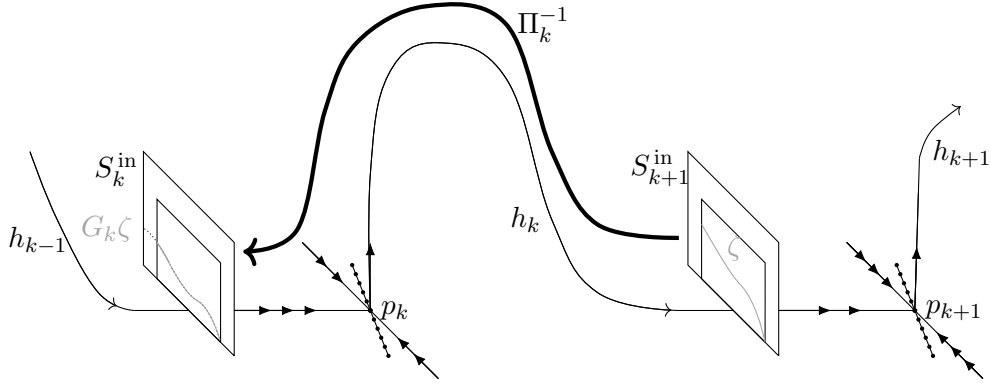
\begin{figure}[H]
						\centering
						\begin{tikzpicture}[scale=0.6]
                              \begin{scope}[yscale=1,xscale=-1]
							%\clip [draw] (1.5,1.5) circle (4.5cm);

\def \InDistLeft {3}
\def \InDistRight {\InDistLeft + 2}
\def \InHeight{2.5}

\definecolor{light-gray}{gray}{0.65}

\coordinate (InBL) at (\InDistLeft,-1);
\coordinate (InBR) at (\InDistRight,1);
\coordinate (InTL) at (\InDistLeft,-1+\InHeight);
\coordinate (InTR) at (\InDistRight,1+\InHeight);

\def \OutHeightBottom {2.5}
\def \OutHeightTop {\OutHeightBottom + 1.7}

\coordinate (OutBL) at (-1.1,\OutHeightBottom);
\coordinate (OutBR) at (1.6,\OutHeightBottom);
\coordinate (OutTL) at (0.9,\OutHeightTop);
\coordinate (OutTR) at (3.6,\OutHeightTop);

\coordinate (Origin) at (0,0);
\coordinate (InOrigin) at ($(InBL)!0.5!(InBR)$);
\coordinate (OutOrigin) at ($(OutBL)!0.5!(OutTL)$);

% origin
\filldraw (Origin) circle (1.2pt) node [anchor=west] {\small$p_k$};
% in/out section
\draw (InBL)--(InBR) -- (InTR) -- (InTL) --(InBL);
%\draw (OutBL)--(OutBR) -- (OutTR) -- (OutTL) --(OutBL);

% delta boxes
% in 
\def \delta {0.7}
\coordinate (InDeltaBL) at ($(InBL)!{0.5-\delta/2}!(InBR)$);
\coordinate (InDeltaBR) at ($(InBL)!{0.5+\delta/2}!(InBR)$);
\path let \p1=(InDeltaBL) in coordinate (InDeltaTL) at (\x1,\y1+\delta*1*\InHeight cm);
\path let \p1=(InDeltaBR) in coordinate (InDeltaTR) at (\x1,\y1+\delta*1*\InHeight cm);
\draw (InDeltaBL) -- (InDeltaBR)--(InDeltaTR)--(InDeltaTL)-- (InDeltaBL);

% out
\def \deltaout {0}
\coordinate (OutDeltaBL) at ($(OutBL)!{0.5-\deltaout/2}!(OutTL)$);
\coordinate (OutDeltaTL) at ($(OutBL)!{0.5+\deltaout/2}!(OutTL)$);
\path let \p1=(OutDeltaBL) in coordinate (OutDeltaBR) at (\x1+\deltaout*1*2.7 cm,\y1);
\path let \p1=(OutDeltaTL) in coordinate (OutDeltaTR) at (\x1+\deltaout*1*2.7 cm,\y1);
%\draw (OutDeltaBL) -- (OutDeltaBR)--(OutDeltaTR)--(OutDeltaTL)-- (OutDeltaBL);

\draw (5.6,2.5)  node [anchor=south] {\small$\insection_k$};
%\draw (4.2,0.5)  node [anchor=south] {\scriptsize$\tinsection_k$};
%\draw (2,3.3)  node [anchor=south] {\scriptsize$\outsection_k$};

%delta sections
\coordinate (InInter) at (intersection of  InBL--InTL and Origin--InOrigin);
%\coordinate (OutInter) at (intersection of  OutDeltaBL--OutDeltaBR and Origin--OutOrigin);
%\coordinate (OutInter) at (intersection of  OutDeltaBL--OutDeltaBR and Origin--OutOrigin);

% line of equilibria
\coordinate (A) at (-0.4,-1);
\coordinate (B) at (0.4,1);
\draw [] (A)--(B); % node[ anchor=west] {$\xc$};
\def \T {10}
\foreach \i in {0,...,\T}
{\coordinate (A\i) at ($(A)!\i/\T!(B)$);
	%	\path let \p1=(A) in coordinate (A\i) at (\i/6*\x1, 2.5/6*\i*\y1); 
	\filldraw [] (A\i) circle (1pt);}

% coordinate system
%\draw (Origin)--(InInter);
%\draw (Origin)--(OutInter);
\draw (-1.5,-1.5)--(1.5,1.5); % node[ anchor=west] {$\xs$};

\coordinate (bla1) at ($(Origin)!1.3!(OutOrigin)$);
\coordinate (bla2) at ($(Origin)!1.3!(InOrigin)$);

%\draw (OutOrigin) -- (bla1); %node[ anchor=east] {$\xu$};
\draw (InOrigin) -- (bla2); % node[ anchor=north] {$\xss$};

% arrows
\draw [] (InInter)--(Origin);
\draw [-Latex, shorten >=1.3cm] (InInter)--(Origin);
\draw [-Latex, shorten >=1cm] (InInter)--(Origin);
\draw [-Latex, shorten >=0.7cm] (InInter)--(Origin);

\draw [-Latex, shorten >=0.5cm] (-1.5,-1.5)--(Origin);
\draw [-Latex, shorten >=0.8cm] (-1.5,-1.5)--(Origin);

\draw [-Latex, shorten >=0.5cm] (1.5,1.5)--(Origin);
\draw [-Latex, shorten >=0.8cm] (1.5,1.5)--(Origin);

\path let \p1=(Origin) in coordinate (OutInter) at (\x1,\y1+2cm);
\draw [-Latex, shorten >=0.3cm] (Origin)--(OutInter);

%\draw [] plot [smooth, tension=0.7] coordinates { (-10.2,1.2) (-10.4,1.8)  (-11,4.4)  (-11.7,5.4)};

\def \InDistLeft {-9}
\def \InDistRight {\InDistLeft + 2}
\def \InHeight{2.5}

\coordinate (InBL1) at (\InDistLeft,-1);
\coordinate (InBR1) at (\InDistRight,1);
\coordinate (InTL1) at (\InDistLeft,-1+\InHeight);
\coordinate (InTR1) at (\InDistRight,1+\InHeight);

\def \OutHeightBottom {2.5}
\def \OutHeightTop {\OutHeightBottom + 1.7}

\coordinate (OutBL1) at (-13.1,\OutHeightBottom);
\coordinate (OutBR1) at (-10.4,\OutHeightBottom);
\coordinate (OutTL1) at (-11.1,\OutHeightTop);
\coordinate (OutTR1) at (-8.4,\OutHeightTop);

\coordinate (Origin1) at (-12,0);
\coordinate (InOrigin1) at ($(InBL1)!0.5!(InBR1)$);
\coordinate (OutOrigin1) at ($(OutBL1)!0.5!(OutTL1)$);

% origin
\filldraw (Origin1) circle (1.2pt) node [anchor=west] {\small$p_{k+1}$};
% in/out section
\draw (InBL1)--(InBR1) -- (InTR1) -- (InTL1) --(InBL1);

% delta boxes
% in 
\def \delta {0.7}
\coordinate (InDeltaBL1) at ($(InBL1)!{0.5-\delta/2}!(InBR1)$);
\coordinate (InDeltaBR1) at ($(InBL1)!{0.5+\delta/2}!(InBR1)$);
\path let \p1=(InDeltaBL1) in coordinate (InDeltaTL1) at (\x1,\y1+\delta*1*\InHeight cm);
\path let \p1=(InDeltaBR1) in coordinate (InDeltaTR1) at (\x1,\y1+\delta*1*\InHeight cm);
\draw (InDeltaBL1) -- (InDeltaBR1)--(InDeltaTR1)--(InDeltaTL1)-- (InDeltaBL1);

% out
\def \deltaout {0.9}
\coordinate (OutDeltaBL1) at ($(OutBL1)!{0.5-\deltaout/2}!(OutTL1)$);
\coordinate (OutDeltaTL1) at ($(OutBL1)!{0.5+\deltaout/2}!(OutTL1)$);
\path let \p1=(OutDeltaBL1) in coordinate (OutDeltaBR1) at (\x1+\deltaout*1*2.7 cm,\y1);
\path let \p1=(OutDeltaTL1) in coordinate (OutDeltaTR1) at (\x1+\deltaout*1*2.7 cm,\y1);

%\draw [blue,<-] plot [smooth, tension=0.7] coordinates { (-9.2,1.2) (-10.4,1.8)  (-11,4.4)  (-11.7,5.4)};

\draw (-6.4,2.5)  node [anchor=south] {\small$\insection_{k+1}$};

%delta sections
\coordinate (InInter1) at (intersection of  InBL1--InTL1 and Origin1--InOrigin1);
\coordinate (OutInter1) at (intersection of  OutDeltaBL1--OutDeltaBR1 and Origin1--OutOrigin1);

% line of equilibria
\coordinate (A1) at (-12.4,-1);
\coordinate (B1) at (-11.6,1);
\draw [] (A1)--(B1); % node[ anchor=west] {$\xc$};
\def \T {10}
\foreach \i in {0,...,\T}
{\coordinate (A1\i) at ($(A1)!\i/\T!(B1)$);
	%	\path let \p1=(A) in coordinate (A\i) at (\i/6*\x1, 2.5/6*\i*\y1); 
	\filldraw [] (A1\i) circle (1pt);}

% coordinate system
\draw (Origin1)--(InInter1);
\draw (Origin1)--(OutInter1);
\draw (-13.5,-1.5)--(-10.5,1.5); % node[ anchor=west] {$\xs$};

\coordinate (bla11) at ($(Origin1)!1!(OutOrigin1)$);
\coordinate (bla21) at ($(Origin1)!1.3!(InOrigin1)$);

\draw (OutOrigin1) -- (bla11); %node[ anchor=east] {$\xu$};
\draw (InOrigin1) -- (bla21); % node[ anchor=north] {$\xss$};

% arrows

\draw [-Latex, shorten >=1.1cm] (InInter1)--(Origin1);
\draw [-Latex, shorten >=0.7cm] (InInter1)--(Origin1);

\draw [-Latex, shorten >=0.4cm] (-13.5,-1.5)--(Origin1);
\draw [-Latex, shorten >=0.65cm] (-13.5,-1.5)--(Origin1);
\draw [-Latex, shorten >=0.9cm] (-13.5,-1.5)--(Origin1);

\draw [-Latex, shorten >=0.4cm] (-10.5,1.5)--(Origin1);
\draw [-Latex, shorten >=0.65cm] (-10.5,1.5)--(Origin1);
\draw [-Latex, shorten >=0.9cm] (-10.5,1.5)--(Origin1);

\draw [-Latex, shorten >=0.7cm] (Origin1)--(OutInter1);

% h_k
\coordinate (x00) at (0,0) ;
\coordinate (x000) at (0,3) ;
\coordinate (x0) at (-0.4,5.2) ;
\coordinate (x1) at (-1.5,5.9) ;
\coordinate (x2) at (-3,5.1) ;
\coordinate (x3) at (-4,2) ;
\coordinate (x4) at (-5,0.4) ;

%Psiglob
\coordinate (anf) at (1,4.4) ;
\coordinate (X0) at (0.3,5.9) ;
\coordinate (X1) at (-1.2,6.7) ;
\coordinate (X2) at (-3,6.3) ;
\coordinate (X3) at (-4,3.5) ;
\coordinate (X4) at (-5,1.9) ;
\coordinate (X5) at (-6.8,1.6) ;

\draw (X2) circle (0pt) node [anchor=west]{\small$\Psi_k^{-1}\qquad\qquad$};

\draw [ultra thick,<-] plot [smooth, tension=0.62] coordinates {(2.8,1.3) (1.6,1.8) (anf) (X0) (X1) (X2) (X3) (X4) (X5)};
\draw (x3) circle (0pt) node [anchor=east]{\small$h_k$};

\draw [] plot [smooth, tension=0.7] coordinates { (x00) (x000) (x0) (x1) (x2) (x3) (x4) (bla21)};
\draw [_->, shorten >=0.1cm] plot [smooth, tension=0.7] coordinates { (x00) (x000) (x0) (x1) (x2) (x3) (x4) (bla21)};

\coordinate (y0) at (-12.4,4) ;
\coordinate (y1) at (-13.3,4.7) ;
\coordinate (y2) at (-14.5,5.5) ;
\coordinate (y3) at (-16,2) ;
\coordinate (y4) at (-17,0.4) ;

\draw (y0) circle (0pt) node [anchor=north]{\small$\qquad h_{k+1}$};
\draw [->, shorten >= 0.2cm] plot [smooth, tension=0.7] coordinates { (bla11) (y0) (y1)}; % (y2) }; %(y2) (y3) (y4)};
%\draw [] plot [smooth, tension=0.7] coordinates {(bla11) (y0) (y1)}; % (y2) };
\draw [] plot [smooth, tension=0] coordinates {(-12,0) (bla11)};

%\draw [] plot [smooth, tension=0.7] coordinates { (bla11) (y0) (y1) (y2) };
%\draw [-Latex, shorten >=0.28cm] plot [smooth, tension=0.7] coordinates { (bla11) (y0) (y1) (y2) };

% h_{k-1}
\coordinate (z2) at (7.5,3.5) ;
\coordinate (z3) at (6.9,2) ;
\coordinate (z4) at (6,0.4) ;

\draw [->, shorten >= 0.2cm] plot [smooth, tension=0.7] coordinates { (z2) (z3) (z4) (bla2) }; %(y2) (y3) (y4)};
\draw [] plot [smooth, tension=0.7] coordinates { (z2) (z3) (z4) (bla2) };

\path let \p1=(x3) in coordinate (LOL) at (\x1+11.7 cm,\y1-1cm);

\draw (LOL) node [anchor=south]{\small$ \,\quad h_{k-1}$};

% graphs
\path let \p1=(InDeltaBL) in coordinate (x0) at (\x1 + 0.3 cm,\y1+ 0.7cm);
\path let \p1=(InDeltaBL) in coordinate (x1) at (\x1 + 0.8 cm,\y1+ 1.2cm);
\path let \p1=(InDeltaBL) in coordinate (x2) at (\x1 + 1.4 cm,\y1+ 2.2cm);
\path let \p1=(InDeltaBL) in coordinate (x3) at (\x1 + 1.7 cm,\y1+ 2.5cm);

\draw [densely dotted] plot [smooth, tension=0.9] coordinates { (InDeltaBL) (x0) (x1) (x2) (x3)};
\draw [light-gray,_-_,shorten >= 0.3cm] plot [smooth, tension=0.9] coordinates { (InDeltaBL) (x0) (x1) (x2) (x3)} node [anchor=east]{\small \, $G_k\lip $};

\path let \p1=(InDeltaBL1) in coordinate (x0) at (\x1 + 0.3 cm,\y1+ 0.9cm);
\path let \p1=(InDeltaBL1) in coordinate (x1) at (\x1 + 0.8 cm,\y1+ 1.6cm);
\path let \p1=(InDeltaBL1) in coordinate (x2) at (\x1 + 1.4 cm,\y1+ 2.6cm);

\draw [light-gray] plot [smooth, tension=0.9] coordinates { (InDeltaBL1) (x0) (x1) (x2)} node [anchor= north west]{\small \, $\lip$};
                              \end{scope}
						\end{tikzpicture} \caption{We apply the inverse map $\Psi_k^{-1}$ to the graph of a Lipschitz continuous function $\lip=\lip_{k+1} \in X$. Its image is the graph of a function  $G_k\lip \in X$. Note that this figure is not precise due to lack of dimensions. The function $\lip$ appears to express $\xu$ in terms of $\xs$, but it expresses $\xc$ in terms of $\xus$. See also \autoref{fig:step3}.} \label{fig:graphtransforms}
					\end{figure}
					For a sequence of Lipschitz functions $\lip = (\lip_k)_{k\in \NN_0} \in X^{\NN_0}$ we define its graph transform $G\lip \in X^{\NN_0}$ pointwise via $(G\lip)_k = G_{k}\lip_{k+1}$, such that
					\begin{align}\label{graph_transform_condition}
						\text{graph}(G_k\lip_{k+1}) = \Psi_k^{-1}(\text{graph}(\lip_{k+1})). \quad 
					\end{align}
					It remains to show that this is well-defined, i.e. that the term on the right side is indeed the graph of a function for all $k \in \NN_0$.
					The contraction and expansion properties in the motivation are satisfied in our situation due to the cone conditions of Lemma~\ref{cones_theorem}.
					%In the following steps we prove these conditions ensure that $G_k\lip_{k+1}$ is well-defined and Lipschitz continuous, and that $G_k$ is a contraction mapping.
					This indicates that in our case the splitting of domain and codomain in contracting (stable) and expanding (unstable) subspaces is given by separating the components of $\x^k \in \tinsection_{k}$ in $\xus^k$ and $\xc^k$. %Before we give a component-wise description of $G_k$, we adapt the notation. 
	   		%		
					
					%\begin{remark}(Notation)
						In the following steps we discuss well-definedness and properties of $G_k$ as a function $X_{k+1}\to X_k$ and of $G_k \lip_{k+1}$ as an element of $X_k$.
						Since $X_k=X$ for all $k \in \NN_0$, we omit the index $k$.
						Whenever we consider a Lipschitz function $\lip$ we think of a function in $X_{k+1}$, i.e. $\lip = \lip_{k+1}$, and we write $G$ instead of $G_k$.
						To distinguish points in $\tinsection_{k+1}$ from those in $\tinsection_{k}$ we denote the former by $y$, and the latter by $x$, see \autoref{fig:step3}. 
					%\end{remark}
					
                        Now to define $G$ let $\lip\in X_{}$ and consider a nonzero point $ (\yus,\lip(\yus)) \in \graph(\lip)$ and map it back with $\Psi^{-1}$. The resulting point is simply
					\begin{equation}
					    \Psi^{-1}(\yus,\lip(\yus)) = \left( \big(\Psi^{-1}(\yus,\lip(\yus))\big)_{u,s}, \big(\Psi^{-1}(\yus,\lip(\yus))\big)_c \right) \,\in \Psi^{-1}(\graph(\lip)).
					\end{equation}
					To achieve \eqref{graph_transform_condition} we define the map $\xus \mapsto G\lip(\xus)$ as
					\begin{equation}\label{graphtransform_def}
						%G\lip: (0,\delta]\times [0,\delta] \longrightarrow [-\delta, \delta] \\
						%\xus\hspace{0.25cm} &\longmapsto G\lip(\xus) = \big(\Psi^{-1}(\xus,\lip(\xus))\big)_c \\
						\big(\Psi^{-1}(\yus, \lip(\yus))\big)_{u,s}\hspace{0.15cm} \longmapsto \big(\Psi^{-1}(\yus,\lip(\yus))\big)_c.
						% &\text{ if }  \yus \neq 0
						% \label{graphtransform_def:1}
						%and hence graph($G\lip$) consists (\textcolor{red}{$\subseteq/\supseteq,..?$}) of all points of the form, ... and therefore $\Psi^{-1}($graph$(\lip))$. \textcolor{red}{why is $\Psi^{-1}$ well defined (domains)?}
						% (0,\xs) &\longmapsto 0, & \text{ else } \label{graphtransform_def:2}
					\end{equation}
					This definition requires some explanation.
					Firstly, note that the preimage of the origin $(0,0,0) \in \text{graph}(\lip)$ is the line $(0,\xs,0)$, due to \eqref{return_map_on_singluar_set}. Thus, $G\lip(0,\xs)=0$.
					%This definition requires some more justification.
					Moreover, for given $\lip \in X_{}$, \begin{itemize}
						\item [(a)] why can each point in the domain of $G\lip$ be written as in \eqref{graphtransform_def}? That is, for given $\xus \in (0,\delta] \times [0,\delta]$, why is there $\yus$ such that $\xus = \big(\Psi^{-1}(\yus, \lip(\yus))\big)_{u,s}$?
						%	/ why can we assume that for each $\xus$ there exists a preimage of $(\xus,\lip(\xus))$ under $\Psi$ with $(u,s)$-component in $[0,\delta]^2$? 
						%/ $(\xus,\lip(\xus)) \in \graph(\lip)$ has a preimage with $(u,s)$-component in $[0,\delta]^2$? 
						\item[(b)] why is $G\lip$ a well-defined function? Assume (a) is settled, and for $\xus$ there are two points $\tyus,\yus\in[0,\delta]^2\setminus \{0\}$ with $\xus = \big(\Psi^{-1}(\tyus, \lip(\tyus))\big)_{u,s} =\big(\Psi^{-1}(\yus, \lip(\yus))\big)_{u,s}$. Then why is $\tyus=\yus$ and therefore $\big(\Psi^{-1}(\tyus, \lip(\tyus))\big)_{c} =\big(\Psi^{-1}(\yus, \lip(\yus))\big)_{c}$?
						\item[(c)]
						why is $G\lip \in X$? Condition \eqref{lipschitz_def2} is satisfied. But why is $G\lip$ Lipschitz continuous with Lipschitz constant $\text{Lip}(G\lip) \le \sigma$?
					\end{itemize}
					
					\smallskip
					\textbf{Step 3: The domain of $G\lip$ is $[0,\delta]^2$.}
					\smallskip
					
					Let $\xus \in (0,\delta]\times [0,\delta] $. We show that there exists $\yus \in [0,\delta]^2\setminus\{0\}$ with 
                    \begin{equation}
                    \big(\Psi^{-1}(\yus, \lip(\yus))\big)_{u,s}=\xus.  
                    \end{equation}
					Consider $\Cus{\sigma}{0}$, the cone in the transverse direction centered at the origin. The line $\{\xus\} \times [-\delta,\delta]$ intersects the preimage $\Psi^{-1}\left(\Cus{\sigma}{0}\right)$ by backward invariance and backward expansion of $\Cus{\sigma}{0}$. Thus, there exists $\x=(\xus,\xc) \in \Pi^{-1}\left(\Cus{\sigma}{0}\right)$, i.e. $\Psi(\x) \in \Cus{\sigma}{0}\subset \tinsection$. Moreover, 
					the line $\{\xus\} \times [-\delta,\delta]$ is contained in $\Cc{\sigma}{\x}$, the center cone at $\x$. By forward invariance \eqref{forward_invariance_center_condition}, the image of that line is contained in the cone $\Cc{\sigma}{\Psi(\x)}$. By forward expansion \eqref{forward_expansion_center_condition}, the image of this line intersected with $\tinsection$ is a curve connecting the planes $\{\xc=\pm\delta\}$. By continuity of $\Psi$, this curve intersects $\graph(\lip)$ in a point $(\yus,\lip(\yus))$. By construction, $\Psi^{-1}(\yus,\lip(\yus)) \in \{\xus\}\times [-\delta,\delta]$ and thus $\big(\Psi^{-1}(\yus,\lip(\yus))\big)_{u,s} = \xus$, see \autoref{fig:step3}.
					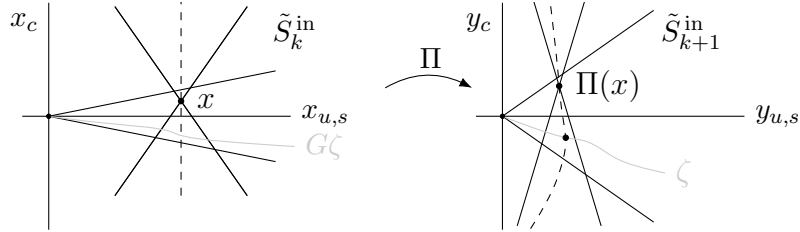
\begin{figure}[H]
						\centering
						\begin{tikzpicture}[scale=0.5]
							\definecolor{light-gray}{gray}{0.65}

\clip (-2,-3) rectangle (20,3);

% left-right shift
\def \s {12}

% length of coordinate system
\def \ly {3}
\def \lx {6.4}

\def \mx {0.7}
\def \my {3}

\coordinate (origin1) at (0,0);
\coordinate (origin2) at (\s,0);

\foreach \i in{1,2}
{\path let \p1=(origin\i) in coordinate (xneg\i) at (\x1-\mx cm ,\y1);
	\path let \p1=(origin\i) in coordinate (xpos\i) at (\x1+\lx cm,  \y1);
	
	\path let \p1=(origin\i) in coordinate (yneg\i) at (\x1,\y1-\my cm);
	\path let \p1=(origin\i) in coordinate (ypos\i) at (\x1,\y1+\ly cm);
	
	%coordinate axes
%	\draw (xneg\i)--(xpos\i) node [anchor=north] {\small $\xus$};
%	\draw (yneg\i)--(ypos\i) node [anchor=north east] {\small $\xc$};
}

\draw (xneg1)--(xpos1) node [anchor=west] {\small $\xus$};
\draw (yneg1)--(ypos1) node [anchor=north east] {\small $\xc$};

\draw (xneg2)--(xpos2) node [anchor=west] {\small $\yus$};
\draw (yneg2)--(ypos2) node [anchor=north east] {\small $\yc$};

\draw (17,3) node [anchor=north] {\small $\tinsection_{k+1}$};
\draw (6.5,3) node [anchor=north] {\small $\tinsection_{k}$};

% cone centers 
\coordinate (center1) at (0,0);
\path let \p1=(center1) in coordinate (center2) at (\x1 +\s cm  ,\y1);

% cone opening
\def \sigma {0.7}

% cone length
\def \conelength {4}
\definecolor{light-gray}{gray}{0.85}

\foreach \i in{1,2}
{\path let \p1=(center\i) in coordinate (coneBL\i) at (\x1-\conelength cm ,\y1- \conelength* \sigma cm);
	\path let \p1=(center\i) in coordinate (coneTL\i) at (\x1-\conelength cm ,\y1+ \conelength* \sigma cm);
	
	\path let \p1=(center\i) in coordinate (coneBR\i) at (\x1+\conelength cm ,\y1- \conelength* \sigma cm);
	\path let \p1=(center\i) in coordinate (coneTR\i) at (\x1+\conelength cm ,\y1+ \conelength* \sigma cm);

}

\draw (center2)--(coneTR2);
\draw (center2)--(coneBR2);

%\path [fill = light-gray] (coneBL2)--(coneTL2)--(center2)--(coneBL2);
%\path [fill = light-gray] (coneBR2)--(coneTR2)--(center2)--(coneBR2);

%\draw (coneBL1)--(coneTR1);
%\draw (coneTL1)--(coneBR1);
%
%\draw (coneBL2)--(coneTR2);
%\draw (coneTL2)--(coneBR2);

\path let \p1=(center1) in coordinate (coneBL1) at (\x1-\conelength cm ,\y1- \conelength* \sigma cm);

\def \conelengthim {6}
\def \sigmaim {0.3}

\foreach \i in{1,2}
{\path let \p1=(center\i) in coordinate (coneBL\i) at (\x1-\conelengthim cm ,\y1- \conelength* \sigmaim cm);
	\path let \p1=(center\i) in coordinate (coneTL\i) at (\x1-\conelengthim cm ,\y1+ \conelength* \sigmaim cm);
	
	\path let \p1=(center\i) in coordinate (coneBR\i) at (\x1+\conelengthim cm ,\y1- \conelength* \sigmaim cm);
	\path let \p1=(center\i) in coordinate (coneTR\i) at (\x1+\conelengthim cm ,\y1+ \conelength* \sigmaim cm);
	
}
%\path [fill = light-gray] (coneBL1)--(coneTL1)--(center1)--(coneBL1);
%\path [fill = light-gray] (coneBR1)--(coneTR1)--(center1)--(coneBR1);

\draw (center1)--(coneTR1);
\draw (center1)--(coneBR1);

%\draw (coneBL2)--(coneTR2);
%\draw (coneTL2)--(coneBR2);

\filldraw (center1) circle (1.5pt);
\filldraw (center2) circle (1.5pt);

%caption
\path let \p1=(center1) in coordinate (center1cap) at (\x1, \y1+0.2 cm);
\path let \p1=(center2) in coordinate (center2cap) at (\x1, \y1+0.2 cm);

%\filldraw (center1cap) circle (0pt) node [anchor=south] {\small $\x$};
%\filldraw (center2cap) circle (0pt) node [anchor=south] {\small $\Psi(\x)$};

%arrow and map
\filldraw (\s-1.9,1) circle (0pt) node [anchor=south] {\small $\Psi$};
\draw [-Latex] (\s-3.1,0.75) to[bend left=30] (\s-0.8,0.75);

 % new centers
\coordinate (center1) at (3.5,0.4);
\coordinate (center2) at (13.5,0.8);

\filldraw (center1) circle (2pt) node [anchor=west] {$\,\x$};
\filldraw (center2) circle (2pt) node [anchor=west] {$\,\Psi(\x)$};

%\filldraw (\s+4,2.8) circle (0pt) node [anchor=west] {\small $\Cus{\sigma}{\Psi(\x)}$};

% cone opening
\def \sigma {0.7}

% cone length
\def \conelength {2.5}

\definecolor{light-gray}{gray}{0.8}

\foreach \i in{1,2}
{\path let \p1=(center\i) in coordinate (coneBL\i) at (\x1-\conelength*\sigma cm ,\y1- \conelength cm);
	\path let \p1=(center\i) in coordinate (coneTL\i) at (\x1-\conelength*\sigma cm ,\y1+ \conelength cm);
	
	\path let \p1=(center\i) in coordinate (coneBR\i) at (\x1+\conelength*\sigma cm ,\y1- \conelength cm);
	\path let \p1=(center\i) in coordinate (coneTR\i) at (\x1+\conelength*\sigma cm ,\y1+ \conelength cm);
	
%	\path [fill = light-gray] (coneBL1)--(coneBR1)--(center1)--(coneBL1);
%	\path [fill = light-gray] (coneTR1)--(coneTL1)--(center1)--(coneTR1);

}

	\draw (coneBL1)--(coneTR1);
\draw (coneTL1)--(coneBR1);

%caption cone centers
\filldraw (center1) circle (1.5pt) ;
\filldraw (center2) circle (1.5pt); 
\path let \p1=(center1) in coordinate (center1cap) at (\x1, \y1+0 cm);
\path let \p1=(center2) in coordinate (center2cap) at (\x1, \y1+0 cm);
%\filldraw (center1cap) circle (0pt) node [anchor=east] {\small $\x$};
%\filldraw (center2cap) circle (0pt) node [anchor=east] {\small $\Psi(\x)$};

%\path let \p1=(center1) in coordinate (coneBL1) at (\x1-\conelength cm ,\y1- \conelength* \sigma cm);
\draw (coneBL1)--(coneTR1);
\draw (coneTL1)--(coneBR1);

\def \conelengthim {3.7}
\def \sigmaim {0.3}

\foreach \i in{1,2}
{\path let \p1=(center\i) in coordinate (coneBL\i) at (\x1-\conelengthim*\sigmaim cm ,\y1- \conelengthim cm);
	\path let \p1=(center\i) in coordinate (coneTL\i) at (\x1-\conelengthim*\sigmaim cm ,\y1+ \conelengthim cm);
	
	\path let \p1=(center\i) in coordinate (coneBR\i) at (\x1+\conelengthim*\sigmaim cm ,\y1- \conelengthim cm);
	\path let \p1=(center\i) in coordinate (coneTR\i) at (\x1+\conelengthim*\sigmaim cm ,\y1+ \conelengthim cm);
}
%\path [fill = light-gray] (coneBL2)--(coneBR2)--(center2)--(coneBL2);
%\path [fill = light-gray] (coneTR2)--(coneTL2)--(center2)--(coneTR2);

\draw (coneBL2)--(coneTR2);
\draw (coneTL2)--(coneBR2);

%vertical line
\path let \p1=(center1) in coordinate (B) at (\x1 ,\y1+ 5 cm);
\path let \p1=(center1) in coordinate (A) at (\x1 ,\y1- 2.5 cm);

\draw [light-gray,xshift=0cm, name path=line1] plot [smooth, tension=0.8] coordinates {(\s,0) (12.8,-0.3) (13.6,-0.55) (14.3,-0.7) (15.2,-1.2) (16.3,-1.5)} node [anchor=west ]{$\lip$};
\draw [dashed, xshift=0cm, name path=line2] plot [smooth, tension=0.8] coordinates {(13,5) (center2) (13.5,-1.5) (12,-4)};
\draw [dashed]  (A) -- (B);

\path [name intersections={of=line1 and line2,by=P1}];
\filldraw (P1) circle (2pt) node [anchor=west] {};

\draw [light-gray,xshift=0cm, name path=line1] plot [smooth, tension=0.8] coordinates {(0,0) (1,-0.13) (2,-0.23) (3,-0.35) (4,-0.6) (6.5,-0.8)} node [anchor=west ]{\small$G\lip$};

\filldraw (origin1) circle (1.5pt);
\filldraw (origin2) circle (1.5pt); 
						\end{tikzpicture} \caption{% Visualization of step 3.
							%	 for all $\xus \in [0,\delta]^2$ there exists $y = (\yus, \lip(\yus))$ with $\xus = (\Psi^{-1}(y))_{u,s}$.
                        The center (transverse) cone at $\x$ (at $0$) is invariant and expanding under $\Psi$ (under $\Psi^{-1}$).}\label{fig:step3}
					\end{figure}
					\smallskip
					\textbf{Step 4: $G\lip$ is a well-defined function.}
					\smallskip
					
					We show that $G\lip$ does not assign two values to a single $\xus$.
					We assume that there exist $\tyus,\yus\in[0,\delta]^2\setminus \{0\}$ with \begin{equation}
					    \xus = \big(\Psi^{-1}(\tyus, \lip(\tyus))\big)_{u,s} =\big(\Psi^{-1}(\yus, \lip(\yus))\big)_{u,s}.
				\end{equation}					
					Two points with the same $\xus$-value are contained in their respective center cones, i.e.,
					\begin{equation}
					    \Psi^{-1}(\tyus,\lip(\tyus)) \in \Cc{\sigma}{\Psi^{-1}(\yus,\lip(\yus))},
					\end{equation} 
                    and thus, by forward invariance \eqref{forward_invariance_center_condition}, also 
                    \begin{equation}
                        (\tyus,\lip(\tyus)) \in \Cc{\sigma}{(\yus,\lip(\yus))}.
                    \end{equation}
					Lipschitz continuity of $\lip$ on the other hand implies
					\begin{equation}
						(\tyus,\lip(\tyus)) \in \Cus{\sigma}{(\yus,\lip(\yus))},
					\end{equation}
					i.e., 
                        $(\tyus,\lip(\tyus)) \in  \Cc{\sigma}{(\yus,\lip(\yus))} \cap  \Cus{\sigma}{(\yus,\lip(\yus))}  = \{(\yus,\lip(\yus))\}
                        $. 
					This implies that the value assigned to $\xus$ under $G\lip$ is unique:
					\begin{equation}
						\big(\Psi^{-1}(\tyus, \lip(\tyus))\big)_{c} =\big(\Psi^{-1}(\yus, \lip(\yus))\big)_{c}.
					\end{equation}
					
					\smallskip
					\textbf{Step 5: $G\lip \in X_{}$.}
					\smallskip
					
					It remains to show Lipschitz continuity of $G\lip$ with Lipschitz constant Lip$(G\lip) \le \sigma$.
					%This follows from ... . More precisely, 
					%We must show that
					%\begin{align}\label{Glip_is_lipschitz1}
					%\| G\lip (\tilde{\big(\Psi^{-1}(\xus, \lip(\xus))\big)_{u,s}}) - G\lip(\big(\Psi^{-1}(\xus, \lip(\xus))\big)_{u,s}) \| \le \sigma \| \tilde{\big(\Psi^{-1}(\xus, \lip(\xus))\big)_{u,s}} - \big(\Psi^{-1}(\xus, \lip(\xus))\big)_{u,s} \|
					%\end{align}
					To that end, take two points $\xus, \txus \in [0,\delta]^2$, the domain of $G\lip$, with
					%If $\xus$ and $\txus$ have nonzero unstable components we can assume that
					\begin{align}
						\txus = \big(\Psi^{-1}(\tyus, \lip(\tyus))\big)_{u,s},   \quad \xus = \big(\Psi^{-1}(\yus, \lip(\yus))\big)_{u,s},
					\end{align}
					for $\tyus, \yus \in [0,\delta]^2$. For Lipschitz continuity of $G\lip$ we must show that
					\begin{align}\label{Glip}
						(\txus,G\lip(\txus)) \in \Cus{\sigma}{\xus,G\lip(\xus)}.
					\end{align}
					%\begin{align}
					%(\txus,(G\lip)(\txus)) \in \Cus{\sigma}{(\xus,\lip(\xus))} 
					%\end{align}
					%for all $\txus, \xus \in [0,\delta]^2$. That is,
					By \eqref{graphtransform_def}, the definition of $G$, this is equivalent to showing that
					\begin{align}
						\Psi^{-1}(\tyus,\lip(\tyus)) \in \Cus{\sigma}{\Psi^{-1}(\yus,\lip(\yus)) }.
					\end{align}
					By Lipschitz continuity of $\lip$ we know that 
                        \begin{equation}
                        (\tyus,\lip(\tyus)) \in \Cus{\sigma}{\yus,\lip(\yus)}.    
                        \end{equation}
					Backward invariance of the transverse cone \eqref{backwards_invariance_transverse_condition} then yields \eqref{Glip}.
					
					\smallskip
					\textbf{Step 6: $G$ is a contraction with unique fixed point $\lipstar$.}
					\smallskip
					
					We prove that  there exists a constant $c < 1$ such that for all $\lip, \tlip \in X_{}$, \begin{equation}\label{G_contraction}
						\| G \tlip- G\lip \|_\infty \le  c\| \tlip - \lip \|_\infty.
					\end{equation}
					We compare the two functions $G\tlip,G\lip$ in an arbitrary point $\xus \in [0,\delta]^2$.
					%When $\xu = 0$ then both, $G\tlip$ and $G\lip$, assign the value zero, i.e. $$\|G\tlip(\xus)-G\lip(\xus)\| = 0.$$
					%
					%When $\xu \neq 0$ 
					We must take into account that the domain has different representations, depending on $\lip$ and $\tlip$. Thus, we obtain two possibly different points $\tyus, \yus$ that satisfy
					\begin{equation}\label{point_in_domain}
						\xus = \big(\Psi^{-1}(\tyus, \tlip(\tyus))\big)_{u,s} = \big(\Psi^{-1}(\yus, \lip(\yus))\big)_{u,s}.
					\end{equation}
					As in the previous steps we use that points with the same transverse component are contained in their respective center cones:
					\begin{equation}
						\Psi^{-1}(\tyus, \tlip(\tyus)) \in \Cc{\sigma}{\Psi^{-1}(\yus, \lip(\yus))}.
					\end{equation}
					By forward invariance of the center cones,
					\begin{equation}\label{randomEQ}
						\big(\tyus, \tlip(\tyus)\big) \in \Cc{\sigma}{\yus, \lip(\yus)},
					\end{equation}
					which means that 
                    \begin{equation}
                     \|\tyus - \yus\| \le \sigma |\tlip(\tyus)- \lip(\yus)|   .
                    \end{equation}
                     The triangle inequality, the Lipschitz continuity of $\lip$ and \eqref{randomEQ}, implies
					\begin{subequations}
						\begin{equation}
							\begin{aligned}
								|\tlip(\tyus) - \lip(\yus) | &\le |\tlip(\tyus) - \lip(\tyus) | + |\lip(\tyus) - \lip(\yus) | \\
								&\le \|\tlip-\lip \|_\infty + \sigma \|\tyus - \yus \|
								\\
								&\le \|\tlip-\lip \|_\infty + \sigma ^2 |\tlip(\tyus)- \lip(\yus)|.
							\end{aligned}
						\end{equation}
					\end{subequations}
					Since $|\tlip(\tyus)- \lip(\yus)|$ appears on both sides of the inequality, rearranging yields
					\begin{equation}\label{rearranged}
						|\tlip(\tyus) - \lip(\yus)| \le \frac{1}{1-\sigma^2} \|\tlip-\lip\|_\infty
					\end{equation}
					Now we can compare $G\tlip$ and $G\lip$.
					Recall that $\tyus$ and $\yus$ yield the same arbitrary point $\xus$ in the domain of $G\tlip$ and $G\lip$. Therefore,
					\begin{subequations}
						\begin{align}
							\| (G \tlip - G\lip) \, \xus \| 
							%=&\| (G \tlip - G\lip) \big(\Psi^{-1}(\yus, \lip(\yus))\big)_{u,s} \| \\
							= &\left\| G \tlip \Big(\big(\Psi^{-1}(\tyus, \tlip(\tyus))\big)_{u,s}\Big)- G \lip \Big(\big(\Psi^{-1}(\yus, \lip(\yus))\big)_{u,s}\Big)\right\| \label{AKI}\\
							=&\left\| \big(\Psi^{-1}(\tyus, \tlip(\tyus))\big)_{c}- \big(\Psi^{-1}(\yus, \lip(\yus))\big)_{c} \right\|.
						\end{align}
						The last term measures the difference in center direction of two points which are contained in the center cone $\Cc{\sigma}{\Psi^{-1}(\yus,\lip(\yus))}$.
						By forward expansion, that amounts to the following inequality, $|\tlip(\tyus) - \lip(\yus) | \ge K_c \| \big(\Psi^{-1}(\tyus, \tlip(\tyus))\big)_{c}- \big(\Psi^{-1}(\yus, \lip(\yus))\big)_{c} \|$, together with \eqref{rearranged}, implies that
						\begin{equation}
							\begin{aligned}
								\| (G \tlip - G\lip) \, \xus \|  &\le 
								\frac{1}{K_c} |\tlip(\tyus) - \lip(\yus)|
								\label{QINEQUASE}\\& \le 
								\frac{1}{K_c(1-\sigma^2)} \|\tlip-\lip\|_\infty.
							\end{aligned}
						\end{equation}
						As $\xus$ is arbitrary, we obtain 
						\begin{equation}\label{GINEQ}
							\| G \tlip - G\lip \|_\infty \le 
							\frac{1}{K_c(1-\sigma^2)} \|\tlip-\lip\|_\infty.
						\end{equation}
					\end{subequations}
					Lemma~\ref{cones_theorem} provides a constant $K_c>1$ for any $\sigma\ (0,1)$. Thus, we choose $\sigma\in (0,1)$ such that $(K_c(1-\sigma^2))^{-1}<1$, which implies that $G$ is a contraction.
                    Note that $K_c\in (1,m(v))$ can be chosen arbitrarily close to $m(v)>1$ for any given $\sigma \in (0,\sqrt{1-1/m(v)})$.%, which is an open set for $v\in (1/2,1)$.
                    \footnote{Since the Cantor set $\mathcal{C}(v)$ is uniformly normally hyperbolic, where the Kasner map has a minimal expansion rate $m(v)>1$ in \eqref{minC}, there is a $\sigma\in (0,1)$ such that the cone properties in Lemma~\ref{cones_theorem} are valid for all heteroclinic chains in $\mathcal{C}(v)$. This is in stark contrast with the critical case $v=1/2$, see \cite[Section 8]{Dutilleul}.} 

                    Since $X_{}$ is a complete metric space, the Banach fixed point theorem yields a unique fixed point $\lipstar \in X$ such that the following invariance of $\lipstar$ holds:
					\begin{equation}\label{invariance_graph}
						\graph(\lipstar) = \graph(G\lipstar) = \Psi^{-1}(\graph(\lipstar)).
					\end{equation}
					%
                    %
                    %We now choose %constants $K_c \in (1, \tilde{K}_c)$, and $\sigma \in \left(0,1\right)$ such that %\begin{align}\label{relation_sigma_K}
				%		K_c> \frac{1}{1-\sigma^2}.
				%	\end{align}
                    %    The constants $K_c$ and $\sigma$ determine the expansion rate in center direction and the opening of the cones, respectively, for which the cone transformation properties are valid.		We employ relation \eqref{relation_sigma_K} in the proof of Theorem~\ref{manifold_theorem}.						The constant $K_{u,s}$ for the contraction in transverse direction depends on the map $\Psi$, and is determined in step 5.
				%	
                    %
					Consequently, $G_k$ is a contraction uniformly in $k \in \NN_0$. Thus, the map $G=(G_k)_{k\in \NN_0}$ is also a contraction on $X^{\NN_0}$ with the metric \eqref{norm_productspace}, with unique fixed point %(a sequence of Lipschitz functions) 
                     $\lipstar = (\lipstar_k)_{k\in \NN_0}$.
					%In the remaining two steps we argue why choosing initial conditions from graph($\lipstar_k$)

                    Since the vector field in \eqref{full:subs} is linear in the parameter $v\in (0,1)$, all the induced maps in this proof are continuous in $v$ and %We have refrained from including the parameter in the proof for relieving the burden of notation, however, one can easily obtain a sequence of graphs 
                    the fixed points $\lipstar = (\lipstar_k)_{k\in \NN_0}$ also depends continuously in $v\in (1/2,1)$. %See \textcolor{red}{XXX Maybe HirschPughShub}.
     
					\smallskip
					\textbf{Step 7: graph$(\lipstar)$ is the stable set of $0 \in \tinsection$ (discrete dynamics).}
					\smallskip
					
					Recall that the origin $0=(0,0,0,0) \in U_k$ represents the point $p_k$, whereas the origin $0=(0,0,0) \in \tinsection_{k}$ has coordinates $(\xuin, \xsin, \xssin, \xcin) = (0,0,1,0)$ in $U_k$ and represents the heteroclinic solution $h_{k-1}$ connecting $p_{k-1}$ with $p_{k}$. Each $U_k$, and in particular each domain $\tinsection_k$ of $\Psi_k$, has the same coordinate representation, and we thereby do not distinguish origins.
					
					For all $n \in \NN_0$, step 6 yields that graph($\lipstar_n$) is a $\Psi_n$-invariant set, i.e. its image is contained in graph$(\lipstar_{n+1})$.
					%Thus, we can consider $\Psi_k=\Psi(k,\cdot)$ as a non-autonomous discrete semi-dynamical system defined on graph$(\lipstar_{k}) \in \tinsection$
					It remains to show that, starting with any graph$(\lipstar_n)$, all points from this graph converge to the origin in $\tinsection$ under action of the sequence $(\Psi_k)_{k\ge n}$. Without loss of generality, we consider $x = (\xus, \lipstar_1(\xus)) \in \tinsection_{1}$ and we abbreviate $\Psi^{k}=\Psi_k \circ ... \circ \Psi_1$.
                        %\begin{equation}
					%    
					%\end{equation}
					%We consider without loss of generality $n=0$ %and we set $$\Psi^k:= \Psi_{k}\circ \dots \Psi_{1}\circ\Psi_1$$
					%More precisely, it converge to the sequence of points $0_k \in \tinsection_k$.
					%represents the heteroclinic chain $(h_k)_{k\in \NN_0}$.
					%
					%
					%
					%We want to have the stable set of the global return map $\Psi := \Psi_n \circ ... \circ \Psi_1$. But so far we have only considered one global excursion map $\Psiglob\circ \Psiloc:\tinsection_k \to \insection_{k+1}$. 
					%However, the analysis from steps (ref) is the same for $\Psi$. Indeed, we find a smaller domain of definition of $\Psi$, as we concatenate only finitely many maps. Moreover, the hyperbolicity is preserved as each individual map is hyperbolic, uniformly in $k$.
					%
					%
					%In this step we show that graph$(\lipstar_n)$ converges to the origin in $\tinsection$ under action of $(\Psi_k)_{k\ge n}$.
					%To show that it converges to the origin under the action of $(\Psi_k)_{k\ge n}$
					The forward contraction of the transverse cone \eqref{forward_contraction_transverse_condition} is uniform in $k$. Thus,
					\begin{equation}
						\left\|(\Psi^k(\x))_{u,s} \right\| \le \frac{1}{(K_{u,s})^{k}} \, \|\xus\| \xrightarrow{k \to \infty}  0.
                        \end{equation}
                        Invariance \eqref{invariance_graph} yields the corresponding center component as
                        \begin{equation}
                            (\Psi^k(\x))_c=\lipstar_{k+1}((\Psi^k(\x))_{u,s}) \xrightarrow{k \to \infty}  0,    
                        \end{equation}	%\intertext{By continuity of $\lipstar_k$ then}
						%(\Psi^{k+N}(\x^k))_{c} = \lipstar((\Psi^{k+N}(\x^k))_{u,s})  \, &\overset{N\to \infty}{\longrightarrow} \lipstar(0,0) = 0.
					%
					where we use that all functions $\lipstar_k$ are Lipschitz continuous with $\lipstar_k(0,0)=0$.
					%
					%Choosing $\graph(\lipstar)$ as the set initial conditions to the flow in continuous time \textcolor{red}{ then what?}
					%\begin{align}
					%\| \big(\Psi^{-1}(\txus, \tlip(\txus))\big)_{c}- \big(\Psi^{-1}(\xus, \lip(\xus))\big)_{c} \| \le K_c^{-1} |\tlip(\txus) - \lip(\xus)|
					%\end{align}

     					\smallskip
					\textbf{Step 8: $\graph(\lipstar)$ yields the stable manifold of $(h_k)_{k\in \NN_0}$ (continuous dynamics).}
					\smallskip
					
					%It remains to show that $\graph(\lipstar_k)$ chosen as initial consitions converges to the sequence of origins in $0_k \in U_k$.As in the previous step we must not distinguish the origins, i.e. we show that the trajectory 
					Let $\Phi_t$ denote the flow of the vector field with an initial condition $\x \in \graph(\lipstar_1) \subset \tinsection_{1}$.
					By the previous step, $\x$ converges to the origin $0 \in \tinsection$ by continually applying the maps $\Psi_k$. Thus, there exists a monotone sequence of times $(\alpha_k)_{k \in \NN_0}$ such that 
                        \begin{equation}
					    \Phi_{\alpha_k}(\x) = \Psi^k(\x) \xrightarrow{k\to\infty} (0,0,1,0).
					\end{equation}
					By smooth dependence on initial conditions for all $\tau\ge 0$, then
					%$(\tau_n)_{n \in \NN}$ such that
					\begin{equation}
					    \Phi_{\tau}(\Phi_{\alpha_k}(\x)) %\Phi_{\tau_n}(\Psi^n(\x))) 
					\xrightarrow{k\to\infty}  \Phi_{\tau }(0,0,1,0).
					\end{equation} 
					As $(0,0,1,0)$ lies on the heteroclinic, the term on the right goes to $0$ as $\tau \to \infty$, and the orbit of $\x$ is closer to $0$ with each iterate of $\Psi_k$. Thus, there is a sequence $(\tau_k)_{k \in \NN_0}$ so that 
                    \begin{equation}
                        \Phi_{\tau_k}(\Phi_{\alpha_k}(\x)) \xrightarrow{k\to\infty} (0,0,0,0).
                    \end{equation}
					The sequence $(t_k)_{k\in \NN_0}$ defined by
					$t_k:=\tau_k + \alpha_k$ yields a sequence of times for which the trajectory of $\x$ converges to the origin.
					We must show that, in the interval $[t_k,t_{k+1}]$, the trajectory of $\x$ converges (in Hausdorff distance) to the heteroclinic solution $h_k(t)$.

					We split the trajectory in three parts.
					First, until the trajectory intersects $\outsection_k$,
					% in the point $\Psiloc_k(\x^k)$,
					we employ the local estimates \eqref{lemma_local_estimates:subs} to see that, for $\xu\to 0$, the trajectory approaches $h_{k-1}(t)$, as it locally coincides with the set $\{\xsss = \xc = 0\}$. Thus, the intersection of the trajectory with $\outsection_k$ converges to $0 \in \outsection_{k}$ with increasing $k$.
					Second, the transition time until the trajectory intersects the next insection $\insection_{k+1}$
					%, i.e. between $\Psiloc_k(\x^k)$ and $\x^{k+1} = \Psi^{k+1}(\x)$,
					is uniformly bounded. By regularity of the vector field, all time-$t$ maps are sufficiently smooth, i.e. the trajectory of $\x$ is located in a tube around $h_k$ whose diameter shrinks with increasing $k$.
					Third, between the intersection
					$\Psi^{k+1}(\x)$ and $ \Phi_{t_{k+1}}(x)$, we again use the vector field to see that the trajectory is close to $h_k$ with respect to Hausdorff distance, since $\left(\Phi_{t_{k+1}}(x)\right)_u$ converges to zero with increasing $k$.

% Step 9 -- replacement for main (20).tex, 2 October 2026.
% Replace everything from the \smallskip before "\textbf{Step 9: ...}" (l. 4345)
% up to, but not including, the \smallskip before "\textbf{Step 10: Bianchi type VIII.}" (l. 4455).
% New labels: cont0, cont1; the label cont is kept.

					\smallskip
					\textbf{Step 9: Continuity of the local stable manifolds in $p$.}
					\smallskip

					We prove that the local stable manifolds depend continuously on $p\in \mathcal{C}(v)$: for every $p\in\mathcal C(v)$ and $\vartheta>0$ there exists $\rho>0$ such that $\|\lipstar_p - \lipstar_{\tilde{p}} \|_\infty \leq \vartheta$ for all $\tilde p\in\mathcal C(v)$ with $|p-\tilde{p}|\leq \rho$, where $|p-\tilde p|$ denotes the arclength distance on $\KC$. Here $\lipstar_p$ and $\lipstar_{\tilde{p}}$ denote the local stable manifolds at $p$ and $\tilde{p}$, i.e. the first components of the fixed points of the graph transforms $G_p$ and $G_{\tilde{p}}$, defined in \eqref{graph_transform_condition} along the heteroclinic chains starting at $p$ and $\tilde p$, respectively. We follow \cite[Proposition 9.21]{Dutilleul} closely; see also \cite{Hirsch2}.

					As in \cite[Section 9.6]{Dutilleul}, we compare graphs in common coordinates. As in \cite[Proposition 3.2]{Dutilleul}, the rectifying coordinates of \autoref{subsec:setting} can be chosen along each half-arc $\A_\alpha^\pm(v)$, with $\xc$ the arclength on $\KC$. Hence, for $p$ and $\tilde p$ in the same half-arc, $\xi_{\tilde{p}}\circ\xi^{-1}_p$ is the translation by $\xc(p)-\xc(\tilde p)$ in the center direction; in particular, it is close to the identity. In this step we write all graphs in these common coordinates, without translating them to the origin: the graph at a point $q\in\mathcal C(v)$ is rooted at $q$, i.e. $\lip_q(0,\xs)=\xc(q)$. Then $\Psi_p$ and $\Psi_{\tilde p}$ are restrictions of the same map.

					Let $\gamma$ be the constant graph family, $\gamma_q\equiv \xc(q)$ for all $q\in\mathcal C(v)$, and denote by $G_p^n\gamma$ the first component of $G_p^n$ applied to $(\gamma_{\KM^k(p)})_{k\in\NN_0}$, and similarly for $\tilde p$. Since $G_p$ and $G_{\tilde{p}}$ are contractions with the constant $c=\frac{1}{K_c(1-\sigma^2)}<1$ of Step 6, there is an integer $n>0$, independent of $\tilde p$, such that $\|\lipstar_p - G_p^n\gamma\|_\infty+\|\lipstar_{\tilde{p}} - G_{\tilde{p}}^n\gamma\|_\infty \leq \vartheta /3$. Thus
					\begin{equation}\label{cont0}
					\begin{aligned}
						\|\lipstar_p - \lipstar_{\tilde{p}} \|_\infty  &\leq \|\lipstar_p - G_p^n\gamma \|_\infty + \|G_p^n\gamma - G_{\tilde{p}}^n\gamma \|_\infty + \|G_{\tilde{p}}^n\gamma - \lipstar_{\tilde{p}} \|_\infty \\
						& \leq \|G_p^n\gamma - G_{\tilde{p}}^n\gamma \|_\infty + \frac{\vartheta}{3}.
					\end{aligned}
					\end{equation}
					Now let $\Lambda$ be any graph family, and denote by $G_p\Lambda$ the graph at $p$ obtained from the graph $\Lambda_{\KM(p)}$ at $\KM(p)$, i.e. $\graph(G_p\Lambda)=\Psi_p^{-1}(\graph(\Lambda_{\KM(p)}))$, and similarly for $\tilde p$. The arguments that led to \eqref{QINEQUASE} and \eqref{GINEQ}, applied to the points $\Psi^{-1}(\yus,\Lambda_{\KM(p)}(\yus))$ and $\Psi^{-1}(\tilde y_{u,s},\Lambda_{\KM(\tilde p)}(\tilde y_{u,s}))$ with the same $(u,s)$-components, yield
					\begin{equation}\label{cont1}
						\| G_p \Lambda - G_{\tilde{p}}\Lambda \|_\infty \le c\, \|\Lambda_{\KM(p)}-\Lambda_{\KM(\tilde{p})}\|_\infty,
					\end{equation}
					provided $p$ and $\tilde{p}$ are sufficiently close, so that the cone properties in Lemma~\ref{cones_theorem} apply to these two points.
					If $\tilde p$ is sufficiently close to $p$, then $\KM^j(p)$ and $\KM^j(\tilde{p})$ are close and lie in the same half-arc for $j=0,\ldots, n$, because $\mathcal C(v)$ is compact and contains neither the boundary points nor the midpoints $\Q_\alpha$ of the arcs (note that $\KM(\Q_\alpha)=\T_\alpha\notin\A(v)$ for $v>\frac12$). Applying \eqref{cont1} $n$ times, and since $\gamma$ is constant, we obtain from \eqref{cont0} that
					\begin{equation}
						\|\lipstar_p - \lipstar_{\tilde p} \|_\infty \leq c^n\|\gamma_{\KM^n(p)} - \gamma_{\KM^n(\tilde{p})} \|_\infty + \frac{\vartheta}{3} = c^n | \KM^n(p) - \KM^n(\tilde{p}) | + \frac{\vartheta}{3}.
					\end{equation}
					Since $|\mathrm{D}\KM|$ is monotone along each half-arc, see \eqref{KM1}, the maximal expansion rate $M(v)$ in \eqref{maxC} bounds $|\mathrm{D}\KM|$ between $\KM^j(p)$ and $\KM^j(\tilde p)$, and thus
					\begin{equation}\label{cont}
						\|\lipstar_p - \lipstar_{\tilde p} \|_\infty \leq \big(c\,M(v)\big)^n \, | p - \tilde{p} | + \frac{\vartheta}{3}.
					\end{equation}
					Choosing $\rho>0$ so small that the above closeness requirements hold and $\rho\,\big(c\,M(v)\big)^n\le\vartheta/3$, we obtain $\|\lipstar_p - \lipstar_{\tilde p} \|_\infty\le\vartheta$.

                    \newpage
     				\smallskip
					\textbf{Step 10: Bianchi type VIII.}
					\smallskip
					
					Each orbit of $\KM$ yields eight possible choices for the signs of the corresponding heteroclinic solutions.
					Two of these choices, $(+,+,+)$ and $(-,-,-)$, correspond to a boundary of a Bianchi type IX subset, and the others correspond to boundaries of the Bianchi type VIII subset.
					The space $X$ from step 1 assumes that the three transverse variables have positive sign. Thus, our setting is tailored for the case of Bianchi type IX. However, the position of the sections $\insection$ and $\outsection$ \eqref{sections} in the local analysis is arbitrary. Indeed, the unstable component grows independently of its sign and the estimates for the stable components are with respect to absolute value. Thus, we obtain the same local results for sections positioned at negative values.
					For the estimates of the transition map, we use general properties which are independent of the signs and the restriction to positive $\xus$ in the definition of $X$ in step 1.
					Thus, for any choice of the signs of the sequence $(h_k(t))_{k \in \NN_0}$ we obtain a sequence of return maps with the properties as described in Lemma~\ref{cones_theorem} and therefore a total of eight local manifolds as described in this proof.
				\end{proof}

				%\subsubsection{General Concatenation}
				%\subsubsection{Examples: period 2, 3 and 4}
				%\subsubsection{The Cantor set}
				             
                \section{Conclusion}\label{chap:conclusion}

                    We discuss our results that were obtained by making use of Ho\v{r}ava-Lifshitz modified gravity, which perturbs\footnote{We emphasize that such perturbation is not generic (in the space of vector fields). Indeed, a generic perturbation of \eqref{full:subs} with $v=1/2$ would destroy the Kasner circle of equilibria, which is not the case by the $v$-perturbation in \eqref{full:subs}. 
                Thus, unfortunately, abstract theorems on the generic behavior of partially hyperbolic dynamical systems do not provide a reasonable description of $v$-perturbations of the Bianchi models. 
                %Even if we do obtain some sort of rigidity for $v \in (1/2,1)$, we expect that the limiting case $v\to 1/2$ is very special (due to degeneracies) and may be the transition to a qualitatively different behavior and lack of continuity (in $v$), even if there are reminiscent structures. For example,  see:  
                } the spatially homogeneous Bianchi models in GR by a parameter $v\in (0,1)$, such that GR occurs at $v=1/2$. We then discuss open issues that ramify from our findings.
                    
				For the supercritical case, $v>1/2$, we have shown that all infinite heteroclinic chains admit a three-dimensional stable manifold. These chains are associated with a Cantor set $\Cantor(v)$ on the Kasner circle. Thereby we have proven Conjectures 7.1 and 7.2 of \cite{HellLappicyUggla}. %, which includes Conjecture 7.1 as a special case, as it only concerns the heteroclinic cycles of length two. 
				One of the key results is Lemma~\ref{cor_supercritical_eigenvalues}, where it is proven that the eigenvalues of the linearization at the Kasner circle are ordered. This allowed us to apply methods developed for general relativity which are used by Liebscher et al. in \cite{Liebscher}. For alternative approaches towards stability of heteroclinic networks which could potentially be applied to general relativity, see \cite{PodviginaAshwin,Podvigina,PeixeAlexandre,CastroPR}.
    
                We emphasize that we have obtained stable manifolds of Lipschitz regularity for $v>1/2$, which are not necessarily the optimal regularity of such manifolds. It would be interesting to investigate if these stable manifolds are of higher regularity, such as the $C^1$-manifolds constructed in \cite{Beguin,Buchner}. Effective methods to obtain higher regularity is to either use the fiber contraction theorem, see \cite{Hirsch}; or a forced smoothness of the embedding of the stable manifold based on the cone field criterion in Lemma~\ref{cones_theorem} and the continuous dependence of the eigenstructure with respect to $p\in \KC$, see \cite[Section 6A]{Hirsch2}. %SEE MARTIN LEGUIL LECTURE 2
                Moreover, Theorem~\ref{manifold_theorem} only states the continuity of the foliation, however, we believe that a higher regularity such as absolute continuity (or stronger) can be achieved by proving a higher regularity of the holonomy map, see \cite[Chapter 11]{Dutilleul} and references therein.
                Even if a higher regularity can be achieved for $v>1/2$, it is not clear if such property would be carried over to the limiting case $v\to 1/2$, along with consequences for the limiting set. % and how this could strengthen our result in Theorem 1: if the dimensionality and graph-property.

				For the subcritical case, $v<1/2$, Conjecture 7.3 in \cite{HellLappicyUggla} proposes that all heteroclinic chains associated with the set $\tilde{\mathcal{C}}(v)$ in \eqref{setsCF} have a stable three-dimensional manifold.
                A first step towards this goal is given by the complete characterization of the linearization at the Kasner circle in Lemma~\ref{cor_subcritical_eigenvalues}, which thereby defines the sectors $\Mid, \Stab, \Unstab$. %, which is .
                A more careful investigation is necessary to examine the relationship of the sectors $\Mid, \Stab, \Unstab$ and the set $\tilde{\mathcal{C}}(v)$ in terms of the discrete dynamics of $\KM$: 
			what happens if an orbit visits different sectors? Is it possible that the overall dynamics is contracting even though there are local passages which are not analogous to the supercritical case and may amount to a few local expansions? %The is related to the investigation of the sets $\Mid, \Stab, \Unstab$ in terms of the discrete dynamics of $\KM$. 
            Which sectors are visited by generic chains? Can one construct sequences visiting the sectors in a prescribed order?
    
			Even the simplest examples of the heteroclinic chains of length three and four are contained in $\tilde{C}(v)$ for some parameter values $v$, however, they %where the local picture is not analogous to the supercritical case: these chains 
                do not admit an ordering of the eigenvalues %such that the unstable eigenvalues has smallest absolute value, which 
                that allows us to apply the methods of \cite{Liebscher}, see \autoref{period3_eigenvalues} and \autoref{period4_to_cycle_figure:full}, and more generally \autoref{sec:periods}.
                This may amount to a change of global dynamical structure. Finding if those correspond to known bifurcations, such as in \cite{Sandstede,Kirk} and references therein, may shed a light in the intricate geometry of phase-space.
                %
				%We expect that the analysis of these chain is more subtle for the following reason.
				%In \cite{Liebscher_magnetic} shadowing in models of Bianchi type $\mathrm{VI}_0$ with magnetic field is investigated.
				To circumvent such issues, one could try to adapt the methods in \cite{Liebscher_magnetic} to obtain a stable manifold of some chains, as that method only requires dominance of the strong stable eigenvalue over the unstable eigenvalue, which is satisfied in $\Mid$. Due to such milder assumption, they also require that the incoming heteroclinic trajectory is the strong stable (non-principal) direction, which is satisfied for the period three cycle, but fails for the period four cycle. 
                %However,
    
                The phenomenon of \emph{switching} describes when solutions of a differential equation follow different heteroclinic chains, for example in case that an equilibrium has two possible outgoing heteroclinics within a heteroclinic chain, see \cite{CastroLohse,Alexandre,Castro} and references therein. This occurs in the set $\tilde{\mathcal{F}}(v)$ in \eqref{setsCF}, and in particular, for the chain of period 4 in \autoref{sec:per4} when $v<v_3$. Unfortunately all such results for switching require hyperbolic equilibria, which is not the case for the Kasner circle. Alternatively, we believe that the center eigendirection of the Kasner circle could trigger a different switching behavior: after following a chain for some time, the drift in the center direction surpasses some threshold and pushes a solution to follow a nearby chain. 
                On one hand, one could possibly obtain solutions that possess more than one heteroclinic chain (or even the whole Mixmaster) in their $\omega$-limit set. This would be a first step towards the \emph{strong Mixmaster attractor conjecture}, see \cite[Section 7]{Mixmaster}. 
                On the other hand, one could possibly construct a solution that does not shadow any orbit of the discrete Kasner map: after shadowing a piece of a heteroclinic chain and reaching sufficiently close to a Taub point (where normal hyperbolicity is arbitrarily weak), the drift in the center direction kicks the solution to a nearby chain. %The $\omega$-limit set of such solutions would consist of several pieces of heteroclinic chains, however, 
                Hence, the continuous dynamics (the Bianchi models) would be poorly approximated by the discrete dynamics (the Kasner map), indicating that the \emph{Kasner map convergence conjecture} in \cite{Mixmaster} is a delicate issue. % - as usual, delicacy surrounds the word 'generic'.

\begin{figure}[H]
		\centering
		\captionsetup[subfigure]{labelformat=empty}
		\subfloat[][]{
			\begin{tikzpicture}[scale=1.2]
				% v= 0.25
% labeling 
\draw (-1.7,1.3) circle(0pt) node[anchor=west] { };
\tiny
% Kasner circle
\draw[thin] (1,0) arc (0:360:1cm and 1cm);

% tangents
%1
\draw[dashed, -, thin] (0.25,0.975)--(3.5,0);
\draw[dashed, -, thin] (0.25,-0.975)--(3.5,0);
%2
\draw[dashed, -, thin] (0.71,-0.72)--(-1.75,-3.03);
\draw[dashed, -, thin] (-0.978,0.25)--(-1.75,-3.03);
%3
\draw[dashed, -, thin] (0.71,0.72)--(-1.75,3.03);
\draw[dashed, -, thin] (-0.978,-0.25)--(-1.75,3.03);

% domains
\draw [Bracket-Bracket] (0.25,-0.9682) arc (-75.5224878:75.5224878:1cm and 1cm);
\draw [Bracket-Bracket] (0.7071,0.7071) arc (45:194.5:1cm and 1cm);
\draw [Bracket-Bracket] (0.7071,-0.7071) arc (-45:-194.5:1cm and 1cm);

%%overlap
\draw [line width = 1.1 pt] (0.7071,0.7071) arc (45:75.52248:1cm and 1cm);
\draw [line width = 1.1 pt] (0.7071,-0.7071) arc (315.52248:285
:1cm and 1cm);
\draw [line width = 1.1 pt] (-0.97, 0.25) arc (165.5:194.5:1cm and 1cm);

%Q/v
\filldraw [black] (3.5,0) circle (0.5pt); \draw  (3.5,0) circle (0pt) node [anchor=west]{$\frac{\Q_1}v$};
\filldraw [black] (-1.75,-3.03) circle (0.5pt); \draw (-1.75,-3.03) circle (0.5pt)  node [anchor=east]{$\frac{\Q_2}v$};
\filldraw [black] (-1.75,3.03) circle (0.5pt); \draw (-1.75,3.03) circle (0.7pt)  node [anchor=east]{$\frac{\Q_3}v$};

%\draw (1,-2.5) circle(0pt) node [text width=3cm] {\footnotesize  $(0 < v < \frac 12$) The Kasner map is multivalued in the intersection of two unstable arcs};

\filldraw  (1.2,-2.5) circle (0pt) node [anchor=north]{\boxed{\text{\scriptsize$v\in (0,1/2)$}}};

%\draw (1.5,-3) circle(0pt) node % [text width=3cm]
%{\fontdimen2\font=2em%
%	\parbox{3.5cm}{\footnotesize \justifying \noindent The Kasner map is multivalued in the intersection of two unstable arcs.}};

%\fontdimen2\font=2em%
%\parbox{3cm}{\RaggedRight This is a ragged right and hyphenated text in a TikZ node.}

\filldraw [black] (0,3) circle (0pt);
\filldraw [black] (0,-3.2) circle (0pt);
				\input{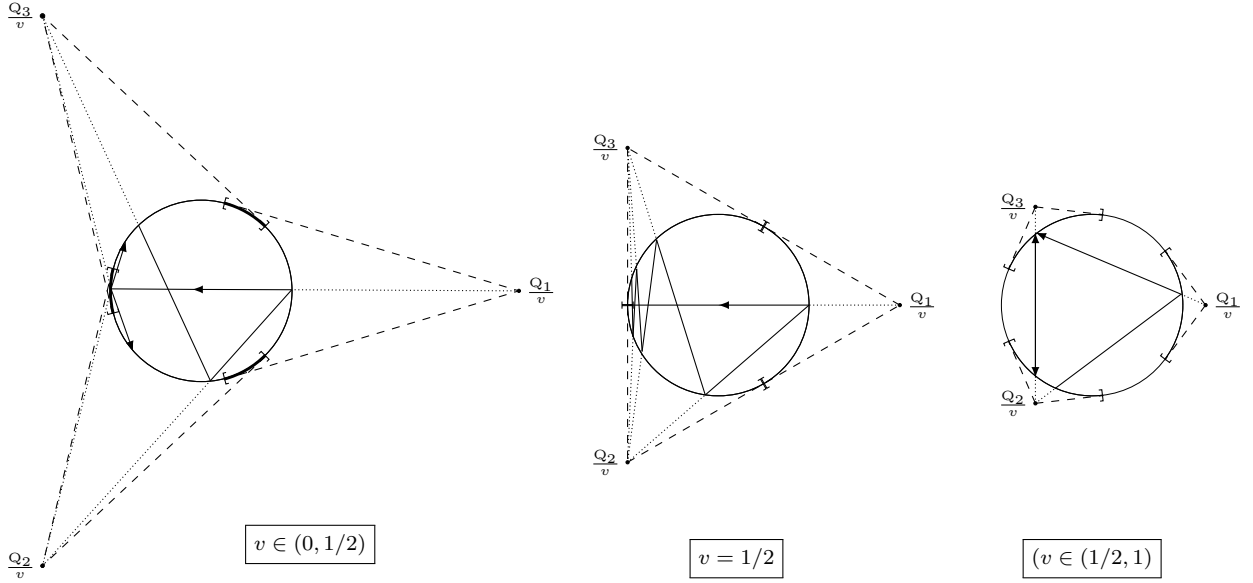}
				% projected outside circle
\draw [densely dotted,  thin] (3.5,0)--(0.997,0.0125);

%\filldraw  (0.997,0.0125) circle (0.7pt) node [anchor=south west]{\scriptsize$p_{\small 0}$};

\draw [densely dotted, thin] (-1.75,3.03)--(-0.69,0.72);

\draw [densely dotted, thin] (-1.75,-3.03)--(0.10, -0.99);

%\draw [green,densely dotted,  thin] (-1.75,3.03) -- ( -0.83, 0.55);
%\draw [densely dotted, very thin] (-1.75,3.03) -- (-0.995,0.022);
%\draw [densely dotted, very thin] (-1.75,3.03) -- ( -0.02, 0.99);

\draw [densely dotted,  thin] (-1.75,-3.03) -- (-0.995,0.022);
%\draw [densely dotted, very thin] (-1.75,-3.03) -- ( -0.75, -0.68);
%\draw [red, densely dotted, very thin] (-1.75,-3.03) -- ( -0.22, -0.975);

%\draw [blue, densely dotted,  thin] (3.5,0) -- (0.86,0.53);
%\draw [densely dotted, very thin] (3.5,0) -- (0.97,-0.24);

% projected inside circle
\draw [thin] (-0.69,0.72)--(0.1,-0.99);
\draw [thin] (0.997,0.0125)--(0.1,-0.99);
\draw [thin,-latex] (0.997,0.0125)--( -0.1, 0.016);
\draw [thin] ( -0.05, 0.0159)--( -0.995, 0.022);
\draw [thin, -latex] (-0.995,0.022)-- ( -0.83, 0.55);
%\draw [thin] ( -0.83, 0.55) -- ( -0.22, -0.975);
%\draw [thin] ( -0.22, -0.975) -- (0.86,0.53);
%\draw [thin,-latex] (0.86,0.53)--( 0.5, 0.6) ;

%\draw [very thin] (-0.995,0.022)-- ( -0.75, -0.67);
%\draw [very thin] ( -0.75, -0.67)-- ( -0.02, 0.99);
%\draw [very thin] ( -0.02, 0.99)-- (0.97,-0.24);
%\draw [very thin,-stealth] (0.97,-0.24)-- ( 0.5, -0.28);
			\end{tikzpicture} \label{conc:figure_subcritical}
		}
%            \qquad \qquad
		\subfloat[][]{
			\begin{tikzpicture}[scale=1.2]
				clip (-1.5,-2) rectangle (2.5,2);
% v= 1/2
\tiny 
% Kasner circle
\draw[thin] (1,0) arc (0:360:1cm and 1cm);
% tangents
\draw[thin,dashed] (2,0)--(-1,1.732) ;
\draw[thin,dashed] (-1,1.732) --(-1,-1.732) ;
\draw[thin,dashed] (-1,-1.732) --(2,0);

%\draw (-1.7,1.3) circle(0pt) node[anchor=west] { };

%% arcs

% A_1
\draw [-Bracket] (1,0) arc (0:60:1cm and 1cm);
\draw [-Bracket] (1, 0) arc (0:-60:1cm and 1cm);

% A_3
\draw [-Bracket] (-0.5,0.866) arc (120:180:1cm and 1cm); 
\draw [-Bracket] (-0.5, 0.866) arc (120:60:1cm and 1cm);

% A_2
\draw [-Bracket] (-0.5,-0.866) arc (240:180:1cm and 1cm);
\draw [-Bracket] (-0.5, -0.866) arc (240:300:1cm and 1cm);

%% Q/v
\filldraw [black] (2,0) circle (0.5pt);
\draw  (2,0) circle (0pt) node [anchor=west]{$\frac{\Q_1}v$};
\filldraw [black] (-1,-1.732) circle (0.5pt);
\draw (-1,-1.732) circle (0pt)  node [anchor=east]{$\frac{\Q_2}v$};
\filldraw [black] (-1,1.732)  circle (0.5pt);
\draw (-1,1.732)  circle (0pt)  node [anchor=east]{$\frac{\Q_3}v$};

\filldraw  (0.2,-2.5) circle (0pt) node [anchor=north]{\boxed{\text{\scriptsize$v=1/2$}}};

%\draw (0,-2.5) circle(0pt) node {\footnotesize $\mathrm{(b)} \, \, v= \frac 12$};
%\draw (0.5,-3) circle(0pt) node % [text width=3cm]
%{\fontdimen2\font=2em%
%	\parbox{3.5cm}{\footnotesize \justifying \noindent The Kasner map is well-defined. Its image $\KM(\KC)$ is a double cover of $\KC$. }};

%align 
%\filldraw [black] (-2,3) circle (0pt);
\filldraw [black] (0,2.5) circle (0pt);
\filldraw [black] (0,-3) circle (0pt);
				clip (-1.5,-2) rectangle (2.5,2);
%\filldraw  (0.98,0.15) circle (0.7pt) node [anchor=south west]{\scriptsize$p_{\small 0}$};

%orbit in correct order
\draw [thin] (-1,0)--(0.1,0);\draw [thin,-latex] (1,0)--(0,0);
\draw [thin] (1,0)--(-0.144,-0.99);
\draw [thin] (-0.144,-0.99)--(-0.68,0.72);
\draw [thin] (-0.68,0.72)--(-0.84,-0.51);
\draw [thin] (-0.84,-0.51)--(-0.9,0.4);
\draw [thin] (-0.9,0.4)--(-0.94,-0.35);
\draw [thin] (-0.94,-0.35)--(-0.955,0.31);

%orbit - dotted
\draw [ thin,densely dotted] (2,0)--(1,0);

\draw [ thin,densely dotted] (-1,-1.732)--(-0.144,-0.99);
\draw [ thin,densely dotted] (-1,-1.732)--(-0.84,-0.51);
\draw [ thin,densely dotted] (-1,-1.732)--(-0.94,-0.35);

\draw [ thin,densely dotted] (-1,1.732)--(-0.68,0.72);
\draw [ thin,densely dotted] (-1,1.732)--(-0.9,0.4);
\draw [ thin,densely dotted] (-1,1.732)--(-0.955,0.31);
			\end{tikzpicture}
		}
            \subfloat[][]{
			\begin{tikzpicture}[scale=1.2]
				clip (-1.5,-2) rectangle (2.5,6);

% 1/2 < v < 1    v=0.8 
\tiny 
\draw (-1.4,1.3) circle(0pt) node[anchor=west] { };

% Kasner circle
\draw[very thin] (1,0) arc (0:360:1cm and 1cm);

% arcs
%A_1
\draw [-Bracket] (1,0) arc (0:36.87:1cm and 1cm);
% node [anchor= west] {$\tac$};
\draw [-Bracket] (1, 0) arc (0:-36.87:1cm and 1cm);
% node [anchor= west] {$\tab$};

%A_3
\draw [-Bracket] (-0.5, 0.866) arc (120:156.87:1cm and 1cm);
% node [anchor= east] {$\tcb$};
\draw [-Bracket] (-0.5, 0.866) arc (120:83.13:1cm and 1cm);
% node [anchor= south] {$\tca$};

%A_2
\draw [-Bracket] (-0.5, -0.866) arc (240:276.87:1cm and 1cm);
\draw [-Bracket] (-0.5, -0.866) arc (240:203.13:1cm and 1cm);

% tangents
\draw[dashed, -, thin] (1.25,0)--(0.8,0.607);
\draw[dashed, -, thin] (1.25,0)--(0.8,-0.607);

\draw[dashed, -, thin] (-0.625,1.0825)--(-0.9257, 0.3893);
\draw[dashed, -, thin] (-0.625,1.0825)--(0.1257, 0.9963);

\draw[dashed, -, thin] (-0.625,-1.0825)--(0.1257, -0.9963);
\draw[dashed, -, thin] (-0.625,-1.0825)--(-0.9257,-0.3893);

%T/v
\filldraw  (1.25,0) circle (0.5pt) node [anchor=west]{$\frac{\Q_1}v$};
\filldraw (-0.625,-1.0825) circle (0.5pt) node [anchor=east]{$\frac{\Q_2}v$};
\filldraw (-0.625,1.0825)  circle (0.5pt)  node [anchor=east]{$\frac{\Q_3}v$};

%\draw (0,-2.5) circle(0pt) node {\footnotesize $ \mathrm{(c)} \, \, \frac 12 < v < 1$};

\filldraw  (0,-2.5) circle (0pt) node [anchor=north]{\boxed{\text{\scriptsize$(v\in (1/2,1)$}}};

%\draw (0.1,-3) circle(0pt) node % [text width=3cm]
%{\fontdimen2\font=2em%
%	\parbox{3.2cm}{\footnotesize \justifying \noindent  The Kasner map is well-defined. Generic orbits eventually enter $\S$.}};

%align 
%\filldraw [black] (-2,3) circle (0pt);
\filldraw [black] (0,3) circle (0pt);
\filldraw [black] (0,-3) circle (0pt);
				%period 2
\draw[densely dotted, thin]  (-0.625,-1.0825)--(-0.625,-0.8);
\draw[densely dotted, thin]  (-0.625,1.0825)--(-0.625,0.8);

\draw[thin,-latex]  (-0.625, 0.8)--(-0.625,-0.8);
\draw[thin,-latex]  (-0.625, -0.8)--(-0.625,0.8);

%stable manifold
\draw[densely dotted,  thin] (1.25,0)--(0.98, 0.12);
\draw[ thin, -latex] (0.98, 0.12)--(-0.625, 0.8);

\draw[densely dotted,  thin] (-0.625,-1.0825)--(-0.41, -0.92);
\draw[thin] (-0.41, -0.92)--(0.98, 0.12);

%\draw[red,  thin] (-0.625,1.0825)--(-0.41, -0.92);
			\end{tikzpicture} \label{conc:figure_supercritical}
		}
		\caption{For $v=1/2$, we call heteroclinic chains of `infinite period' by chains that end at the Taub point and its backwards images under the Kasner map only alternates between two different maps that get arbitrarily close to such Taub point. The existence of a stable manifold for such chain is not known. For $v<1/2$, this chain becomes a multivalued heteroclinic chain. For $v>1/2$, the period two heteroclinic chain converges to the Taub point, as $v\to 1/2$, however, it is not clear what happens with its stable manifold in the limit. %\textcolor{red}{it seems that the period 2 has a homoclinic, in terms of the kasner map, no? look at all pre images alternating the maps 2 and 3. Perhaps GR is a homoclinic bifurcation?} %Moreover, we do not know what is the role of the chains which are arbitrarily close to the Taub points across the bifurcation $v=1/2$. 
        } \label{conc:figures_TaubChain}
	\end{figure}

The most relevant bifurcation occurs at the GR value $v=1/2$, and it concerns the
heteroclinic chains that end at the Taub points. For $v=1/2$, the chains that come
arbitrarily close to the Taub points form a generic subset of the Kasner circle, and
among them there are a few special ones: there exist chains of ``infinite period'',
which take infinite time, in the backwards direction, to return to the Taub point,
see \autoref{conc:figures_TaubChain}. Since the heteroclinic chains of period two
degenerate at the Taub points as $v\to 1/2$, such chains of infinite period should be
related to the period two chains and to their stable manifolds, with respect to the
discrete dynamics of the Kasner map and thereby to the continuous dynamics. This
suggests that a crisis occurs at the crossing $v=1/2$, akin to
\cite{grebogi82,grebogi83}.

These degeneracies are also what prevents our construction from being carried to the
limit. Indeed, from the absolute continuity of the foliation of stable manifolds over
a generic subset of the Kasner circle, the authors of \cite{Dutilleul} obtain a set of
positive Lebesgue measure of Bianchi type $\mathrm{VIII}$ and $\mathrm{IX}$ solutions
converging to the Mixmaster, and we expect the limiting set of the stable manifolds of
Theorem~\ref{manifold_theorem}, as $v\to 1/2$, to contain it. However, most of the
construction carried out in this paper degenerates in that limit, such as the
contraction rates and the size of the sections. Nonetheless, the unfolding of the GR vector field through
the parameter $v\neq 1/2$ indicates a fruitful direction towards rigorous results in
the degenerate GR case, through approximation by modified gravity theories.
%
                %We mention a remark on such chains that gets arbitrarily close to a Taub point and their respective stable manifolds of Bianchi type IX solutions. Due to closedness of the $\omega$-limit set (of the continuous dynamics), the Taub points should also be in the $\omega$-limit set of such solutions within a stable manifold. This proves that the union of such a Taub point with such a given chain (that gets arbitrarily close to the Taub point) is contained in the $\omega$-limit set of the type IX stable manifold was proven. However, the other inclusion is not clear to me. Are there other points in the $\omega$-limit set, e.g., pre-images of a Taub point? Probably not, because of continuity...
 
				Lastly, it remains a challenge to look for results of the exceptional Bianchi type $\mathrm{VI}_{-1/9}$, see \cite{hewetal03}. Can any of the results in \cite{Liebscher,Liebscher_magnetic,Beguin,Buchner} be replicated? The first rigorous heteroclinic chains for type $\mathrm{VI}_{-1/9}$ were constructed in \cite{LappicyUggla}. We emphasize that a stable manifold for a heteroclinic chain of type $\mathrm{VI}_{-1/9}$ would be a first step towards proving such stability results (of certain codimension) for spatially inhomogeneous cosmologies: in a suitable gauge, the equations near a singularity can be rewritten as a system of PDEs and ODEs such that the type $\mathrm{VI}_{-1/9}$ is an invariant boundary. %Secondly, for such (or others) heteroclinic chains, the linearization and the ordering with a spectral gap condition may not be always satisfied, which should be an indication that the local and global passages can be modified appropriately to obtain a stable manifold in these models. Johannes claims that some class B models allow such Takens linearization, however, never proves the theorem for them... 
				
                %\textcolor{red}{Ringstrom also gives an overview of stefan/tom, and mention that this initial condition ideas might be the parametrization of the stable manifold of the kasner circle points, however, this depends on a gauge choice. That should be doable, no? Look into claes gauge choice, where the spatially homogeneous (SH) models form an invariant boundary. Linearize at the kasner cirlce! the linearization should consist of one block (with 2 stable, 1 unstable, 1 center directions, coming from the SH models), and another block which is the 'PDE linearization'. In a sense, this is similar to the wave equation... All we need is a spectral gap to 'isolate all the stable eigenmodes'! Can we find the codimension?}
				%https://arxiv.org/abs/2202.04919

				\textbf{Acknowledgments.} We are grateful for inspiring discussions (live or via e-mails) with François Béguin, Bernold Fiedler, Hans Ringström, Alexandre Rodrigues and Claes Uggla.  % for interesting discussions and valuable suggestions. %Some over short e-mails, others with enduring sessions with wine. 
                This paper is the outcome of HS's thesis defended at Freie Universität Berlin in early 2023. 
                PL was financially supported by Marie Skłodowska-Curie Actions, H2020 Cofund, 847635, UNA4CAREER, % (UNA Europa, an alliance of universities FOR the emergence of talent and the development of research CAREERs)  
                with the project DYNCOSMOS. 

                \textbf{AI Statement.} We used Claude Opus 5.5 for language and spelling corrections, as well as for auditing the proofs of this manuscript.

				\bibliographystyle{plain}

			\end{document}